\documentclass[12pt,letterpaper]{article}

\usepackage[margin=15pt,font=small,labelfont={bf,sf}]{caption}
\usepackage{epsfig}
\usepackage{amsmath}
\usepackage{amssymb}
\usepackage{amsthm}
\usepackage{indentfirst}
\usepackage{xspace}
\usepackage{multirow}
\usepackage{hyperref}
\usepackage{cleveref}
\usepackage[dvipsnames]{xcolor}
\usepackage{braket}
\definecolor{darkred}{rgb}{0.5,0.15,0.15}
\hypersetup{colorlinks=true,urlcolor=darkred,linkcolor=darkred,citecolor=darkred}
\usepackage{verbatim}
\usepackage[letterpaper,margin=1in,headheight=15pt]{geometry}
\usepackage{mathpazo}
\usepackage{microtype}
\usepackage{authblk}
\usepackage{tikz-cd}
\usepackage{silence}
\usepackage[missing=]{gitinfo2}
\usepackage{enumitem}
\usepackage{blkarray}
\usepackage{subcaption}
\usepackage{pdfpages}

\usepackage{CJKutf8}
\font\maljapanese=dmjhira at 2ex

\usepackage{tikz}
\usepackage{circuitikz}
\usetikzlibrary{calc}   
\usetikzlibrary{topaths}
\usetikzlibrary{decorations}
\usetikzlibrary{decorations.pathmorphing}
\usetikzlibrary{braids}

\tikzset{
branchpoint/.pic={code ={
\draw[very thick, orange] (-0.1,-0.1)--(0.1,0.1);
\draw[very thick, orange] (-0.1,0.1)--(0.1,-0.1);
}
}
}

\newcommand{\Z}{\ensuremath{\mathbb{Z}}}
\newcommand{\R}{\ensuremath{\mathbb{R}}}
\newcommand{\C}{\ensuremath{\mathbb{C}}}
\newcommand{\CP}{\ensuremath{\mathbb P}}

\newcommand{\cF}{\ensuremath{\mathcal{F}}}
\newcommand{\cH}{\ensuremath{\mathcal H}}
\newcommand{\cV}{\ensuremath{\mathcal V}}

\newcommand{\FN}{\mathrm{FN}}
\newcommand{\MCG}{\mathrm{MCG}}

\newcommand{\SkAlg}{\mathrm{SkAlg}}
\newcommand{\SkMod}{\mathrm{SkMod}}
\newcommand{\SkCat}{\mathrm{SkCat}}
\newcommand{\OpSk}{\mathrm{OpSk}}

\DeclareMathOperator{\frakp}{\mathfrak p}
\DeclareMathOperator{\fraka}{\mathfrak a}

\theoremstyle{definition}
\newtheorem{remark}{Remark}[section]

\numberwithin{equation}{section}

\title{Liouville Blocks from Spectral Networks Part I: Moore-Seiberg parallel transport from q-nonabelianisation}
\date{{{\tiny \color{gray} \tt}} 
{{\tiny \color{gray} \tt}}} 
\author{Lotte Hollands \\  \href{mailto:l.hollands@hw.ac.uk}{l.hollands@hw.ac.uk} 
   \and Subrabalan Murugesan \\  \href{mailto:sm2160@hw.ac.uk}{sm2160@hw.ac.uk} }
\affil{ \small Department of Mathematics, Heriot-Watt
  University, and\\ Maxwell Institute for Mathematical Sciences, Edinburgh}

\begin{document}

\maketitle
\begin{abstract}
This is part I of a series of papers in which we investigate the role of spectral networks in quantum Liouville theory, with particular emphasis on spectral networks of Fenchel–Nielsen type. In this part, we focus on the computation of $SL(2,\mathbb{C})$ quantum parallel transports on any (punctured) Riemann surface $C$, which in quantum Liouville theory has a well-known prescription through the Moore-Seiberg formalism. Our main achievement is rederiving this prescription through an extension of the $q$-nonabelianisation formalism applied to spectral networks of Fenchel-Nielsen type. One of the spin-offs is a proposal for a quantum version of the NRS proposal.
\end{abstract}

\tableofcontents

\newpage

\section{Introduction}

Liouville CFT has played a central role in two-dimensional quantum field theory for several decades. Beginning with the seminal papers \cite{Dorn:1994xn, Zamolodchikov:1995aa}, which proposed exact expressions for its three-point structure constants, Liouville theory emerged as a rare example of an interacting CFT solvable by analytic bootstrap methods \cite{Teschner:1995yf, Teschner:2001rv}. Subsequent developments connected Liouville theory to geometry and topology, most notably through its relation to the quantisation of Teichm\"uller space \cite{Teschner:1995dt, Chekhov:1999tn}, placing it at the intersection of conformal field theory, low-dimensional geometry, and quantum topology. More recently, Liouville theory has been given a rigorous mathematical foundation through probabilistic methods, culminating in a proof of the DOZZ three-point function \cite{kupiainen2019integrabilityliouvilletheoryproof}.

Toda CFTs arise as higher-rank extensions of Liouville theory, in which the single exponential interaction is replaced by a set of interacting fields organised by a Lie algebra. Their properties have been explored using tools such as the higher-rank conformal bootstrap \cite{Fateev:2007ab}, the quantisation of moduli spaces of flat $G$-connections \cite{fock2006moduli}, and the AGT correspondence relating four-dimensional $\mathcal{N}=2$ QFTs to two-dimensional CFTs~\cite{Alday:2009aq, Wyllard:2009hg} (reviewed for instance in \cite{LeFloch:2020uop}). Despite substantial progress, several foundational questions remain open. In particular, a universally accepted closed-form expression for three-point functions of arbitrary primary fields is still lacking (but see \cite{Huang:2011qx,Aganagic:2014oia,Mitev:2017jqj,Coman:2017qgv, Fucito:2023txg} for various very promising approaches). 

In the mean-time, the notion of a spectral network was introduced in the influential works \cite{Gaiotto:2009hg, Gaiotto:2012rg}. Aside from being a powerful tool to enumerate the various types of BPS states in four-dimensional $\mathcal{N}=2$ theories \cite{Gaiotto:2010be, Gaiotto:2011tf, Hollands:2016kgm, Hao:2019ryd}, spectral networks were soon found to be essential for understanding coordinate systems and quantisation in higher Teichm\"uller theory \cite{Gaiotto:2012db, Hollands:2013qza}. See for instance \cite{Kineider:2024ttx} for a thorough and pedagogical review. In particular, $\mathcal{W}$-abelianisation provides a systematic method to construct systems of Darboux coordinates on moduli spaces of flat connections, which are known as spectral coordinates \cite{Gaiotto:2012rg, Hollands:2013qza}. One of the main advantages of spectral coordinates is their good WKB properties. This is made precise by the relation of $\mathcal{W}$-abelianisation to the exact WKB analysis \cite{Hollands:2019wbr}. In particular, this leads to a practical approach of computing the generating function of oper connections \cite{Hollands:2017ahy, Hollands:2021itj}, even for higher rank theories \cite{Hollands:2019wbr}. Through the NS-NRS proposals~\cite{Nekrasov:2009rc, Nekrasov:2011bc}, this generating function is related to the the semi-classical limit of Toda conformal blocks.

One of the motivations behind this series of papers is to lift the above relation between spectral networks and Toda conformal blocks to the quantum level, with the ultimate goal of computing Toda three-point functions with arbitrary insertions. For this, we take inspiration from the free-field approach to Toda CFTs \cite{Dotsenko:1984ad, Dotsenko:1984nm}, since its screening contours resemble quantised spectral network trajectories. The relation of the free-field formalism to the theory of matrix models~\cite{Dijkgraaf:2009pc} prompted us to consider free-field correlators with insertions of screening charges along the spectral network trajectories. A similar, but novel, strategy that we refer to as CFT-nonabelianisation was developed in \cite{Hao:2024vlg} for Liouville theory at $c=1$. In contrast to the standard approach of considering a singular spectral curve in the large-$N$ limit, CFT-nonabelianisation sets out with a spectral network on a smooth spectral curve. In part II of this series \cite{HollandsMurugesanII}, we compare both approaches, before arguing that the framework of \cite{Hao:2024vlg} is ultimately the right one, and we initiate its extension to the case $c \neq 1$ for general Toda theories.

$\mathcal{W}$-abelianisation generates various types of spectral coordinates, including the celebrated Fock-Goncharov cluster coordinates \cite{fock2006moduli, Gaiotto:2012db}, but it is well-known that Fenchel-Nielsen type spectral coordinates play an elevated role in the context of Toda CFTs. In particular, only when expressed in terms of complexified Fenchel-Nielsen coordinates, the generating function of oper connections reproduces the Yang-Yang function \cite{Nekrasov:2011bc}. Yet, it was found more recently that the perspective of cluster coordinates provides a more complete and unifying picture. Indeed, the generating function of oper connections in terms of cluster coordinates defines a non-perturbative completion (in the sense of \cite{Alim:2021mhp}) of the Yang-Yang function. From the perspective of the four-dimensional $\mathcal{N}=2$ theory, the Yang-Yang function computes the four-dimensional instanton partition function in the Nekrasov-Shatashvili limit \cite{Nekrasov:2009rc}; in addition, the non-perturbative Yang-Yang function encodes the complete four-dimensional BPS particle spectrum \cite{Hollands:2021itj, Grassi:2021wpw}. At present, however, the role of generic cluster coordinates in Toda CFTs is not very well-understood (see \cite{goncharov2019quantum} for some progress in this direction). This series of papers also attempts to remedy this.

The main results in part I are as follows:
\begin{itemize}
\item In \S\ref{sec:spectral_network}, we construct semi-classical $SL(2, \mathbb{C})$ monodromy representations in terms of complexified Fenchel-Nielsen coordinates for arbitrary Riemann surfaces $C$ (with regular punctures), by computing parallel transport matrices via $\mathcal{W}_\text{FN}$-abelianisation with respect to Fenchel-Nielsen type spectral networks $\mathcal{W}_\text{FN}$ on the gauged three-punctured sphere. 

\item It is well-known that quantum $SL(2, \mathbb{C})$ parallel transports on any surface $C$ (with regular punctures) can be computed in quantum Liouville theory through the Moore-Seiberg formalism \cite{Moore:1988qv}. In \S\ref{sec:liouville-cft}, we bring the Moore-Seiberg matrices in the $\mathcal{W}_\text{FN}$-abelianised gauge, uncovering a simple relation between ordinary and quantum parallel transports on arbitrary surfaces. (This is established in \S\ref{subsec:bpz-monodromies} on the gauged 3-punctured sphere and in  \S\ref{sec:QPgeneralC} in general.) In particular, we verify that the complexified Fenchel-Nielsen coordinates are quantised in terms of difference operators, when acting on conformal blocks (as already known from \cite{Drukker:2009id, Alday:2009fs}). 

\item Furthermore, in \S\ref{sec:liouville-cft} we work out the NRS proposal \cite{Nekrasov:2011bc} (in the formulation \cite{Hollands:2021itj}) for the gauged three-punctured sphere and  develop a quantum NRS proposal. The quantum NRS proposal says that the quantum Fenchel-Nielsen twist coordinates, when acting on the Nekrasov partition function, compute derivatives of the Yang-Yang function. (The action of the quantum spectral coordinates on Liouville blocks is defined in the sense of CFT-abelianisation, as developed further in Part II \cite[\S3]{HollandsMurugesanII}.) 

\item Finally, in \S\ref{sec:symmdefects}, we argue that Fenchel-Nielsen type spectral networks may be interpreted as symmetry defects (as formulated in  \cite{Gaiotto:2014kfa}) in Liouville theory. 

\item An alternative approach for calculating quantum $SL(2,\mathbb{C})$ monodromies with respect to Fock-Goncharov type spectral networks $\mathcal{W}_\text{FG}$ was developed in \cite{Neitzke:2020jik} and called $q$-nonabelianisation. In \S\ref{sec:q-nonabelianisation}, we extend this framework to compute $q$-parallel transport with respect to Fenchel-Nielsen type spectral networks $\mathcal{W}_\text{FN}$ on arbitrary surfaces $C$ (with regular punctures). We find agreement with Moore-Seiberg parallel transports up to (somewhat trivial) winding factors. (This is established in \S\ref{sec:q-FN-nonab} for the gauged three-punctured sphere and in \S\ref{subsec:gluing-q-nonab} for arbitrary surfaces.)

\item Whereas $q$-nonabelianisation defines a homomorphism from a nonabelian skein algebra on $C$ into an abelian skein algebra on $\Sigma$, which may be identified with a quantum torus $\mathbf{T}_\Gamma$ generated by the quantum spectral coordinates $X_\gamma$, the $q$-parallel transport map takes values in a module $\mathbf{A}$ for the quantum torus $\mathbf{T}_\Gamma$ that is generated by open abelian skeins $Y_{\widetilde{\mathfrak{p}}}$. At the end of \S\ref{subsec:quantum-FN-coordinates-2}, we explain how to extract the quantum spectral coordinates $X_\gamma$ from the open skeins $Y_{\widetilde{\mathfrak{p}}}$ using cup and cap operations. 

\item Also, in \S\ref{subsec:quantum-FN-coordinates-2}, we deduce the abelianisation of Liouville conformal blocks with degenerate insertions (in the spirit of Part II \cite[\S3]{HollandsMurugesanII}) by interpreting $q$-nonabelianisation in the background of the AGT correspondence.


\end{itemize}

\subsection{Outline}

The outline of part I is as follows:

Let $C$ be a (possibly punctured) Riemann surface. 
Section~\ref{sec:spectral_network} starts with a quick refresher on rank 2 spectral networks $\mathcal{W}^\vartheta(\phi_2)$, which are defined in terms of a quadratic differential~$\phi_2$ together with a phase $\vartheta$. In particular, we introduce the class of Fock-Goncharov type spectral networks $\mathcal{W}^\text{FG}$, which are dual to ideal triangulations of $C$, and the class of Fenchel-Nielsen type spectral networks $\mathcal{W}^\text{FN}$, which are dual to a pants decomposition of $C$ and can be built out of copies of the so-called Fenchel-Nielsen molecules. 

We then review the $\mathcal{W}$-abelianisation method, which brings any $SL(2,\mathbb{C})$ flat connection $\nabla$ on $C$ in an almost diagonal form with respect to the network $\mathcal{W}$, and the $\mathcal{W}$-nonabelianisation method, which computes the non-abelian parallel transport of an $SL(2,\mathbb{C})$ flat connection $\nabla$ in terms of the abelian parallel transport of an $SL(2,\mathbb{C})$-equivariant $\C^\times$-connection $\nabla^\text{ab}$ on the covering $\Sigma$, defined by $\phi_2$. $\mathcal{W}$-nonabelianisation with respect to a Fock-Goncharov type spectral network yields monodromy representations in terms of Fock-Goncharov cluster coordinates, while $\mathcal{W}$-nonabelianisation with respect to a Fenchel-Nielsen type spectral network yields monodromy representations in terms of complexified Fenchel-Nielsen coordinates. 

We emphasise that both the $\mathcal{W}$-abelianisation as well as the $\mathcal{W}$-nonabelianisation methods commute with the operation of gluing pairs of pants when the network $\mathcal{W}$ is of Fenchel-Nielsen type. Therefore, it is sufficient to determine the monodromy representation of any $SL(2,\mathbb{C})$ flat connection on the pair of pants (or more precisely, the gauged 3-punctured sphere) to compute the monodromy representation of any $SL(2,\mathbb{C})$ flat connection on an arbitrary surface $C$. We bring the non-abelian parallel transports across the pair of pants, from a marked point on boundary component 1 to a marked point on boundary component 2, in the simple form (see \cref{eq:parallel-transport-FN})\footnote{Eq.~\eqref{eq:parallel-transport-FN} is the analogous formula for $P^\text{FN}_a$ and therefore has permuted masses compared to \cref{{eq:PTb-FN}}. }
\begin{align}
\begin{aligned}\label{eq:PTb-FN}
 P^\textrm{FN}_b =  \begin{pmatrix}  0 & 1 \\  -1 & 0 \end{pmatrix} \begin{pmatrix}  \gamma & \alpha \, \frac{T_a T_b}{T_c} \\ \beta \, \frac{T_c}{T_a T_b} & \gamma \end{pmatrix} \begin{pmatrix} T_b & 0 \\  0 & \frac{1}{T_b} \end{pmatrix}, 
   \end{aligned}
\end{align} 
in terms of the exponentiated, complexified Fenchel-Nielsen length coordinates $M_{1,2,3}$ and twist coordinates $T_{a,b,c}$. The combinations
\begin{align}
\begin{aligned}
     \gamma^2 
     & =   \frac{(M_2 M_3-M_1^{-1})(M_1 M_3^{-1} - M_2^{-1})}{(M_1-M_1^{-1}) (M_2-M_2^{-1})},
\end{aligned}
\end{align}
and
\begin{align}
    \alpha \, \beta = \frac{ ( M_2^{-1} M_3  - M_1^{-1}) ( M_1 M_3^{-1}  - M_2  )} { (M_1-M_1^{-1}) (M_2-M_2^{-1}) }
\end{align}
are invariant under abelian (more precisely, $SL(2,\mathbb{C})$-equivariant) gauge transformations at the marked points.
The non-abelian parallel transport on any other surface $C$ can then be easily computed by concatenating the non-abelian parallel transports $P^\textrm{FN}$.

Section~\ref{sec:liouville-cft}
starts with a quick introduction to Liouville CFT. In particular, we review how to compute Moore-Seiberg parallel transports on $C$ using the BPZ equation. We then argue that the BPZ equation transforms as a quantum oper connection since it acts on differential forms of degree 
\begin{align}
\Delta_{1,2} = -\frac{1}{2}-\frac{3 b^2}{4}, 
\end{align}
and thus reduces to an oper connection in the limit $b \to 0$. We are therefore able to interpret Moore-Seiberg parallel transports as quantum parallel transports.

We bring the Moore-Seiberg parallel transport matrices on the 3-punctured sphere in the abelianised form (see \cref{eq:MSP})
\begin{align}
 P^\mathrm{MS}_{b} =\, q^{\frac{1}{2}} \sqrt{-\frac{m_1}{m_2}} \,
         \begin{pmatrix}
            0 & 1 \\ -1  & 0 \end{pmatrix}
            \begin{pmatrix}
            \gamma & \alpha \, \frac{T^\textrm{\,Lax}_a T^\textrm{\,Lax}_b}{T^\textrm{\,Lax}_c} \\  \beta \, \frac{T^\textrm{\,Lax}_c}{T^\textrm{\,Lax}_a T^\textrm{\,Lax}_b} & \gamma \end{pmatrix} 
            \begin{pmatrix}
           \frac{i T_b^\textrm{\,Lax}}{\sqrt{M_2}} & 0 \\ 0  &   \frac{\sqrt{M_2}}{i T_b^\textrm{\,Lax}}   \end{pmatrix},
\end{align}
where $T_{b}^\text{Lax}$ (and cyclically permuted for $T_{a,c}^\text{Lax}$) can be expressed in terms of $m_{1,2,3}(\alpha_{1,2,3})$ as (see \cref{eq:T-BPZ-1})
\begin{align}\label{eq:TLax}
   & \left( \frac{iT_b^\textrm{Lax}}{\sqrt{M_2}} \right)^2 = \, \frac{\Gamma [1-m_1] \Gamma [1-m_2]}{\Gamma [1+m_1] \Gamma [1+m_2] } \frac{\Gamma [\frac{1}{2} (1+m_1+m_2 \pm m_3)]}{\Gamma [\frac{1}{2} (1-m_1-m_2 \pm m_3)] },
\end{align}
and where $M_k = - \exp(-i \pi m_k)$ and $q = \exp(-i \pi b^2)$.

We point out that \cref{eq:TLax}, in the semi-classical limit $b \to 0$, determines the Yang-Yang function $\widetilde{W}_\mathrm{eff}$ on the 3-punctured sphere through the NRS relations 
\begin{align}\label{eqn:oper-Lagrangian-3sphere}
  \log T^\text{Lax}_b = - b \left( \frac{\partial \widetilde{W}_\mathrm{eff}(\vec{m})}{\partial m_1} + \frac{\partial \widetilde{W}_\mathrm{eff}(\vec{m})}{\partial m_2} \right),
\end{align}
together with its cyclic permutations.
Here, $T^\text{Lax}_b$ should be interpreted as the evaluation of the exponentiated Fenchel-Nielsen twist coordinate~$T_b$ on the semi-classical limit of the BPZ connection (i.e.~the Lax connection). 

Moore-Seiberg parallel transport along open paths on arbitrary surfaces $C$ can be obtained by inserting a matrix of shift operators at every pants curve (see \cref{sec:QPgeneralC} for details). Suppose that we denote the Liouville momentum through the pants curve by $\alpha$. Then, through a twist flow argument (similar as in the semi-classical case), we argue that the difference operator
\begin{align}
    \hat{T}^\text{BPZ}_\alpha = e^{b \partial_\alpha}
\end{align}
represents a quantum exponentiated Fenchel-Nielsen twist coordinate $\hat{T}_\alpha$ acting on the  BPZ connection (see \cref{eq:quantum-coordinates}). 

By gluing gauged 3-spheres the NRS relation~\eqref{eqn:oper-Lagrangian-3sphere}  extends to arbitrary surfaces $C$.  That is, the complex Lagrangian $\mathcal{L}^\text{oper}$ of oper connections (inside the moduli space of flat $SL(2,\mathbb{C})$-connections) can be parametrised by the system of equations
\begin{align}\label{eqn:oper-Lagrangian}
  \log T_k (\nabla^\text{Lax}) =  b \frac{\partial \widetilde{W}_\mathrm{eff}(\vec{a})}{\partial a_k},
\end{align}
where $k$ labels the pants cycles of $C$ and $\{ a_k, \log T_k \}$ is a system of semi-classical Fenchel-Nielsen length-twist coordinates.

We argue that the quantum twist coordinates $\hat{T}_\alpha$ should be thought of, in the sense of CFT-abelianisation \cite{Hao:2024vlg}, as computing the Liouville momentum through associated B-cycles of a Heisenberg conformal block on $\Sigma$. This leads us to propose a quantum version of the NRS proposal, which says that the operator (see \cref{eqn:quantumNRS})
\begin{align}\label{eqn:quantum-oper-Lagrangian}
 \hat{T}_\alpha - e^{b \left( \frac{\partial{\widetilde{W}_\textrm{eff}(\vec{a})}}{\partial F a} + \mathcal{O}(b^2) \right)}
\end{align}
acts trivially (in the sense of CFT-abelianisation) on conformal blocks. In other words, the Lagrangian $\mathcal{L}^\text{oper}$ of oper connections is quantised by the operator~\eqref{eqn:quantum-oper-Lagrangian} acting on BPZ connections. 

Our last result in section~\ref{sec:liouville-cft} identifies Fenchel-Nielsen spectral networks as symmetry defects in the Liouville CFT. Indeed, since the Moore-Seiberg parallel transport on any surface $C$ can be brought in $\mathcal{W}$-abelianised gauge with respect to a Fenchel-Nielsen spectral network, the spectral network itself acts on degenerate operators in the CFT as a symmetry defect. 

In section~\ref{sec:q-nonabelianisation}, we relate the Moore-Seiberg formalism with a new  quantisation of $SL(2, \C)$-parallel transports on the surface $C$ using skein theory and $q$-nonabelianisation. We start with an introduction to the $q$-nonabelianisation formalism \cite{Neitzke:2020jik}, which quantises ordinary $\mathcal{W}$-nonabelianisation into a skein algebra homomorphism 
\begin{equation}
    \psi^q_\mathcal W : \SkAlg(C, SL(2, \C)) \to \SkAlg(\Sigma, GL(1, \C)).
\end{equation}
We then introduce the space of open skeins on a surface $C$: this is the space of all open, upward-travelling paths inside $C \times I$ with boundaries on a fixed set $\mathbb X \subset C$ (see \cref{eq:def-open-skein-module}),
\begin{equation}
    \OpSk\big(C\times I, \mathbb X, SL(2, \C)\big) = \mathrm{Span}\{\text{open paths } \frakp \subset C \times I : \partial \frakp \subset \mathbb X \}/\sim.
\end{equation}
This allows us to define the $q$-parallel transport map as a linear map (see \cref{eq:q-parallel-def})
\begin{equation}
    P^q_\mathcal W : \OpSk\big(C \times I, \mathbb X, SL(2, \C)\big) \to \OpSk\big(\Sigma \times I, \pi^{-1}(\mathbb X), GL(1, \C)\big)
\end{equation}
obtained by restricting $q$-nonabelianisation to open paths.

After defining a suitable averaging-procedure for Fenchel-Nielsen type spectral networks, we are able to bring the basic $q$-parallel transport with respect to a Fenchel-Nielsen network into the form (see \cref{eq:quantum-parallel-transport-FN})
\begin{equation}
    P^{q\FN}_{\frakp_b} = \begin{pmatrix}
        0 & Y^{12}_{\iota}\\ Y^{21}_{\iota} & 0
    \end{pmatrix} \begin{pmatrix}  \gamma  Y^{11}_{2, L} & \alpha  Y^{12}_{2, L} \\ - q^{-1}\beta  Y^{21}_{2, L} & \gamma  Y^{22}_{2, L} \end{pmatrix} 
    \begin{pmatrix}
        q Y^{11}_{f} & 0 \\ 0 & Y^{22}_{f},
    \end{pmatrix}
\end{equation}
independently using an isotopy argument and a first-principle computation. We interpret the composite objects (see \cref{eq:quantum-open-coordinates})
\begin{align}
\begin{aligned}
& Y^{11}_{b, L} = Y^{12}_\iota Y^{21}_{2, L} Y^{11}_f \qquad Y^{12}_{b, L} = Y^{12}_{\iota}Y^{22}_{2,L}Y^{22}_f \\
& Y^{21}_{b, L} = Y^{21}_\iota Y^{11}_{2,L}Y^{11}_f \qquad Y^{22}_{b, L} = Y^{21}_\iota Y^{12}_{2,L}Y^{22}_f
\end{aligned}
\end{align}
as (compositions of) quantum exponentiated Fenchel-Nielsen coordinates. In particular, $Y_{b,L}^{21}$ may be identified with the quantum twist coordinate $\hat{T}_\alpha$. 

More precisely, we underscore that the quantum $Y$-coordinates are valued in a different quantum algebra than the usual quantum cluster $X$-coordinates (since they are defined with respect to open instead of closed paths on $\Sigma$).  In \cref{subsubsec:xy-coordinates}, we identify the precise relation between these two types of coordinates. We introduce a family of wedge operators (see \cref{eq:wedge-op})
\begin{equation}
    \iota^\textrm{ab}_\alpha : \widetilde F_\alpha \longrightarrow \widetilde F_\textrm{deg} \otimes\widetilde F_{\alpha'} \quad \text{ and } \quad \pi^\textrm{ab}_\alpha :\widetilde F_\textrm{deg} \otimes \widetilde F_{\alpha} \longrightarrow \widetilde F_{\alpha'},
\end{equation}
so that for any open path $\widetilde {\mathfrak b}_p$, we have the relation (see \cref{eq:YtoX-open})
\begin{align}
 X^{ij}_{\mathfrak{b}_p} = \pi^\text{ab}_{\alpha_i} \circ Y^{ij}_{\mathfrak{b}_p} \circ \iota^\text{ab}_{\alpha_f}.
\end{align}

Physically, we remind ourselves in \cref{sec:CStheory} that $q$-nonabelianisation is a version of a UV-IR map between an $SL(2,\C)$-Chern-Simons theory on $C \times I$ and a $GL(1,\C)$-Chern-Simons theory on $\Sigma \times I$. Equivalently, this can also be phrased as a UV-IR map between a boundary Liouville theory on $C$ and a boundary free boson theory on $\Sigma$ (we make this precise through the CFT-abelianisation map in \cite{HollandsMurugesanII}.). Then, the quantum $Y$-coordinates can be realised as open abelian Verlinde line operators in a free boson theory on $\Sigma$ (equivalently, as non-compact Wilson lines (labelled by the fundamental representation) in a $GL(1,\mathbb{C})$-Chern-Simons theory on $\Sigma \times I$).

The physical interpretation of $q$-nonabelianisation allows us to evaluate the quantum $Y$-coordinates on the gauged $3$-sphere in terms of Heisenberg conformal blocks on the cover~$\Sigma$ (see \cref{subsubsec:y-coordinates-physics}). We find that all the quantum coordinates must act multiplicatively on these Heisenberg conformal blocks, where the multiplicative constant is the semi-classical spectral coordinate. Moreover, assuming that $q$-nonabelianisation and CFT-abelianisation commute, we also deduce the abelianisation of the space of Liouville conformal blocks on the gauged $3$-sphere with one degenerate insertion.

We extend this construction to arbitrary surfaces $C$ in \cref{subsec:gluing-q-nonab}. We argue that it is sufficient to compute $q$-parallel transports restricted to individual pairs of pants in a pants decomposition of $C$, and glue the resulting $q$-parallel transports by inserting the matrix (see \cref{eq:shift-matrix-q-nonab})
\begin{equation} 
    \begin{pmatrix}
        \Delta_q^{-1} & 0 \\ 0 & \Delta_q \end{pmatrix} \bigg( \dots \bigg)^\textrm{op} ,
\end{equation}
where $\Delta_q^{\pm 1} (M) = q^{\pm 1} M$ is a difference operator acting on the pants cycle coordinate and the superscript "op" denotes orientation reversal. We conclude section~\ref{sec:q-nonabelianisation} by comparing the resulting monodromy representations to those computed using the Moore-Seiberg formalism on the 4-punctured sphere. We find agreement up to (somewhat trivial) winding factors.

\subsection{Outlook}

We conclude by outlining several open problems and directions for future research:

\begin{itemize}

\item  In \S\ref{sec:q-nonabelianisation} we have given a description of $q$-parallel transport in terms of quantum Fenchel-Nielsen coordinates, through the $q$-nonabelianisation approach of Neitzke and Yan in \cite{Neitzke:2020jik}. In particular, we found agreement with the Moore-Seiberg approach in Liouville CFT. Yet another approach for computing quantum parallel transport is given by Schrader and Shapiro in \cite{schrader2025algebraic}. It would be worthwhile to analyse the similarities and differences between both approaches, and compare computations in terms of Fock-Goncharov as well as Fenchel-Nielsen coordinates. 

\item In \cref{subsubsec:quantum-nrs-proposal}, we propose a quantisation of the Nekrasov-Rosly-Shatashvili proposal from \cite{Nekrasov:2011bc} (in the language of \cite{Hollands:2021itj}). The verification of this proposal relies on a detailed formulation of the CFT-abelianisation map. With the progress made in~\cite{HollandsMurugesanII}, it is possible to work out the quantum NRS proposal in examples. In particular, it is tempting to think that this proposal leads to (a refinement of) the Baxter TQ-equation (see for instance \cite[eq.~7.11]{Nekrasov:2013xda}) on the 4-punctured sphere for the pure $SU(2)$ theory.  

\item Originally in \cite{Neitzke:2020jik}, the $q$-nonabelianisation map was only defined for generic Fock-Goncharov networks. Although we are able to define $q$-parallel transports for Fenchel-Nielsen networks, we have not addressed the $q$-nonabelianisation of arbitrary closed links with respect to a Fenchel-Nielsen network in this paper.\footnote{\textit{Note added in v2}: Recently, exciting progress in this direction was achieved in~\cite{Gupta:2026vxz}.} It is now known \cite{panitch20243dquantumtracemap} that $q$-nonabelianisation for Fock-Goncharov networks agrees with other triangulation-dependent quantisation procedures in the literature. The final piece of this puzzle is to extend the $q$-nonabelianisation map to Fenchel-Nielsen networks, and compare it with other pants decomposition-dependent quantisation procedures such as the one described in \cite{Detcherry_2025}.

\end{itemize}

\subsection{Acknowledgements}

First of all, we would like to thank the \href{https://sites.google.com/view/griftseminar} {GRIFT seminar} for creating an inspiring and supportive research environment. We thank Ioana Coman for introducing us to the Coulomb gas formalism and collaborating with us in an early stage of this project. We thank Andy Neitzke for multiple beneficial discussions and, in particular, for suggesting to look into the Maulik-Okounkov R-matrix. We thank Misha Bernstein for introducing us to free-field representations more generally. Many other enlightening discussions were had with Murad Alim, Jennifer Brown, Nitin Chidambaram, Tudor Dimofte, Veronica Fantini, David Jordan, Jasper Kager, Pietro Longhi, Fabrizio del Monte, Sasha Shapiro, Joerg Teschner, Campbell Wheeler, Yegor Zenkevich, amongst others. Specifically, we thank Nitin Chidambaram, David Jordan, Andy Neitzke and Joerg Teschner for their comments on a previous draft. Furthermore, we thank the organisers of the CAMP in the Lakes workshop, the organisers of the MFO Mini-Workshop on Resurgence, Difference Equations and Quantum Modularity, and the organisers of the Chiralisation and QFT workshop, at each of which parts of this work were presented.

\section{Gauging and gluing Fenchel-Nielsen molecules}
\label{sec:spectral_network}

This section is mostly a synopsis of material in \cite{Gaiotto:2009hg, Gaiotto:2012rg, Hollands:2013qza, Hollands:2017ahy, Hollands:2019wbr, Hollands:2021itj} that will be built upon in later sections. We start this section in \S\ref{sec:spectral-networks} with a quick intro to spectral networks~$\mathcal{W}$ on a (possibly punctured) Riemann surface $C$, with special emphasis on those of Fenchel-Nielsen type. In \S\ref{sec:nonab} we summarise the abelianisation and non-abelianisation methods that bring the parallel transport of flat $SL(2,\mathbb{C})$-connections on $C$ in an almost-diagonal gauge with respect to a fixed spectral network $\mathcal{W}$. We emphasise that both methods commute with the operation of gluing pairs of pants when the network~$\mathcal{W}$ is of Fenchel-Nielsen type. In \S\ref{subsubsec:gaugedsphere} we derive nonabelian parallel transport matrices on the \emph{gauged 3-punctured sphere} in terms of Fenchel-Nielsen length-twist coordinates, while in \S\ref{sec:gluing-nonab} we discuss how to glue gauged 3-punctured spheres to compute parallel transport on arbitrary Riemann surfaces (possibly with regular punctures) in terms of Fenchel-Nielsen length-twist coordinates.

\subsection{Basics of spectral networks}\label{sec:spectral-networks}

 Consider a degree 2 spectral cover $\Sigma \subset T^* C$ of a (possibly punctured) Riemann surface~$C$ cut out by an equation of the type\footnote{Note that $\phi_2$ is defined here with an additional sign compared to (for instance) \cite{Gaiotto:2009hg,Hollands:2013qza}, yet agrees with the conventions of (for instance) \cite{Gaiotto:2012rg} and Andy Neitzke's swn plotter \cite{swn-plotter}.} 
 \begin{align}
 \Sigma: \quad    \lambda^2 + \phi_2 = 0, \label{eq:def-spectral-cover}
 \end{align}
where $\phi_2$ is a quadratic differential on $C$ and $\lambda$ is the tautological 1-form on $T^*C$. Such a spectral covering is in general branched, and for the moment we will assume that the ramification points are simple. We shall choose a system of branch-cuts on the base curve~$C$ to trivialise the covering. All final results are independent of the choice of trivialisation, however, and can be presented in a trivialisation-independent way.

\paragraph{Spectral network}

Let $\lambda_i$ denote the restriction of $\lambda$ to the $i$-th sheet of $\Sigma$. Choose a phase $\vartheta \in \R/2\pi \Z$. We define an $ij$-\emph{trajectory} on $C$ to be a (real) curve $w(t) : \mathbb{R} \to C$ such that locally,  for each point $z = w(t)$ on the curve,
\begin{equation}
    e^{-i\vartheta}(\lambda_i - \lambda_j)(w^\prime(t)) \in \mathbb R_{\ge 0}. \label{eq:ij-trajectory}
\end{equation}
It is easy to check that every simple branch-point of type $(ij)$ emits three trajectories, which are either of type $ij$ or $ji$. The set of all $ij-$trajectories defines a foliation of the surface $C$. This is called the \emph{WKB foliation}.

\begin{remark}
    Leaves of the WKB foliation may also be defined as (real) curves $w(t) : \R \to C$ such that
    \begin{equation}
        \textrm{Im}~e^{-i\vartheta}\int_{w} \left( \lambda_i - \lambda_j\right) = \text{ constant}.
    \end{equation}
    Since $\lambda_i = - \lambda_j$, all trajectories are flow lines for the Hamiltonian function $\textrm{Im}(e^{-i\vartheta} \sqrt{\phi_2})$. Equivalently, $12$-trajectories ($21$-trajectories) are steepest ascent (descent) trajectories for the Morse function $\textrm{Re}(e^{-i\vartheta} \sqrt{\phi_2})$. This fact is used in \cite{HollandsMurugesanII} to define matrix model partition functions with spectral network contours.
\end{remark}

\begin{figure}[h]
   \centering
  \includegraphics[width=0.6\linewidth]{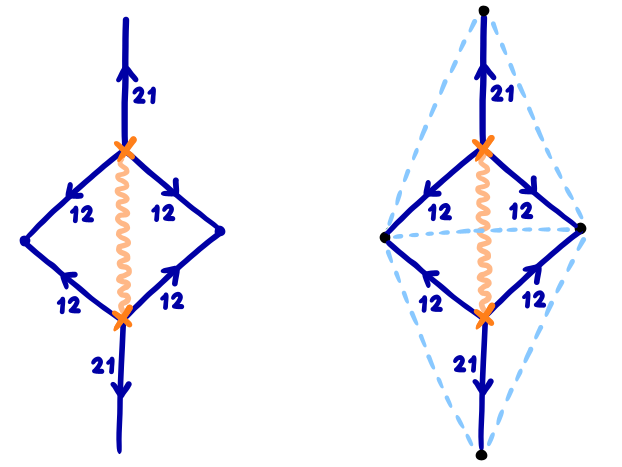}
    \caption{Left: FG network on the 3-punctured sphere, with punctures at $z=\pm 1$ and $z=\infty$, for $\phi_2 = \frac{(z^2 + 1)}{(z^2 - 1)^2} (dz)^2$ at $\vartheta = \frac{\pi}{2}$. Labellings are fixed with respect to the choice of trivialisation $\lambda_1 = \sqrt{-z^2-1}/(z^2-1)$ and $\lambda_2 = - \sqrt{-z^2-1}/(z^2-1)$. Right: same network together with the dual ideal triangulation (in dashed light blue).}
    \label{fig:FG-example}
\end{figure}

An $ij$-trajectory is said to be \emph{critical} if one of its end-points lies on a branch-point.  Critical trajectories are sometimes called \emph{walls} as well. We orient critical trajectories away from the branch-point. The union of all critical trajectories is called the critical graph. 
The critical graph together with a choice of branch-cuts, orientations and labellings is called a \emph{spectral network} $\mathcal W^\vartheta(\phi_2)$ (or more precisely, a WKB spectral network \cite{Gaiotto:2009hg}). See \cref{fig:FG-example} for an example. We will refer to the connected components of $C\backslash\mathcal W$ as cells. 

\paragraph{Degenerate networks}
\label{par:degenerate-network}

We will also be interested in the degenerate limit in which $\phi_2$ becomes the square of a 1-form $\omega$ on $C$. In this case, the spectral cover 
\begin{align}
\Sigma: \quad \lambda^2 + \omega^2 = 0
\end{align}
degenerates into two copies of the curve $C$ that are glued transversely at the zeroes of $\omega$. In particular, this implies that the ramification points of the covering $\Sigma \to C$ are no longer simple, and now emit four trajectories. We shall refer to the resulting spectral network, defined by $\phi^\text{sing}_2 = \omega^2$, as a \emph{degenerate network} $\mathcal{W}^\vartheta(\phi_2^\text{sing})$. See fig.~\ref{fig:degenerate-FN-example} for an example of a degenerate network of Fenchel-Nielsen type (explained next).

\paragraph{Fock-Goncharov and Fenchel-Nielsen}

Critical trajectories generically end at a puncture, but may also run between two branch-points. Such trajectories are called \emph{saddle trajectories} or \emph{double walls}. When a network has no saddle trajectories, the network is said to be of \emph{Fock-Goncharov type} (FG-type) \cite{Gaiotto:2012db}. An FG network is dual to an ideal triangulation of the base curve $C$. (That is, the edges of the triangulation do not intersect the trajectories of the spectral network, but instead run parallel to them.) See fig.~\ref{fig:FG-example}. 

\begin{figure}[h]
   \centering
  \includegraphics[width=0.85\linewidth]{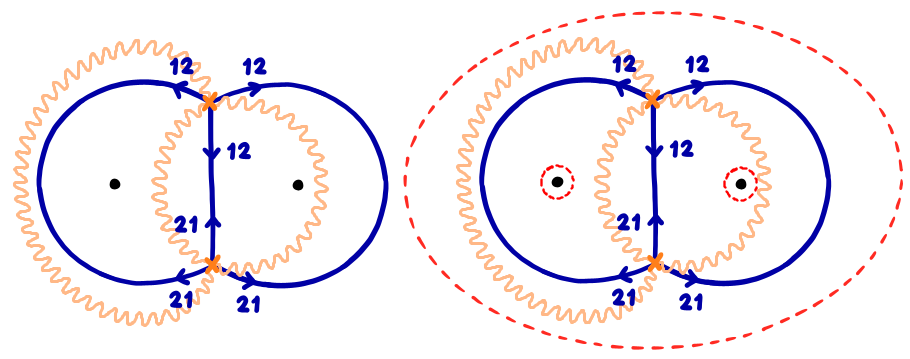}
    \caption{Left: Picture of a Fenchel-Nielsen type spectral network on the 3-punctured sphere, with punctures at $z=\pm 1$ and $z=\infty$, for $\phi_2 = \frac{(z^2 + 1)}{(z^2 - 1)^2} (dz)^2$ at $\vartheta = 0$. Right: same spectral network together with the dual pair of pants decomposition (in dashed red).}
    \label{fig:FN-example}
\end{figure}

\begin{figure}[h]
   \centering
  \includegraphics[width=0.75\linewidth]{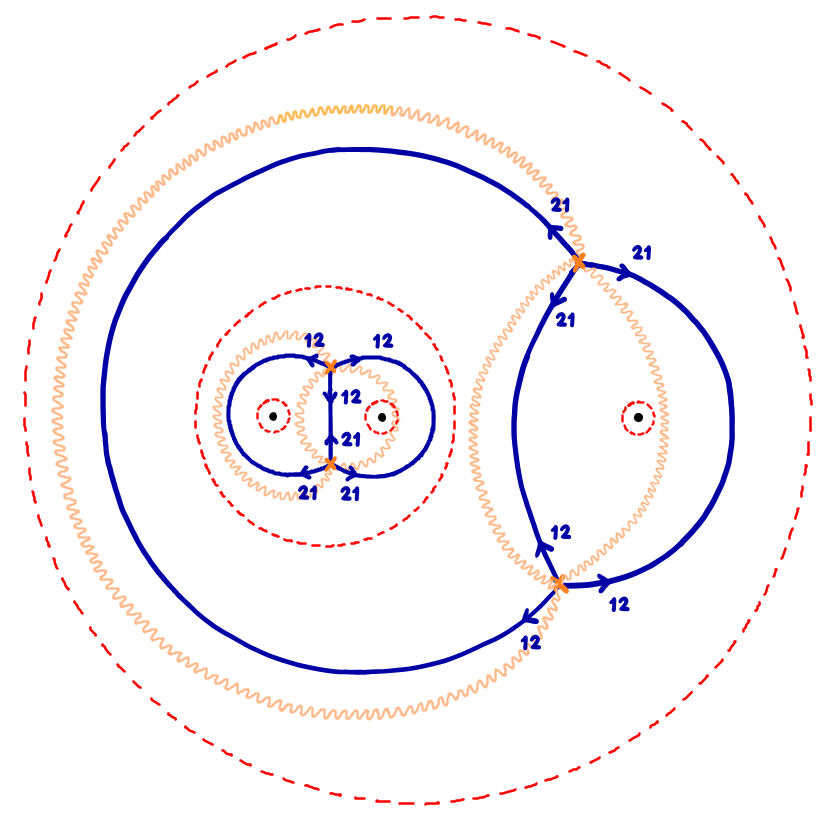}
    \caption{Picture of a Fenchel-Nielsen type spectral network (in dark blue) on the 4-punctured sphere, with punctures at $z=0,~ 0.2,~ 1$ and $\infty$, for $\phi_2 = \frac{2.6 - 2.94 z + 15.04 z^2 - 14.7 z^3 + 6.5 z^4}{z^2 (z-0.2)^2 (z-1)^2 }$ at $\vartheta_\mathrm{FN}=0$. The dual pants decomposition (in dashed red) separates the punctures at $z=0$ and $0.2$ from those at $z=1$ and $\infty$. Note that we can align the orientation of saddle trajectories in both molecules by adding an additional branch-cut around the pants cycle.  }
    \label{fig:FN-4pt-sphere-example}
\end{figure}

On the other hand, for certain phases, all $ij-$trajectories (critical and non-critical) may become compact curves on $C$. When the network is dual to a pants decomposition of~$C$, we instead call it of \emph{Fenchel-Nielsen type} (FN-type) \cite{Hollands:2013qza}. See fig.~\ref{fig:FN-example} and \ref{fig:FN-4pt-sphere-example} for examples. The appearance of pants decompositions is the first indication that spectral networks play an important role in the description of conformal field theories such as Liouville theory.

Let us for the moment restrict ourselves to surfaces $C$ with regular punctures.\footnote{This is our main class of examples, where we usually also assume that $C$ is hyperbolic (i.e.~$3g+n-3>0$), but all discussions can be easily extended to irregular punctures. See for instance \cite{Hollands:2021itj}.} Suppose that we fix a pants decomposition $\mathcal{P}$ of the surface $C$, with pants-cycles $\mathfrak{a}_p$ and loops $\underline{\mathfrak{a}}_k$ around the punctures $z_k$. (A pair of pants in $\mathcal{P}$ may thus have up to three regular punctures as its boundary components.) Given any collection of positive numbers $a_p$ and $\mu_k$, we can find a differential $\phi_2$ and a phase $\vartheta_\mathrm{FN}$ such that \cite{Liu:JS}
\begin{align}\label{eq:Strebelform}
   e^{-i \vartheta_\textrm{FN}} \oint_{\gamma_p} \sqrt{\phi_2} =  a_p \quad 
\textrm{and} \quad  
   e^{-i \vartheta_\textrm{FN}}  \oint_{\underline{\gamma}_k} \sqrt{\phi_2} = \mu_k,
\end{align}
for suitable lifts $\gamma_p$, $\underline{\gamma}_k$ of the 1-cycles $\mathfrak{a}_p$, $\underline{\mathfrak{a}}_k$ to $\Sigma$.
The differential $\phi_2^\text{Str} = e^{-2i \vartheta_\textrm{FN}} \, \phi_2$ is known as a Jenkins-Strebel differential.
The resulting spectral network $\mathcal{W}_{\textrm{FN}} = \mathcal{W}^{\vartheta_\mathrm{FN}}(\phi_2)$ is of Fenchel-Nielsen type and dual to the pants decomposition~$\mathcal{P}$. Each of the saddle trajectories in the network  $\mathcal{W}_{\textrm{FN}}$ in fact encodes a family of $12$ and $21$-trajectories. These families become visible when we slightly change the phase $\vartheta_\textrm{FN}$ to obtain the resolutions $\mathcal{W}^\pm_\text{FN} = \mathcal{W}^{\vartheta_\mathrm{FN} \pm \eta}(\phi_2)$, for infinitesimal $\eta>0$. See \cref{fig:FN-4pt-sphere-trajectories-molI.png} for an example.

\paragraph{Fenchel-Nielsen molecules}

\begin{figure}[h]
   \centering
  \includegraphics[width=0.8\linewidth]{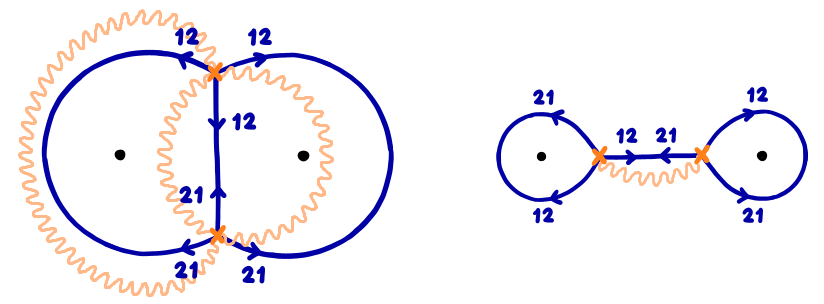}
    \caption{Fenchel-Nielsen type networks on the 3-punctured sphere with quadratic differential from \cref{eq:diffformolecules} with $\mu_k>0$. Left: molecule I with $\mu_0+\mu_1>\mu_\infty$. Right: molecule II with $\mu_0+\mu_1<\mu_\infty$. There are two versions of each molecule, related by exchanging the labels $1 \leftrightarrow 2$.}
    \label{fig:molecule1+2}
\end{figure}

FN networks on the 3-punctured sphere $\mathbb{P}^1_{0,1,\infty}$ can be  generated by the quadratic differential
\begin{align}\label{eq:diffformolecules}
    \phi_2 = -\frac{\mu^2_{\infty} z^2 - (\mu^2_{\infty} + \mu^2_0 - \mu^2_1)z + \mu^2_0}{z^2 (z - 1)^2} \, (dz)^2.
\end{align}
The two possible isotopy classes are illustrated in  \cref{fig:molecule1+2}. These networks were called \emph{molecule I and II} in \cite{Hollands:2013qza}. (More precisely, since the punctures are distinguishable, there are four possible isotopy classes really: one with the topology of molecule I and three with the topology of molecule II. All of these can be obtained by suitable choices of the parameters $\mu_k \in \mathbb{R}$. ) 

Note that when $\mu_0 + \mu_1 = \mu_\infty$, the numerator of $\phi_2$ factorises as 
\begin{align}
(\mu_0+\mu_1)^2 \left( z- \frac{\mu_0}{\mu_0+\mu_1} \right)^2.
\end{align}
This implies that the two simple branch-points have merged into a double branch-point. The resulting network is therefore a degenerate network of FN type. See \cref{fig:degenerate-FN-example}.

\begin{figure}[h]
   \centering
  \includegraphics[width=0.3\linewidth]{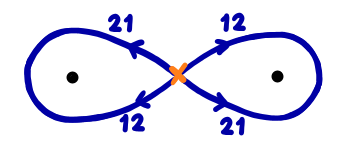}
    \caption{A degenerate spectral network of Fenchel-Nielsen type on the three-punctured sphere, with punctures at $z=\pm 1$ and $z=\infty$, for $\phi_2 = \frac{z^2}{(z^2 - 1)^2} (dz)^2$ at $\vartheta = 0$.}
    \label{fig:degenerate-FN-example}
\end{figure}

\paragraph{Gluing Fenchel-Nielsen molecules}
The restriction of any FN network $\mathcal{W}_{\textrm{FN}}$ to a pair of pants in $\mathcal{P}$ is isotopic to one of the two FN~molecules. This operation is somewhat subtle though: the restriction of the spectral network $\mathcal{W}^{\vartheta_\text{FN} \pm \eta}$, for finite $\eta>0$, to a pair of pants is \emph{not} isotopic to a spectral network on the 3-punctured sphere obtained by changing the phase of a FN molecule by finite $\eta>0$. Let us explain this in more detail. 

The resulting network on the 3-punctured sphere consists of just 6 trajectories that start at one of the two branch-points and wind into one of the three punctures. (This network is dual to an ideal triangulation with two triangles.) Instead, the trajectories of the network $\mathcal{W}^{\vartheta_\text{FN} \pm \eta}$ on generic $C$, which originate at the branch-points of a given pair of pants in $\mathcal{P}$, will either wind into a puncture or cross a pants tube.  This implies that the restriction of the spectral network $\mathcal{W}^{\vartheta_\text{FN} \pm \eta}$ to any pair of pants will have additional incoming trajectories (compared to the network on the 3-punctured sphere) from each boundary component that is not a puncture. See \ref{fig:FN-4pt-sphere-trajectories-molI.png} for an example on the 4-punctured sphere.

 \begin{figure}
   \centering
  \includegraphics[width=0.75\linewidth]{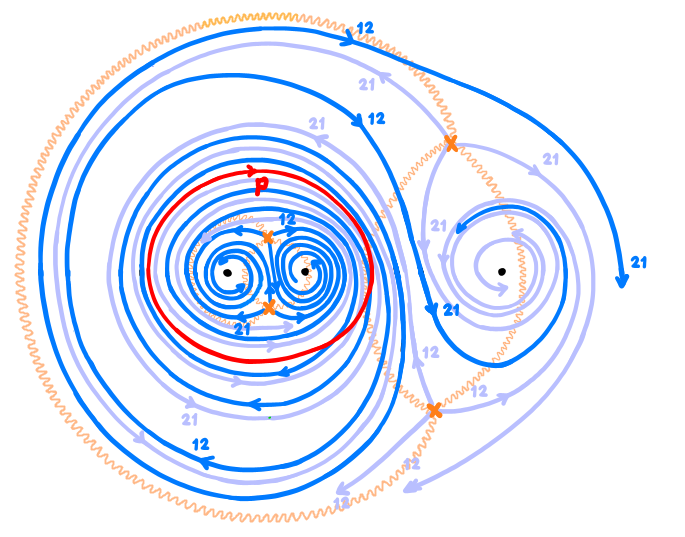}
    \caption{Cartoon of spectral network on the 4-punctured sphere for the same quadratic differential $\phi_2$ as in \cref{fig:FN-4pt-sphere-example}, but now with $\vartheta = \eta >0 $ deviating slightly from the Fenchel-Nielsen phase. The red loop is a choice of pants cycle. The cartoon highlights all trajectories (in bright blue) that start in the left pair of pants and de-emphasises all trajectories (in light blue) that start in the right pair of pants. The light (bright) blue trajectories are additional incoming trajectories into the left (right) pair of pants.}
    \label{fig:FN-4pt-sphere-trajectories-molI.png}
\end{figure}

Now, when we tune $\eta \to 0$, the trajectories of $\mathcal{W}^{\vartheta_\text{FN} \pm \eta}$ start to wind more and more frequently around the punctures and pants tubes. Precisely at the critical phase $\vartheta_\text{FN}$ all trajectories are compact, so that no trajectory traverses any pants tube any longer. The restriction of the critical network $\mathcal{W}_{\textrm{FN}}$ to the pair of pants therefore \emph{does} have the same topology as that of an FN molecule. As a consequence, we can simply glue FN molecules on pairs of pants to obtain FN networks on generic surfaces $C$. 

In the following, we will mostly consider FN networks that are built out of copies of molecule I. This is mostly for presentation purposes, since molecule I does not distinguish between punctures. We expect that the main results hold for any other configuration of molecules as well.

\paragraph{Higher rank spectral networks}

In this paper, we mostly focus on spectral networks of rank 2. Nonetheless, we briefly introduce spectral networks of higher rank here.

Spectral networks of type $A_{N-1}$ are defined with respect to a degree $N$ spectral cover $\Sigma \subset T^*C$ cut out by an equation of the type \cite{Gaiotto:2012rg}
\begin{align}
    \Sigma: \quad \lambda^N + \phi_2 \lambda^{N-2} + \ldots + \phi_N\,,
\end{align}
where $\underline{\phi}=(\phi_2, \ldots, \phi_N)$ is a tuple of $k$-differentials on $C$ and $\lambda$ is the tautological 1-form on $T^*C$. As previously, the family of spectral networks $\mathcal{W}^\vartheta(\underline{\phi})$, for $\vartheta \in \mathbb{R}/2 \pi \mathbb{Z}$, is defined as the collection of critical $ij$-trajectories, oriented away from the branch-points.

\subsection{Recap of abelianisation and nonabelianisation}\label{sec:nonab}

Spectral networks encode a lot of information, for instance about the BPS spectrum of $\mathcal{N}=2$ theories of class S \cite{Gaiotto:2009hg, Gaiotto:2010be, Gaiotto:2012rg, Hollands:2016kgm}. But they can also be employed to construct Darboux coordinates on moduli spaces of flat connections on $C$, through a procedure that is called $\mathcal{W}$-abelianisation \cite{Gaiotto:2012rg, Hollands:2013qza}.   We briefly summarise $\mathcal{W}$-abelianisation here for spectral networks $\mathcal{W}= \mathcal{W}^\vartheta(\phi_2)$ of rank 2. 

Note that the following discussion is valid only for non-degenerate spectral networks. To be precise, $\mathcal{W}$-abelianisation is not even defined at any value~$\vartheta$ for which the network~$\mathcal{W}$ has saddle trajectories. If the spectral network does have saddles, we define its $\mathcal{W}$-abelianisation as the average of the $\mathcal{W}$-abelianisations of its two resolutions. We will go through explicit examples of this type in \S\ref{sec:nonab-pants}.

\subsubsection{\texorpdfstring{$\mathcal{W}$-abelianisation}{W-abelianisation}}

Fix a rank 2 vector bundle $E \to C$ and consider any flat $SL(2,\mathbb C)$-connection $\nabla$ on $E$ together with a  \emph{$\mathcal{W}$-framing}. The $\mathcal{W}$-framing is a discrete choice on $\nabla$ that depends only on the isotopy class of $\mathcal{W}$. Its purpose is to render the $\mathcal{W}$-abelianisation procedure (as explained below) 1-1. If $\mathcal{W}$ is a spectral network of FG type, the $\mathcal{W}$-framing consists of a choice of eigenline $\ell_k$ at each puncture. If $\mathcal{W}$ is a spectral network of FN type, the $\mathcal{W}$-framing consists of a choice of eigenline $\ell^\pm_n$ in each direction of going around a ring domain \cite[\S5.3]{Hollands:2013qza}. We require $\ell^+_n \neq \ell^-_n$, so that the $\mathcal{W}$-framing for an FN network only exists when the monodromy around each ring domain is diagonalisable.

The task in $\mathcal{W}$-abelianisation is to bring $\nabla$ into an almost-diagonal form. More precisely, we want to find a gauge in which $\nabla$ is diagonal in each cell of $C\backslash \mathcal{W}$ and in which the diagonal bases across $ij$-trajectories are related by unipotent transformations. That is, we look for a basis $(s_1, s_2)$ of local sections of $E$ in each cell of $C\backslash \mathcal{W}$ such that 
\begin{align}
\nabla s_i = d_i \otimes s_i
\end{align}
for some $d_i \in T^*C$, and such that two local bases across an $ij$-trajectory are related by the transformation
\begin{align}
s_i &\mapsto s_i^\prime = s_i + c_{ij}s_j, \notag \\
s_k &\mapsto s_k^\prime = s_k, \quad k\neq i. \label{eq:change-of-basis}
\end{align}
We also impose that the local sections are  shuffled as
\begin{align}
s_i &\mapsto s_j\notag\\
s_j &\mapsto -s_i, 
\end{align}
 when crossing a branch-cut labelled by $(ij)$. With the above choice of $\mathcal{W}$-framing for spectral networks of FG and FN type, we can find a unique almost-diagonal form of this kind for every $\mathcal{W}$-framed flat connection~$\nabla$ (up to $SL(2,\mathbb{C})$-equivalence) \cite{Hollands:2013qza}.

Given the almost-diagonal form of $\nabla$, we can define a $\mathbb{C}^*$-connection $\nabla^\text{ab}$ on the cover~$\Sigma$ through the equation 
\begin{equation}
    \nabla^\text{ab}s_i = d_i \otimes s_i. \label{eq:abelian-connection}
\end{equation}
The abelian connection $\nabla^\text{ab}$ is indeed well-defined globally precisely because the gauge transformations between the cells can be brought into unipotent form. It is almost-flat, in the sense that its holonomy around the lift of a branch-point is $-1$ instead of $1$.

The resulting procedure  defines the $\mathcal{W}$-abelianisation map 
\begin{equation} \psi_{\mathcal W}^\textrm{ab} : \mathcal M_\mathcal{W}(C, SL(2,\mathbb C)) \to \mathcal M^\text{al}(\Sigma, GL(1,\mathbb C))\end{equation}
between $\mathcal{W}$-framed flat $SL(2,\mathbb{C})$-connections on the base~$C$ and almost-flat $\mathbb{C}^\times$-connections  on the cover~$\Sigma$. In fact, the almost-flat connections in the image automatically have additional structure. This was formulated in \cite[\S 4.2]{Hollands:2013qza} in terms of the existence of a $\nabla^\text{ab}$-invariant pairing, and referred to as being $SL(2,\mathbb{C})$-\emph{equivariant}.
The $\mathcal{W}$-abelianisation map is 1-1 between flat $SL(2,\mathbb{C})$ connections and equivariant $\mathbb{C}^\times$-connections for any spectral network of FG or FN-type \cite[\S5]{Hollands:2013qza}. Let us emphasise that the definition of the $\mathcal{W}$-abelianisation map  $ \psi_{\mathcal W}^\text{ab} $ depends only on the isotopy class of the spectral network $\mathcal{W}$. 

\subsubsection{Spectral coordinates}

Given any flat $\mathbb{C}^\times$-connection on $\Sigma$, we can compute its holonomies along the 1-cycles $\gamma  \in \Gamma = H_1(\Sigma, \mathbb{Z})$. Almost-flat $GL(1,\mathbb C)$-connections $\nabla^\text{ab}$ require a $\mathbb{Z}_2$-extension $\hat{\Gamma}$ of the charge lattice $\Gamma$, to incorporate the non-trivial loops around branch-points on $\Sigma$. Using the $\mathcal{W}$-abelianisation map $\psi_{\mathcal W}^\text{ab}$, we may thus associate to any $\mathcal{W}$-framed flat $SL(2,\mathbb{C})-$connection $\nabla$ the $\mathbb{C}^\times$-numbers
\begin{align}
    \mathcal{X}^\mathcal{W}_{\hat{\gamma}}(\nabla) = \mathrm{Hol}_{\hat{\gamma}} \nabla^\text{ab} \in \mathbb{C}^\times,
\end{align}
for $\hat{\gamma} \in \hat\Gamma$.
The objects $\mathcal{X}^\mathcal{W}_{\hat{\gamma}}$ thus define a set of coordinates on $\mathcal M_\mathcal{W}(C, SL(2,\mathbb C))$ and are known as the \emph{spectral coordinates}. 

More precisely, spectral coordinates are usually defined with respect to the original charge lattice $\Gamma$. This is possible because one can define a sign $\tau(\hat{\gamma})$ (in terms of a spin structure on $C$) such that \cite[Appendix B]{Grassi:2021wpw}
\begin{align}
\mathcal{X}^\mathcal{W}_\gamma = \tau(\hat{\gamma}) \mathcal{X}^\mathcal{W}_{\hat{\gamma}} \label{eq:holonomy-function}
\end{align}
only depends on $\gamma \in \Gamma$ while obeying the product rule 
\begin{align}\label{eq:Xalgebra}
\mathcal{X}^\mathcal{W}_\gamma \mathcal{X}^\mathcal{W}_{\gamma'} = \mathcal{X}^\mathcal{W}_{\gamma+\gamma'}.
\end{align}

Note that since $\nabla$ is an $SL(2,\mathbb{C})-$flat connection, the spectral coordinate $\mathcal{X}^\mathcal{W}_{\gamma}(\nabla)$ is only nonzero when $\gamma$ is an odd 1-cycle on $\Sigma$. Moreover, 
\begin{align}
    \mathcal{X}^\mathcal{W}_{\gamma^\text{odd}}(\nabla) = \mathcal{X}^\mathcal{W}_{\gamma}(\nabla),
\end{align}
where $\gamma^\text{odd} = \frac{1}{2} (\gamma + \sigma_* \gamma)$ is defined with respect to the covering involution $\sigma$ on $\Sigma$. 
Also note that spectral coordinates are automatically (exponential) Darboux with respect to the canonical Poisson structure on $\mathcal M_\mathcal{W}(C, SL(2,\mathbb C))$. Indeed, the Poisson bracket between any two spectral coordinates is given by 
\begin{equation}
    \{\mathcal X^\mathcal{W}_\gamma, \mathcal X^\mathcal{W}_{\gamma'} \} = \langle \gamma, \gamma' \rangle \, \mathcal X^\mathcal{W}_\gamma \mathcal X^\mathcal{W}_{\gamma'},
\end{equation}
where $\langle \_, \_\rangle$ is the homological intersection form.

\begin{figure}[h]
   \centering
  \includegraphics[width=0.5\linewidth]{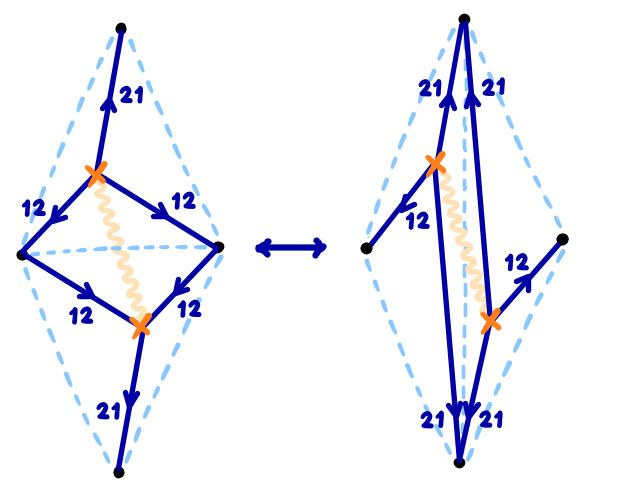}
    \caption{Example of a flip transition of an FG-type spectral network.  }
    \label{fig:flip}
\end{figure}

 The spectral coordinates $\{ \mathcal{X}_\gamma^\text{FG} \}$ associated to a spectral network of FG-type reproduce the Fock-Goncharov coordinates attached to its dual ideal triangulation \cite{Gaiotto:2012db, Hollands:2013qza}. Under a flip transition of the spectral network (see \cref{fig:flip}), the coordinates $\{ \mathcal{X}_\gamma^\text{FG} \}$ transform through the cluster mutation
 \begin{align}
 \begin{aligned}
    &  {\mathcal X^\text{FG}_{\gamma}}'' \mapsto 
1/ \mathcal  X^{\text{FG}}_\gamma  \\
& {\mathcal X^\text{FG}_{\gamma'}}'' \mapsto  \mathcal X^{\text{FG}}_{\gamma'} \left( 1 + \mathcal X^{\text{FG}}_\gamma \right)^{\langle \gamma, \gamma' \rangle},
\end{aligned}
 \end{align}
for $\gamma' \neq \gamma$. The spectral coordinates $\{ \mathcal{X}_\gamma^\text{FN} \}$ associated to a spectral network of FN-type can be identified with (the exponential of) complexified Fenchel-Nielsen length-twist coordinates \cite{Hollands:2013qza}. Indeed, the $SL(2,\mathbb{C})$-monodromy around a pants cycle $\mathfrak{a}$ is diagonal in terms of the spectral coordinates $\mathcal{X}_{A}^\text{FN}$, where $A$ is a lift of $\mathfrak{a}$, since the $\mathcal{W}$-abelianisation map brings $\nabla$ in diagonal form on $C\backslash \mathcal{W}$ and thus in a neighborhood of the pants cycle.
 
\paragraph{Exact WKB analysis}

There is a close relation between $\mathcal{W}$-abelianisation and the exact WKB analysis \cite{Hollands:2019wbr}. This relation states that for any oper $\epsilon$-connection $\nabla_\epsilon^\mathrm{oper}$ we have the relation\footnote{To be precise, on the RHS we limit ourselves to the even part of the Riccati expansion. Including the odd part introduces a minus-sign which turns the Borel sum into a function on the extended lattice $\hat{\Gamma}$ \cite[\S3.6]{Grassi:2021wpw}. }
\begin{align}
\mathcal{X}^{\mathcal{W}_\textrm{WKB}}_\gamma (\nabla^\mathrm{oper}_\epsilon) = \mathcal{B}_{\vartheta_\textrm{WKB}}  \exp \left( \frac{1}{\epsilon} \int_\gamma \lambda^\textrm{WKB}_\epsilon \right),
\end{align}
where $\mathcal{B}_{\vartheta_\textrm{WKB}}$ denotes the Borel resummation, in the direction 
\begin{align}
\vartheta_\textrm{WKB} = \arg(\epsilon)
\end{align} 
of the exponentiated Riccati differential $\lambda^\textrm{WKB}_\epsilon$ determined by the oper $\epsilon$-connection $\nabla^\mathrm{oper}_\epsilon$. The spectral network $\mathcal{W}_\textrm{WKB}$ is generated by the quadratic differential $\varphi^\textrm{WKB}_2$ that is obtained from $\nabla^\mathrm{oper}_\epsilon$ in the semi-classical limit $\epsilon \to 0$ at phase $\vartheta_\textrm{WKB}$. 

In particular, this implies that for any phase $\vartheta$, in a half-plane centered around $\vartheta_\mathrm{WKB}$, we may approximate 
\begin{align}\label{eq:WKBapprox}
\mathcal{X}^{\mathcal{W}_\textrm{WKB}^\vartheta}_\gamma (\nabla^\mathrm{oper}_\epsilon) \sim c_\gamma  \exp \left( \frac{1}{\epsilon} \int_\gamma \lambda \right) \text{ for } \epsilon \to 0,
\end{align}
where $\mathcal{W}_\textrm{WKB}^\vartheta$ is the WKB spectral network with respect to the quadratic differential  $\varphi^\textrm{WKB}_2$ at phase $\vartheta$ and $\lambda = \sqrt{\varphi^\textrm{WKB}_2}$.

\subsubsection{\texorpdfstring{$\mathcal{W}$-nonabelianisation}{W-nonabelianisation}}

There also exists a $\mathcal{W}$-nonabelianisation map 
\begin{equation} 
\psi_{\mathcal W}^\textrm{nab} : \mathcal M^\text{al}(\Sigma, GL(1,\mathbb C))\to  \mathcal M_\mathcal{W}(C, SL(2,\mathbb C)) \label{eq:nonab-def}
\end{equation}
that in some sense reverses this entire procedure \cite{Gaiotto:2012rg, Hollands:2013qza}.\footnote{Let us emphasize that $\mathcal{W}$-nonabelianisation is the inverse of $\mathcal{W}$-abelianisation for those types of spectral networks $\mathcal{W}$ for which the latter map is 1-1. This is true for rank 2 spectral networks of Fenchel-Nielsen or Fock-Goncharov type \cite{Hollands:2013qza}, but not true in general (see \cite{Hollands:2019wbr} for an example). In general, $\mathcal{W}$-abelianisation and $\mathcal{W}$-nonabelianisation should be treated as self-standing procedures.} The map $\psi_{\mathcal W}^\textrm{nab}$ can be used to compute the parallel transport of a flat $SL(2, \mathbb{C})$-connection $\nabla$ (in its image) along an open path $\mathfrak{p} \subset C$ in terms of the abelian parallel transport of its pre-image $\nabla^\text{ab}$ along a collection of paths $\mathfrak{p}^{ij}$ on the cover $\Sigma$. This collection consists of \emph{direct lifts} of the path $\mathfrak{p}$ to the cover, as well as \emph{detour paths} that are allowed to take detours along the lifts of the $ij$-trajectories of the network~$\mathcal W$ to the cover $\Sigma$. The map $\psi_{\mathcal W}^\textrm{nab}$ is known to be a bijection for spectral networks $\mathcal{W}$ of Fock-Goncharov and Fenchel-Nielsen type \cite{Hollands:2013qza}. Non-abelian parallel transport can then be described by a map
\begin{equation}\label{eq:non-ab-par-transport}
 P^\text{nab}_\mathcal{W}: M_\mathcal{W}(C, SL(2,\mathbb C)) \to \mathrm{Mat}_{2 \times 2}\big( \mathcal M^\textrm{al}(\Sigma, GL(1, \C))\big)/\mathcal{G}_\text{ab},
\end{equation}
where abelian gauge transformations in $\mathcal{G}_\text{ab}$ act on end-points of open paths $\mathfrak{p}$ by multiplication by a diagonal $SL(2,\mathbb{C})$-matrix. 

\begin{figure}[h]
   \centering
  \includegraphics[width=0.22\linewidth]{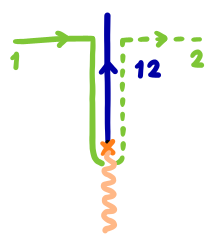}
    \caption{Example of a detour path (in light green), starting on sheet 1 and ending on sheet~2 of the covering $\Sigma$. }
    \label{fig:detour-path}
\end{figure}

Consider, for example, an open path $\mathfrak{p}$ on $C$ that crosses an $ij$-trajectory of the spectral network $\mathcal{W}$. Then the collection of lifts of $\mathfrak{p}$ to the cover contains the two canonical lifts of $\mathfrak{p}$ to each sheet of the covering, i.e.~the direct lifts, as well as precisely one detour path that runs from sheet $i$ to sheet $j$ along the $ij$-trajectory, as illustrated in \cref{fig:detour-path}.  
That is, suppose that the path~$\mathfrak p$ crosses the $ij-$trajectory at a point $w \in C$. Then the detour path starts on sheet $i$ until it hits the $ij-$trajectory, when it follows the trajectory from the point $w^{(i)}$ on sheet $i$ to the point $w^{(j)}$ on sheet $j$, encircles the branch-point, and then continues on sheet~$j$. 

Note that the sheet labellings at the start and end of the detour path are chosen in such a way that the abelian parallel transport $a_{ij}$ along the detour path $\delta_{ij}$ is exponentially small in the WKB approximation~\eqref{eq:WKBapprox}. 
Indeed, since $e^{-i \vartheta} (\lambda_i-\lambda_j) \ge 0$ along the $ij$-trajectory, the abelian holonomy $a_{ij}$ is exponentially small in the limit $\epsilon \to 0$ with the above conventions.

The non-abelian parallel transport $P(\mathfrak{p})$ of the flat connection $\nabla$ along the base path $\mathfrak{p}$ is then simply the sum of the abelian parallel transports of the almost-flat connection $\nabla^\text{ab}$ along the collection of lifts to $\Sigma$. In particular, note that the monodromy along a closed path on $C$ can receive contributions from open paths on the cover $\Sigma$.

\subsection{Gauging the 3-punctured sphere}\label{subsubsec:gaugedsphere}

As we have explained above, an FN network is isotopic to one of the two FN molecules when it is restricted to a pair of pants. Yet, the moduli space of flat connections restricted to any pair of pants (when it is embedded in a larger surface $C$ and thus contains fewer than three regular punctures as its boundary components) is different from the moduli space of flat connections on the 3-punctured sphere (where we fix the conjugacy class of the monodromies around the punctures). Indeed, the complex dimension of the moduli spaces of flat connections of the larger Riemann surface $C$ increases by two, and not just one, for every additional internal pants cycle. Whereas the conjugacy class of the monodromy around an internal pants cycle parametrises one complex degree of freedom, each internal pants cycle is also associated with an additional complex degree of freedom.  

\begin{figure}[h]
   \centering
  \includegraphics[width=0.35\linewidth]{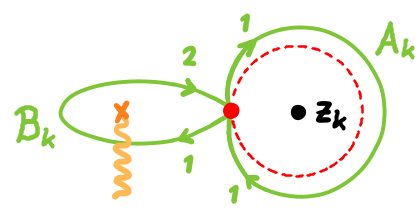}
    \caption{When gauging a puncture we cut out a small open disc around the puncture and fix the trivialisation of the bundle $E$ at a marked point on the boundary of the open disc. As a result, the abelian holonomies along the paths $A_k$ and $B_k$ on $\Sigma$ turn into new spectral coordinates. }
    \label{fig:gaugedpuncture}
\end{figure}

We can replicate these additional degrees of freedom on the pair of pants by cutting out a small open disc around each puncture of the 3-punctured sphere and fixing a trivialisation of the rank 2 bundle $E$ at a marked point on each new boundary component. 
We refer to the above operation as \emph{gauging} the puncture.\footnote{The nomenclature "gauging" originates from the AGT correspondence \cite{Alday:2009fs}, in which vertex operators $V_\alpha$ correspond to flavour symmetries in the dual $\mathcal{N}=2$ theory, and where summing over the complete basis of descendants $\sum_Y \mathcal L_{Y} V_{\alpha}$ corresponds to gauging this flavour symmetry (see \cite{Hollands:2011zc} for more details). } From the abelianisation perspective, gauging a puncture on $C$ introduces two additional spectral coordinates on $C$: the abelian holonomy (which we refer to as $M_k$) along a lift $A_k$ of the loop $\underline{\mathfrak{a}}_k$ around the puncture, and the abelian holonomy (which we refer to as $T_k$) along a dual open path $B_k$ on the cover~$\Sigma$ that connects the two lifts of the marked point. This is illustrated in fig.~\ref{fig:gaugedpuncture}. To be precise, we choose the lift $A_k$ such that $|M_k|<1$ and the lift $B_k$ such that the intersection product $A_k \cdot B_k =1$. This implies that $\log M_k$ is a complexified FN length coordinate. Similarly, $\log T_k$ is a complexified FN twist coordinate \cite[\S8.4]{Hollands:2013qza}. 

Note that we can phrase the choice of FN length coordinate in terms of the $\mathcal{W}$-framing as follows. If the $\mathcal{W}$-framing is chosen such that the small eigenvalue $|M_k|<1$ (versus the large eigenvalue $|M_k|>1$) is assigned to the orientation of the loop $\mathfrak{a}_k$, and if the orientation of $\mathfrak{a}_k$ agrees with the orientation of the $ij$-trajectories in this ring domain, then we define the exponentiated FN length coordinate as the abelian holonomy along the lift~$A_k$ of the loop~$\mathfrak{a}_k$ to the $j$th sheet (versus the $i$th sheet). 

Vice versa, if we instead decide to fix the lift~$A_k$ once and for all, we have to specify the $\mathcal{W}$-framing such that $|M_k|<1$. This choice of $\mathcal{W}$-framing is called the  \emph{WKB framing} in \cite{Hollands:2021itj}, since in the setting of $\epsilon$-connections and the WKB approximation~\eqref{eq:WKBapprox}, it implies that the exponentiated FN length coordinate is exponentially small in the limit $\epsilon \to 0$. 

We name the object that is obtained after gauging all three punctures of the 3-punctured sphere a \emph{gauged 3-punctured sphere}, or gauged 3-sphere in short. Note that the gauged 3-sphere is equivalent to the \emph{gauged pair of pants}. If the conjugation class of the monodromy along one boundary component (of a gauged 3-sphere) is the same as along another boundary component (of possibly a different gauged 3-sphere), these boundary components can be glued by identifying the fixed trivialisations at their marked points. Gluing of gauged 3-spheres can thus be defined on the level of moduli spaces of flat $SL(2,\mathbb{C})$-connections. 

Moreover, it can be shown that this gluing operation commutes with $\mathcal{W}$-abelianisation for spectral networks of FN type \cite{Hollands:2013qza}. The reason why this works is that none of the trajectories of the FN network cross the new pants cycle. Abelianisation thus brings the monodromies along the new pants cycle into diagonal form.

\paragraph{Nonabelianising Fenchel-Nielsen molecules}\label{sec:nonab-pants}

\begin{figure}[h]
   \centering
  \includegraphics[width=0.55\linewidth]{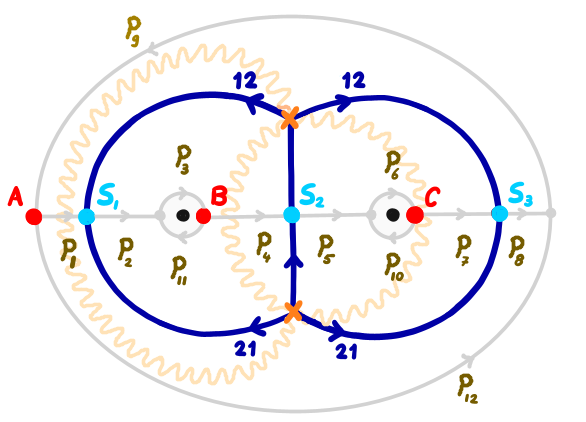}
    \caption{Choice of a trivialisation of $\Sigma$ (in light orange) and a path groupoid (in light grey) with respect to the Fenchel-Nielsen type molecule I. We gauge each puncture by fixing a trivialisation at the marked points labelled by $A$, $B$ and $C$ (in red). The S-matrices (in light blue), attributed to each crossing with a double trajectory (in dark blue) are labelled by $S_1$, $S_2$ and $S_3$.}
    \label{fig:molecule1-nonab}
\end{figure}

In order to make concrete the above discussion,  we apply the nonabelianisation method (as formulated in \cite[\S7]{Hollands:2013qza}\footnote{The only difference is a change in conventions: in line with the Moore-Seiberg computations later, we write $\mathfrak{p}^{ij}$ for a path that starts at sheet $i$ and ends at sheet $j$ and thus adopt the left-to-right convention for path composition. 
}) to FN molecule I (as illustrated in fig.~\ref{fig:molecule1-nonab}) on the gauged 3-punctured sphere.

Consider the path groupoid in fig.~\ref{fig:molecule1-nonab} relative to molecule I, where we fix a trivialisation of the bundle $E$ at the three marked points (in dark green) to the left of the punctures. This implies that the abelian connection on the covering $\Sigma$ is completely fixed in terms of six spectral coordinates: the eigenvalues $M_{1,2,3}$ of the abelian holonomy around the punctures, and the abelian parallel transport coefficients $T_{a,b,c}$ along lifts of the paths  
\begin{align}\label{eq:pathspABC}
\mathfrak{p}_a &= \mathfrak{p}_1 \, \mathfrak{p}_2 \, \mathfrak{p}_3, \quad
\mathfrak{p}_b = \mathfrak{p}_4 \, \mathfrak{p}_5 \, \mathfrak{p}_6, \quad
\mathfrak{p}_c = \mathfrak{p}_7 \, \mathfrak{p}_8 \, \mathfrak{p}_9, 
\end{align}
to the covering $\Sigma$ (starting from sheet 2 and ending on sheet 1).

As we mentioned previously, the nonabelianisation method cannot be applied to FN-type networks precisely at the critical phase. Therefore, the best we can do is apply the nonabelianisation method to their resolutions and take an average. The families of 12 and 21-trajectories in the resolutions are then encoded in the so-called S-matrices, which describe the unipotent gauge transformations across the saddle trajectories in the given resolution.

Since we want to relate the resulting $SL(2,\mathbb{C})$-connections for both resolutions, rather than starting with equivalent abelian connection on $\Sigma$, we introduce the notation $P^\pm$ for the abelian parallel transports in either resolution $\pm$. Fix the (clockwise for punctures 1 and 2, and anti-clockwise for puncture 3) abelian holonomies around the punctures to be of the form
\begin{align}\label{eq:localmonodromy}
    \underline{M}_1 = \begin{pmatrix} M_1 & 0 \\ 0 & \frac{1}{M_1} \end{pmatrix}, ~~\underline{M}_2 = \begin{pmatrix} M_2  & 0 \\ 0 & \frac{1}{M_2} \end{pmatrix},~~\underline{M}_3 = \begin{pmatrix} M_3  & 0 \\ 0 & \frac{1}{M_3} \end{pmatrix},
\end{align}
with conventions such that $|M_k|<1$ (i.e.~such that $M_k$ is exponentially small in the WKB approximation).\footnote{Note that swapping the labels $1 \leftrightarrow 2$ in the molecule inverts the WKB masses $M_{1,2,3}$.}
Then we find that the non-abelian parallel transport $P^{\mathrm{FN}}_a$ along the path~$\mathfrak{p}_{a}$ can be brought in the form 
\begin{align}
\begin{aligned}
\label{eq:PTAgauged3-puncturedsphere}
P_a^{\textrm{FN},-} &=   \begin{pmatrix} 0 & -1 \\  1 & 0 \end{pmatrix}  \begin{pmatrix}  1 &   \frac{ ( M_1^{-1} M_2 -  M_3^{-1} )}{ (M_3^{-1}-M_3) } \frac{T_a^- T_c^-}{T_b^- } \\ 0 & 1 \end{pmatrix}  \begin{pmatrix}  1 & 0 \\ \frac{( M_2^{-1} M_3 - M_1) }{(M_1^{-1} - M_1) } \frac{T_b^- }{T_a^- T_c^-}  & 1 \end{pmatrix} \begin{pmatrix}  T^-_a & 0 \\  0 & \frac{1}{T^-_a} \end{pmatrix} ,
\end{aligned}
\end{align}
in the American resolution, and 
\begin{align}
\begin{aligned}
\label{eq:PTBgauged3-puncturedsphere}
P_a^{\textrm{FN},+} &=   \begin{pmatrix}  0 & -1 \\  1 & 0 \end{pmatrix}  \begin{pmatrix}  1 &  0 \\  \frac{ (M_1 M_2^{-1} - M_3) }{(M_3^{-1}-M_3) } \frac{T_b^+ } {T_a^+ T_c^+ }  & 1 \end{pmatrix} \begin{pmatrix}  1 & \frac{ (M_2 M_3^{-1}- M_1^{-1})}{ (M_1^{-1}-M_1)  } \frac{ T_a^+ T_c^+ }{T_b^+ } \\  0 & 1 \end{pmatrix}  \begin{pmatrix} T^+_a & 0 \\  0 & \frac{1}{T^+_a} \end{pmatrix}   , 
\end{aligned}
\end{align}
in the British resolution, whereas the remaining parallel transports $P_{b,c}^{\textrm{FN},\pm}$ can be found by cyclically permuting the indices.

\begin{figure}[h]
   \centering
  \includegraphics[width=0.85\linewidth]{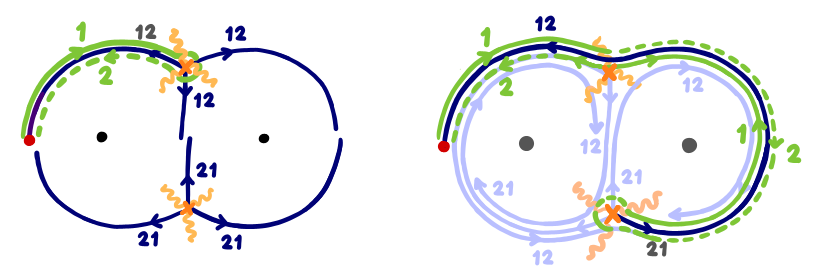}
    \caption{From left to right: spectral network $\mathcal{W}_\textrm{FN}^-$ (in dark blue) at increasing values of the (real) mass filtration parameter $\mu= e^{-i \vartheta} \int_w \sqrt{\phi_2} $ \cite{Gaiotto:2009hg}. Lightest detour path (in light green) on the left, and next-to-lightest detour path (in light green) on the right. The two infinite families of detour paths described in the text appear when increasing $\mu$ further. }
    \label{fig:detour-paths-mol1}
\end{figure}

The second and third matrices in each of the above equations are S-matrices associated to crossing a family of either 12 or 21-trajectories in the network. These families are encoded in the off-diagonal element of the S-matrix through the abelian parallel transport along their associated detour paths. For instance, the first off-diagonal element in \eqref{eq:PTAgauged3-puncturedsphere} can be formally expanded as\footnote{This expansion is well-defined in the WKB approximation, when $M_k$ is exponentially small.}
\begin{align}
 - \frac{ T_a^- T_c^-}{  T_b^- } \frac{ (1- M_1^{-1} M_2 M_3)}{ (1-M_3^2) } =  - \frac{ T_a^- T_c^-}{  T_b^- } \left(  \sum_{k >0 } (M_3^2)^{k}  -  M_1^{-1} M_2 M_3 \sum_{k \ge 0} (M_3^2)^{k} \right),
\end{align}
where the $k$th contribution to the first sum can be matched with the abelian parallel transport along a detour path that starts from the base-point on the double trajectory labelled by $S_1$ and winds $k$ times in the clock-wise direction around the molecule before encircling the upper branch-point, whereas the $k$th contribution to the second sum can be matched with the abelian parallel transport along a detour path that starts from the same base-point and winds $k$ times in the clock-wise direction around the molecule before encircling the lower branch-point. This is illustrated in \cref{fig:detour-paths-mol1}. (See the discussion around fig.~36 in \cite{Hollands:2013qza} for more details.)

We find the relation between the spectral coordinates $T^\pm$ by comparing the traces of the $P^{\textrm{FN},\pm}$'s. This yields
\begin{align}
\begin{aligned}\label{eq:Tabcrelation}
\frac{T^+_a}{T_b^+} &=  \frac{ \left(M_2 - M_2^{-1} \right) (M_2 - M_1 M_3)}{\left(M_3 - M_3^{-1} \right) (M_3 - M_1 M_2)} \, \frac{T^-_a}{T^-_b} \quad \textrm{and} \\
\frac{T^+_b}{T_c^+}&=   \frac{ \left(M_3 - M_3^{-1} \right) (M_3 - M_1 M_2)}{ \left(M_1 - M_1^{-1} \right) (M_1 - M_2 M_3)}  \,\frac{T^-_b}{T_c^-}.
\end{aligned}
\end{align}
In terms of the \emph{average} twist coordinates
\begin{align}
T_{a,b,c} = \sqrt{T^+_{a,b,c} T^-_{a,b,c}}\,
\end{align}
we thus have $\text{Tr} \, P^{\textrm{FN},+}_{a}[T^+_{a,b,c}(T_{a,b,c})] =  \text{Tr} \, P^{\textrm{FN},-}_{a}[T^-_{a,b,c}(T_{a,b,c})]$.

The relations~\eqref{eq:Tabcrelation} only determine the average coordinates $T_{a,b,c}$ up to an overall normalisation. This freedom can be fixed to equate the  matrices $P^{\textrm{FN},\pm}_{a}[T^\pm_{a,b,c}(T_{a,b,c})]$ themselves. This results in the average parallel transport matrix
\begin{align}\label{eq:parallel-transport-FN}
\begin{aligned}
P^\text{FN}_a= P^\textrm{FN} \begin{bmatrix} M_3 & M_1 & M_2 \\ T_a & T_b & T_c \end{bmatrix} =  \begin{pmatrix}  0 & 1 \\  -1 & 0 \end{pmatrix} \begin{pmatrix}  \gamma & \alpha \, \frac{T_a T_c}{T_b} \\ \beta \, \frac{T_b}{T_a T_c}& \gamma \end{pmatrix} \begin{pmatrix} T_a & 0 \\  0 & \frac{1}{T_a} \end{pmatrix},
   \end{aligned}
\end{align}
where
\begin{align}
\begin{aligned}\label{eq:gammaA} 
     \gamma^2 := \left( \frac{T_a^+}{T_a} \right) 
     & =  1+  \frac{ ( M_1^{-1}  M_2 - M_3^{-1}) ( M_2^{-1} M_3  - M_1  )} { (M_1-M_1^{-1}) (M_3-M_3^{-1}) } \\
     & =   \frac{(M_1 M_2-M_3^{-1})(M_2^{-1} M_3  - M_1^{-1})}{(M_1-M_1^{-1}) (M_3-M_3^{-1})},
\end{aligned}
\end{align}
and
\begin{align}
\begin{aligned}
\label{eq:alphabetaA}
    \beta =  - \frac{M_1 M_3}{M_2} \, \alpha &= \sqrt{ \frac{ ( M_1 M_2^{-1}  - M_3) ( M_2^{-1} M_3  - M_1  )} { (M_1-M_1^{-1}) (M_3-M_3^{-1})} }, \\
    \alpha \beta & =   \frac{ ( M_1^{-1} M_2  - M_3^{-1}) ( M_2^{-1} M_3  - M_1  )} { (M_1-M_1^{-1}) (M_3-M_3^{-1})} .
\end{aligned}
\end{align}
We refer to the matrix 
\begin{align}\label{eq:average-S-matrix}
    S^\textrm{FN}_\mathrm{av} \begin{bmatrix} M_3 & M_1 & M_2 \\ T_a & T_b & T_c \end{bmatrix} := \begin{pmatrix}  \gamma & \alpha \, \frac{T_a T_c}{T_b} \\ \beta\, \frac{T_b} {T_a T_c} & \gamma \end{pmatrix} 
\end{align}
as the \emph{average} S-matrix. Note that the S-matrix is of the same form in the $\mathcal{W}$-abelianisation formalism (see for instance eq.~(2.12) in \cite{Hollands:2019wbr}).

We observe that the average S-matrix~\eqref{eq:average-S-matrix} may just as well be obtained as the entry-wise average of the S-matrices in the American and British resolutions.  Geometrically, this implies that the total abelian parallel transport on each sheet at the FN phase, i.e.
\begin{equation}
    \big[S^\FN_\textrm{av}\big]_{ij} = \sqrt{\big[ S^+\big]_{ij} \big[ S^-\big]_{ij}}, \label{eq:Smatrix-average}
\end{equation}
is the geometric mean of the abelian parallel transports in either resolution. This will inspire our definition for the quantum FN S-matrix in \S\ref{sec:q-FN-nonab}.

The same holds for the matrices $P^{\textrm{FN},\pm}_{b}[T^+_{a,b,c}(T_{a,b,c})] $ and $P^{\textrm{FN},\pm}_{c}[T^-_{a,b,c}(T_{a,b,c})]$, which can be brought in the form 
\begin{align}
P^\textrm{FN}_b = P^\textrm{FN} \begin{bmatrix} M_1 & M_2 & M_3 \\ T_b & T_c & T_a \end{bmatrix} \quad \mathrm{and} \quad P^\textrm{FN}_c = P^\textrm{FN} \begin{bmatrix} M_2 & M_3 & M_1 \\ T_c & T_a & T_b \end{bmatrix}, \label{eq:parallel-transport-FN-2}
\end{align}
respectively.

Since we have fixed an ($SL(2,\mathbb{C})$-equivariant) trivialisation at all ends of the paths $\mathfrak{p}_{a,b,c}$, the entries of the matrices~$P_{a,b,c}^\text{FN}$ are invariant under the remaining abelian (or rather equivariant) gauge transformations. Yet, under a change of trivialisation at the end-points, the parallel transports~$P_{a,b,c}^\text{FN}$ transform as 
\begin{align}
    \widetilde{P}_{a/b/c}^\text{FN}  = G_k^{-1} \, P_{a/b/c}^\text{FN} \, G_l,
\end{align}
where $G_{k/l} = \text{diag}(g_{k/l}, g^{-1}_{k/l})$. The resulting parallel transports $\widetilde{P}^\text{FN}$ can still be written in the form \eqref{eq:parallel-transport-FN}, but now in terms of the new twist coordinates 
\begin{align}
\widetilde{T}_{a/b/c} =   T_{a/b/c} \, g_k \,  g_l.
\end{align}
Therefore, only the combinations
\begin{align}
\gamma^2 ~~\text{and} ~~ \alpha \beta, \label{eq:invariants-s-matrix}
\end{align}
with $\gamma^2 - \alpha \beta = 1$, remain invariant under changes of trivialisation. These remarks will be important when "ungauging" the punctures, for instance, when gluing boundary components.

Independently, note that the inverse of $P^\text{FN}$ can be brought in the form 
\begin{align}
\begin{aligned}
 \left( P^\textrm{FN} \right)^{-1} \begin{bmatrix} M_3 & M_1 & M_2 \\ T_a & T_b & T_c \end{bmatrix} =   \begin{pmatrix}  0 & -1 \\  1 & 0 \end{pmatrix} \begin{pmatrix}  \gamma & \beta \, \frac{T_a T_b}{T_c} \\ \alpha \, \frac{T_c}{T_a T_b} & \gamma \end{pmatrix} \begin{pmatrix} T_a & 0 \\  0 & \frac{1}{T_a} \end{pmatrix}, 
   \end{aligned}
\end{align}
and thus has the same equivariant invariants $\gamma^2$ and $\alpha \beta$.

It is also useful to know that the traces of the matrices $P^{\textrm{FN},\pm}_{a,b,c}[T^\pm_{a,b,c}(T_{a,b,c})]$ are equal to
\begin{align}
\begin{aligned}
  \mathrm{Tr}\, P^\textrm{FN}_a &= \sqrt{\frac{(M_1 M_2-M_3) (M_2 M_3-M_1)}{M_1 M_3 (1-M_1^2) (1-M_3^2)}} \left( \frac{T_c}{T_b} +  \frac{M_1 M_3}{M_2} \frac{T_b}{T_c} \right) , \\
   \mathrm{Tr} \,P^\textrm{FN}_b &= \sqrt{\frac{(M_2 M_3-M_1) (M_3 M_1-M_2)}{M_2 M_1 (1-M_2^2) (1-M_1^2)}} \left( \frac{T_a}{T_c} +  \frac{M_2 M_1}{M_3} \frac{T_c}{T_a} \right) ,\\
    \mathrm{Tr} \,P^\textrm{FN}_c &= \sqrt{\frac{(M_3 M_1-M_2) (M_1 M_2-M_3)}{M_3 M_2 (1-M_3^2) (1-M_2^2)}} \left( \frac{T_b}{T_a} +  \frac{M_3 M_2}{M_1} \frac{T_a}{T_b} \right).
    \label{eq:trP}
\end{aligned}
\end{align}

Finally, a basis of paths on the gauged 3-sphere is given by the paths $\mathfrak{p}_{a,b,c}$ together with the paths
\begin{align}\label{eq:pathspprime}
\mathfrak{p}'_a &=  \mathfrak{p}_1 \, \mathfrak{p}_2 \,\mathfrak{p}^{-1}_{11}, \quad
\mathfrak{p}'_b = \mathfrak{p}_4 \, \mathfrak{p}_5 \,\mathfrak{p}^{-1}_{10} , \quad
\mathfrak{p}'_c =  \mathfrak{p}_7 \, \mathfrak{p}_8 \,\mathfrak{p}_{12}^{-1}. 
\end{align}
The non-abelian parallel transport along the $\mathfrak{p}'_{a,b,c}$ in the resolution $\pm$ is simply given by 
\begin{align}\label{eq:parallel-transport-FN-prime}
    P'^{\textrm{FN},\pm}_{a,b,c}  =  P^{\textrm{FN},\pm}_{a,b,c} \, \underline{M}_{1,2,3}^{-1},
\end{align}
respectively.

\subsection{Gluing through nonabelianisation}\label{sec:gluing-nonab}

Also, $\mathcal{W}$-nonabelianisation for an FN network $\mathcal{W}_\mathrm{FN}$  (defined with respect to a pants decomposition~$\mathcal{P}$) commutes with dissecting the surface~$C$ into pairs of pants (in the pants decomposition $\mathcal{P}$). Just as before, this $\mathcal{W}$-nonabelianisation is defined as the average of both resolutions $\mathcal{W}_\mathrm{FN}^\pm$. 

The reader might be concerned about what happens to the additional incoming trajectories into each pair of pants at phase $\vartheta_\textrm{FN} \pm \eta$, with $\eta>0$ finite, in this process. (Recall that fig.~\ref{fig:FN-4pt-sphere-trajectories-molI.png} illustrates such trajectories in the 4-punctured sphere example.) It turns out that these additional trajectories decouple in the limit $\eta \to 0^+$, so that the S-matrices for the resolutions $\mathcal{W}^\pm_{\textrm{FN}}$ reproduce the S-matrices on the gauged 3-sphere, computed in \S\ref{sec:nonab-pants}. We have included the precise argument in Appendix~\ref{appendix-nonab}. 

As a concrete example, we compute the nonabelian parallel transport matrices on the 4-punctured sphere.

\subsubsection{Example: 4-punctured sphere}

Consider the 4-punctured sphere together with the FN-type spectral network in \cref{fig:FN-4pt-sphere-example}. In the following we detail how to construct the monodromy representation on this sphere by gluing two gauged 3-spheres and then "ungauging" the 4 remaining punctures. Note that the moduli space of flat $SL(2,\mathbb{C})$-connections on the gauged 4-punctured sphere is complex 10-dimensional, whereas the moduli space of flat $SL(2,\mathbb{C})$-connections on the (ungauged) 4-punctured sphere is complex 2-dimensional.

\begin{figure}[h!]
   \centering
  \includegraphics[width=0.65\linewidth]{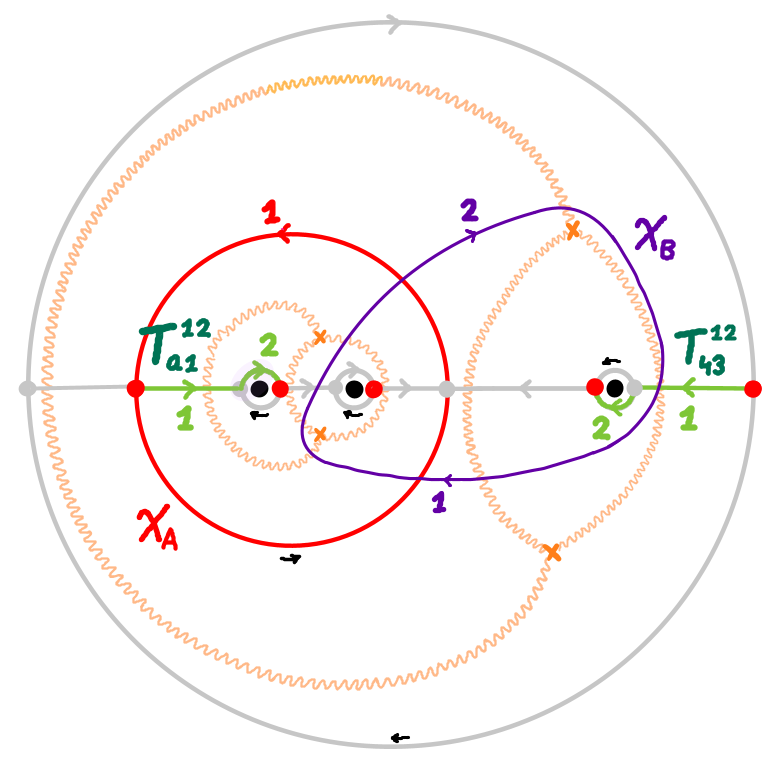}
    \caption{ Choice of path groupoid (in light grey and red) on 4-punctured sphere. The little black arrows indicate the orientations of 1-cycles around punctures and pants curve.
    $\mathcal{X}_A$ and $\mathcal{X}_B$ are the abelian parallel transports along the A-cycle (in red) and the B-cycle (in purple) on $\Sigma$, respectively. $T^{ij}_{kl}$ is the abelian parallel transport along an open path $\mathfrak{p}^{ij}_{kl} \subset \Sigma$ which runs from the lift of the marked point for puncture $k$ to sheet $i$ to the lift of the marked point for puncture $l$ to sheet $j$. Examples are the light green paths on the left and right.}
    \label{fig:FN-4pt-sphere-path-groupoid}
\end{figure}

Consider the path groupoid as illustrated in \cref{fig:FN-4pt-sphere-path-groupoid}. Nonabelianisation with respect to an FN network implies that the resulting $SL(2,\mathbb{C})$-monodromies around the four punctures~$z_k$, as well as the pants-curve~$\mathfrak{a}$, are diagonal. We introduce parameters $M_k$ and $A$ such that these monodromies equal
\begin{align}
    \underline{M}_k = \begin{pmatrix} M_k & 0  \\ 0 & M_k^{-1} \end{pmatrix} \quad \textrm{and} \quad  M_\mathfrak{a} = \begin{pmatrix} A  & 0  \\ 0 &  A^{-1} \end{pmatrix},
\end{align}
respectively, in the directions indicated by little black arrows in \cref{fig:FN-4pt-sphere-path-groupoid}. The gauged 3-spheres admit an additional six dual spectral coordinates $T_{L/R,\, a/b/c}$, which we will denote here as 
\begin{align}
    \{ T^{ij}_{kl} : \, k,l \in \{\mathfrak{a},1,\ldots,4\},\, ij \in \{12,21\}, \, T^{ij}_{kl} = -1/T^{ji}_{kl} = 1/T^{ji}_{lk} \},
\end{align}
where $T^{ij}_{kl}$ measures the abelian parallel transport along an open path $\mathfrak{p}^{ij}_{kl} \subset \Sigma$ which runs from the lift of the marked point for puncture $k$ to sheet $i$ to the lift of the marked point for puncture $l$ to sheet $j$. (The marked points are the red dots in \cref{fig:FN-4pt-sphere-path-groupoid}.)

Gluing the gauged 3-spheres reinstates the abelian gauge transformations at the marked point on the pants cycle. 
Parallel transport on the resulting gauged 4-punctured sphere can be described in terms of the four masses $M_k$, the exponentiated FN length coordinate $\mathcal{X}_A = A$, and the five dual gauge-invariant coordinates
\begin{align}
    \{ T^{12}_{12}, \, T^{12}_{34}, \, T^{12}_{1\mathfrak{a}} \, T^{21}_{\mathfrak{a}3}, \, T^{12}_{2\mathfrak{a}} \, T^{21}_{\mathfrak{a}3}, \, T^{12}_{1\mathfrak{a}} \, T^{21}_{\mathfrak{a}4} \}.
\end{align}
Nonabelianisation thus tells us that the $SL(2,\mathbb{C})$ monodromy representations on the gauged 4-punctured sphere is indeed parametrised by 10 independent spectral coordinates. 

If we "ungauge" the 4 punctures $z_k$, the resulting $SL(2,\mathbb{C})$ monodromy representations can be solely expressed in terms of the two left-over spectral coordinates, which may be parametrised as 
\begin{align}
\mathcal{X}_A = A \quad \mathrm{and} \quad \mathcal{X}_B = -M_1 M_4 \, \frac{T_{12}}{T_{\mathfrak{a}1} \, T_{2\mathfrak{a}}} \, \frac{T_{34}}{T_{\mathfrak{a}3} \, T_{4\mathfrak{a}}},
\end{align}
where all upper labels of $T$'s are chosen as $ij=12$. These expressions match with the abelian parallel transport along the 1-cycles $A$ and $B$ on the cover $\Sigma$ illustrated as well in \cref{fig:FN-4pt-sphere-path-groupoid}. The masses $M_k$ have now turned into "frozen" parameters and the additional $T$'s into gauge-dependent degrees of freedom. 

Let us go back to the definition of the twist flow in \cite[\S8.4]{Hollands:2013qza} to see that $\mathcal{X}_B$ defines an FN twist coordinate. Suppose that we cut open the 4-punctured sphere along the pants curve. This fixes trivialisations at the marked points of both copies of the pants curve. Suppose that we now change the trivialisation (just) on the left pair of pants, using the transformation $G = \text{diag}(g,g^{-1})$, and then glue together the two pair of pants again. This operation defines the twist flow. Note that under the twist flow $T_{\mathfrak{a}1}  \mapsto g \, T_{\mathfrak{a}1} $ and $T_{2 \mathfrak{a}} \mapsto g \, T_{2 \mathfrak{a}}$, so that  
\begin{align}\label{eqn:FNtwist}
    \mathcal{X}_B \mapsto g^{-2} \mathcal{X}_B.
\end{align}
The transformation above implies that $\mathcal{X}_B$ is indeed an exponentiated (complexified) FN twist coordinate. (Recall from \cite[\S8.4]{Hollands:2013qza} that there is an ambiguity in the FN twist coordinate $\mathcal{X}_B$ parameterised by multiplying by a power of the FN length coordinate $\mathcal{X}_A$.)

The nonabelian parallel transport along any concatenation of paths on $C$ can be expressed as a product of nonabelian parallel transport matrices, similar to $P_{a/b/c}^\textrm{FN}$ and ${P'}_{a/b/c}^{\textrm{FN}}$ defined in eq.~\eqref{eq:parallel-transport-FN} and eq.~\eqref{eq:parallel-transport-FN-prime}, respectively. More precisely, a basis of nonabelian parallel transports on the gauged 4-punctured sphere is given by the matrices\footnote{In our nonabelianisation computation we moved the branch-cuts so that we first cross the branch-point and then the S-matrices, when moving along the open paths $\mathfrak{p}_{kl}$. } 
\begin{align}
    \begin{aligned}
P_{L,\mathfrak{a} 1}^{\textrm{FN}}  =  P_L^{\textrm{FN}} \begin{bmatrix} M_1 & M_2 & A \\ T_{12} & T_{2\mathfrak{a}} & T_{\mathfrak{a}1} \end{bmatrix}, 
& \quad P_{R,\mathfrak{a} 4}^{\textrm{FN}}  =  P_R^{\textrm{FN}} \begin{bmatrix} A & M_4 & M_3 \\ T_{4 \mathfrak{a}} & T_{34} & T_{\mathfrak{a} 3} \end{bmatrix}, \\
P_{L,12}^{\textrm{FN}}  =  P_L^{\textrm{FN}} \begin{bmatrix} M_2 & A & M_1 \\ T_{2\mathfrak{a}} & T_{\mathfrak{a} 1} & T_{12} \end{bmatrix}, 
& \quad P_{R,43}^{\textrm{FN}}  =  P_R^{\textrm{FN}} \begin{bmatrix} M_3 & A & M_4 \\ T_{\mathfrak{a}3} & T_{4 \mathfrak{a}} & T_{34} \end{bmatrix}, \\ 
P_{L,2\mathfrak{a}}^{\textrm{FN}}  =  P_L^{\textrm{FN}} \begin{bmatrix} A & M_1 & M_2 \\ T_{\mathfrak{a} 1} & T_{12} & T_{2\mathfrak{a}} \end{bmatrix},
& \quad P_{R,3\mathfrak{a}}^{\textrm{FN}}  =  P_R^{\textrm{FN}} \begin{bmatrix} M_4 & M_3 & A \\ T_{34} & T_{\mathfrak{a} 3} & T_{4 \mathfrak{a}} \end{bmatrix},
    \end{aligned}
\end{align}
as well as their inverses, where $P^\textrm{FN}_L = P^\textrm{FN}$ and $P_R^\textrm{FN}$ differs from $P_L^\textrm{FN}$ by inverting the diagonal matrix diag$(T, T^{-1})$ as well as the S-matrix $S^\textrm{FN}_\textrm{av}$, defined in \eqref{eq:average-S-matrix}. (This is because the orientation of the trajectories in the right molecule is opposite to that of the left molecule for the FN network. It might be possible to resolve the difference between $P^\textrm{FN}_L$ and $P^\textrm{FN}_R$ by adding an additional branch-cut along the pants cycle.) The form of the equivariant invariant~$\gamma^2$, as well as of course $\alpha \beta = \gamma - 1$, is the same for both $P^\text{FN}_L$ and $P^\text{FN}_R$.

The nonabelian parallel transport around the loop $\beta$, defined as the projection of the 1-cycle~$B$ to the surface $C$ (oriented in the direction of the little black arrow near the puncture at $z=1$), is for instance given by
\begin{align}
\begin{aligned}\label{eq:semi-classicalMbeta}
P^\text{FN}_\beta &= P_L^\textrm{FN} \begin{bmatrix} M_1 & M_2 & A \\ T_{12} & T_{2\mathfrak{a}} & T_{\mathfrak{a} 1} \end{bmatrix}  \, \underline{M}_1 \, \left( P_L^\textrm{FN} \begin{bmatrix} M_1 & M_2 & A \\ T_{12} & T_{2 \mathfrak{a}} & T_{\mathfrak{a} 1} \end{bmatrix}  \right)^{-1} \times \\
& \qquad  P_R^{\textrm{FN}} \begin{bmatrix} A & M_4 & M_3 \\ T_{4 \mathfrak{a}} & T_{34} & T_{\mathfrak{a} 3} \end{bmatrix}  \, \underline{M}_4^{-1} \, \left( P_R^{\textrm{FN}} \begin{bmatrix} A & M_4 & M_3 \\ T_{4 \mathfrak{a}} & T_{34} & T_{\mathfrak{a} 3} \end{bmatrix} \right)^{-1},
\end{aligned}
\end{align}
reproducing invariants already computed in \cite{Hollands:2013qza}.\\

Using the $\mathcal{W}$-nonabelianisation method, it is straight-forward to compute the monodromy representation of any surface $C$ (even with any type of punctures) in terms of FN-type spectral coordinates. 

Let us restrict ourselves here to a surface $C$ with only regular punctures, together with a pants decomposition $\mathcal{P}$. Consider the  FN spectral network whose restriction to each pair of pants is isotopic to molecule I. (Let us add branch-cuts around all pants cycles, so that the orientations of the saddle trajectories in each molecule are aligned.)

We start the $\mathcal{W}$-nonabelianisation procedure by fixing a path groupoid on $C$ that, when restricted to any pair of pants in $\mathcal{P}$, is equivalent (up to homotopy) to the one illustrated in \cref{fig:molecule1-nonab}. Any path $\mathfrak{p}$ in this groupoid can thus be decomposed as
\begin{align}
    \mathfrak{p} = \mathfrak{p}^\textrm{\maljapanese\char"4E}\mathfrak{p}^\textrm{\maljapanese\char"1A} \cdots \mathfrak{p}^\textrm{\maljapanese\char"4A}, 
\end{align}
where each $\mathfrak{p}^\textrm{\maljapanese\char"4D}$ is (a connected component of) the restriction of the path $\mathfrak{p}$ to a single pair of pants labelled by $\textrm{\maljapanese\char"4D}$. The non-abelian parallel transport along $\mathfrak{p}$ can be written in the form
\begin{align}\label{eq:nonab-P}
    P = P^\textrm{\maljapanese\char"4E}P^\textrm{\maljapanese\char"1A} \cdots P^\textrm{\maljapanese\char"4A},
\end{align}
where each $P^\textrm{\maljapanese\char"4D}$ is equal to a non-abelian parallel transport $P_{a/b/c}$ as computed in \S\ref{sec:nonab-pants}. 

After gluing the pants boundaries, the associated abelian parallel transports $T_{L/R}$ combine into abelian parallel transports $\mathcal{X}_\gamma$ along 1-cycles $\gamma$ on $\Sigma$, which cross the lifts of the pants cycles. If we choose a basis of A-cycles on $\Sigma$ consisting of lifts $A_p$ of each pants cycle $\mathfrak{a}_p$, a basis of B-cycles~$B_p$ on $\Sigma$ can be formed by additional closed 1-cycles~$\gamma$. Through nonabelianisation, the nonabelian parallel transports are then expressed in terms of the abelian holonomies 
\begin{align}
\mathcal{X}_{A_p}, \mathcal{X}^\pm_{B_p}, ~ \text{and}~ M_k,
\end{align}
 with $p$ labelling the pants cycles and $k$ the regular punctures on $C$. By a similar argument as presented around \cref{eqn:FNtwist}, the coordinates $\mathcal{X}_{A_p}$ and $\mathcal{X}^\pm_{B_p}$ have the characteristics of exponentiated (complexified) FN length and twist coordinates, respectively.

\section{Liouville CFT and quantum Fenchel-Nielsen coordinates}\label{sec:liouville-cft}

The goal of this section is to describe a quantisation of the Fenchel–Nielsen parallel transports derived in the previous section. We achieve this using the Moore-Seiberg transformations of conformal blocks in the framework of Liouville conformal field theory. By studying the transformation properties of the BPZ connection for Liouville theory, we are able to show that the Moore-Seiberg transformations quantise the Fenchel-Nielsen parallel transports.

In \S\ref{sec:basicsLiouville}, we provide the necessary background by introducing the Lax connection, conformal blocks, (degenerate) vertex operators, and the BPZ connection. In \S\ref{subsec:bpz-monodromies}, we summarise how to compute quantum parallel transports using the Moore-Seiberg groupoid, and how to bring the Moore-Seiberg (MS) parallel transport in the Fenchel-Nielsen gauge, relative to a chosen pants decomposition. In all this, we emphasize that results on the gauged three-punctured sphere can easily be generalised to an arbitrary (punctured) Riemann surface through a gluing construction. We furthermore show how the semi-classical conformal blocks are encoded in the evaluation of the Fenchel-Nielsen coordinates on the Lax connection, and how the Moore-Seiberg monodromies quantise this result. Finally, in \S\ref{sec:symmdefects}, we observe that these results lead to an interpretation of Fenchel-Nielsen type spectral networks as topological defects in Liouville theory.

\subsection{Basics of Liouville CFT}\label{sec:basicsLiouville}

Liouville theory is a quantum field theory described by a scalar field $\varphi(z,\bar{z})$ on a Riemann surface $C$ with an exponential potential. It is governed by the (renormalised) action  \cite{Alvarez-Gaume:1991joj},
\begin{equation}
    S = \frac 1 {4\pi} \int_C d^2 z\, \sqrt{\widehat{g}} \left(\widehat{g}^{\alpha \beta}\partial_\alpha \varphi \partial_\beta \varphi + Q \widehat{R} \varphi + 4\pi \mu e^{2b\varphi} \right), \label{eq:liouvilleaction}
\end{equation}
where $\widehat{g}$ is a constant-curvature background metric\footnote{Remember that, according to the uniformisation theorem, there is a unique such metric (up to diffeomorphisms) for every conformal class on $C$.}, the background charge
\begin{align}
    Q= b + \frac{1}{b},
\end{align}
measures the coupling of the quantum Liouville field $\varphi$ to the curvature~$\widehat{R}$, and $\mu$ is called the cosmological constant.\footnote{See, for example, \cite{Ribault:2016sla, Teschner:2017del} for very helpful lecture notes on Liouville theory and CFT's in general.} 

In the semi-classical limit $b \to 0$, 
the equations of motion of the Liouville theory describe constant-curvature deformations  
\begin{equation}
    g = e^{2 b \varphi} \widehat g
\end{equation}
of the background metric $\widehat{g}$. 
Another way to phrase this is to say that the Liouville equation of motion, in the semi-classical limit $b \to 0$, is equivalent to the rank-$2$ linear differential equation 
\begin{equation}
    (d + A)\Psi = 0, \label{eq:Lax}
\end{equation}
where the flat $SL(2, \R)$ connection~$A$ is known as the \emph{Lax connection} \cite{DHoker:1991yue}. Locally, the Lax equation~\eqref{eq:Lax} can be brought into the form \cite[Section 3]{Drukker:2009id}
\begin{equation}
    \left( \partial_z^2 + b^2 T_\textrm{cl}(z)\right) \psi(z) = 0 \label{eq:null-equation-cl},
\end{equation}
where $T_\textrm{cl}$ is the semi-classical energy-momentum tensor of the  Liouville theory. As we detail in \S\ref{subsec:bpz-monodromies}, the Lax connection transforms as an $SL(2, \mathbb{R})$-oper connection. 

At the quantum level, Liouville theory is a two-dimensional conformal field theory, whose symmetry algebra is the left-right Virasoro algebra $\textbf{Vir}_c \otimes \overline{\textbf{Vir}}_c$ with central charge 
\begin{equation}
    c = 1 + 6 Q^2.
\end{equation}
The quantum theory is invariant under the symmetry $b \leftrightarrow 1/b$ (when simultaneously rescaling the cosmological constant $\mu$), which is obviously broken in the semi-classical limit $b \to 0$.

The basic local operators in Liouville theory are the primary vertex operators $V_{\Delta_\alpha}(z, \overline z)$ which are characterised by their conformal weights $\Delta_\alpha$. In particular, $V_{\Delta_\alpha}(z,\overline{z})$ transforms as a $(\Delta_\alpha,\Delta_\alpha)$-form under holomorphic/anti-holomorphic changes of variables.  In what follows, it is useful to write the conformal weight as
\begin{equation}
    \Delta_\alpha = \alpha(Q-\alpha),
\end{equation}
where $\alpha$ is called the \emph{Liouville momentum} (or simply momentum), and label the primary vertex operators by their momentum instead, i.e~$V_\alpha$.

Through the state-operator correspondence
\begin{align}
   \lim_{z, \overline{z} \to 0} V_\alpha(z, \overline{z}) | 0 \rangle = | v_\alpha \rangle,
\end{align}
primary vertex operators correspond to highest weight states (or primary states) $v_\alpha \otimes \overline v_\alpha$ of left-right Virasoro modules $\cV_\alpha \otimes \overline{\cV}_\alpha$ with conformal dimension $(\Delta_\alpha, \Delta_\alpha)$. We denote the vertex operators that correspond to a generic state $v \otimes \overline{v} \in \cV_\alpha \otimes \overline{\cV}_\alpha$ by
\begin{align}\label{def:genericV}
V_\alpha(v \otimes \overline{v};z,\overline{z}).
\end{align}

For unitarity, it is essential that $c \in \mathbb{R}$ -- only then is the canonical inner product defined by $(L_n)^\dagger = L_{-n}$ positive-definite. This implies that $b$ is either real (i.e.~$c\ge25$) or purely imaginary (i.e.~$c\le1$). A Virasoro module $\cV_\alpha$ is unitary iff 
\begin{align}\label{eq:unitarity}
    \Delta_\alpha \ge \frac 1 4 Q^2 \quad \Longleftrightarrow \quad \alpha = \frac{Q}{2} + i\mathit{p} \in \frac Q 2 + i \R \quad \text{with} ~ \mathit{p} \in \mathbb{R},
\end{align}
and the theory has a continuous operator spectrum
\begin{equation}\label{eq:Liouville-spectrum}
    \cH = \int_{\alpha \in \frac{Q}{2} + i \R_+} d\alpha \, \cV_\alpha \otimes \overline \cV_\alpha.
\end{equation}

\begin{remark}
    The only unitary theories with $b \in i \mathbb{R}_+$, and thus central charge $c\le 1$, are part of the minimal model series which have $b=i \sqrt{p'/p}$ for coprime integers $p'>p$. These minimal models have a finite spectrum, listed in the famous Kac table. Nonetheless, extensions of Liouville theory to $c\le 1$ have been considered, for instance, in~\cite{McElgin:2007ak}.
\end{remark}

Although unitarity restricts the Liouville momentum to 
\(\alpha = \frac{Q}{2} + i \mathit{p} \), it is customary to analytically continue results to allow \(\alpha \in \mathbb{C}\). In particular, we will often consider Verma modules \(\mathcal{V}_\alpha\) for arbitrary complex momenta \(\alpha \in \mathbb{C}\), and will consider
complex-valued background charges \(Q\). Within this generalization, the Lax connection is naturally a flat $SL(2,\mathbb{C})$-connection.

Liouville theory was originally "solved" in \cite{Dorn:1994xn, Zamolodchikov:1995aa}, where it was proposed that the basic building blocks of Liouville theory, the three-point functions 
\begin{align}
\begin{aligned}
    &\mathcal{C}(\alpha_1,\alpha_2,\alpha_3) = \lim_{|z|\to\infty} |z|^{\Delta_{\alpha_3}} \, \langle V_{\alpha_1}(0) V_{\alpha_2}(1) V_{\alpha_3}(z) \rangle,  
\end{aligned}
\end{align}
have the analytic expression
\begin{align}
\begin{aligned}\label{eq:DOZZ}
    &\mathcal{C}(\alpha_1,\alpha_2,\alpha_3) = \left( \pi \mu \gamma(b^2) b^{2-2b^2} \right)^{(Q-\alpha_1-\alpha_2 - \alpha_3)/b} \times\\
    & \quad \times  \, \frac{\Upsilon_b'(0)\Upsilon_b(2 \alpha_1)\Upsilon_b(2 \alpha_2) \Upsilon_b(2 \alpha_3)}{\Upsilon_b(\alpha_1 + \alpha_2 + \alpha_3 -Q)\Upsilon_b(\alpha_1 + \alpha_2 - \alpha_3)\Upsilon_b(\alpha_1 - \alpha_2 + \alpha_3) \Upsilon_b(-\alpha_1 + \alpha_2 + \alpha_3)},
\end{aligned}
\end{align}
with 
\begin{align}\label{eq:ell-upsilon-gammab}
\gamma(x) = \frac{\Gamma(x)}{\Gamma(1-x)}    \quad \textrm{and} \quad \, \Upsilon_b(x) = \frac{1}{\Gamma_b(x)\Gamma_b(Q-x)}, 
\end{align}
in terms of the double Barnes functions $\Gamma_b(x) = \Gamma_{1/b}(x)$. Eq~\eqref{eq:DOZZ} is now known as the DOZZ formula \cite{Teschner:2001rv}. The DOZZ proposal was put on a much more solid basis in \cite{Teschner:1995yf}, and recently proved using probabilistic methods in \cite{kupiainen2019integrabilityliouvilletheoryproof}. 

\subsubsection{Conformal blocks}

The correlation functions of Liouville theory factorise into holomorphic (chiral) and anti-holomorphic (anti-chiral) contributions. More precisely, given a pair of pants decomposition of the surface $C$, and given that the OPE between two primary fields necessarily has the form
\begin{align}
\begin{aligned}
 &V_{\alpha_1}(z_1, \bar z_1) V_{\alpha_2}(z_2, \bar z_2) =\\
 & \qquad \int d\alpha \, \mathcal{C}(\alpha_1,\alpha_2,\alpha^*) \, |z_1-z_2|^{2 (\Delta_\alpha - \Delta_1-\Delta_2)} \, \left( V_{\alpha}(z_1, \bar{z}_1) + \mathcal{O}(|z_1-z_2|) \right),
 \end{aligned}
\end{align}
where $\alpha^* = Q-\alpha$ is the conjugate momentum, and where the subleading terms are corrections from descendant fields, it follows that any Liouville correlation function can be written as an integral over internal momenta through the pants cycles, with the integrand consisting of a product of three-point functions $\mathcal{C}$, one for each pair of pants, times a factorised product $\mathcal{F} \overline \cF$ over descendant contributions. 

The factors $\mathcal{F}$ and $\overline \cF$ are known as holomorphic and anti-holomorphic \emph{conformal blocks}, respectively. Whereas the three-point functions $\mathcal{C}$ encode the dynamics of Liouville theory, the conformal blocks $\mathcal{F}$ and $\overline{\mathcal{F}}$ are entirely determined by the Virasoro algebra.

For example, the four-point correlation function of primary vertex operators may be written as
\begin{equation}
\begin{aligned}\label{eq:Liouvillecorrelator}
   &  \langle V_{\alpha_1}(z_1, \bar z_1) V_{\alpha_2}(z_2, \bar z_2)V_{\alpha_3}(z_3, \bar z_3) V_{\alpha_4}(z_4, \bar z_4)\rangle = \\
   & \qquad \quad \int d\nu(\alpha) \cF_\alpha(z_1, z_2, z_3, z_4) \overline \cF_\alpha(\bar z_1, \bar z_2, \bar z_3, \bar z_4).
\end{aligned}
\end{equation}
where the measure
\begin{align}
    d \nu(\alpha) = \int \frac{d \alpha}{2 \pi} \, |x^{\Delta_\alpha - \Delta_{1} - \Delta_{2}}|^2 \, \mathcal{C}(\alpha_1,\alpha_2,\alpha^*) \, \mathcal{C}(\alpha, \alpha_3,\alpha_4),
\end{align}
contains a product of two 3-point functions. The conformal block $\mathcal{F}_\alpha$ is labelled by an internal momentum $\alpha$ and has an expansion of the form
\begin{align}\label{eq:4-pt-block-in-q}
\mathcal{F}_{\alpha}(z_1, z_2, z_3, z_4) =  1 +  \frac{(\Delta_\alpha - \Delta_1 + \Delta_2)(\Delta_\alpha + \Delta_3 - \Delta_4) }{2 \Delta_\alpha } \, x + \mathcal{O}(x^2)
\end{align}
in terms of the cross-ratio
\begin{align}
    x = \frac{(z_1-z_2)(z_3-z_4)}{(z_1-z_3)(z_2-z_4)}.
\end{align}

\begin{remark}\label{remark:1-loop}
The three-point function $\mathcal{C}(\alpha_1,\alpha_2, \alpha_3)$ may also be brought into a factorised form
\begin{align}
    \mathcal{C}(\alpha_1,\alpha_2, \alpha_3) = |\mathcal{C}_\text{hol}(\alpha_1,\alpha_2, \alpha_3) |^2,
\end{align}
although there is no canonical way of doing this. One possibility is to split each $\Upsilon_b$-factor as in eq.~\eqref{eq:ell-upsilon-gammab} \cite{Alday:2009aq}. Another possibility is to consider each $\Upsilon_b$ as the product~\cite{Teschner:2013tqy}
\begin{align}\label{eq:CFTsquareroot}
    \Upsilon_b(x) = \sqrt{\Upsilon_b(x) \Upsilon_b(Q-x)}.
\end{align}
We refer to the former normalisation as the topological string normalisation, and to the latter one as the CFT normalisation. In contrast to the \emph{Virasoro} conformal blocks~$\mathcal{F}^\text{Vir} = \mathcal{F}$ that are independent of the dynamics, we will define \emph{Liouville} conformal blocks as the combination 
\begin{align}\label{eqn:Liou-vs-Vir-blocks}
\mathcal{F}^\text{Liou} = \left(\prod \mathcal{C}_\text{hol} \right) \mathcal{F}^\text{Vir},
\end{align}
where the product runs over all pair of pants in the chosen pants decomposition of $C$.

\end{remark}

In the following, we restrict ourselves to holomorphic Virasoro blocks~$\mathcal{F}^\text{Vir}$, and drop the prefixes ``holomorphic'' and "Virasoro". Conformal blocks are typically defined on genus zero surfaces and extended to higher genus surfaces through a procedure known as gluing. We denote the space of conformal blocks on a surface $C$ by $\mathrm{CB}(C)$. 

On the Riemann sphere $\mathbb{P}^1$, with punctures at $z_1, \ldots, z_n$ labelled by Virasoro modules $\cV_{\alpha_1}, \ldots, \cV_{\alpha_n}$, a \emph{conformal block} is defined as a map
\begin{equation}\label{eq:conformalblock}
    \cF^\text{Vir} : \cV_{\alpha_1}(z_1) \otimes \ldots \otimes \cV_{\alpha_n} (z_n) \to \C,
\end{equation}
that satisfies the so-called conformal Ward identities. The latter identities are derived by inserting a copy of the stress-energy tensor $T(z)$ into the conformal block, or more precisely,
\begin{align*}
\oint_\gamma \eta(z) T(z),
\end{align*}
where $\eta(z)$ is a meromorphic function that is holomorphic away from the punctures and $\gamma$ is a contour that does not encircle any of the punctures. The Ward identities thereby describe how the conformal blocks transform under infinitesimal conformal symmetries. 

At times, it is useful to rewrite the $n-$point conformal block as a map between Virasoro modules
\begin{equation}
    \cF^\text{Vir} : \cV_{\alpha_1}(z_1) \otimes \ldots \otimes \cV_{\alpha_{n-1}} (z_{n-1})\to \cV_{\alpha_n}^*(z_n).
\end{equation}
The dual Virasoro module $\cV_\alpha^*$ is defined with respect to the standard inner product on~$\cV_\alpha$.\footnote{The Verma module $\cV_\alpha$ carries a well-defined inner product only when $\alpha \in\frac Q 2 + i \R$. Nevertheless, we can extend this to all values of $\alpha$ through analytic continuation. This is discussed in detail in \cite[Section 4]{Teschner:2001rv}.} 

We shall refer to conformal blocks on genus zero surfaces with $n$ insertions as \emph{$n-$point conformal blocks}. The Virasoro symmetry ensures that an $n$-point conformal block $\mathcal F^\textrm{Vir}$ is fully determined by its action on the primary states $v_{\alpha_k} \in \mathcal V_{\alpha_k}$. So, without any loss in generality, from here onwards, we restrict our attention to conformal blocks evaluated on primary vertex operators,
\begin{equation}
    \mathcal F^\textrm{Vir}(V_{\alpha_1}(z_1), \ldots, V_{\alpha_n}(z_n)). \label{eq:conformal-block-primary}
\end{equation}

\paragraph{\texorpdfstring{Gluing conformal blocks\\}{Gluing conformal blocks:}}

A conformal block on an arbitrary (possibly punctured) Riemann surface can be built out of conformal blocks on $3$-punctured spheres. This construction builds on the plumbing construction that may be used to sew Riemann surfaces geometrically. This is particularly useful because, up to normalisation, there is a unique $3$-point conformal block,
\begin{equation}
    \mathcal F^\textrm{Vir} : \mathcal V_{\alpha_1}(z_1) \otimes \mathcal V_{\alpha_2}(z_2) \to \mathcal V_{\alpha_3}^*(z_3).
\end{equation}

That is, suppose we want to glue the Riemann surfaces $C_1$ and $C_2$ at the punctures (or generic marked points) $p_1 \in C_1$ and $p_2 \in C_2$ (where $C_1$ may be equal to $C_2$). Choose local coordinates $z$ and $w$ on $C_1$ and $C_2$, respectively, such that \[z(p_1) = 0 \text{ and } w(p_2) = \infty.\]
Also, fix a sewing parameter $\kappa \in \C$. Then, cut open small discs around $p_1$ and $p_2$ of radius $r > |\kappa|$ such that the chosen coordinates are well-defined in this neighbourhood. To glue the two surfaces together, we simply identify the regions,
\begin{equation}
    \{ \frac{|\kappa|}{r} < |z| <r, \frac{|\kappa|}{r} < |w| <r\}
\end{equation}
through the condition 
\begin{equation}
    wz = \kappa.
\end{equation}
We denote the newly obtained Riemann surface by $C$.

Now, suppose that both surfaces $C_1$ and $C_2$ carry conformal blocks $\cF^\text{Vir}_{C_1}$ and $\cF^\text{Vir}_{C_2}$ respectively, and we would like to extend this to a conformal block on $C$. The vertex operators inserted outside the sewing region remain unaffected, but the Virasoro modules at $p_1$ and $p_2$ are required to be conjugate to each other. We collectively denote the vertex operators away from $p_1$ on $C_1$ by $V_{C_1}$, and the vertex operators away from $p_2$ on $C_2$ by $V_{C_2}$. 

Let $\cV_\alpha$ and $\cV_{\alpha}^*$ be two conjugate representations at the points $p_1$ and $p_2$ respectively. Then, the conformal block $\cF^\text{Vir}_{C,\alpha}$ on $C$ is defined as 
\begin{equation}
    \cF^\text{Vir}_{C,\alpha} \left(V_{C_1}, V_{C_2}\right) = \sum_{v, v'} \cF^\text{Vir}_{C_1} \left(V_{C_1}, V_\alpha(v; 0) \right) K_{\alpha; v, v'}^{-1}\, \cF^\text{Vir}_{C_2} \, \left((V_{\alpha^*}(v';\infty),\, V_{C_2} \right), \label{eq:gluing}
\end{equation}
where the right-hand side is a sum over all possible states $v, v'$ in the representations $\cV_\alpha$ and $\cV_\alpha^*$, respectively, and $K_{\alpha; v, v'}$ is the Kac matrix defined by two-point conformal locks
\begin{equation}
    K_{\alpha; v, v'} = \cF^\text{Vir}_{\CP^1} \left(V_{\alpha^*}(v; \infty), V_{\alpha}(v'; 0) \right), \label{eq:kac-matrix}
\end{equation}
which can be computed systematically. 

\smallskip

In principle, it is then possible to construct conformal blocks on any arbitrary Riemann surface $C$ from the knowledge of the three-point conformal blocks for all possible Virasoro representations. This motivates the definition of the space $\mathrm{gCB}(C_{0,3})$ of \emph{gauged} $3$-point conformal blocks as the space 
\begin{equation}
    \mathrm{gCB}(C_{0,3}) := \bigoplus_{\alpha_1,\, \alpha_2,\, \alpha_3} \hom\bigg(\mathcal V_{\alpha_1} (z_1) \otimes \mathcal V_{\alpha_2} (z_2) \otimes \mathcal V_{\alpha_3} (z_3), \, \C\bigg)^\text{Ward id} \label{eq:gauged-conformal-block-Liouville}
\end{equation}
of conformal blocks on the gauged 3-sphere $C_{0, 3}$,
where the sum is over $\mathcal V_{\alpha_k}$ for \emph{generic} momenta $\alpha_k$ (as opposed to degenerate momenta that will be introduced in \cref{eq:degenerate-momenta}). Observe that $\mathrm{gCB}(C_{0,3})$ is an infinite-dimensional vector space since all three momenta $\alpha_1$, $\alpha_2$ and $\alpha_3$ can take arbitrary (generic) complex values. Instead, its subspace $\mathrm{CB}(C_{0,3})$ of conformal blocks on the bare ("non-gauged") 3-punctured sphere is 1-dimensional. That is, its only freedom is the choice of normalisation. 

The space of conformal blocks on an arbitrary Riemann surface $C$ is, similarly, infinite-dimensional in general. Furthermore, we can also define the space of gauged conformal blocks $\mathrm{gCB}(C)$ on any surface $C$ by summing over all possible vertex operator representations at its punctures.

As we vary the complex structure on $C$, the space of conformal blocks organises itself into a holomorphic vector bundle $\mathcal{CB}$ over the moduli space $\mathcal{M}$ of complex structures on $C$. The variation of the conformal block in the direction of a Beltrami differential $\mu$ can be expressed in terms of a \emph{projectively flat} (or twisted) connection on $\mathcal{M}$, whose connection 1-form is defined by inserting 
\begin{align}
    \int_C \mu \, T(z)
\end{align}
into the conformal block. More details can be found for instance in \cite[Section 2]{Hao:2024vlg}. 

Suppose $C$ is a genus $g$ surface with $n$ punctures. Given any pants decomposition $\mathcal{P}$ of~$C$, we can construct a canonical basis 
\begin{equation} \label{eq:basisF}
    \cF^\text{Vir}_{\vec \alpha}
\end{equation}
of the (possibly infinite-dimensional) space of conformal blocks on $C$ that is obtained by gluing $2g-2+n$ three-point conformal blocks, and labelled by a choice of $3g-3+n$ internal momenta 
\begin{align}
\vec \alpha = (\alpha_1, \ldots, \alpha_{3g-3+n}),
\end{align}
i.e.~one for each pants cycle in $\mathcal P$.
The bases associated with the different pairs of pants decompositions $\mathcal P$ are known as channels in the physics literature.

\begin{remark}
   Note that even though the expressions for these bases depend non-trivially on the choice of pants decomposition, the resulting correlation functions are independent of this choice by definition (since any two-dimensional CFT ought to be modular invariant).  
\end{remark}

\subsubsection{Null-vector decoupling}

Generically, a Verma module $\mathcal V_\alpha$ forms an irreducible representation of the Virasoro algebra. However, for special values
\begin{equation}
    \alpha_{p,q} = (1-p) \, \frac 1 {2b} + (1-q) \, \frac b 2  \quad \text{ with }~p,q \in \Z_+, \label{eq:degenerate-momenta}
\end{equation}
of the Liouville momentum, the corresponding Virasoro module is reducible and contains a sub-module generated by a null state (or null vector). All states in this sub-module have vanishing norm, and may thus be consistently quotiented out. The resulting finite-dimensional module $\mathcal{V}_{p,q}$ is known as a \emph{degenerate} representation, the associated primary vertex operator $V_{p,q} := V_{\alpha_{p,q}}$ is known as a degenerate primary vertex operator, and the corresponding momentum~$\alpha_{p,q}$, as a degenerate momentum.\footnote{From a representation-theoretic perspective, an infinite-dimensional reducible Verma module and its finite-dimensional irreducible quotient are two starkly different objects. It will also be a consistent theme in this paper that the degenerate operators $V_{p,q}$ often behave differently when compared to generic primary operators $V_\alpha$.}

The degenerate momenta~$\alpha_{p,q}$ are closely related to the fundamental weights of $SL(2, \mathbb C)$. Indeed, we may rewrite \cref{eq:degenerate-momenta} as
\begin{equation}
        \alpha_{p, q} = -\frac 1 b\, \omega_{p-1} - b\,\omega_{q-1}, \label{eq:degenerate-sl2-weights}
\end{equation}
where $\omega_{p-1}$ and $\omega_{q-1}$ are the highest weights of the $p$th and $q$th symmetric tensor representations of $SL(2, \C)$, respectively. In this sense, a degenerate operator with momentum $\alpha_{1,q}$ \emph{quantises} the $q$-dimensional representation of $SL(2,\C)$. This can be made more precise through skein theory (see \S\ref{sec:CStheory}). For quantum monodromy representations, the degenerate operator $V_{1,2}$ corresponding to the $2$-dimensional or defining representation of $SL(2,\C)$ is the most relevant to us.

If a degenerate vertex operator $V_{p,q}(z)$ is inserted in a conformal block $\cF^\text{Vir}(\ldots)$, the algebraic null-vector constraint turns into a differential operator of degree $pq$ acting on the block $\cF^\text{Vir}(V_{p,q}(z), \ldots)$. The resulting differential equation is known as the null-vector decoupling equation or the \emph{BPZ equation} \cite{Belavin:1984vu}. For instance, for the degenerate vertex operator $V_{1,2}(z)$ we obtain the degree 2 differential equation\footnote{The detailed equation can for instance be found in \cite[Appendix D]{Ashok:2015gfa}.}
\begin{equation}\label{eq:BPZ12}
    (\partial_z^2 + b^2 T(z)) \cF^\text{Vir}(V_{1,2}(z), \ldots) = 0.
\end{equation}
The space of conformal blocks $\cF^\text{Vir}(V_{p,q}(z), \ldots)$ may thus be identified with the space of solutions to the BPZ equation, and is therefore $pq$-dimensional. 

The rank $pq$ matches with the number of fusion channels in the OPE between the degenerate vertex operator $V_{p,q}(z)$ and another generic vertex operator $[V_{\alpha_k}] \in \cV_{\alpha_k}$,
\begin{align}\label{eq:OPE12alpha}
\lim_{z \to z_k} V_{p,q}(z) \left[V_{\alpha_k}\right](z_k) = \sum_{r,s} c_{r,s} \, z^{\Delta_{ \alpha_k - \frac{r}{2b} - \frac{sb}{2}} -\Delta_{\alpha_k} - \Delta_{ p,q}} \, \left[V_{\alpha_k - \frac{r}{2b} - \frac{sb}{2}}\right](z_k)
\end{align}
where
\[r \in \{-p+1, -p+3, \ldots, p-1 \} \text{ and } s \in \{ -q+1, -q+3, \ldots, q-1 \}.\]
Then, letting the insertion point $z$ approach a puncture $z_k$, the resulting conformal blocks
\begin{align}
\cF^\text{Vir} (V_{\alpha - \frac{r}{2b} - \frac{sb}{2}}(z_k), \ldots),
\end{align}
generate the $pq$ independent local solutions to the BPZ differential equation near $z_k$. 

Note that the OPE~\eqref{eq:OPE12alpha} also implies constraints on the internal momenta $\vec{a}$ of the conformal block $\mathcal{F}^\text{Vir}$, which are generically unrestricted. For instance, the three-point conformal block $\mathcal{F}^\text{Vir}_{\mathbb{P}^1}(V_{\alpha_1} V_{\alpha_2} V_{\alpha_3})$ with degenerate insertion $V_{\alpha_2}=V_{1,2}$ is non-vanishing only when the non-degenerate momenta $\alpha_1$ and $\alpha_3$ are related by 
\begin{equation}
    \alpha_3 = \alpha_1 - \, s \frac{b}{2},
\end{equation}
with $s = \pm$. This restriction continues to hold when gluing three-point functions to construct conformal blocks on arbitrary surfaces. 

Comparing the BPZ equation~\eqref{eq:BPZ12} with the Lax equation~\cref{eq:null-equation-cl} suggests that the BPZ equation for $V_{1,2}$ should be viewed as a \emph{quantisation} of the Lax equation. This implies that the degenerate vertex operator~$V_{1,2}(z)$ should be viewed as the quantisation of the semi-classical field $\psi(z)$. More precisely, the BPZ equation~\eqref{eq:BPZ12} should be interpreted as a "quantum oper" connection on a rank 2 vector bundle over $C$, associated with the fundamental representation of $SL(2)$. We clarify what we mean by this in \S\ref{subsec:bpz-monodromies}, where we discuss the global transformation properties of the BPZ equation~\eqref{eq:BPZ12} on a general Riemann surface. 

More generally, the BPZ equation for a degenerate operator $V_{1,p}(z)$ should be thought of as a quantum connection in a rank $p$ vector bundle, associated with the $p$th symmetric tensor representation of $SL(2)$, whereas the BPZ equation for $V_{p,q}(z)$ is associated with a component of the tensor product of the $p$th and the $q$th symmetric tensor representations of $SL(2)$.

\subsubsection{AGT correspondence and NRS proposal} \label{subsubsec:agt-nrs}

We conclude the introduction to Liouville theory with a brief discussion on the AGT correspondence between 4d $\mathcal N =2$ $SU(2)$ gauge theories and 2d Liouville CFT \cite{Alday:2009aq}. The AGT correspondence states that the supersymmetric partition function $\mathcal Z[S^4_b]$ of a 4d $\mathcal N = 2$ $SU(2)$ theory $T[C]$ on the squashed 4-sphere 
\[S^4_b := \{(x_1, x_2, x_3, x_4, x_5) \in \R^5 \: : \: x_5^2 + b^2 \left(x_1^2 + x_2^2\right) + b^{-2} \left(x_3^2 + x_4^2\right) = R^2 \}, \]
can be identified with a Liouville correlation function whose vertex operator insertions are determined by the matter content of the 4d theory. This follows from defining a 6d $(2,0)$ $A_1$-theory in the background $S^4_b \times C$.

Similar to the factorisation of Liouville correlation functions into holomorphic and anti-holomorphic conformal blocks, the 4d partition function $\mathcal Z[S^4_b]$ can also be (schematically) factorised as
\begin{equation}\label{eqn:AGQS4}
    \mathcal Z[S^4_b] = \int da \, |\mathcal Z^\textrm{1-loop}(a)|^2 \, \mathcal Z[\R^4_{\epsilon_1, \epsilon_2}](a) \, \overline{ \mathcal Z[\R^4_{\epsilon_1, \epsilon_2}]}(a),
\end{equation}
given a weak-coupling description of $T[C]$ (which is equivalent to a choice of pants decomposition of $C$ \cite{Gaiotto:2009we}). The Coulomb parameters are collectively denoted by $a$, the $\mathcal Z^\textrm{1-loop}$ (or $\overline{\mathcal Z^\textrm{1-loop}}$) encodes the 1-loop contribution to the Nekrasov partition function in the $\Omega-$background $\R^4_{\epsilon_1, \epsilon_2}$, whereas $\mathcal Z[\R_{\epsilon_1, \epsilon_2}^4]$ (or $\overline{\mathcal Z[\R_{\epsilon_1, \epsilon_2}^4]}$) contains its tree-level and instanton (or anti-instanton) contribution. The instanton parameters of $T[C]$ may be identified with the complex structure parameters of $C$. 

Comparing the factorisation~\eqref{eqn:AGQS4} with the factorisation of Liouville correlation functions (see for instance \cref{eq:Liouvillecorrelator}), the AGT correspondence proposes the identification of instanton partition functions $\mathcal Z[\R^4_{\epsilon_1, \epsilon_2}]$ with the holomorphic conformal blocks $\mathcal F$, where the deformation parameters are related as
\begin{equation}
    \epsilon_1 = \frac 1 {b R}  \quad \text{ and } \quad \epsilon_2 = \frac b R.
\end{equation}
Additionally, $|\mathcal Z^\textrm{1-loop}|^2$ may be identified with the product of Liouville three-point functions $\mathcal C(\alpha_1,\alpha_2,\alpha_3)$ contributing to the Liouville correlation function. It is customary to set $R = 1$.

The semi-clasical limit of the AGT correspondence can be understood via the Nekrasov-Rosly-Shatashvili (NRS) proposal~\cite{Nekrasov:2011bc}, which states that the generating function of the family of opers, when expressed in terms of complex FN-type coordinates, computes the effective twisted superpotential $\widetilde{W}^\mathrm{eff}$ for the associated class S theory~$T[C]$. 

The easiest way to understand this proposal is to consider the theory $T[C]$ on a background $\R^2 \times D_\epsilon$ where $D_\epsilon$ is the cigar geometry with an $\Omega-$deformation.\footnote{Starting from the $\Omega-$background $\R^2_{\epsilon_2} \times \R^2_{\epsilon_1}$, we obtain the background $\R^2 \times D_\epsilon$ by first, sending $\epsilon_2 \to 0$, and then replacing the standard metric on $\R^2_{\epsilon = \epsilon_1}$ with the cigar metric $ds^2 = dr^2 + f(r) d\theta^2$, where $f(r) \sim r^2$ as $r \to 0$ and $f(r) = \rho^2$ for a constant radius $\rho$ as $r \to \infty$. } At low energies, we can reduce the theory along $D_\epsilon$ to an effective 2d $\mathcal N = (2, 2)$ theory of twisted chiral superfields (sourced by the 4d vector multiplets)  on $\R^2$, coupled to an effective twisted superpotential $\widetilde{W}_\textrm{eff}$. In \cite{Nekrasov:2009rc}, Nekrasov and Shatashvili proposed that the effective twisted superpotential and the instanton partition function for the theory $T[C]$ are related as
\begin{equation}
    \widetilde W_\textrm{eff}(\epsilon) = \lim_{\epsilon_2 \to 0} \epsilon_2 \log \mathcal Z[\R^4_{\epsilon_1 = \epsilon, \, \epsilon_2}].
\end{equation}
Furthermore, they argued that the effective twisted superpotential plays the role of a Yang-Yang functional in a quantum integrable system underlying the Seiberg-Witten geometry of $T[C]$. This is now known as the Bethe-gauge correspondence.

Through the AGT-correspondence for general $C$, the exponent of the one-loop contribution to $\widetilde{W}^\mathrm{eff}$ is closely related to the square-root of the three-point contribution to the associated Liouville correlator (see the discussion around \cref{eq:Weff3pt} and \eqref{eq:Csc} for the precise statement), and the exponent of the instanton contribution to $\widetilde{W}^\mathrm{eff}$ is equal to a corresponding conformal block~$\mathcal{F}$, all in the limit $\epsilon_2 \sim b \to 0$. This establishes that the semi-classical ($b \to 0$) limit of Liouville conformal blocks is captured by the Yang-Yang function of a quantum integrable system determined by $T[C]$.

At the same time, it was argued by Nekrasov, Rosly and Shatashvili in \cite{Nekrasov:2011bc} that the superpotential~$\widetilde W_\textrm{eff}$ computes the difference of two generating functions of holomorphic Lagrangian submanifolds inside the moduli space of flat $\epsilon$-connections. This can be seen by, instead of reducing the 4d theory on all of $D_\epsilon$ to get a 2d theory, compactifying along the circle direction of the cigar (i.e.~write $D_\epsilon = \R_+ \times S^1$ where the $S^1$ degenerates at the tip of the cigar) to obtain a 3d $\mathcal N = 4$ theory on $\R^2 \times \R_+$. Then, the two holomorphic Lagrangian submanifolds correspond to a choice of two $\frac 1 2$-BPS boundary conditions at the two ends of~$\R^+$ \cite{Nekrasov:2010ka}.

The boundary condition at the tip of $\R_+$ is purely dictated by the geometry, and is described by the Lagrangian submanifold of $\epsilon$-opers inside the moduli space. The boundary condition at the other end is less restrictive. It was argued in \cite{Hollands:2021itj} that for any choice of holomorphic Darboux coordinates $\{x_i, y_i\}$ on the moduli space of flat $\epsilon$-connections, the second boundary condition is given by fixing all $y_i \equiv 0 \,(\text{mod } 2\pi)$. We will come back to this in \S\ref{sec:nrs-FN-gauge}.

\subsection{Moore-Seiberg monodromies on the gauged 3-sphere}\label{subsec:bpz-monodromies}

We have learned in \S\ref{sec:basicsLiouville} that the space of Liouville conformal blocks forms a vector bundle~$\mathcal{CB}$ over the moduli space $ \mathcal{M}_{g,n} $ of genus $g$ Riemann surfaces $C_{g,n}$ with $n$ punctures. In particular, when $g=0$ we have that 
\begin{align}
\mathcal{M}_{0,n} &= \{(z_1, \ldots, z_n) \in (\mathbb{CP}^1)^n : z_i \neq z_j \text{ for } i, j \in \{1, \ldots, n\} \} / \text{PSL}(2, \mathbb{R}),
\end{align}
where the group $ \text{PSL}(2, \mathbb{R}) $ acts via Möbius transformations. The vector bundle~$\mathcal{CB}$ is equipped with a projectively flat connection, and parallel transport around loops in $ \mathcal{M}_{g,n} $ thus induces nontrivial transformations in the space of conformal blocks. In fact, since
\begin{align}
\mathcal{M}_{g,n} &= \mathcal{T}_{g,n} / \text{MCG}(C_{g,n}),
\end{align}
where the Teichmüller space $ \mathcal{T}_{g,n} $ is the universal cover of the moduli space $ \mathcal{M}_{g,n}$ and the mapping class group $ \text{MCG}(C_{g,n})$ parametrises non-trivial loops in $ \mathcal{M}_{g,n}$, the space of conformal blocks forms a projective representation of the mapping class group \cite{Moore:1988qv}.

We have also explained in \S\ref{sec:basicsLiouville} that any pair of pants decomposition of the surface~$C_{g,n}$ induces a basis for the space of Liouville blocks on $C_{g,n}$. For instance, on the 4-punctured sphere, there are three standard choices known as the s-channel, the t-channel and the u-channel basis corresponding to the three different pants decomposition. See~\cref{fig:conformal-block-bases}. The generators of each basis are labelled by a single intermediate momentum $\alpha$.

\begin{figure}[ht]
\centering
    \begin{tikzpicture}
        \node[xshift = -4.4cm, yshift = 0.5cm] at (0,0){s-channel:};
        \node[xshift = -1.7cm, yshift = 0.5cm] at (0,0){$\cF_\alpha^{(s)}\begin{bmatrix}
        \alpha_3 & \alpha_2 \\ \alpha_4 & \alpha_1
    \end{bmatrix} =$};
        \draw (0,0)--(2,0);
        \node[yshift = -0.2cm] at (0,0) {$\alpha_4$};
        \node[yshift = -0.2cm] at (2,0) {$\alpha_1$};
        \node[yshift = 0.2cm] at (0.5,1) {$\alpha_3$};
        \node[yshift = 0.2cm] at (1.5,1) {$\alpha_2$};
        \node[yshift = -0.2cm] at (1,0) {$\alpha$};
        \draw (0.5,0)-- (0.5,1);
        \draw (1.5,0)--(1.5,1);
    \end{tikzpicture} \\
    \begin{tikzpicture}
        \node[xshift = -4.4cm, yshift = 0.5cm] at (0,0){t-channel:};
        \node[xshift = -1.7cm, yshift = 0.5cm] at (0,0){$\cF_\alpha^{(t)}\begin{bmatrix}
        \alpha_3 & \alpha_2 \\ \alpha_4 & \alpha_1
    \end{bmatrix} =$};
        \draw (0,0)--(2,0);
        \node[yshift = -0.2cm] at (0,0) {$\alpha_4$};
        \node[yshift = -0.2cm] at (2,0) {$\alpha_1$};
        \node[yshift = 0.2cm] at (0.5,1) {$\alpha_3$};
        \node[yshift = 0.2cm] at (1.5,1) {$\alpha_2$};
        \node[xshift = 0.2cm, yshift = 0.25cm] at (1,0) {$\alpha$};
        \draw (1,0.5)-- (0.5,1);
        \draw (1,0.5)--(1.5,1);
        \draw (1,0)--(1,0.5);
    \end{tikzpicture} \\
    \begin{tikzpicture}
        \node[xshift = -4.4cm, yshift = 0.5cm] at (0,0){u-channel:};
        \node[xshift = -1.7cm, yshift = 0.5cm] at (0,0){$\cF_\alpha^{(u)}\begin{bmatrix}
        \alpha_3 & \alpha_2 \\ \alpha_4 & \alpha_1
    \end{bmatrix} =$};
        \draw (0,0)--(2,0);
        \node[yshift = -0.2cm] at (0,0) {$\alpha_4$};
        \node[yshift = -0.2cm] at (2,0) {$\alpha_1$};
        \node[yshift = 0.2cm] at (0.5,1) {$\alpha_3$};
        \node[yshift = 0.2cm] at (1.5,1) {$\alpha_2$};
        \node[yshift = -0.2cm] at (1,0) {$\alpha$};
        \draw (0.5,0)--(1.5,1);
        \draw (1.5,0)--(1.1,0.4);
        \draw (0.9,0.6)--(0.5,1);
    \end{tikzpicture} 
    \caption{The three standard choices of bases for a four-point conformal block.}
    \label{fig:conformal-block-bases}
\end{figure}

The mapping class group $\MCG(C_{g,n})$ has a presentation with three generators known as fusion, braiding, and modular (or S) moves. Given a pair of pants decomposition $\mathcal P$ of the surface~$C_{g,n}$, the action of each basic move on $\mathcal P$ produces a new pair of pants decomposition $\mathcal P'$. This structure may be captured in a groupoid whose vertices are all different pair of pants decompositions of $C_{g,n}$ and the edges/morphisms are actions of the generators. This groupoid is known as the \emph{Moore-Seiberg groupoid}.  

Starting with a given pair of pants decomposition $\mathcal{P}$, there may be several different sequences of basic moves leading to another decomposition $\mathcal{P}'$. Such consistency relations between the three basic moves correspond to closed 2-cells in the Moore-Seiberg groupoid. We refer the reader to \cite{bakalov1998legoteichmullergame} for more details. Since the S-move can be expressed in terms of the other generators (and is not relevant anyhow when discussing genus $0$ surfaces), we will focus on the fusion and braiding moves from now on.

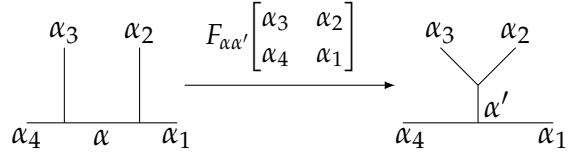
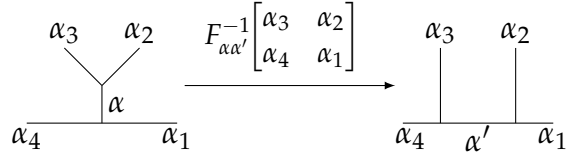
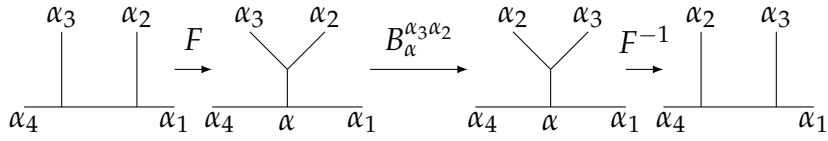
\begin{figure}[h]
\centering
    \begin{subfigure}{0.8\textwidth}
        \centering
        \begin{tikzpicture}
        \draw (0,0)--(2,0);
        \node[yshift = -0.2cm] at (0,0) {$\alpha_4$};
        \node[yshift = -0.2cm] at (2,0) {$\alpha_1$};
        \node[yshift = 0.2cm] at (0.5,1) {$\alpha_3$};
        \node[yshift = 0.2cm] at (1.5,1) {$\alpha_2$};
        \node[yshift = -0.2cm] at (1,0) {$\alpha$};
        \draw (0.5,0)-- (0.5,1);
        \draw (1.5,0)--(1.5,1);
        \draw [-latex] (2.1,0.5)--(4.9,0.5);
        \node[yshift = 0.65cm, xshift = 3.4cm] at (0,0.5) {\small $ F_{\alpha \alpha'}  \!\!\begin{bmatrix}
        \alpha_3 & \alpha_2 \\ \alpha_4 & \alpha_1
    \end{bmatrix}$};
        \draw[xshift = 1cm] (4,0)--(6,0);
        \draw[xshift = 1cm] (5,0.5)-- (4.5,1);
        \draw[xshift = 1cm] (5,0.5)--(5.5,1);
        \draw[xshift = 1cm] (5,0)--(5,0.5);
        \node[yshift = -0.2cm, xshift = 5.1cm] at (0,0) {$\alpha_4$};
        \node[yshift = -0.2cm, xshift = 5cm] at (2,0) {$\alpha_1$};
        \node[yshift = 0.2cm, xshift = 5cm] at (0.5,1) {$\alpha_3$};
        \node[yshift = 0.2cm, xshift = 5cm] at (1.5,1) {$\alpha_2$};
        \node[yshift = 0.25cm, xshift = 5.25cm] at (1,0) {$\alpha'$};
    \end{tikzpicture}
     \caption{The fusion move expresses an s-channel block in terms of t-channel blocks.}
    \label{fig:F-move}
    \end{subfigure}
    \begin{subfigure}{0.8\textwidth}
    \centering
        \begin{tikzpicture}
        \draw (0,0)--(2,0);
        \node[yshift = -0.2cm] at (0,0) {$\alpha_4$};
        \node[yshift = -0.2cm] at (2,0) {$\alpha_1$};
        \node[yshift = 0.2cm] at (0.5,1) {$\alpha_3$};
        \node[yshift = 0.2cm] at (1.5,1) {$\alpha_2$};
        \node[xshift = 0.2cm, yshift = 0.25cm] at (1,0) {$\alpha$};
        \draw (1,0.5)-- (0.5,1);
        \draw (1,0.5)--(1.5,1);
        \draw (1,0)--(1,0.5);
        \draw [-latex] (2.1,0.5)--(4.9,0.5);
        \node[yshift = 0.65cm, xshift = 3.4cm] at (0,0.5) {\small $ F^{-1}_{\alpha \alpha'}  \!\!\begin{bmatrix}
        \alpha_3 & \alpha_2 \\ \alpha_4 & \alpha_1
    \end{bmatrix}$};
        \draw[xshift = 1cm] (4,0)--(6,0);
        \draw[xshift = 1cm] (4.5,0)-- (4.5,1);
        \draw[xshift = 1cm] (5.5,0)--(5.5,1);
        \node[yshift = -0.2cm, xshift = 5.1cm] at (0,0) {$\alpha_4$};
        \node[yshift = -0.2cm, xshift = 5cm] at (2,0) {$\alpha_1$};
        \node[yshift = 0.2cm, xshift = 5cm] at (0.5,1) {$\alpha_3$};
        \node[yshift = 0.2cm, xshift = 5cm] at (1.5,1) {$\alpha_2$};
        \node[yshift = -0.2cm, xshift = 5cm] at (1,0) {$\alpha'$};
    \end{tikzpicture}
     \caption{The inverse fusion move expresses a t-channel block in terms of s-channel blocks.}
    \label{fig:F-move-inv}  
    \end{subfigure}
    \begin{subfigure}{0.8\textwidth}
        \centering
    \begin{tikzpicture}
        \draw (0,0)--(2,0);
        \node[yshift = -0.2cm] at (0,0) {$\alpha_4$};
        \node[yshift = -0.2cm] at (2,0) {$\alpha_1$};
        \node[yshift = 0.2cm] at (0.5,1) {$\alpha_3$};
        \node[yshift = 0.2cm] at (1.5,1) {$\alpha_2$};
        \draw (0.5,0)-- (0.5,1);
        \draw (1.5,0)--(1.5,1);
        \draw [-latex] (2,0.5)--(2.5,0.5);
        \node[yshift = 0.4cm, xshift = 2.25cm] at (0,0.5) {$F$};
        \draw (2.5,0)--(4.5,0);
        \draw (3.5,0.5)-- (3,1);
        \draw (3.5,0.5)--(4,1);
        \draw (3.5,0)--(3.5,0.5);
        \node[yshift = -0.2cm, xshift = 2.6cm] at (0,0) {$\alpha_4$};
        \node[yshift = -0.2cm, xshift = 2.5cm] at (2,0) {$\alpha_1$};
        \node[yshift = 0.2cm, xshift = 2.5cm] at (0.5,1) {$\alpha_3$};
        \node[yshift = 0.2cm, xshift = 2.5cm] at (1.5,1) {$\alpha_2$};
        \node[yshift = -0.2cm, xshift = 2.5cm] at (1,0) {$\alpha$};
        \draw [-latex] (4.6,0.5)--(5.9,0.5);
        \node[yshift = 0.4cm, xshift = 5.25cm] at (0,0.5) {$B_\alpha^{\alpha_3 \alpha_2}$};
        \draw[xshift = 1cm] (5,0)--(7,0);
        \draw[xshift = 1cm] (6,0.5)-- (5.5,1);
        \draw[xshift = 1cm] (6,0.5)--(6.5,1);
        \draw[xshift = 1cm] (6,0)--(6,0.5);
        \node[yshift = -0.2cm, xshift = 6.1cm] at (0,0) {$\alpha_4$};
        \node[yshift = -0.2cm, xshift = 6 cm] at (2,0) {$\alpha_1$};
        \node[yshift = 0.2cm, xshift = 6 cm] at (0.5,1) {$\alpha_2$};
        \node[yshift = 0.2cm, xshift = 6 cm] at (1.5,1) {$\alpha_3$};
        \node[yshift = -0.2cm, xshift = 6 cm] at (1,0) {$\alpha$};
        \draw [xshift = 1cm, -latex] (7,0.5)--(7.5,0.5);
        \node[yshift = 0.4cm, xshift = 8.25cm] at (0,0.5) {$F^{-1}$};
        \draw[xshift = 1cm] (7.5,0)--(9.5,0);
        \draw[xshift = 1cm] (8,0)-- (8,1);
        \draw[xshift = 1cm] (9,0)--(9,1);
        \node[yshift = -0.2cm, xshift = 8.6cm] at (0,0) {$\alpha_4$};
        \node[yshift = -0.2cm, xshift = 8.5cm] at (2,0) {$\alpha_1$};
        \node[yshift = 0.2cm, xshift = 8.5cm] at (0.5,1) {$\alpha_2$};
        \node[yshift = 0.2cm, xshift = 8.5cm] at (1.5,1) {$\alpha_3$};
    \end{tikzpicture} 
    \caption{The braiding move (in our notation) expresses a t-channel block in terms of permuted t-channel blocks.}
    \label{b-move}
    \end{subfigure}
  \caption{The action of the Moore-Seiberg moves on Liouville blocks. Figures $(a)$ and $(b)$ illustrate the fusion and inverse fusion moves, respectively. Figure~$(c)$ illustrates the braiding move.}   
\end{figure}

The \emph{fusion} move expresses an s-channel conformal block in terms of t-channel conformal blocks, as in
\begin{align}
    \label{eq:fusion-transformation}
     \cF_\alpha^{(s)}\begin{bmatrix}
        \alpha_3 & \alpha_2 \\ \alpha_4 & \alpha_1
    \end{bmatrix} &= \int d \alpha' \;F_{\alpha \alpha'} \begin{bmatrix}
        \alpha_3 & \alpha_2 \\ \alpha_4 & \alpha_1
    \end{bmatrix} \cF_{\alpha'}^{(t)}\begin{bmatrix}
        \alpha_3 & \alpha_2 \\ \alpha_4 & \alpha_1
    \end{bmatrix},
\end{align}
with inverse
\begin{align}
    \cF_\alpha^{(t)}\begin{bmatrix}
        \alpha_3 & \alpha_2 \\ \alpha_4 & \alpha_1
    \end{bmatrix} &= \int d \alpha' \;F^{-1}_{\alpha \alpha'} \begin{bmatrix}
        \alpha_3 & \alpha_2 \\ \alpha_4 & \alpha_1
    \end{bmatrix} \cF_{\alpha'}^{(s)}\begin{bmatrix}
        \alpha_3 & \alpha_2 \\ \alpha_4 & \alpha_1
    \end{bmatrix}.
\end{align}
This is illustrated in \cref{fig:F-move,fig:F-move-inv}. The integral kernel of the fusion transformation is called the fusion matrix $F_{\alpha \alpha'}$ (or also the $6j$-symbol). The consistency conditions in the Moore-Seiberg groupoid lead to the following symmetry properties for the fusion matrices:
\begin{equation}
    \label{eq:fusion-matrices-symmetry} F_{\alpha  \alpha'}\begin{bmatrix}
        \alpha_3 & \alpha_2 \\ \alpha_4 & \alpha_1
    \end{bmatrix} = F_{\alpha \alpha'}\begin{bmatrix}
        \alpha_2 & \alpha_3 \\ \alpha_1 & \alpha_4
    \end{bmatrix} = F_{\alpha \alpha'}\begin{bmatrix}
        \alpha_4 & \alpha_1 \\ \alpha_3 & \alpha_2
    \end{bmatrix}, \quad F^{-1}_{\alpha \alpha'}\begin{bmatrix}
        \alpha_3 & \alpha_2 \\ \alpha_4 & \alpha_1
    \end{bmatrix} = F_{\alpha \alpha'}\begin{bmatrix}
        \alpha_3 & \alpha_4 \\ \alpha_2 & \alpha_1
    \end{bmatrix}.
\end{equation}

The \emph{braiding} move can be expressed as a sequence of three moves, in which we first use the fusion move to express an s-channel block into t-channel conformal blocks, then twist the two "upward-facing" legs of the t-channel blocks, and finally express the result in terms of s-channel blocks by using the inverse fusion move. This is illustrated in \cref{b-move}. The twisting action $B$ on the t-channel blocks is simply given by the diagonal transformation
\begin{align}
    \label{eq:braiding-transformation}
     \cF_\alpha^{(t)}\begin{bmatrix}
        \alpha_3 & \alpha_2 \\ \alpha_4 & \alpha_1
    \end{bmatrix} &= B_\alpha^{\alpha_3 \alpha_2}\,\cF_{\alpha}^{(t)}\begin{bmatrix}
        \alpha_2 & \alpha_3 \\ \alpha_4 & \alpha_1
    \end{bmatrix},
\end{align}
where the coefficient
\begin{equation}
    \label{eq:braiding-move} B_{\alpha}^{\alpha_3 \alpha_2} = \exp(-i \pi (\Delta(\alpha) - \Delta(\alpha_2) - \Delta(\alpha_3)))
\end{equation}
is merely a complex number. The twisting action should be thought of as transporting $\alpha_3$ around $\alpha_2$ in the \emph{clockwise} direction.

The fusion matrices are generally complicated, infinite-dimensional objects. They simplify drastically, though, if we pick one of the operators to be degenerate. 

\subsubsection{Computing Moore-Seiberg parallel transports}

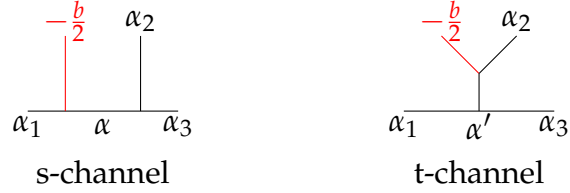
\begin{figure}[h]
    \centering
    \begin{tikzpicture}
        \draw (0,0)--(2,0);
        \node[yshift = -0.2cm] at (0,0) {$\alpha_1$};
        \node[yshift = -0.2cm] at (2,0) {$\alpha_3$};
        \node[yshift = 0.2cm][red] at (0.5,1) {$-\frac b 2$};
        \node[yshift = 0.2cm] at (1.5,1) {$\alpha_2$};
        \node[yshift = -0.2cm] at (1,0) {$\alpha$};
        \draw[draw=red] (0.5,0)-- (0.5,1);
        \draw (1.5,0)--(1.5,1);
        \node[yshift = -0.8cm] at (1,0) {s-channel};
        \draw[xshift = 5cm] (0,0)--(2,0);
        \node[xshift = 5cm, yshift = -0.2cm] at (0,0) {$\alpha_1$};
        \node[xshift = 5cm, yshift = -0.2cm] at (2,0) {$\alpha_3$};
        \node[xshift = 5cm, yshift = 0.2cm][red] at (0.5,1) {$-\frac b 2$};
        \node[xshift = 5cm, yshift = 0.2cm] at (1.5,1) {$\alpha_2$};
        \node[xshift = 5cm, yshift = -0.2cm] at (1,0) {$\alpha'$};
        \draw[xshift = 5cm, red] (1,0.5)-- (0.5,1);
        \draw[xshift = 5cm] (1,0.5)--(1.5,1);
        \draw[xshift = 5cm] (1,0)--(1,0.5);
        \node[yshift = -0.8cm, xshift = 5cm] at (1,0) {t-channel};
    \end{tikzpicture} 
    \caption{A 4-point conformal block with one degenerate operator (in red) inserted is shown in both s- and t-channel. Here, $\alpha = \alpha_1 \pm b/2$ and $\alpha' = \alpha_2 \pm b/2$. }
    \label{fig:four-point-with-degenerate}
\end{figure}

Consider a four-point conformal block $\mathcal{F}^\text{Vir}_{1,2}(z) = \mathcal{F}^\text{Vir}(V_{1,2}(z), \ldots)$ with one degenerate insertion~$V_{1,2}$. This block is illustrated in \cref{fig:four-point-with-degenerate} in the s-channel as well as the t-channel. Since the internal momenta in these channels are restricted to $\alpha = \alpha_1 \pm b/2$ and $\alpha' = \alpha_2 \pm b/2$, respectively, it follows that the corresponding fusion and braiding matrices are two-dimensional. 

In particular, if we define
\begin{align}
       F_{ss'}\begin{bmatrix}
        \alpha_2 & -b/2 \\ \alpha_3 & \alpha_1
    \end{bmatrix} :=  F_{\alpha_1 -  s\frac{b}{2}
    ,\alpha_2- s'\frac{b}{2}} \begin{bmatrix}
        \alpha_2 & -b/2 \\ \alpha_3 & \alpha_1
    \end{bmatrix}, 
\end{align}
we have that \cite{Belavin:1984vu}\footnote{These expressions can also be found in \cite[Appendix B]{Drukker:2009id}.} 
\begin{equation}
    \label{eq:fusion-matrices-can} 
   F_{++}\begin{bmatrix}
        \alpha_2 & -b/2 \\ \alpha_3 & \alpha_1
    \end{bmatrix}  = \frac{\Gamma [b (2 \alpha_1-b) ] \Gamma [b (Q-2 \alpha_2)]}{\Gamma [b (\alpha_1-\alpha_2+\alpha_3-\frac{b}{2})] \Gamma [b(\alpha_1-\alpha_2-\alpha_3+Q-\frac{b}{2})]}\,,
\end{equation} 
and, 
\begin{align}
\begin{aligned}
    F_{-+}\begin{bmatrix}
        \alpha_2 & -b/2 \\ \alpha_3 & \alpha_1
    \end{bmatrix}  &= F_{++}\begin{bmatrix}
        \alpha_2 & -b/2 \\ \alpha_3 & Q-\alpha_1
    \end{bmatrix}, \\
    F_{+-}\begin{bmatrix}
        \alpha_2 & -b/2 \\ \alpha_3 & \alpha_1
    \end{bmatrix} &= F_{++}\begin{bmatrix}
        Q-\alpha_2 & -b/2 \\ \alpha_3 & \alpha_1
    \end{bmatrix}, \text{ and }\\
    F_{--}\begin{bmatrix}
        \alpha_2 & -b/2 \\ \alpha_3 & \alpha_1
    \end{bmatrix} &= F_{++}\begin{bmatrix}
        Q-\alpha_2 & -b/2 \\ \alpha_3 & Q-\alpha_1
    \end{bmatrix} .
\end{aligned}
\end{align}

It is convenient to introduce the notation\footnote{Later, around eq.~\eqref{eq:Znek3-sphere}, we will introduce masses $\widetilde{m}_k$ such that $b \widetilde{m}_k = m_k$. In terms of the $\Omega$-deformation parameters, this means that $\widetilde m_k = \epsilon_1 m_k$ (with $R = 1$). Even though the $m_k$s make the notation most efficient here, the $\widetilde{m}_k$s are closest to the gauge theory masses. \label{footnote:hbar}} 
\begin{align}\label{eq:AGQnotation}
    m_k := b(2 \alpha_k - Q),
\end{align}
so that 
\begin{equation}
    \label{eq:fusion-matrices-can-M} 
   F_{ss'}\begin{bmatrix}
        \alpha_2 & -b/2 \\ \alpha_3 & \alpha_1
    \end{bmatrix}  = \frac{\Gamma [1+ s m_1 ] \Gamma [-s'  m_2]}{\Gamma [\frac{1}{2}(1+ s m_1 - s' m_2 + m_3 )] \Gamma [\frac{1}{2}(1+ s m_1 - s' m_2 -  m_3 )]}\,.
\end{equation}
In this form, it is easy to see that the fusion matrices are invariant under the simultaneous exchanges
\begin{equation}
    (s \leftrightarrow -s,\: m_1 \leftrightarrow -m_1),  \quad \text{ as well as } \quad  (s' \leftrightarrow -s',\: m_2 \leftrightarrow -m_2).
\end{equation} 
The first exchange in each parenthesis swaps rows/columns, and is thus equivalent to multiplication by an off-diagonal matrix. The second exchange in each bracket conjugates the momentum $\alpha_{1,2} \leftrightarrow Q-\alpha_{1,2}$. Hence,
\begin{align}
\begin{aligned}
    F_{(-s) s'}\begin{bmatrix}
        \alpha_2 & -b/2 \\ \alpha_3 & \alpha_1
    \end{bmatrix} &= F_{s s'}\begin{bmatrix}
        \alpha_2 & -b/2 \\ \alpha_3 & Q-\alpha_1
    \end{bmatrix} \quad \text{ and } \\
    F_{s (-s')}\begin{bmatrix}
        \alpha_2 & -b/2 \\ \alpha_3 & \alpha_1
    \end{bmatrix} &= F_{s s'}\begin{bmatrix}
        Q-\alpha_2 & -b/2 \\ \alpha_3 & \alpha_1
    \end{bmatrix}. \label{eq:Fchangerows}
\end{aligned}
\end{align}
These identities will come in handy when we glue 3-punctured spheres in \S\ref{sec:QPgeneralC}.

In the notation~\eqref{eq:AGQnotation}, the weights $\Delta_{\alpha_k} = \alpha_k (Q-\alpha_k)$ are given by
\begin{align}
    \Delta_{\alpha_k} = \frac{(1 + b^2)^2 - m_k^2}{4 b^2},
\end{align}
so that the \emph{clockwise} braiding matrix is computed by
\begin{align} \label{eq:braiding-matrix-can-M} 
    B^{-\frac{b}{2},\alpha_k}_{ss'} :=  B^{-\frac{b}{2},\alpha_k}_{\alpha_k - s\frac{b}{2}} \, \delta_{s s'}= e^{-\frac{i \pi}{2} (b^2+1 + s m_k)}  \delta_{s s'} = q^{\frac{1}{2}} e^{ -\frac{i \pi }{2} (1+ s m_k)} \delta_{s s'},
\end{align}
where $q=e^{-i \pi b^2}$.

\begin{figure}[h]
   \centering
  \includegraphics[width=0.35\linewidth]{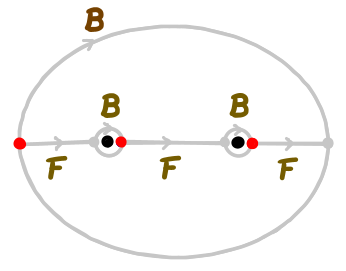}
    \caption{The fusion and braiding matrices correspond to parallel transport of the $q$-hypergeometric equation on the 3-punctured sphere. 
    } 
    \label{fig:FBtoPT}
\end{figure}

You may recognise the fusion matrices $F_{ss'}$ in equation~\eqref{eq:fusion-matrices-can-M} as the connection matrices for the hypergeometric differential equation. The braiding matrices $B_{ss'}$ in eq.~\eqref{eq:braiding-matrix-can-M} agree with this interpretation up to the factor of $q^{1/2}$. Indeed, the Lax equation on the 3-punctured sphere is described by a hypergeometric (or rather Riemann's) differential equation. This implies that the Moore-Seiberg parallel transport is identical to its semi-classical limit, up to the overall factor of $q$.

\subsubsection{BPZ equation and quantum oper connections} \label{sec:quantumoper}

In the next few paragraphs, we interpret the BPZ equation as a \emph{quantum oper connection}, and argue that the fusion and braiding matrices~\eqref{eq:fusion-matrices-can-M} are computing the monodromies of the quantum oper connection on the $3$-punctured sphere. The paths corresponding to the individual fusion and braiding matrices are illustrated in \cref{fig:FBtoPT}.

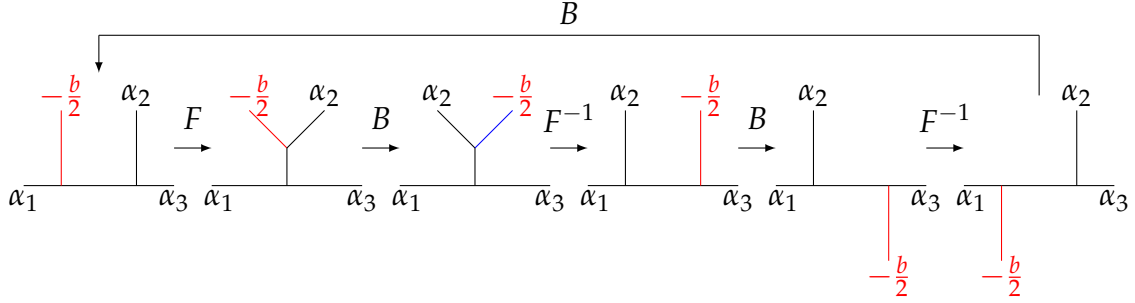
\begin{figure}[h]
   \centering
  \begin{subfigure}{\textwidth}
  \centering
      \begin{tikzpicture}
        \draw (0,0)--(2,0);
        \node[yshift = -0.2cm] at (0,0) {$\alpha_1$};
        \node[yshift = -0.2cm] at (2,0) {$\alpha_3$};
        \node[yshift = 0.2cm][red] at (0.5,1) {$-\frac b 2$};
        \node[yshift = 0.2cm] at (1.5,1) {$\alpha_2$};
        \draw[red] (0.5,0)-- (0.5,1);
        \draw (1.5,0)--(1.5,1);
        \draw [-latex] (2,0.5)--(2.5,0.5);
        \node[yshift = 0.4cm, xshift = 2.25cm] at (0,0.5) {$F$};
        \draw (2.5,0)--(4.5,0);
        \draw[red] (3.5,0.5)-- (3,1);
        \draw (3.5,0.5)--(4,1);
        \draw (3.5,0)--(3.5,0.5);
        \node[yshift = -0.2cm, xshift = 2.6cm] at (0,0) {$\alpha_1$};
        \node[yshift = -0.2cm, xshift = 2.5cm] at (2,0) {$\alpha_3$};
        \node[yshift = 0.2cm, xshift = 2.5cm][red] at (0.5,1) {$- \frac b 2$};
        \node[yshift = 0.2cm, xshift = 2.5cm] at (1.5,1) {$\alpha_2$};
        \draw [-latex] (4.5,0.5)--(5,0.5);
        \node[yshift = 0.4cm, xshift = 4.75cm] at (0,0.5) {$B$};
        \draw (5,0)--(7,0);
        \draw (6,0.5)-- (5.5,1);
        \draw[blue] (6,0.5)--(6.5,1);
        \draw (6,0)--(6,0.5);
        \node[yshift = -0.2cm, xshift = 5.1cm] at (0,0) {$\alpha_1$};
        \node[yshift = -0.2cm, xshift = 5 cm] at (2,0) {$\alpha_3$};
        \node[yshift = 0.2cm, xshift = 5 cm] at (0.5,1) {$\alpha_2$};
        \node[yshift = 0.2cm, xshift = 5 cm][red] at (1.5,1) {$-\frac b 2$};
        \draw [-latex] (7,0.5)--(7.5,0.5);
        \node[yshift = 0.4cm, xshift = 7.25cm] at (0,0.5) {$F^{-1}$};
        \draw (7.5,0)--(9.5,0);
        \draw (8,0)-- (8,1);
        \draw[red] (9,0)--(9,1);
        \node[yshift = -0.2cm, xshift = 7.6cm] at (0,0) {$\alpha_1$};
        \node[yshift = -0.2cm, xshift = 7.5cm] at (2,0) {$\alpha_3$};
        \node[yshift = 0.2cm, xshift = 7.5cm] at (0.5,1) {$\alpha_2$};
        \node[yshift = 0.2cm, xshift = 7.5cm][red] at (1.5,1) {$-\frac b 2$};
        \draw [-latex] (9.5,0.5)--(10,0.5);
        \node[yshift = 0.4cm, xshift = 9.75cm] at (0,0.5) {$B$};
        \draw (10,0)--(12,0);
        \draw (10.5,0)-- (10.5,1);
        \draw[red] (11.5,0)--(11.5,-1);
        \node[yshift = -0.2cm, xshift = 10.1cm] at (0,0) {$\alpha_1$};
        \node[yshift = -0.2cm, xshift = 10cm] at (2,0) {$\alpha_3$};
        \node[yshift = 0.2cm, xshift = 10cm] at (0.5,1) {$\alpha_2$};
        \node[yshift = -0.2cm, xshift = 10cm][red] at (1.5,-1) {$-\frac b 2$};
        \draw [-latex] (12,0.5)--(12.5,0.5);
        \node[yshift = 0.4cm, xshift = 12.25cm] at (0,0.5) {$F^{-1}$};
        \draw (12.5,0)--(14.5,0);
        \draw[red] (13,0)-- (13,-1);
        \draw (14,0)--(14,1);
        \node[yshift = -0.2cm, xshift = 12.6cm] at (0,0) {$\alpha_1$};
        \node[yshift = -0.2cm, xshift = 12.5cm] at (2,0) {$\alpha_3$};
        \node[yshift = -0.2cm, xshift = 12.5cm][red] at (0.5,-1) {$-\frac b 2$};
        \node[yshift = 0.2cm, xshift = 12.5cm] at (1.5,1) {$\alpha_2$};
        \draw (13.5, 1.2)-- (13.5, 2);
        \draw (13.5, 2)-- (1, 2);
        \draw[-latex] (1, 2)-- (1, 1.5);
        \node[yshift = 2.3cm, xshift = 7.25cm] at (0,0) {$B$};
    \end{tikzpicture}
  \end{subfigure}
  \caption{Conformal block computation corresponding to moving the degenerate vertex operator $V_{1,2}(z)$ along a trivial loop on the 3-punctured sphere.}
    \label{fig:FBFBFB}
\end{figure}

We previously noted that the fusion matrices $F_{ss'}$ in equation~\eqref{eq:fusion-matrices-can-M} are equal to the connection matrices for the hypergeometric differential equation. However, observe that the sequence of Moore-Seiberg moves (illustrated in \cref{fig:FBFBFB}) transporting the degenerate operator along a path trivial in homology (we have suppressed the internal momenta for brevity),
    \begin{align}\label{eq:qPtrivialloop}
        F\begin{bmatrix}
        -b/2 & \alpha_2 \\ \alpha_1 & \alpha_3
        \end{bmatrix} \cdot \left(B^{-\frac b 2, \alpha_2}\right) \cdot F^{-1} \begin{bmatrix}
        \alpha_2 & -b/2 \\ \alpha_1 & \alpha_3
        \end{bmatrix} \cdot \left( B^{-\frac b 2, \alpha_3}\right) \cdot F^{-1} \begin{bmatrix}
        \alpha_3 & -b/2 \\ \alpha_2 & \alpha_1
        \end{bmatrix} \cdot \left( B^{-\frac b2, \alpha_1}\right),
    \end{align}
is equal to $q^{3/2}$ times the identity\footnote{This is possible because the braiding and fusion matrices define a \emph{projective representation} of the mapping class group on conformal blocks.}; for the hypergeometric equation this monodromy is simply the identity. So, let us first understand the difference between the standard hypergeometric equation and the BPZ equation.

Indeed, suppose we consider the BPZ equation (or null-vector decoupling equation) for the four-point conformal block $\mathcal{F}^\text{Vir}_{1,2}(z)$ 
\begin{align}\label{eq:BPZ3pt}
\left( \frac{z(1 - z)}{b^2} \partial_z^2 + (2z-1) \partial_z +  \Delta_{1,2} + \frac{\Delta_1}{z} - \Delta_4 + \frac{\Delta_2}{1-z} \right) \mathcal{F}^\text{Vir}_{1,2}(z) = 0
\end{align}
as a \emph{standard} Fuchsian ODE on the 3-punctured sphere. Then, it can be brought in the form of Riemann's differential equation -- a slight generalisation of the hypergeometric differential equation -- with local exponents 
\begin{align}
\begin{aligned}\label{eq:localexponentsBPZ}
    \lambda_1^\pm &= \frac{b^2 + 1}{2} \pm \frac {m_1} 2, \\
    \lambda_2^\pm &= \frac{b^2 + 1}{2} \pm \frac {m_2} 2, \\
    \lambda_3^\pm &=  - \frac{2 b^2 + 1}{2} \pm \frac {m_3} 2,
\end{aligned}
\end{align}
at the punctures $z=0,1$ and $\infty$ respectively. Check that these exponents add up to 1, as they should for any second-order Fuchsian differential equation on the 3-punctured sphere. This ensures that the monodromies around trivial loops are indeed trivial.

Also, note that the local exponents at $z=0,1$ agree with the eigenvalues 
\begin{align}
    - q \, e^{\pm i \pi m_k}
\end{align}
of the square of the braiding matrix~\eqref{eq:braiding-matrix-can-M}. Yet, the local exponents at $z=\infty$ do not. This explains why the monodromy of the differential equation~\eqref{eq:BPZ3pt} around a trivial loop is trivial, but the monodromy in terms of braiding and fusion matrices could be non-trivial.

In a standard ODE, the local exponent at $z=\infty$ is computed through the M\"obius transformation $z \mapsto 1/z$, where $\mathcal F_{1,2}(z)$ is treated as a function on the complex plane. But, the four-point function $\mathcal{F}_{1,2}(z)$ transforms as a $\Delta_{1,2}$-differential, with 
\begin{align}
    \Delta_{1,2} = - \frac{1}{2} - \frac{3}{4} b^2.
\end{align}
This property, in particular, implies that the M\"obius transformation $z \mapsto 1/z$ will act non-trivially on the BPZ equation~\eqref{eq:BPZ3pt} for generic values of $b$. Taking this into account, we can reproduce the ``correct'' local exponents at all three punctures $z=0, 1$ and $\infty$.

\paragraph{\texorpdfstring{Oper connections\\}{Oper connections}}
Before we work this out in detail, note that $\mathcal{F}_{1,2}(z)$ transforms as a $-\frac{1}{2}$-differential when $b \to 0$. As we explain in the following, this implies that the BPZ connection reduces to an $SL(2)$-oper connection in the semi-classical limit. Indeed, it is well known that the Lax equation transforms as an oper connection (see for instance \cite{Teschner:2013tqy}). Whereas oper connections are defined in such a way that their defining equation stays invariant under holomorphic coordinate transformations\footnote{See, for instance, \cite[\S8.1]{Hollands:2017ahy} for a discussion on oper connections in accessible terms.}, this is certainly not the case for the BPZ connection at general values of $b$. 

Indeed, let us find out how the following general, second-order differential equation
\begin{align}
y''(z) + q_1(z) y'(z) + q_2(z) y(z) = 0 
\end{align}
transforms when $y(z)$ is a $\Delta_{1,2}$-differential. Under a holomorphic coordinate transformation $z \mapsto z(w)$, we have that 
\begin{align}
    \widetilde{y}(w) = y(z(w)) (z'(w))^{\Delta_{1,2}},
\end{align}
and thus 
\begin{align}
\begin{aligned}
    \widetilde{y}'(w) &= (z'(w))^{\Delta_{1,2}+1} \left( y'(z(w))  + \Delta_{1,2} \, \frac{z''(w)}{z'(w)^2} \, y(z(w)) \right), \\
     \widetilde{y}''(w) &= (z'(w))^{\Delta_{1,2}+ 2} \left( y''(z(w))  +  (2 \Delta_{1,2}+ 1) \, \frac{z''(w) }{ z'(w)^{2}} \, y'(z(w)) \right) \\
     & \quad + (z'(w))^{\Delta_{1,2}} \Delta_{1,2}  \left( (\Delta_{1,2}- 1)\, \frac{z''(w)^2}{z'(w)^2} + \frac{z'''(w)}{z'(w)} \right) y(z(w)).
\end{aligned}
\end{align}
This implies that the transformed differential equation
\begin{align}
    \widetilde{y}''(w) + \widetilde{q}_1(w) \widetilde{y}'(w) + \widetilde{q}_2(w) \widetilde{y}(w) = 0 
\end{align}
can be written in terms of the original variables as
\begin{align}\label{eq:qoper}
  & 0 \, = \, y''(z(w))  + \left( \frac{\widetilde{q}_1(w)}{z'(w)} + (2 \Delta_{1,2}+ 1) \, \frac{z''(w) }{ z'(w)^{2}} \right)  y'(z(w)) \, + \\
  &+  \left( \frac{\widetilde{q}_2(w)}{z'(w)^2} + \widetilde{q}_1(w) \Delta_{1,2} \, \frac{z''(w)}{z'(w)^3}
  + \frac{\Delta_{1,2}}{z'(w)^2} \left( (\Delta_{1,2}- 1)\, \frac{z''(w)^2}{z'(w)^2} + \frac{z'''(w)}{z'(w)} \right) \right) y(z(w)),  \notag
\end{align}
where we have not yet specified how the coefficients $q_1(z)$ and $q_2(z)$ transform. 

In particular, for weight $\Delta_{1,2} = - 1/2$ and coefficient $q_1(z)=0$, the transformed differential equation reads 
\begin{align}
  & 0 \, = \, y''(z(w))  +  \left(
  \frac{\widetilde{q}_2(w)}{z'(w)^2} +  \frac{1}{2 z'(w)^2}\left( \frac{3}{2} \frac{z''(w)^2}{z'(w)^2} - \frac{z'''(w)}{z'(w)} \right) \right) y(z(w))\\
  & \qquad = y''(z(w))  +  \left(
  \frac{\widetilde{q}_2(w)}{z'(w)^2} + \frac{1 }{2 } \{ w, z\} \right) y(z(w)),
  \notag
\end{align}
where the brackets $\{ \,.\,,\,.\,\}$ denote the Schwarzian derivative.
This implies that $q_2(z)$ should transform as a projective connection, i.e.~as
\begin{align}
    \widetilde{q}_2(w) = (z'(w))^2 \left( q_2(z(w)) - \frac{1}{2} \{ w,z \} \right),
\end{align}
in order to preserve the form of the original equation under holomorphic coordinate transformations. The resulting object is called an $SL(2)$-oper connection. (Note that the Schwarzian $\{\,.\,,\,.\,\}$ vanishes for M\"obius transformations, in which case all complications disappear, so that $q_2(z)$ just transforms as a quadratic differential.)

\paragraph{\texorpdfstring{Local exponents for $\Delta_{1,2}$-differentials\\}{Local exponents for \Delta-differentials}}
In our more general setup, we have $\Delta_{1,2} = - \frac{1}{2} - \frac{3}{4} b^2$ and $q_1(z) \neq 0$ instead. For instance, for large values of $z \gg 1$, we have
\begin{align}
  q_1(z) &= \frac{1+ \lambda^+_3 + \lambda^-_3}{z}  = -\frac{2 b^2}{z},\\
   q_2(z) & = \frac{\lambda^+_3 \lambda^-_3}{z^2} = \frac{b^4 +  b^2 + \frac{1}{4}}{z^2} - \frac{m_3^2}{4z^2}. \notag
\end{align}
In this generality, the form of the differential equation is not even preserved by M\"obius transformations. Indeed, for $z(w) = -1/w$, we compute that eq.~\eqref{eq:qoper} turns into 
\begin{align}
\begin{aligned}
  & 0 \, = \, y''(z(w))  + \left( \widetilde{q}_1(w) \, w^2  + 3 b^2 w \right)  y'(z(w)) \, + \\
  &+  \left(\widetilde{q}_2(w) \, w^4 +  \widetilde{q}_1(w) \left( 1 + \frac{3}{2} b^2 \right) w^3
  +  \frac{3}{2} b^2 \left(1+\frac{3}{2} b^2 \right) w^2  \right) y(z(w)).
  \end{aligned}
\end{align}
If we assume that the $q_k(z)$ transform as $k$-differentials under M\"obius transformations, i.e.~that $\widetilde{q}_k(w) = (z'(w))^k\, q_k(z(w))$, we find that the new coefficients $\overline{\underline{q}}_k(w)$ of the transformed BPZ equation, locally around $z(w)=0$, are given by
\begin{align}
  \overline{\underline{q}}_1(w) &= -2b^2 w + 3 b^2 w = - \frac{b^2}{z(w)}  \\
  \overline{\underline{q}}_2(w) &= \left( b^4 + b^2 + \frac{1}{4} -\frac {m_3^2}{4} \right) w^2 - 2 b^2 \left( 1 + \frac{3}{2} b^2 \right) w^2 +  \frac{3}{2} b^2 \left(1+\frac{3}{2} b^2 \right) w^2 \notag \\
  &= \left( \frac{(b^2+1)^2}{4}- \frac {m_3^2}{4} \right) \frac{1}{z(w)^2}.
\end{align}
This implies that the local exponents are given by 
\begin{align}
    \overline{\underline{\lambda}}_3^\pm = \frac{b^2+1}{2} \pm \frac {m_3}2,
\end{align}
restoring the symmetry between the local exponents at the three punctures! 

This confirms our claim that the asymmetry in the local exponents~\eqref{eq:localexponentsBPZ} is merely an artefact of not carefully taking into account the transformation properties of the BPZ equation. That is, the monodromies of the BPZ equation \emph{are} honestly computed by the braiding and fusion matrices. 

Moreover, we also see that the BPZ equation should be considered a \emph{quantum oper connection} on the holomorphic bundle of conformal blocks $\cF_{1,2}(z)$, that only reduces to a genuine oper connection in the semi-classical limit $b \to 0$. It then immediately follows that any monodromy (or parallel transport, in general) computed through the braiding and fusion matrices has an interpretation as a quantum monodromy (or quantum parallel transport). We call this the \emph{Moore-Seiberg (MS) parallel transport}.

\subsubsection*{Fenchel-Nielsen gauge}

So far, we have learned that quantum parallel transports can be computed using fusion and braiding matrices, that fusion matrices~$F$ correspond to connection coefficients for the hypergeometric differential equation, and that braiding matrices~$B$ match with local monodromy matrices up to a simple factor of $q$. In this subsection, we relate these observations to parallel transports in terms of FN coordinates, as calculated in \S\ref{sec:spectral_network} using $\mathcal{W}$-nonabelianisation.  

We introduce the MS parallel transports
\begin{align}
\begin{aligned}
 P^\mathrm{MS}_{a,b,c}&=    F\begin{bmatrix}
        \cdot & -b/2 \\ \cdot & \cdot \end{bmatrix} B^{-\frac{b}{2}, \cdot} \quad \textrm{and} \\
 (P')^\mathrm{MS}_{a,b,c} &=    F\begin{bmatrix}
        \cdot & -b/2 \\ \cdot & \cdot \end{bmatrix} \left( B^{-\frac{b}{2},\cdot} \right)^{-1}.
\end{aligned}
\end{align}
along the paths $\mathfrak{p}_{a,b,c}$ and $\mathfrak{p}'_{a,b,c}$ on the 3-punctured sphere (which were introduced in \cref{eq:pathspABC} and \cref{eq:pathspprime}).\footnote{Unfortunately, there is a clash between the subscript $b$ used to label the open paths on the 3-punctured sphere and the quantum parameter $b$ appearing in arguments and exponents. }

Since the expressions \cref{eq:fusion-matrices-can-M} for the fusion matrices in terms of the $m_k$ are devoid of any $b$-factors, and the only $b$-factors in the braiding matrices are in the form of a uniform, overall factor of $q^{\pm 1/2}$ in the front, we can conclude that the MS parallel transport matrices within the 3-punctured sphere are ``identical'' to their semi-classical limits upto an overall power of $q$. We summarise this as 
\begin{align}\label{eq:PMS}
 P^\mathrm{MS}_{a,b,c} =  q^{\frac{1}{2}}  \, P^\mathrm{Rie}_{a,b,c} \quad \textrm{and/or} 
 \quad  (P')^\mathrm{MS}_{a,b,c}  =  q^{-\frac{1}{2}}  \, (P')^\mathrm{Rie}_{a,b,c},
\end{align}
where $P^\textrm{Rie}$ computes the same parallel transport matrix for the semi-classical Riemann equation.

The above MS parallel transports can be brought in a form very similar to that of the parallel transport matrices in FN coordinates computed in \cref{eq:parallel-transport-FN}. Indeed,\footnote{Here we have labelled the first column/row with $+$ and the second column/row with $-$.} 
\begin{align}
\begin{aligned}
\label{eq:MSP}
 P^\mathrm{MS}_{b} & \begin{bmatrix}
        \alpha_2 & -b/2 \\ \alpha_3 & \alpha_1 \end{bmatrix} 
        \, = \\
        &=\, q^{\frac{1}{2}} \sqrt{-\frac{m_1}{m_2}} \,
         \begin{pmatrix}
            0 & 1 \\ -1  & 0 \end{pmatrix}
            \begin{pmatrix}
            \gamma & \alpha \, \frac{T^\textrm{\,Rie}_a T^\textrm{\,Rie}_b}{T^\textrm{\,Rie}_c} \\ \beta \, \frac{T^\textrm{\,Rie}_c}{T^\textrm{\,Rie}_a T^\textrm{\,Rie}_b} & \gamma \end{pmatrix} 
            \begin{pmatrix}
           \frac{i T_b^\textrm{\,Rie}}{\sqrt{M_2}} & 0 \\ 0  &   \frac{\sqrt{M_2}}{i T_b^\textrm{\,Rie}}   \end{pmatrix},
\end{aligned}
\end{align}
and similarly for $P^\mathrm{MS}_{a,c}$. Here, $\alpha$, $\beta$ and $\gamma$ are in agreement with (the cyclically permuted version of) equations~\eqref{eq:gammaA} and~\eqref{eq:alphabetaA}, under the identification
\begin{equation}
    M_{k}=-\exp (-i \pi m_{k}) = q \exp (- 2 \pi i b \alpha_k)
\end{equation}
between the abelian holonomies and the external momenta.
In contrast to \cref{eq:parallel-transport-FN}, the objects $T_{a,b,c}^\text{Rie}$ are not abstract coordinates, but have explicit expressions
\begin{align}\label{eq:T-BPZ-1}
   & \left( \frac{i T_b^\textrm{Rie}}{\sqrt{M_2}} \right)^2  =  \, \frac{\Gamma [1-m_1] \Gamma [1-m_2]}{\Gamma [1+m_1] \Gamma [1+m_2] } \frac{\Gamma [\frac{1}{2} (1+m_1+m_2 \pm m_3)]}{\Gamma [\frac{1}{2} (1-m_1-m_2 \pm m_3)] },
\end{align}
where the product is over all four choices of signs.

Comparing \cref{eq:MSP} with \cref{eq:parallel-transport-FN}, we note two key differences. One of the differences is the prefactor $q^{1/2}\sqrt{-m_1/m_2}$. While the $q^{1/2}$ term has its origin in the braiding matrix, the factor $\sqrt{-m_1/m_2}$ is due to the determinant of the fusion matrix 
\begin{align} \label{eq:detF}
 \det F \begin{bmatrix}
        \alpha_2 & -b/2 \\ \alpha_3 & \alpha_1
    \end{bmatrix}= \det \begin{pmatrix} F_{--} & F_{-+} \\ F_{+-} & F_{++} \end{pmatrix} = - \frac{m_1}{m_2}
\end{align}
not being equal to 1. The latter can be eliminated through a diagonal (but non-equivariant) scaling of the basis of conformal blocks at the marked points. 

The second difference is the factor $i/\sqrt{M_2}$ in front of the twist coordinate $T_b^{\text{Rie}}$. This factor is rather harmless, as it can be gauged away by a diagonal (and equivariant) $SL(2, \C)$-transformation on the basis of conformal blocks at the marked points. Instead, we could also define 
\begin{align}
 \widetilde{T}_b^\textrm{\,Rie} = \frac{i T_b^\textrm{\,Rie}}{\sqrt{M_2}}, 
\end{align}
so that the coefficients $\alpha$ and $\beta$ change into 
\begin{align}
\begin{aligned}
    \widetilde{\alpha} = -i \, \sqrt{\frac{M_1 M_2}{M_3}} \, \alpha & \quad \text{and} \quad
    \widetilde{\beta} = +i \, \sqrt{\frac{M_3}{M_1 M_2}} \, \beta = -i \, \sqrt{\frac{M_1 M_2}{M_3}} \, \alpha\,.
\end{aligned}
\end{align}
This makes explicit that the Moore-Seiberg parallel transport prefers a slightly different gauge compared to the one arising from the averaging procedure in $\mathcal{W}$-nonabelianisation.

Comparing eq.~\eqref{eq:MSP} with eq.~\eqref{eq:parallel-transport-FN}, we conclude that in the semiclassical limit, where the quantum monodromy reduces to the Lax monodromy, the FN twist coordinates on the gauged 3-sphere \emph{evaluate} to a product of gamma-functions, as expressed in eq.~\eqref{eq:T-BPZ-1}.

We may thus phrase the relation between eq.~\eqref{eq:MSP} and eq.~\eqref{eq:parallel-transport-FN} as follows. The monodromy representation of the semi-classical limit of the BPZ equation (i.e.~Riemann's differential equation) on the 3-punctured sphere is equivalent to the monodromy representation in FN coordinates on the gauged 3-sphere, when the latter coordinates are evaluated on the oper connection $\nabla^\textrm{Rie}$ associated to Riemann's differential equation. That is,
\begin{align}\label{comparison-3pt-BPZ-FN}
    T_{a,b,c} ( \nabla^\textrm{Rie} ) = T_{a,b,c}^\textrm{Rie}.
\end{align}
(Recall that the FN twist coordinate, defined through the twist-flow argument, is only determined up to adding a complex multiple of the FN length coordinate. Since $\log T$ and $\log \widetilde{T}$ differ by such a factor, the choice of $T$ versus $\widetilde{T}$ as exponentiated FN twist coordinates is not too important.)

\subsubsection{Relation to the NRS proposal} \label{sec:nrs-FN-gauge}

\begin{figure}[h]
   \centering
  \includegraphics[width=0.5\linewidth]{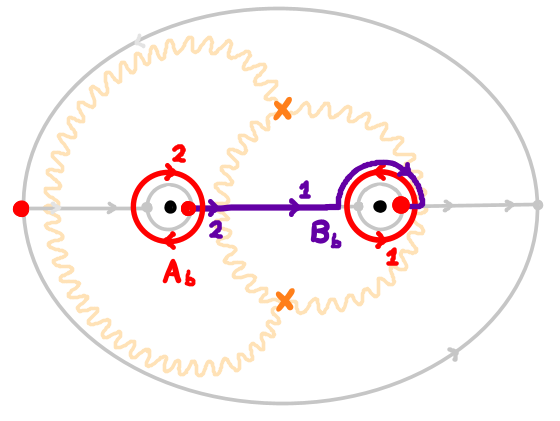}
    \caption{The 1-cycles $A_b$ and $B_b$ on the covering $\Sigma$ of the 3-punctured sphere. }
    \label{fig:A+Bcycles-3pt}
\end{figure}

The above expression~\eqref{comparison-3pt-BPZ-FN} is closely related to the NRS proposal introduced in \S\ref{subsubsec:agt-nrs}. Recall that the NRS proposal states that the effective twisted superpotential $\widetilde W_\textrm{eff}$ of a class S theory $T[C]$ can be identified with the generating function of the Lagrangian submanifold of (Lax) opers for a choice of holomorphic Darboux coordinates. For (non-exponentiated) FN-type coordinates $\{a_k, t_k\}$, this proposal can be rephrased as 
\begin{align}\label{eq:NRS}
   t_k (\nabla^\textrm{Lax}) = b \, \frac{\partial \widetilde{W}_\mathrm{eff}(\vec{a})}{\partial a_k}.
\end{align}
To write down the precise form of the NRS proposal on the gauged 3-punctured sphere, consider the symplectic basis of A and B-cycles
\begin{align}
\begin{aligned}\label{eq:ABcycles3-sphere}
A_a = \frac{1}{2} \left(\underline{\gamma}^{2}_3 - \underline{\gamma}^{1}_1 \right), \qquad & B_a = \mathfrak{p}^{21}_a, \\
A_b = \frac{1}{2} \left(\underline{\gamma}^{2}_1 - \underline{\gamma}^{1}_2 \right), \qquad & B_b = \mathfrak{p}^{21}_b, \\
A_c = \frac{1}{2} \left(\underline{\gamma}^{2}_2 - \underline{\gamma}^{1}_3 \right), \qquad & B_c = \mathfrak{p}^{21}_c,
\end{aligned}
\end{align}
on its cover $\Sigma$. This basis is illustrated in \cref{fig:A+Bcycles-3pt}.  In the above set of equations, the 1-cycle~$\underline{\gamma}^{i}$ is the lift to sheet $i$ of the 1-cycle $\underline{\mathfrak{a}}$, while the open paths $\mathfrak{p}^{ij}$ are the lifts, starting at sheet $i$ and ending at sheet $j$, of the open paths $\mathfrak{p}$. Then, the NRS proposal, in the form presented in \cite{Hollands:2021itj}\footnote{Be aware of some notational differences with \cite{Hollands:2021itj}. For instance, their $x$ would correspond to our $\frac{M}{b}$.}, tells us that the effective twisted superpotential $\widetilde{W}^\mathrm{eff}$ can be extracted from the relations
\begin{align}
\begin{aligned}
\label{eq:recipe3pt}
    t_a (\nabla^\textrm{Lax}) & =   b \left( \frac{\partial \widetilde{W}_\textrm{eff}}{\partial m_3} +\frac{\partial \widetilde{W}_\textrm{eff}}{\partial m_1} \right),  \\
    t_b (\nabla^\textrm{Lax}) & =  b \left( \frac{\partial \widetilde{W}_\textrm{eff}}{\partial m_1} +\frac{\partial \widetilde{W}_\textrm{eff}}{\partial m_2} \right),  \\
         t_c (\nabla^\textrm{Lax}) & =   b  \left( \frac{\partial \widetilde{W}_\textrm{eff}}{\partial m_2} +\frac{\partial \widetilde{W}_\textrm{eff}}{\partial m_3} \right).
\end{aligned}
\end{align}

\begin{remark}
    We should be mindful about the following. The Lax connection $\nabla^\textrm{Lax}$ on the gauged 3-sphere \emph{reduces} to Riemann's connection $\nabla^\textrm{Rie}$ on the 3-punctured sphere, but is \emph{not} strictly equal to it. In the limit where we "ungauge" the punctures, the superpotential on the gauged 3-sphere simplifies to the superpotential on the 3-punctured sphere, and thus only retains its perturbative (i.e.~classical + one-loop) part. If we therefore evaluate the FN twist coordinates $t$ on $\nabla^\textrm{Rie}$, instead of on the more general $\nabla^\textrm{Lax}$, we only expect to reproduce the perturbative part of the complete $\widetilde{W}_\textrm{eff}$ on the gauged 3-sphere. We will get back to the full statement in \S\ref{sec:quantumFN}.
\end{remark}

From the equations~\eqref{eq:T-BPZ-1} and~\eqref{comparison-3pt-BPZ-FN}, we read off that
\begin{align}
\begin{aligned}
  t_a(\nabla^\textrm{Rie}) & \sim  \log \sqrt{\frac{\Gamma [1-m_3] \Gamma [1-m_1]}{\Gamma [1+m_3] \Gamma [1+m_1] } \frac{\Gamma [\frac{1}{2} (1+m_3+m_1 \pm m_2)]}{\Gamma [\frac{1}{2} (1-m_3-m_1 \pm m_2)] }},\\
   t_b (\nabla^\textrm{Rie}) & \sim \log \sqrt{ \frac{\Gamma [1-m_1] \Gamma [1-m_2]}{\Gamma [1+m_1] \Gamma [1+m_2] } \frac{\Gamma [\frac{1}{2} (1+m_1+m_2 \pm m_3)]}{\Gamma [\frac{1}{2} (1-m_1-m_2 \pm m_3)] }},\\
     t_c (\nabla^\textrm{Rie}) & \sim \log \sqrt{ \frac{\Gamma [1-m_2] \Gamma [1-m_3]}{\Gamma [1+m_2] \Gamma [1+m_3] } \frac{\Gamma [\frac{1}{2} (1+m_2+m_3 \pm m_1)]}{\Gamma [\frac{1}{2} (1-m_2-m_3 \pm m_1)] }},
     \end{aligned}
\end{align}
where all products are over all combinations of signs, and where $\sim$ means that this equation holds up to multiplicative factors in the masses $M_k$. Making use of the identities
\begin{align}
    \frac{\Gamma[1+X]^2}{\Gamma[1-X]^2} = - \frac{\Gamma[1+X]}{\Gamma[-X]}  \frac{\Gamma[X]}{\Gamma[1-X]}  , 
\end{align}
and 
\begin{align}
 \partial_X \Upsilon(\kappa + \lambda X) = \gamma \log \frac{\Gamma[\kappa + \lambda X]}{\Gamma[1-\kappa - \lambda X]}, 
\end{align}
for the $\Upsilon$-function defined by
\begin{align}
\Upsilon(X) = \int_{\frac{1}{2}}^X \log \frac{\Gamma(X')}{\Gamma(1-X')} dX',
\end{align}
we find that 
\begin{align}
\begin{aligned}\label{eq:Weff3pt}
\widetilde{W}_\textrm{eff}(m_1,m_2,m_3)&=  \frac{1}{2b} \, \Upsilon(\pm m_1) + \frac{1}{2b} \,\Upsilon(\pm m_2) + \frac{1}{2b} \, \Upsilon(\pm m_3) \\
& \qquad - \frac{1}{2b} \, \Upsilon \left( \frac{1}{2}(1-m_1 \pm m_2 \pm m_3) \right),
\end{aligned}
\end{align}
where implicit sums are taken over all combinations of signs, 
up to quadratic terms in the masses $m_k$ (i.e.~up to the classical contributions to $\widetilde{W}_\textrm{eff}$). 

This correctly reproduces the one-loop contribution to $\widetilde{W}_\textrm{eff}$ associated with the gauged 3-sphere (up to quadratic terms in the masses $m_k$). Indeed, the 4d $\mathcal{N}=2$ field theory corresponding to the gauged 3-punctured sphere is the $SU(2)^3$ gauge theory coupled to a trifundamental half-hypermultiplet. The first line in the expression~\eqref{eq:Weff3pt} enumerates the one-loop contributions for the three gauge multiplets (with Coulomb parameters $m_{1,2,3}$). The second line computes the one-loop contribution for the trifundamental half-hypermultiplet (with masses $\pm m_1 \pm m_2 \pm m_3$).\footnote{See \cite{Hollands:2011zc} for more details on this theory and half-hypermultiplets.} 

Note that the expression~\eqref{eq:Weff3pt} is also closely related to the \emph{square-root} of the semi-classical limit of the Liouville three-point function $\mathcal C(\alpha_1, \alpha_2, \alpha_3)$ from eq.~\eqref{eq:DOZZ}
 \cite{Zamolodchikov:1995aa, Teschner:2013tqy}. Indeed, we have that 
 \begin{align}
 \begin{aligned}
      \frac{\partial}{\partial x} \left( \lim_{b \to 0} b \log \Gamma_b(x) \right) =  \log \sqrt{ \frac{b^{-b x}}{2 \pi b} } \,\Gamma(b x),
\end{aligned}
 \end{align}
(see for instance \cite[Appendix A]{Jeong:2018qpc}), so that 
\begin{align}\label{eq:Upsilon-clas}
    \lim_{b \to 0}  \log \Upsilon_b(x)   
     = \int dx \log  b^{-2-2 bx} + \frac{1}{b} \, \Upsilon(bx) ,
 \end{align}
The semi-classical limit of the Liouville three-point function~\eqref{eq:DOZZ} is therefore given by 
\begin{align}
\begin{aligned}\label{eq:Csc}
  \lim_{b \to 0} \log\mathcal  C(\alpha_1, \alpha_2, \alpha_3) &= 
 \frac{1}{b} \, \Upsilon(-m_1) + \frac{1}{b} \, \Upsilon(-m_2) + \frac{1}{b} \, \Upsilon(-m_3) \\
& \quad - \frac{1}{b} \, \Upsilon \left( \frac{1}{2} (1- m_1 \pm m_2 \pm m_3) \right) + \mathcal{O}(\log b).
\end{aligned}
\end{align}

Relations similar to eq.~\eqref{eq:recipe3pt} came up independently in the work of Lisovyy and Naidiuk~\cite{Lisovyy_2022}, where they were proven and named "Trieste formulae", as a reference to analogous equations in~\cite{Bonelli:2022ten}. Here, we want to emphasise that these relations are simply an example of more general equations that can be extracted from the "geometric recipe"~\cite{Hollands:2019wbr}, based on the NRS proposal \cite{Nekrasov:2011bc}.

\subsection{Moore-Seiberg monodromies on arbitrary surfaces}\label{sec:QPgeneralC}

Thus far, we have only considered MS parallel transports within a gauged 3-punctured sphere. In the following, we construct MS parallel transports on arbitrary surfaces through the gluing construction.

Fix an arbitrary (possibly punctured) Riemann surface $C$ together with a pants decomposition~$\mathcal{P}$. Choose a path groupoid on $C$ that respects the pants decomposition $\mathcal{P}$ and is homotopic to the path groupoid illustrated in fig.~\ref{fig:molecule1-nonab} when restricted to  any pair of pants.\footnote{If a pair of pants has punctures instead of boundary components, we may choose to "gauge" the punctures.} Then, the MS parallel transport along any path $\mathfrak{p}$ on $C$ may be computed by decomposing the path $\mathfrak{p}$ with respect to this groupoid, as
\begin{align}\label{eq:generalpathC}
    \mathfrak{p} = \mathfrak{p}^\textrm{\maljapanese\char"4E}\mathfrak{p}^\textrm{\maljapanese\char"1A} \cdots \mathfrak{p}^\textrm{\maljapanese\char"4A}, 
\end{align}
where each $\mathfrak{p}^\textrm{\maljapanese\char"4D}$ is an open path, embedded in a single pair of pants, that connects two boundary components, and is thus homotopic to either $\mathfrak{p}_{a,b,c}$ or $\mathfrak{p}'_{a,b,c}$. The MS parallel transport along $\mathfrak{p}$ is then a concatenation of the MS parallel transports 
\begin{align}
\begin{aligned}
 P^\mathrm{MS}_{\textrm{\maljapanese\char"4D}}&=    F\begin{bmatrix}
        \cdot & -b/2 \\ \cdot & \cdot \end{bmatrix} B^{-\frac{b}{2}, \cdot} \quad \textrm{and} \\
 (P')^\mathrm{MS}_{\textrm{\maljapanese\char"4D}} &=    F\begin{bmatrix}
        \cdot & -b/2 \\ \cdot & \cdot \end{bmatrix} \left( B^{-\frac{b}{2},\cdot} \right)^{-1},
\end{aligned}
\end{align}
and the braiding matrices.

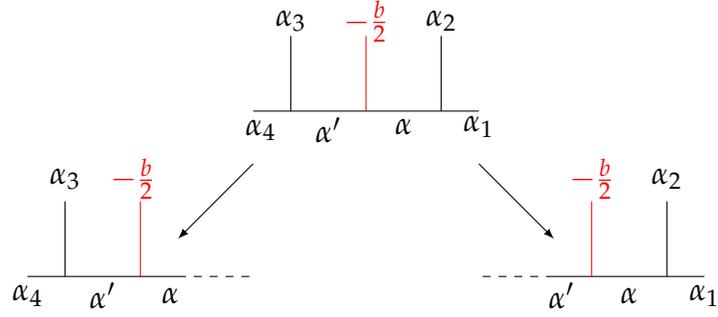
\begin{figure}
    \centering
    \begin{subfigure}{0.8\textwidth}
        \centering
        \begin{tikzpicture}
        \draw[yshift = 0.2cm] (0,0)--(3,0);
        \node[xshift = 0.1cm, yshift = -0.05cm] at (0,0) {$\alpha_4$};
        \node at (2,0) {$\alpha$};
        \node[yshift = 0.4cm] at (0.5,1) {$\alpha_3$};
        \node[yshift = 0.4cm][red] at (1.5,1) {$-\frac b 2$};
        \node[yshift = -0.05cm] at (1,0) {$\alpha'$};
        \node at (3,0) {$\alpha_1$};
        \node[yshift = 0.4cm] at (2.5,1) {$\alpha_2$};
        \draw[yshift = 0.2cm] (0.5,0)-- (0.5,1);
        \draw[yshift = 0.2cm, red] (1.5,0)--(1.5,1);
        \draw[yshift = 0.2cm] (2.5,0)--(2.5,1);
        
        \draw[-latex] (0, -0.5)--(-1, -1.5);
        \node[xshift=-0.15cm, yshift = 0.25cm] at (-0.5, -1) {\,};

        \draw[xshift = -0.5cm, yshift = -2cm] (-0.5, 0)--(-2.5, 0);
        \draw[xshift = -0.5cm, yshift = -2cm, red] (-1, 0)--(-1,1);
        \draw[xshift = -0.5cm, yshift = -2cm] (-2, 0)--(-2, 1);
        \node[xshift = -0.5cm, yshift = -1.7cm][red] at (-1.1, 1) {$-\frac b 2$};
        \node[xshift = -0.5cm, yshift = -1.7cm] at (-2, 1) {$\alpha_3$};
        \node[xshift = -0.6cm, yshift = -2.25cm] at (-0.5, 0) {$\alpha$};
        \node[xshift = -0.5cm, yshift = -2.25cm] at (-2.5, 0) {$\alpha_4$};
        \node[xshift = -0.5cm, yshift = -2.25cm] at (-1.5, 0) {$\alpha'$};
        \draw[xshift = -0.5cm, yshift = -2cm, dashed] (-0.5, 0)--(0.5,0);
        
        \draw[-latex] (3, -0.5)--(4, -1.5);
        \node[xshift=0.25cm, yshift = 0.25cm] at (3.5, -1) {\,};

        \draw[xshift = 0.5cm, yshift = -2cm] (3.5, 0)--(5.5, 0);
        \draw[xshift = 0.5cm, yshift = -2cm, red] (4, 0)--(4, 1);
        \draw[xshift = 0.5cm, yshift = -2cm] (5, 0)--(5, 1);
        \node[xshift = 0.5cm, yshift = -1.7cm][red] at (4, 1) {$-\frac b 2$};
        \node[xshift = 0.5cm, yshift = -1.7cm] at (5, 1) {$\alpha_2$};
        \node[xshift = 0.6cm, yshift = -2.25cm] at (3.5, 0) {$\alpha'$};
        \node[xshift = 0.5cm, yshift = -2.25cm] at (5.5, 0) {$\alpha_1$};
        \node[xshift = 0.5cm, yshift = -2.25cm] at (4.5, 0) {$\alpha$};
        \draw[xshift = 0.5cm, yshift = -2cm, dashed] (3.5, 0)--(2.5,0);
        
        \end{tikzpicture}
    \end{subfigure}
    
    \caption{A five-point conformal block including a single degenerate insertion on the 4-punctured sphere with a marked point. The internal momenta $\alpha$ and $\alpha'$ differ by $\pm b/2$. The internal momentum through the pants cycle depends on which side of the pants tube the degenerate operator is inserted. If it is inserted to the left the internal momentum through the pants cycle is $\alpha$, whereas if it is moved to the right of the pants tube, the internal momentum through the pants cycle is given by $\alpha'$.}
    \label{fig:five-point-conformal-block}
\end{figure}

As explained in \S\ref{subsec:bpz-monodromies}, the MS parallel transport on 3-punctured spheres is identical to its semi-classical limit, up to a factor of $q$. 
The concatenated MS parallel transport on~$C$, however, deviates from its semi-classical expression by the appearance of \emph{shifts} in the internal momenta (see for instance \cite{Drukker:2009id}). The underlying reason is that the momenta $\alpha$ and $\alpha'$ on either side of a degenerate operator $V_{1,2}$ differ by a factor $\pm b/2$. This means that the internal momentum through a pants cycle shifts by this factor when the degenerate operator is moved from one pair of pants to the other. This is illustrated in \cref{fig:five-point-conformal-block} in the example of a 4-punctured sphere. So, depending on whether the degenerate operator is to the left or the right of the pants tube, the quantum monodromy around the pants cycle is either
\begin{equation}\label{eqn:qmonA}
P^{\mathrm{MS}}_\mathfrak{a} = q \,\mathrm{Diag}\left(A, \frac 1 A\right),
     \quad \textrm{or} \quad
P^{\mathrm{MS}}_{\mathfrak{a}'} = q^{2}\, \mathrm{Diag}\left(\frac 1 A, A \right),
\end{equation}
where $A = - e^{- i \pi a} = q \, e^{-2 i \pi b \alpha  } $.

\begin{figure}[h]
     \centering
     \includegraphics[width=0.5\linewidth]{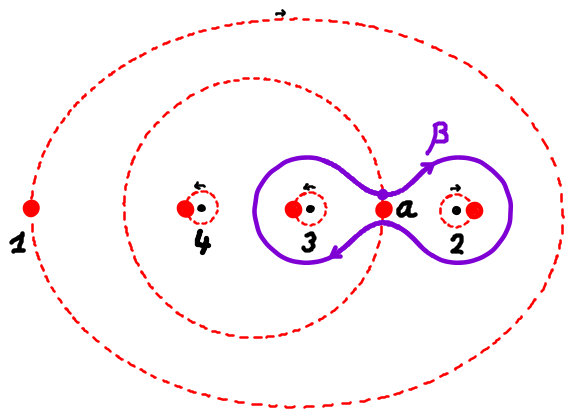}
     \caption{Path $\beta$ on the 4-punctured sphere, with base-point just above the marked point labelled $a$.}
    \label{fig:CFT-4pt-beta}
 \end{figure}

\subsubsection{Example: 4-punctured sphere}

Let us, for instance, compute the Moore-Seiberg monodromy along the path 
\begin{align}
\beta = \mathfrak{p}_{a2} \circ \mathfrak{a}_{2} \circ \mathfrak{p}^{-1}_{a2} \circ \mathfrak{p}_{a3} \circ \mathfrak{a}^{-1}_{3} \circ \mathfrak{p}^{-1}_{a3}
\end{align}
on the 4-punctured sphere, as illustrated in \cref{fig:CFT-4pt-beta}. Semi-classically, this monodromy is computed from Heun's equation, which takes up the role of the Lax connection in this example. This monodromy was explicitly calculated in a series expansion in $x$, the complex structure parameter on the 4-punctured sphere, in \cite[eq. (9.71)]{Hollands:2017ahy}. It has the leading order behaviour
\begin{align}\label{Pbetaleadingx}
  &  P^{\textrm{Heu}}_{\beta} = P_{ \mathfrak{a} 2}^\textrm{Rie} \, P^\textrm{MS}_2 \left( P_{\mathfrak{a}2}^\textrm{Rie} \right)^{-1} \mathbf{T}_\text{cl} \,P_{\mathfrak{a}3}^\textrm{Rie} \, \left(P^\textrm{MS}_3\right)^{-1} \left( P_{\mathfrak{a} 3}^\textrm{Rie} \right)^{-1}  \mathbf{T}_\text{cl}^{-1} \left( 1 + \mathcal{O}(x) \right), 
\end{align}
where 
\begin{align}
   P_{ \mathfrak{a} 2}^\textrm{Rie} = q^{-\frac{1}{2}} F \begin{bmatrix}
         \alpha_2 & -b/2 \\  \alpha_1 & -\alpha 
    \end{bmatrix}  B^{-\frac{b}{2},\alpha_2} , \quad P_{ \mathfrak{a} 3}^\textrm{Rie} = q^{-\frac{1}{2}} F \begin{bmatrix}
         \alpha_3 & -b/2 \\  \alpha_4 & \alpha 
    \end{bmatrix}  B^{-\frac{b}{2},\alpha_3} 
\end{align}
and 
\begin{align}
    \textbf{T}_\text{cl} = \text{diag} \left( x^{-\frac{1+a}{2}},x^{-\frac{1-a}{2}} \right),
\end{align}
if we make the identifications 
\begin{align}
m_{1,2} &= b(2 \alpha_{1,2} - Q) \quad \textrm{and} \quad m_{3,4} = b(Q - 2 \alpha_{3,4}),
\end{align}
consistent with our conventions in \S\ref{sec:spectral_network}.

\begin{figure}[h]
    \centering
    \begin{tikzpicture}
       \draw (0,0)--(3,0);
        \draw (0.5,0)-- (0.5,1);
        \draw[red] (1.5,0)--(1.5,1);
        \draw (2.5,0)--(2.5,1);
        \node[xshift = 0.1cm, yshift = -0.25cm] at (0,0) {$\alpha_4$};
        \node[yshift = 0.2cm] at (0.5,1) {$\alpha_3$};
        \node[yshift = -0.2cm] at (1,0) {$\alpha'$};
        \node[yshift = -0.25cm] at (2,0) {$\alpha$};
        \node[yshift = 0.2cm] at (2.5,1) {$\alpha_2$};
        \node[xshift = -0.1cm, yshift = -0.25cm] at (3,0) {$\alpha_1$};

        \draw[-latex] (3,0.5)--(3.5,0.5);
        \node[yshift = 0.3cm] at (3.25,0.5) {$F$};

        \draw[xshift = 3.5cm] (0,0)--(2.5,0);
        \draw[xshift = 3.5cm] (0.5,0)-- (0.5,1);
        \draw[xshift = 3.5cm, red] (1.5,0.5)--(1,1);
        \draw[xshift = 3.5cm] (1.5,0.5)--(2,1);
        \draw[xshift = 3.5cm] (1.5,0)--(1.5,0.5);
        \node[xshift = 3.6cm, yshift = -0.25cm] at (0,0) {$\alpha_4$};
        \node[xshift = 3.5cm, yshift = 0.2cm] at (0.5,1) {$\alpha_3$};
        \node[xshift = 3.5cm, yshift = -0.2cm] at (1,0) {$\alpha'$};
        \node[xshift = 3.5cm, yshift = 0.2cm] at (2,1) {$\alpha_2$};
        \node[xshift = 3.4cm, yshift = -0.25cm] at (2.5,0) {$\alpha_1$};

        \draw[xshift = 3.5cm, -latex] (2.5, 0.5)--(3, 0.5);
        \node[xshift = 3.5cm, yshift = 0.25cm] at (2.75, 0.5) {$B^2$};

        \draw[xshift = 6.5cm] (0,0)--(2.5,0);
        \draw[xshift = 6.5cm] (0.5,0)-- (0.5,1);
        \draw[xshift = 6.5cm, red] (1.5,0.5)--(1,1);
        \draw[xshift = 6.5cm] (1.5,0.5)--(2,1);
        \draw[xshift = 6.5cm] (1.5,0)--(1.5,0.5);
        \node[xshift = 6.6cm, yshift = -0.25cm] at (0,0) {$\alpha_4$};
        \node[xshift = 6.5cm, yshift = 0.2cm] at (0.5,1) {$\alpha_3$};
        \node[xshift = 6.5cm, yshift = -0.2cm] at (1,0) {$\alpha'$};
        \node[xshift = 6.5cm, yshift = 0.2cm] at (2,1) {$\alpha_2$};
        \node[xshift = 6.4cm, yshift = -0.25cm] at (2.5,0) {$\alpha_1$};
    
        \draw[xshift = 6.5cm, -latex] (2.5, 0.5)--(3, 0.5);
        \node[xshift = 6.65cm, yshift = 0.25cm] at (2.75, 0.5) {$F^{-1}$};

        \draw[xshift = 9.5cm] (0,0)--(3,0);
        \draw[xshift = 9.5cm] (0.5,0)-- (0.5,1);
        \draw[xshift = 9.5cm, red] (1.5,0)--(1.5,1);
        \draw[xshift = 9.5cm] (2.5,0)--(2.5,1);
        \node[xshift = 9.6cm, yshift = -0.25cm] at (0,0) {$\alpha_4$};
        \node[xshift = 9.5cm, yshift = 0.2cm] at (0.5,1) {$\alpha_3$};
        \node[xshift = 9.5cm, yshift = -0.2cm] at (1,0) {$\alpha'$};
        \node[xshift = 9.5cm, yshift = -0.2cm] at (2,0) {$\alpha''$};
        \node[xshift = 9.5cm, yshift = 0.2cm] at (2.5,1) {$\alpha_2$};
        \node[xshift = 9.5cm, yshift = -0.25cm] at (3,0) {$\alpha_1$};

        \draw[xshift = 9.5cm] (1.5, -0.5)--(1.5, -2);
        \draw[xshift = 9.5cm, -latex] (1.5, -2)--(1, -2);
        \node[xshift = 9.75cm] at (1.5, -1.5) {$F$};

        \draw[xshift = 8cm, yshift = -2.5cm] (0,0)--(2.5,0);
        \draw[xshift = 8cm, yshift = -2.5cm] (1,0.5)-- (0.5,1);
        \draw[xshift = 8cm, yshift = -2.5cm, red] (1,0.5)--(1.5,1);
        \draw[xshift = 8cm, yshift = -2.5cm] (1,0)--(1,0.5);
        \draw[xshift = 8cm, yshift = -2.5cm] (2,0)--(2,1);
        \node[xshift = 8.1cm, yshift = -2.75cm] at (0,0) {$\alpha_4$};
        \node[xshift = 8cm, yshift = -2.3cm] at (0.5,1) {$\alpha_3$};
        \node[xshift = 8cm, yshift = -2.75cm] at (1.5,0) {$\alpha''$};
        \node[xshift = 8cm, yshift = -2.3cm] at (2,1) {$\alpha_2$};
        \node[xshift = 7.9cm, yshift = -2.75cm] at (2.5,0) {$\alpha_1$};

        \draw[xshift = 5cm, yshift = -2.5cm, latex-] (2.5, 0.5)--(3, 0.5);
        \node[xshift = 5cm, yshift = -2.25cm] at (2.75, 0.5) {$B^2$};

        \draw[xshift = 5cm, yshift = -2.5cm] (0,0)--(2.5,0);
        \draw[xshift = 5cm, yshift = -2.5cm] (1,0.5)-- (0.5,1);
        \draw[xshift = 5cm, yshift = -2.5cm, red] (1,0.5)--(1.5,1);
        \draw[xshift = 5cm, yshift = -2.5cm] (1,0)--(1,0.5);
        \draw[xshift = 5cm, yshift = -2.5cm] (2,0)--(2,1);
        \node[xshift = 5.1cm, yshift = -2.75cm] at (0,0) {$\alpha_4$};
        \node[xshift = 5cm, yshift = -2.3cm] at (0.5,1) {$\alpha_3$};
        \node[xshift = 5cm, yshift = -2.75cm] at (1.5,0) {$\alpha''$};
        \node[xshift = 5cm, yshift = -2.3cm] at (2,1) {$\alpha_2$};
        \node[xshift = 4.9cm, yshift = -2.75cm] at (2.5,0) {$\alpha_1$};

        \draw[xshift = 2cm, yshift = -2.5cm, latex-] (2.5, 0.5)--(3, 0.5);
        \node[xshift = 2cm, yshift = -2.25cm] at (2.75, 0.5) {$F^{-1}$};

        \draw[xshift = 1.5cm, yshift = -2.5cm] (0,0)--(3,0);
        \draw[xshift = 1.5cm, yshift = -2.5cm] (0.5,0)-- (0.5,1);
        \draw[xshift = 1.5cm, yshift = -2.5cm, red] (1.5,0)--(1.5,1);
        \draw[xshift = 1.5cm, yshift = -2.5cm] (2.5,0)--(2.5,1);
        \node[xshift = 1.6cm, yshift = -2.75cm] at (0,0) {$\alpha_4$};
        \node[xshift = 1.5cm, yshift = -2.3cm] at (0.5,1) {$\alpha_3$};
        \node[xshift = 1.5cm, yshift = -2.75cm] at (1,0) {$\alpha'''$};
        \node[xshift = 1.5cm, yshift = -2.75cm] at (2,0) {$\alpha''$};
        \node[xshift = 1.5cm, yshift = -2.3cm] at (2.5,1) {$\alpha_2$};
        \node[xshift = 1.5cm, yshift = -2.75cm] at (3,0) {$\alpha_1$};

        \draw (1, -0.6)--(1, -2);
        \draw[-latex] (1, -2)--(1.5, -2);
        \draw (1, -0.6)--(1, -2);
        \draw[-latex] (1, -2)--(1.5, -2);
        \node[xshift = -0.5cm] at (0.8, -1.5) {$P^{\mathrm{MS}}_{\beta}$};
    \end{tikzpicture}
    \caption{Computation of the quantum parallel transport $P^{\mathrm{MS}}_{\beta}$ of a degenerate operator (in red) around the 1-cycle $\beta$ on the four-punctured sphere, illustrated in \cref{fig:CFT-4pt-beta}. }
    \label{fig:B-cycle-monodromy}
\end{figure}
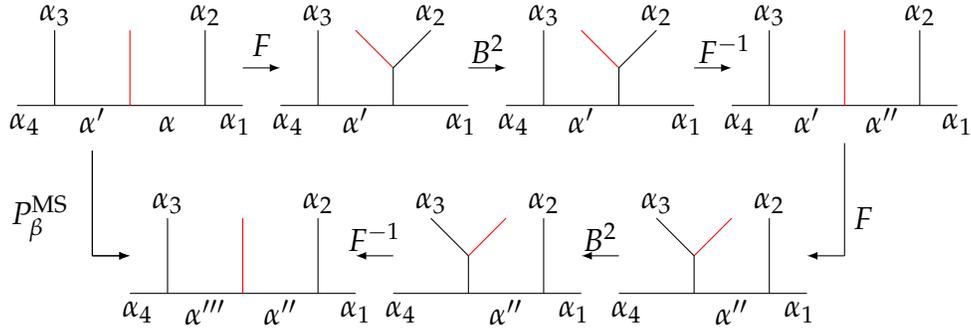

The quantum result can instead be computed through the sequence of moves shown in \cref{fig:B-cycle-monodromy}. Assuming that the degenerate operator is initially inserted to the left of the pants tube, with internal momentum $\alpha$, the quantum parallel transport along the path $\beta$ can be factorised as
\begin{align}\label{eq:quantumMbeta}
    \left[P^{\mathrm{MS}}_{\beta}\right]_{ss'} = \sum_{t = \pm 1} \left[P^{\mathrm{MS}}_{\beta_R} \right]_{-s t}\begin{bmatrix}
        -b/2 & \alpha_2 \\ \alpha' & \alpha_1
    \end{bmatrix}   \left[P^{\mathrm{MS}}_{\beta_L}\right]_{-t s'}\begin{bmatrix}
        \alpha_3 & -b/2  \\ \alpha_4 & \alpha''
    \end{bmatrix} ,
\end{align} 
where
\begin{align}
\begin{aligned}
    P^{\mathrm{MS}}_{\beta_R} \begin{bmatrix}
         \alpha_2 & -b/2  \\ \alpha_1 & \alpha' 
    \end{bmatrix} &= F\begin{bmatrix}
        \alpha_2 & -b/2  \\ \alpha_1 & \alpha'
    \end{bmatrix} \cdot \left( B^{-b/2, \, \alpha_2
    }\right)^2 \cdot F^{-1}\begin{bmatrix}
        \alpha_2 & -b/2  \\ \alpha_1 & \alpha' 
    \end{bmatrix}, \\
    P^{\mathrm{MS}}_{\beta_L} \begin{bmatrix}
        \alpha_3 & -b/2  \\ \alpha_4 & \alpha''
    \end{bmatrix} &= F\begin{bmatrix}
        \alpha_3 & -b/2  \\ \alpha_4 & \alpha''
    \end{bmatrix} \cdot \left( B^{-b/2, \, \alpha_3
    }\right)^2 \cdot F^{-1}\begin{bmatrix}
        \alpha_3 & -b/2  \\ \alpha_4 & \alpha''
    \end{bmatrix},
\end{aligned}
\end{align}
compute the quantum parallel transport around the restriction of the path $\beta$ to the right and left pair of pants, respectively, with
\begin{align}
\begin{aligned}
    \alpha' &= \alpha - s \,\frac b 2, \\
    \alpha'' &= \alpha' - t b/2 = \alpha - (s + t) \,\frac b 2,\\
    \alpha''' &= \alpha'' - s' b/2 = \alpha - (s + t + s')\, \frac b 2.
    \end{aligned}
\end{align}

We can express the result~\eqref{eq:quantumMbeta} in terms of the initial internal momentum $\alpha$ by formulating the shifted internal momenta $\alpha'$ and $\alpha''$ in terms of difference operators as
\begin{align}
    \alpha' = e^{-s \frac{b}{2} \partial_\alpha} \, \alpha \quad \textrm{and} \quad \alpha'' = e^{-(s+t) \frac{b}{2} \partial_\alpha} \, \alpha.
\end{align}
This implies that
\begin{align}
\begin{aligned}\label{eq:qmon4ptbeta}
    \left[P^{\mathrm{MS}}_{\beta} \right]_{ss'} & =  \sum_{u, t, v = \pm 1 }
     e^{u \frac{b}{2}  \partial_\alpha} \, \delta_{s, -u} \,\left[P^{\mathrm{MS}}_{\beta_R} \right]_{u t}\begin{bmatrix}
        \alpha_2 & -b/2  \\ \alpha_1 & \alpha 
    \end{bmatrix} \, e^{v \frac{b}{2}  \partial_\alpha} \,\delta_{t, -v}  \, \left[P^{\mathrm{MS}}_{\beta_L} \right]_{v s'}\begin{bmatrix}
        \alpha_3 & -b/2  \\ \alpha_4 & \alpha
    \end{bmatrix} \\
&=    \sum_{u, t, v = \pm 1 } e^{-u \frac{b}{2}  \partial_\alpha} \, \delta_{s, u} \,\left[P^{\mathrm{MS}}_{\beta_R} \right]_{u t}\begin{bmatrix}
         \alpha_2 & -b/2 \\  \alpha_1 & Q-\alpha
    \end{bmatrix} \, e^{ v \frac{b}{2}  \partial_\alpha} \,\delta_{t, v}  \, \left[P^{\mathrm{MS}}_{\beta_L}\right]_{v s'}\begin{bmatrix}
    \alpha_3 & -b/2  \\ \alpha_4 &  \alpha
    \end{bmatrix},
    \end{aligned}
\end{align}
where the difference operators act on all terms to their right. To get to the second line, we have used the relations~\eqref{eq:Fchangerows} for the fusion matrices $F$. Note that the product
\begin{align}
\begin{aligned}
P^{\mathrm{MS}}_{\beta_R} \begin{bmatrix}
         \alpha_2 & -b/2 \\  \alpha_1 & Q-\alpha 
    \end{bmatrix}  P^{\mathrm{MS}}_{\beta_L}
 \begin{bmatrix}
        \alpha_3 & -b/2  \\ \alpha_4 & \alpha
    \end{bmatrix} 
    \end{aligned}
\end{align}
indeed agrees with the one-loop contribution (that is, leaving out the matrices $\mathbf{T}_\text{cl}$ in \cref{Pbetaleadingx}) of the leading behaviour in $x$ of $P_{\beta}^\textrm{Heu}$, up to a factor $q^2$.  Higher order corrections in $x$ to $P_{\beta}^\textrm{Heu}$ originate from the action of the difference operators in $P^\textrm{MS}_{\beta}$. \newline

Let us return to the general surface $C$ now. The above computation generalises to the arbitrary path 
\begin{align}
    \mathfrak{p} = p^\textrm{\maljapanese\char"4E}p^\textrm{\maljapanese\char"1A} \cdots p^\textrm{\maljapanese\char"4A}
\end{align}
from eq.~\eqref{eq:generalpathC} as follows. Each path $\mathfrak{p}^\textrm{\maljapanese\char"4D}$ contributes the usual factor 
\begin{align}\label{eq:MSpathP}
P^\textrm{MS}_\textrm{\maljapanese\char"4D} = F \begin{bmatrix}
         \alpha_3 & -b/2 \\  \alpha_1 & \alpha_2 
    \end{bmatrix}  B^{-\frac{b}{2},\alpha_{2}}   
\end{align}
to the quantum parallel transport, where $\alpha_{1,2,3}$ are the external momenta of the corresponding pair of pants. But additionally, each time a pants tube with internal momentum $\alpha$ is crossed, we insert the matrix of difference operators\footnote{Remember that the first column/row is indexed by $+$ and the second column/row by $-$.}
\begin{equation}
    \underline{\widehat{\mathbf{T}}}_\alpha = \begin{pmatrix}
        0 & e^{-\frac{b}{2} \partial_\alpha} \\ e^{\frac{b}{2} \partial_\alpha} & 0
    \end{pmatrix}.
\end{equation}

\begin{figure}[h]
    \centering
    \begin{tikzpicture}
        \draw[xshift = -3cm] (0,0)--(2,0);
        \draw[xshift = -3cm] (1,0)--(1,1);
        \draw[xshift = -3cm] (0.4,0.1)--(0.5,0)--(0.4,-0.1);
        \draw[xshift = -3cm] (1.6,0.1)--(1.5,0)--(1.6,-0.1);
        \draw[xshift = -3cm] (0.9,0.6)--(1,0.5)--(1.1,0.6);
        \node[xshift = -3cm] at (0,-0.3) {$\alpha_1$};
        \node[xshift = -3cm] at (2,-0.3) {$Q-\alpha$};
        \node[xshift = -3cm] at (1, 1.2) {$\alpha_2$};

        \node at (-0.5, 0.75) {\small{and}};

        \draw[xshift = 0cm] (0,0)--(2,0);
        \draw[xshift = 0cm] (1,0)--(1,1);
        \draw[xshift = 0cm] (0.4,0.1)--(0.5,0)--(0.4,-0.1);
        \draw[xshift = 0cm] (1.6,0.1)--(1.5,0)--(1.6,-0.1);
        \draw[xshift = 0cm] (0.9,0.6)--(1,0.5)--(1.1,0.6);
        \node[xshift = 0cm] at (0,-0.3) {$\alpha$};
        \node[xshift = 0cm] at (2,-0.3) {$\alpha_4$};
        \node[xshift = 0cm] at (1, 1.2) {$\alpha_3$};
        
        \draw[-latex] (2,0.5)--(3,0.5);
        \node at (2.5, 0.75) {\small{conj.}};
        
        \draw[xshift = 3cm] (0,0)--(2,0);
        \draw[xshift = 3cm] (1,0)--(1,1);
        \draw[xshift = 3cm] (0.4,0.1)--(0.5,0)--(0.4,-0.1);
        \draw[xshift = 3cm] (1.4,0.1)--(1.5,0)--(1.4,-0.1);
        \draw[xshift = 3cm] (0.9,0.6)--(1,0.5)--(1.1,0.6);
        \node[xshift = 3cm] at (0,-0.3) {$\alpha_1$};
        \node[xshift = 3cm] at (2,-0.3) {$\alpha$};
        \node[xshift = 3cm] at (1, 1.2) {$\alpha_2$};
        
        \node[xshift = 3cm] at (2.5, 1) {\small{glue}};
        \node[xshift = 3cm] at (2.5, 0.60) {\small{with}};

        \draw[xshift = 6cm] (0,0)--(2,0);
        \draw[xshift = 6cm] (1,0)--(1,1);
        \draw[xshift = 6cm] (0.4,0.1)--(0.5,0)--(0.4,-0.1);
        \draw[xshift = 6cm] (1.6,0.1)--(1.5,0)--(1.6,-0.1);
        \draw[xshift = 6cm] (0.9,0.6)--(1,0.5)--(1.1,0.6);
        \node[xshift = 6cm] at (0,-0.3) {$\alpha$};
        \node[xshift = 6cm] at (2,-0.3) {$\alpha_4$};
        \node[xshift = 6cm] at (1, 1.2) {$\alpha_3$};

        \draw[-latex, xshift = 6cm] (2,0.5)--(3,0.5);

        \draw[xshift = 9cm] (0,0)--(3,0);
        \draw[xshift = 9cm] (1,0)--(1,1);
        \draw[xshift = 9cm] (2,0)--(2,1);
        \draw[xshift = 9cm] (0.4,0.1)--(0.5,0)--(0.4,-0.1);
        \draw[xshift = 9cm] (1.4,0.1)--(1.5,0)--(1.4,-0.1);
        \draw[xshift = 9cm] (0.9,0.6)--(1,0.5)--(1.1,0.6);
        \draw[xshift = 9cm] (1.9,0.6)--(2,0.5)--(2.1,0.6);
        \draw[xshift = 9cm] (2.6,0.1)--(2.5,0)--(2.6,-0.1);
        \node[xshift = 9cm] at (0,-0.3) {$\alpha_1$};
        \node[xshift = 9cm] at (1.5,-0.3) {$\alpha$};
        \node[xshift = 9cm] at (1,1.2) {$\alpha_2$};
        \node[xshift = 9cm] at (2,1.2) {$\alpha_3$};
        \node[xshift = 9cm] at (3,-0.3) {$\alpha_4$};
    \end{tikzpicture}
    \caption{Two punctures (possibly within the same surface) can be glued together if they carry conjugate momenta $\alpha$ and $Q - \alpha$, respectively. This is because after reversing the orientation of one of the momenta, this momentum is flowing in the same direction with the same magnitude as the other momentum.}
    \label{fig:orientation-reversal}
\end{figure}
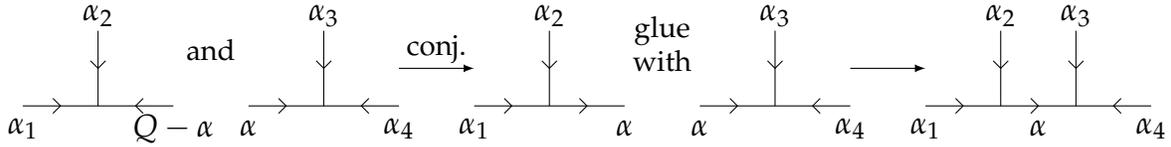

Alternatively, we can rewrite this matrix as
\begin{align} \label{eq:shift-matrix-S3}
\begin{aligned}
    \underline{\widehat{\mathbf{T}}}_\alpha &= \widehat{\mathbf{T}}^{-1}_\alpha \, \begin{pmatrix} 0 & 1 \\ 1 & 0\end{pmatrix} = \begin{pmatrix} 0 & 1 \\ 1 & 0\end{pmatrix} \, \widehat{\mathbf{T}}_\alpha  \quad \textrm{with} \quad  \widehat{\mathbf{T}}_\alpha = \begin{pmatrix}
        e^{\frac{b}{2} \partial_\alpha} & 0 \\ 0 & e^{-\frac{b}{2} \partial_\alpha}
    \end{pmatrix},
    \end{aligned}
\end{align}
and note that multiplication by the off-diagonal matrix conjugates the momentum of the adjacent three-punctured parallel transports, as in
\begin{align}
 P^{\mathrm{MS}}_\textrm{\maljapanese\char"4D} \begin{bmatrix}
         \alpha_2 & -b/2  \\ \alpha_1 & \alpha 
    \end{bmatrix} \, \begin{pmatrix} 0 & 1 \\ 1 & 0\end{pmatrix} =  \begin{pmatrix} 0 & 1 \\ 1 & 0\end{pmatrix} \, P^{\mathrm{MS}}_\textrm{\maljapanese\char"4D} \begin{bmatrix}
         \alpha_2 & -b/2  \\ \alpha_1 & Q-\alpha 
    \end{bmatrix}.
\end{align}
That is, we may also choose to insert the diagonal difference operator $\widehat{\mathbf{T}}_\alpha$ instead, while conjugating the momentum on either side of the pants tube. Here, we assign an orientation to the momentum $\alpha$ in each pants tube, and conjugation $\alpha \mapsto Q - \alpha$ reverses this orientation. This is clear from the gluing construction discussed in the previous section, and illustrated in \cref{fig:orientation-reversal}.

As an example, we can write the parallel transport $P^\textrm{MS}_{\beta}$ on the 4-punctured sphere as
\begin{equation}
    P^\textrm{MS}_\beta = \widehat{\mathbf{T}}^{-1}_{\alpha} \, P^\textrm{MS}_{\beta_R, Q-\alpha} \, \widehat{\mathbf{T}}_\alpha \, P^\textrm{MS}_{\beta_L, \alpha} \label{eq:gluing-ms-beta}
\end{equation}
where we have suppressed all the irrelevant labels in the monodromy. Furthermore, if we eliminate the diagonal difference operators $\widehat{\mathbf{T}}_\alpha$ from the resulting expression, we immediately find agreement (up to a simple factor in $q$) with the one-loop contribution to the semi-classical parallel transport computed by the Lax connection. For this reason, we work with the diagonal difference operators $\widehat{\mathbf{T}}_\alpha$ in the following.

\subsubsection*{Aside: parallel transport versus monodromy invariants}
\label{sec:parallel-vs-loop-1}

\begin{figure}[h]
    \centering
    \begin{tikzpicture}
        \draw (0,0)--(1,0);
        \draw (0.5,0)--(0.5,1);
        \node[yshift = -0.2cm] at (0,0) {$\alpha_1$};
        \node[yshift = -0.2cm, xshift = -0.1cm] at (1,0) {$\alpha_3$};
        \node[yshift = 0.2cm] at (0.5,1) {$\alpha_2$};
        
        \draw[-latex] (1,0.5)--(1.5,0.5);
        \node[yshift = 0.2cm] at (1.25, 0.5) {$\iota$};

        \draw[xshift = 1.5cm] (0,0)--(2,0);
        \draw[xshift = 1.5cm] (1.5,0)--(1.5,1);
        \draw[xshift = 1.5cm, dashed] (0.5,0)--(0.5,0.5);
        \draw[xshift = 1.5cm, red] (0.5,0.5)--(0,1);
        \draw[xshift = 1.5cm, red] (0.5,0.5)--(1,1);
        \node[yshift = -0.2cm, xshift = 1.6cm] at (0,0) {$\alpha_1$};
        \node[yshift = -0.2cm, xshift = 1.4cm] at (2,0) {$\alpha_3$};
        \node[yshift = 0.2cm, xshift = 1.5cm] at (1.5,1) {$\alpha_2$};

        \draw[xshift = 1.5cm, -latex] (2,0.5)--(2.5,0.5);
        \node[xshift = 1.5cm, yshift = 0.25cm] at (2.25, 0.5) {$F^{-1}$};

        \draw[xshift = 4cm] (0,0)--(2,0);
        \draw[xshift = 4cm] (1.5,0)--(1.5,1);
        \draw[xshift = 4cm, red] (0.5,0)--(0.5,0.75);
        \draw[xshift = 4cm, red] (1,0)--(1,0.75);
        \node[yshift = -0.2cm, xshift = 4.1cm] at (0,0) {$\alpha_1$};
        \node[yshift = -0.2cm, xshift = 3.9cm] at (2,0) {$\alpha_3$};
        \node[yshift = 0.2cm, xshift = 4cm] at (1.5,1) {$\alpha_2$};

        \draw[xshift = 4cm, -latex] (2,0.5)--(2.5,0.5);
        \node[xshift = 4cm, yshift = 0.25cm] at (2.25, 0.5) {$F$};

        \draw[xshift = 6.5cm] (0,0)--(2,0);
        \draw[xshift = 6.5cm] (1.25,0.5)--(1.5,1);
        \draw[xshift = 6.5cm, red] (0.5,0)--(0.5,0.75);
        \draw[xshift = 6.5cm, red] (1.25,0.5)--(1,1);
        \draw[xshift = 6.5cm] (1.25,0)--(1.25,0.5);
        \node[yshift = -0.2cm, xshift = 6.6cm] at (0,0) {$\alpha_1$};
        \node[yshift = -0.2cm, xshift = 6.4cm] at (2,0) {$\alpha_3$};
        \node[yshift = 0.2cm, xshift = 6.5cm] at (1.5,1) {$\alpha_2$};

        \draw[xshift = 6.5cm, -latex] (2,0.5)--(2.5,0.5);
        \node[xshift = 6.5cm, yshift = 0.25cm] at (2.25, 0.5) {$B$};

        \draw[xshift = 9cm] (0,0)--(2,0);
        \draw[xshift = 9cm, red] (1.25,0.5)--(1.5,1);
        \draw[xshift = 9cm, red] (0.5,0)--(0.5,0.75);
        \draw[xshift = 9cm] (1.25,0.5)--(1,1);
        \draw[xshift = 9cm] (1.25,0)--(1.25,0.5);
        \node[yshift = -0.2cm, xshift = 9.1cm] at (0,0) {$\alpha_1$};
        \node[yshift = -0.2cm, xshift = 8.9cm] at (2,0) {$\alpha_3$};
        \node[yshift = 0.2cm, xshift = 9cm] at (1,1) {$\alpha_2$};

        \draw[xshift = 9cm, -latex] (2,0.5)--(2.5,0.5);
        \node[xshift = 9cm, yshift = 0.25cm] at (2.25, 0.5) {$F^{-1}$};

        \draw[xshift = 11.5cm] (0,0)--(2,0);
        \draw[xshift = 11.5cm, red] (1.5,0)--(1.5,0.75);
        \draw[xshift = 11.5cm, red] (0.5,0)--(0.5,0.75);
        \draw[xshift = 11.5cm] (1,0)--(1,1);
        \node[yshift = -0.2cm, xshift = 11.6cm] at (0,0) {$\alpha_1$};
        \node[yshift = -0.2cm, xshift = 11.4cm] at (2,0) {$\alpha_3$};
        \node[yshift = 0.2cm, xshift = 11.5cm] at (1,1) {$\alpha_2$};

        \draw[xshift = 11.5cm, -latex] (1,-0.25)--(1,-0.75);
        \node[xshift = 11.5cm, xshift = 0.25cm] at (1, -0.5) {$B$};

        \draw[xshift = 11.5cm, yshift = -2cm] (0,0)--(2,0);
        \draw[xshift = 11.5cm, yshift = -2cm, red] (1.5,0)--(1.5,-0.75);
        \draw[xshift = 11.5cm, yshift = -2cm, red] (0.5,0)--(0.5,0.75);
        \draw[xshift = 11.5cm, yshift = -2cm] (1,0)--(1,1);
        \node[yshift = -2.2cm, xshift = 11.6cm] at (0,0) {$\alpha_1$};
        \node[yshift = -2.2cm, xshift = 11.4cm] at (2,0) {$\alpha_3$};
        \node[yshift = -2.2cm, xshift = 11.8cm] at (1,1) {$\alpha_2$};

        \draw[xshift = 9cm, yshift = -2cm, latex-] (2,0.5)--(2.5,0.5);
        \node[xshift = 9.1cm, yshift = -1.75cm] at (2.25, 0.5) {$F^{-1}$};

        \draw[xshift = 9cm, yshift = -2cm] (0,0)--(2,0);
        \draw[xshift = 9cm, yshift = -2cm, red] (1,0)--(1,-0.75);
        \draw[xshift = 9cm, yshift = -2cm, red] (0.5,0)--(0.5,0.75);
        \draw[xshift = 9cm, yshift = -2cm] (1.5,0)--(1.5,1);
        \node[yshift = -2.2cm, xshift = 9.1cm] at (0,0) {$\alpha_1$};
        \node[yshift = -2.2cm, xshift = 8.9cm] at (2,0) {$\alpha_3$};
        \node[yshift = -1.8cm, xshift = 9cm] at (1,1) {$\alpha_2$};

        \draw[xshift = 6.5cm, yshift = -2cm, latex-] (2,0.5)--(2.5,0.5);
        \node[xshift = 6.6cm, yshift = -1.75cm] at (2.25, 0.5) {$F^{-1}$};

        \draw[xshift = 6.5cm, yshift = -2cm] (0,0)--(2,0);
        \draw[xshift = 6.5cm, yshift = -2cm] (1.5,0)--(1.5,1);
        \draw[xshift = 6.5cm, yshift = -2cm, red] (0.5,0)--(0.5,-0.75);
        \draw[xshift = 6.5cm, yshift = -2cm, red] (1,0)--(1,0.75);
        \node[yshift = -2.2cm, xshift = 6.6cm] at (0,0) {$\alpha_1$};
        \node[yshift = -2.2cm, xshift = 6.4cm] at (2,0) {$\alpha_3$};
        \node[yshift = -1.8cm, xshift = 6.5cm] at (1.5,1) {$\alpha_2$};

        \draw[xshift = 4cm, yshift = -2cm, latex-] (2,0.5)--(2.5,0.5);
        \node[xshift = 4cm, yshift = -1.75cm] at (2.25, 0.5) {$B$};

        \draw[xshift = 4cm, yshift = -2cm] (0,0)--(2,0);
        \draw[xshift = 4cm, yshift = -2cm] (1.5,0)--(1.5,1);
        \draw[xshift = 4cm, yshift = -2cm, red] (0.5,0)--(0.5,0.75);
        \draw[xshift = 4cm, yshift = -2cm, red] (1,0)--(1,0.75);
        \node[yshift = -2.2cm, xshift = 4.1cm] at (0,0) {$\alpha_1$};
        \node[yshift = -2.2cm, xshift = 3.9cm] at (2,0) {$\alpha_3$};
        \node[yshift = -1.8cm, xshift = 4cm] at (1.5,1) {$\alpha_2$};

        \draw[xshift = 1.5cm, yshift = -2cm, latex-] (2,0.5)--(2.5,0.5);
        \node[xshift = 1.5cm, yshift = -1.75cm] at (2.25, 0.5) {$F$};

        \draw[xshift = 1.5cm, yshift = -2cm] (0,0)--(2,0);
        \draw[xshift = 1.5cm, yshift = -2cm] (1.5,0)--(1.5,1);
        \draw[xshift = 1.5cm, yshift = -2cm, dashed] (0.5,0)--(0.5,0.5);
        \draw[xshift = 1.5cm, yshift = -2cm, red] (0.5,0.5)--(0,1);
        \draw[xshift = 1.5cm, yshift = -2cm, red] (0.5,0.5)--(1,1);
        \node[yshift = -2.2cm, xshift = 1.6cm] at (0,0) {$\alpha_1$};
        \node[yshift = -2.2cm, xshift = 1.4cm] at (2,0) {$\alpha_3$};
        \node[yshift = -1.8cm, xshift = 1.5cm] at (1.5,1) {$\alpha_2$};
        
        \draw[latex-, yshift = -2cm] (1,0.5)--(1.5,0.5);
        \node[yshift = -1.8cm] at (1.25, 0.5) {$\pi$};

        \draw[yshift = -2cm] (0,0)--(1,0);
        \draw[yshift = -2cm] (0.5,0)--(0.5,1);
        \node[yshift = -2.2cm] at (0,0) {$\alpha_1$};
        \node[yshift = -2.2cm, xshift = -0.1cm] at (1,0) {$\alpha_3$};
        \node[yshift = -1.8cm] at (0.5,1) {$\alpha_2$};
    \end{tikzpicture}
    \caption{Comb diagrams for the computation of the action of a Verlinde operator supported on a trivial loop inside a three-punctured sphere. The dashed lines represent the identity operator and red lines represent the degenerate operators.}
    \label{fig:trivial-Verlinde}
\end{figure}
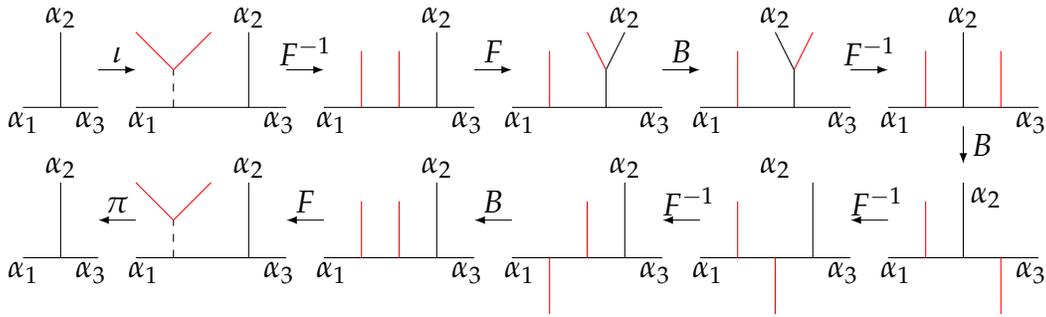

Our emphasis in this paper is on the quantum parallel transport of vertex operators, in contrast to its associated quantum monodromy invariants, which are also known as \emph{Verlinde operators} \cite{Alday:2009fs, Drukker:2009id}. In the semi-classical limit, Verlinde operators are simply given by the conventional traces of the monodromies of the Lax connection. Quantum-mechanically, though, they are computed by extending the quantum parallel transport computation with two additional steps. First, we start with inserting the identity operator into a given conformal block $\mathcal F$, while resolving it into a pair of degenerate operators $V_{1,2}(z)$ using the OPE
\begin{equation}
    \lim_{z\to 0} V_{1,2}(z) V_{1,2}(0) \sim V_{1,3}(0) + V_{1,1}(0),
\end{equation}
where $V_{1,1}(0)$ is the identity operator. Then, we take the degenerate operator $V_{1,2}(z)$ around a given loop $\gamma$ as before when computing the quantum parallel transport along $\gamma$. And finally, we fuse the two degenerate vertex operators together, and project to the channel of the identity operator. Altogether, this may be represented as a map    
\begin{equation}
    \mathcal L_\gamma := \pi \circ P^{\mathrm{MS}}_\gamma \circ \iota : \mathrm{CB}(C_{g, n}) \to \mathrm{CB}(C_{g, n}), \label{eq:def-verlinde-operator}
\end{equation}
where $\iota$ denotes the inclusion (and resolution) of the identity operator, and $\pi$ denotes the (fusion and) projection back to the identity operator (see \cref{fig:trivial-Verlinde}). The map $\mathcal L_\gamma$ is called a \emph{Verlinde operator} supported on $\gamma$, and it indeed computes a genuine quantum monodromy invariant.

\begin{remark}
    Previously, we had noted that the quantum parallel transport around a trivial loop may not be trivial because of the projective nature of the BPZ connection. Nevertheless, since Verlinde operators are quantum invariants, they \emph{are} necessarily trivial when supported on a trivial loop.

    As an example, we compute the Verlinde operator supported on a trivial loop on a three-punctured sphere corresponding to the computation~\eqref{eq:qPtrivialloop}. The sequence of moves computing this operator is illustrated in \cref{fig:trivial-Verlinde}. After some straightforward algebra, we verify that this indeed evaluates to $1$.\footnote{The loop operator is scaled by a factor of $- (q+ q^{-1})$ \cite{Drukker:2009id} for agreement with the skein algebra.}
\end{remark}

\subsubsection{Quantum Fenchel-Nielsen coordinates}\label{sec:quantumFN}

Our aim in this section is to find out how the relation~\eqref{comparison-3pt-BPZ-FN} between MS parallel transport and parallel transport expressed in FN-type coordinates on the 3-sphere generalises to an arbitrary surface $C$.

Even though the MS parallel transport on the 3-sphere only differs from its semi-classical limit by a factor of $q$, this is not the case for a general surface $C$. Therefore, an extension of relation~\eqref{comparison-3pt-BPZ-FN} can only be formulated at the semi-classical level, where we know how to evaluate FN-type spectral coordinates on oper connections such as~$\nabla_C^\textrm{Lax}$. Indeed, given any symplectic basis of closed 1-cycles on $\Sigma$, respecting a given pants decomposition, the NRS proposal \cite{Nekrasov:2011bc} can be rephrased as \cite{Hollands:2021itj} 
\begin{align}\label{eq:NRS-lengthtwist}
  t_{k} \left(\nabla^\textrm{Lax}_C \right)  = t_k^\text{Lax} = b \left( \frac{\partial{\widetilde{W}_\textrm{eff}(\vec{a})}}{\partial a_k} \right),
\end{align}
where $a_k$ and $t_k = \log T_k$ are (non-exponentiated) FN-type spectral coordinates on the surface $C$.\footnote{Note that eq.~\eqref{eq:NRS-lengthtwist} is the kind of relation we find for closed $B$-cycles, for open $B$-cycles we instead find relations of the form eq.~\eqref{eq:recipe3pt}.} This can be verified on a case by case basis as a perturbation series in the complex couplings~$\vec{x}$ for any surface~$C$, either by evaluating the expressions for the FN-type coordinates in terms of the framing data for $\nabla^\textrm{Lax}_C$ (as in \cite{Hollands:2021itj}), or by 
comparing the specific monodromy representation of the Lax connection~$\nabla^\textrm{Lax}_C$ to the generic monodromy representation in terms of the FN-type coordinates (as in \cite{Hollands:2017ahy}). 

If we send all complex structure parameters $x_k \to 0$, the Lax connection $\nabla^\textrm{Lax}_C$ degenerates into a collection of Riemann's differential equations on the individual pairs of pants. In the following, we denote the resulting ``degenerate'' Lax connection as
\begin{align}
    \nabla^\textrm{Rie}_C = \lim_{\vec{x} \to 0} \nabla^\textrm{Lax}_C.
\end{align}
As a result of the degeneration, only the perturbative contribution to the effective twisted superpotential $\widetilde{W}_\textrm{eff}(\vec{a})$ survives. We thus find that
\begin{align}
  t_{k}\left(\nabla^\textrm{Rie}_C\right)  = b \left( \frac{\partial \widetilde{W}^\textrm{pert}_\textrm{eff} (\vec{a})}{\partial a_k} \right).
\end{align}
where $\widetilde{W}^\textrm{pert}_\textrm{eff} (\vec{a})$ sums the classical and 1-loop contributions~\eqref{eq:Weff3pt} for each pair of pants in the pants decomposition of $C$. 

In the quantum setting, it is natural to define quantum FN coordinates~$X^{\mathrm{FN}}_\gamma$, and more generally any quantum spectral coordinates~$X^{\mathcal W}_\gamma$, as Heisenberg-Verlinde operators in the context of the CFT-abelianisation formalism of \cite{Hao:2024vlg}, at least for $c = 1$. That is, for any $\mathcal{W}$-abelianisation of a Liouville conformal block on $C$ in terms of a Heisenberg conformal block on $\Sigma$, the operators $X^\text{FN}_\gamma$ should be defined as computing the internal momentum through the \( \mathcal W \)-abelianised conformal block. A systematic treatment of quantum FN coordinates therefore requires an extension of CFT-abelianisation formalism to the case \( c \neq 1 \), which is the subject of~\cite{HollandsMurugesanII}. Here, we will assume that such a formalism exists and derive some implications.

We define the quantum coordinates~$X^{\mathrm{FN}}_\gamma$ with respect to the same basis of A and B-cycles on $\Sigma$, so that we still have two coordinates $X^{\mathrm{FN}}_A$ and $X^{\mathrm{FN}}_B$ for every pants cycle on $C$. Since the FN network does not cross the pants cycles, the quantum length coordinate~$X^{\mathrm{FN}}_A$ does not just measure the Heisenberg momentum, but in fact even the Liouville 
momentum $\alpha$ through the pants cycle. More precisely,
\begin{align}
X^\text{FN}_A (\nabla_C^\text{BPZ}) = - e^{i \pi a} = q \, e^{- 2 i \pi b  \alpha }. 
\end{align}
Note that this also implies that Liouville blocks on $C$, defined through the (reversed) CFT-nonabelianisation procedure with respect to an FN network, will be diagonal with respect to the quantum length coordinates. (And that this will not be the case for any other choice of spectral network.)

The quantum FN twist coordinate can be defined through the twist flow, similar as in the semi-classical set-up (see \S\ref{sec:gluing-nonab}). The twist flow acts on the BPZ connection~$\nabla_C^\text{BPZ}$ by cutting the surface $C$ along the pants cycle, then applying an automorphism to $\nabla_C^\text{BPZ}$ restricted to either half, and finally gluing back.  

In fact, since the quantum parallel transport of the BPZ connection  on $C$, calculated through MS parallel transport matrices in \S\ref{sec:QPgeneralC}, respects the chosen pants decomposition of $C$, it is reasonable to stipulate that the MS parallel transport matrices compute quantum monodromies in terms of quantum FN-type coordinates 
\begin{align}
X^{\mathrm{FN}}_A(\nabla^\text{BPZ})~~\text{and}~~ X^{\mathrm{FN}}_B(\nabla^\text{BPZ})
\end{align}
evaluated on the BPZ connection. This means that 
we can simply read off the quantum FN-twist coordinates, evaluated on the BPZ connection, from the MS parallel transports computed in \S\ref{sec:QPgeneralC}. 

Recall that MS parallel transport on a general surface $C$ differs from MS parallel transport on the 3-punctured sphere by the additional insertions of matrices~$\widehat{\mathbf{T}}_\alpha$ of difference operators
\begin{align}
\widehat{T}^\text{BPZ}_\alpha = e^{\frac{b}{2} \partial_\alpha} 
\end{align}
at the pants cycles. Since the quantum twist flow, with parameter~$\lambda$, acts by inserting the diagonal matrix $D_\lambda = \mathrm{Diag} (\lambda, \lambda^{-1})$ at the cut pants cycle, it maps
\begin{align}
    \widehat{T}^\text{BPZ}_\alpha \mapsto \lambda \widehat{T}^\text{BPZ}_\alpha.
\end{align}
This implies that $X^\text{FN}_B$ defined by
\begin{align}
\begin{aligned}
X^\text{FN}_B (\nabla_C^\text{BPZ}) =  \left( \widehat{T}^\text{BPZ}_\alpha \right)^2 = e^{b \partial_{\alpha}} \label{eq:quantum-coordinates}
\end{aligned}
\end{align}
is a quantum twist coordinate. Note that this relation indeed reproduces the correct quantum commutation relation
\begin{align}
    X^\text{FN}_A X^\text{FN}_B  = q^{-2} \, X^\text{FN}_B  X^\text{FN}_A.
\end{align}

\subsubsection{Quantum NRS proposal}
\label{subsubsec:quantum-nrs-proposal}

The family of $SL(2,\mathbb{C})$-oper connections $\nabla^\text{Lax}_C$ on a Riemann surface $C$ forms a complex Lagrangian subvariety $\mathcal{L}^\text{oper}$ of the moduli space $\mathcal{M}^\textrm{flat}(C, SL(2,\mathbb{C})$ of flat $SL(2,\mathbb{C})$-connections on $C$. As already discussed in \S\ref{sec:nrs-FN-gauge} and \S\ref{sec:quantumFN}, the NRS proposal \cite{Nekrasov:2011bc} relates the generating function of this Lagrangian subvariety, when expressed in terms of FN-type coordinates $\{a_k,t_k\}$, to the effective twisted superpotential $\widetilde{W}_\textrm{eff}(\vec{a})$ associated with the 4d $\mathcal{N}=2$ theory $T[C]$, as in  
\begin{align}\label{eq:NRS-exp}
  T_\mathfrak{a} \left(\nabla^\textrm{Lax}_C \right)  = \exp \left( b \, \frac{\partial{\widetilde{W}_\textrm{eff}(\vec{a})}}{\partial a} \right),
\end{align}

It would be exciting if we could introduce the quantum parameter $b$ in the above equation. Let us note that there are various perspectives on quantising complex Lagrangians. In this specific instance, quantising the oper Lagrangian $\mathcal{L}^\text{oper}$ was formulated in \cite{Gukov:2008ve} as computing the open string Hilbert space between the so-called "brane of opers" -- an ABA-brane supported on $\mathcal{L}^\text{oper}$-- and the "canonical coisotropic brane" -- a BBB-brane supported on the whole of $\mathcal{M}^\textrm{flat}(C, SL(2,\mathbb{C})$, both introduced in \cite{Kapustin:2006pk}. It was argued that this Hilbert space is the space of conformal blocks $\mathcal{F}$ on $C$.

Here, we want to approach this question from the CFT-abelianisation perspective of \S\ref{sec:quantumFN}. The quantum version of \cref{eq:NRS-exp} must impose that the quantum operator
\begin{align}\label{eqn:quantumNRS}
 \widehat{T}_\mathfrak{a}  - e^{b  \left( \frac{\partial{\widetilde{W}_\textrm{eff}(\vec{a})}}{\partial a} \right) + \mathcal{O}(b^2) }
\end{align}
acts trivially on the BPZ connection. Not literally, but in the sense of CFT-abelianisation, in which the quantum FN twist coordinate computes the Liouville momentum of the $\mathcal{W}$-abelianised block through the B-cycle. We should then interpret the operator~\eqref{eqn:quantumNRS} as defining a difference connection on the Hitchin section.

Note that if we bring a degenerate vertex operator $V_{1,2}(z)$ close to a puncture, we can use the OPE~\eqref{eq:OPE12alpha} between $V_{1,2}(z)$ and a generic vertex operator to write the conformal block $\mathcal{F}_{1,2}(z)$, i.e.~the BPZ connection, as a generic conformal block $\mathcal{F}$ on the surface $C$ with shifted momenta. Using the AGT correspondence, we thus conclude that a quantum operator of the form~\eqref{eqn:quantumNRS} annihilates the Nekrasov partition function $Z^\text{Nek}(\vec{a})$.

As before, we should interpret this relation at the level of CFT-abelianisation. Interestingly though, this relation does literally turn into a difference equation in case $C$ is the 3-punctured sphere. Indeed, recall that the Nekrasov partition function on the 3-punctured sphere is given by its 1-loop contribution
\begin{align}
\begin{aligned}\label{eq:Znek3-sphere}
     Z_\text{Nek}(\widetilde{m}_1,\widetilde{m}_2,\widetilde{m}_3) &= \sqrt{\frac{\Upsilon_b \left( \pm \widetilde{m}_1 \right) \Upsilon_b \left( \pm \widetilde{m}_2 \right) \Upsilon_b \left( \pm \widetilde{m}_3 \right) }{\Upsilon_b \left( \frac{Q}{2} - \frac{\widetilde{m}_1}{2} \pm \frac{\widetilde{m}_2}{2} \pm \frac{\widetilde{m}_3}{2} \right)}},
\end{aligned}
\end{align}
in the CFT normalisation (see remark~\ref{remark:1-loop}), where we defined $\widetilde{m} = m/b$. Now, the $\Upsilon_b$-functions have the property 
\begin{align}\label{eq:Upsilonb}
    \Upsilon_b(x+b) = \frac{\Gamma(bx)}{\Gamma(1-bx)} \, b^{1-2bx} \, \Upsilon_b(x), 
\end{align}
where 
\begin{align}
\lim_{b \to 0} \, \frac{\partial}{\partial x} \left( b \log \Upsilon_b(x) \right) = \log  \left( \frac{\Gamma(bx)}{\Gamma(1-bx)} \, b^{-2-2bx} \right) ,
\end{align}
as spelled out in eq.~\eqref{eq:Upsilon-clas}. This implies that (up to some overall factors of $b$)
\begin{align}\label{eqn:quantumNRS-3pt}
    \frac{Z_\text{Nek}(\widetilde{m}_1+b,\widetilde{m}_2,\widetilde{m}_3)}{Z_\text{Nek}(\widetilde{m}_1,\widetilde{m}_2,\widetilde{m}_3)} = \exp  \left( b \, \frac{\partial \widetilde{W}_\text{eff} (m_1, m_2, m_3 )}{\partial m_1}   \right), 
\end{align}
where $\widetilde{W}_\text{eff} (m_1,m_2,m_3) $ (in the CFT normalisation) is computed in eq.~\eqref{eq:Weff3pt}. In fact, the same equation holds for either normalisation of the 1-loop factors, and is similar to the quantum curve for closed moduli as introduced in \cite{Alim:2022oll}. 

It would be exciting to find a generalisation of this difference equation to other surfaces~$C$. Analogous to the known fact that 2d $\mathcal{N}=(2,2)$ theories in the $\frac{1}{2} \Omega$-background define oper equations on the UV curve $C$, we expect that 4d $\mathcal{N}=2$ theories of type $A_1$ in the full $\Omega$-background can be described in terms of a difference equation of (in general) degree $2$ defined on the base of the Hitchin integrable system (i.e.~the 4d Coulomb branch). This difference equation might just be (a refinement of) the Baxter TQ-relation \cite[eq.~7.11]{Nekrasov:2013xda} for the 4-punctured sphere. The latter is a three-term relation amongst the objects $Z_\text{Nek}(\alpha \pm b/2)$ and $Z_\text{Nek}(\alpha)$ that can be obtained from the leading contribution in $z$ of the degenerate block $\mathcal{F}_{1,2}(z)$.

\subsection{Spectral network as symmetry defect}\label{sec:symmdefects}

In the final section of this chapter we comment on the physical role of spectral networks in Liouville theory. We argue that the FN networks can be interpreted as topological defects in the CFT.

\subsubsection{Symmetry defects in QFT}

A topological defect $\mathcal{D}$ is an (extended) operator in a quantum field theory whose expectation value is invariant under small deformations of the support of the operator. In the groundbreaking work \cite{Gaiotto:2014kfa}, it was shown that generalized symmetries of a quantum field theory correspond one-to-one with topological defects. We therefore also refer to a topological defect as a \emph{symmetry defect}. 

\begin{figure}[h]
    \centering
    \includegraphics[width=0.6\linewidth]{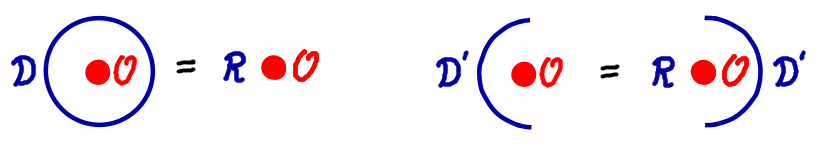}
    \caption{On the left: the symmetry defect $\mathcal{D}$ (in dark blue) acts on the (extended) operator~$\mathcal{O}$ (in red) by wrapping the support of the operator. On the right: the symmetry defect~$\mathcal{D}'$ acts on the (extended) operator $\mathcal{O}$ by moving through it.}
    \label{fig:sym-defect}
\end{figure}

A symmetry defect $\mathcal{D}$ of co-dimension $d+1$ acts on an extended operator $\mathcal{O}$ of dimension $d$ by wrapping the support of the operator around the defect. When the support of the operator is shrunk to zero size, the relation between the defect and the operator can be written in the form
\begin{align}
    \mathcal{D} \cdot \mathcal{O} = \mathcal{R} \cdot \mathcal{O},
\end{align}
where $\mathcal{R}$ is the representation with which the generalised symmetry acts on $\mathcal{O}$. Alternatively, this can be expressed as
\begin{align}
    \mathcal{D}' \cdot \mathcal{O} = \mathcal{R} \cdot \mathcal{O} \cdot \mathcal{D}', \label{eq:transport-sym-defect}
\end{align}
where the original defect $\mathcal{D}$ is split into two defects $\mathcal{D}'$, each supported on half of the original support of $\mathcal{D}$, so that the operator $\mathcal{O}$ is sandwiched between the two operators $\mathcal{D}'$ (see \cref{fig:sym-defect}). We should think of this as transporting the defect $\mathcal D$ across the operator $\mathcal O$.

Conventional $0$-form symmetries of an $n$-dimensional quantum field theory correspond to topological defects with an $(n-1)$-dimensional support. The topological defect acts on local operators according to the representation of the $0$-form symmetry in which the operator transforms.

\subsubsection{Fenchel-Nielsen networks as symmetry defects}
We now argue that Fenchel-Nielsen type networks act as $0$-form symmetry defects in Liouville theory, associated with its underlying $SL(2,\C)$ symmetry. Surely, spectral networks are real 1-dimensional objects on a Riemann surface $C$. Furthermore, the abelianisation map $\psi_\mathcal{W}$, and more generally the $q$-nonabelianisation map $\psi^q_\mathcal{W}$ (as defined in \cite{Neitzke:2020jik}), is only sensitive to the isotopy class of the spectral network $\mathcal{W}$. In the following section \S\ref{sec:q-nonabelianisation}, we also show that the MS monodromies can be reproduced through the $q$-nonabelianisation procedure with respect to an FN spectral network evaluated on conformal blocks (i.e.~applied to the BPZ connection).

Our aim is to then show that an FN network $\mathcal{W}_\mathrm{FN}$ acts on the degenerate vertex operators $V_{p,q}$ as 
\begin{align}
    \mathcal{W}_\mathrm{FN} \cdot V_{p,q} = \mathcal{R} \cdot V_{p,q} \cdot \mathcal{W}_\mathrm{FN},
\end{align}
where $\mathcal{R}$ is (a component of) the tensor product representation of the $p$th and the $q$th symmetric tensor representations of $SL(2)$ under which the degenerate vertex operators $V_{p,q}$ transform. Yet, note that acting with the spectral network $\mathcal{W}_\mathrm{FN}$ on $V_{p,q}$ is equivalent to moving the degenerate vertex operator $V_{p,q}$ through $\mathcal{W}_\mathrm{FN}$, and this multiplies the operator with the corresponding quantum parallel transport. 

As we have seen throughout this section (in the example of $(p,q)=(1,2)$), this parallel transport can be naturally brought in the FN gauge~\eqref{eq:MSP} with respect to a pants decomposition of~$C$. Hence, this allows us to conclude that FN spectral networks indeed act as symmetry defects on the vertex operators $V_{1,2}$, where the action is given by the quantum parallel transport, in FN gauge, defined by the BPZ equation for $V_{1,2}$. We expect that this statement can be easily generalised to all degenerate vertex operators $V_{p,q}$.

\section{\texorpdfstring{Quantum parallel transport from $q$-nonabelianisation}{Quantum parallel transports from q-nonabelianisation}}\label{sec:q-nonabelianisation}

As argued in the previous section, the Moore-Seiberg formalism realises a quantisation of the moduli space $\mathcal{M}(C, SL(2,\C))$ of flat connections in terms of difference operators~$\widehat{T}^\text{BPZ}$. Furthermore, these difference operators may be interpreted as quantum exponentiated FN twist coordinates $\widehat{T}$ acting on conformal blocks. 

More abstractly, the algebra of functions on the moduli space $\mathcal{M}(C, SL(2,\C))$ is quantised by the \emph{skein algebra}. To obtain a concrete realisation as an operator algebra acting on a Hilbert space, one typically embeds the skein algebra
\begin{equation}
F : \SkAlg(C,G) \longrightarrow \mathbf T
\end{equation}
into a quantum torus algebra $\mathbf T$. For non-abelian $G$, such an embedding is not canonical and requires additional structure on the surface $C$, such as a triangulation \cite{Fock:2003xxy, Bonahon_2011, Douglas:2021fyw}, a pants decomposition \cite{Detcherry_2025}, or a spectral network \cite{Gabella:2016zxu, Neitzke:2020jik, Neitzke:2021gxr}. The latter approach is known as the \emph{$q$-nonabelianisation map} \cite{Neitzke:2020jik}, and defines the quantum analogue of the (ordinary) nonabelianisation map. (See also~\cite{Galakhov:2014xba, Longhi:2016wtv} for a related construction.)

In this section we extend the 
$q$-nonabelianisation map to FN-type networks. Our aim is to find the precise relation between the Moore–Seiberg formalism and $q$-nonabelianisation. Achieving this requires addressing two important issues. First, $q$-nonabelianisation in~\cite{Neitzke:2020jik} was primarily studied for spectral networks of FG-type, so that we need to extend the formalism to spectral networks of FN-type. In particular, we need to determine how to average over resolutions. Second, the intention of \cite{Neitzke:2020jik} was to calculate quantum invariants, whereas in this paper we are mainly interested in computing quantum parallel transports.

In \S\ref{sec:basics-q-nonab}, we review the $q$-nonabelianisation map as defined in \cite{Neitzke:2020jik}, while extending it into a $q$-parallel transport map for open paths. In \S\ref{sec:q-FN-nonab}, we determine the average quantum S-matrices for FN-type spectral networks on the gauged three-punctured sphere and find agreement with the Moore-Seiberg formalism. In \S\ref{subsec:quantum-FN-coordinates-2}
we show how to extract quantum FN  coordinates from the $q$-parallel transport map. In \S\ref{subsec:gluing-q-nonab}, we extend the identification between Moore-Seiberg parallel transport and $q$-parallel transport to arbitrary surfaces~$C$. We conclude the section with an example.

\subsection{\texorpdfstring{Basics of $q$-nonabelianisation}{Basics of q-nonabelianisation}}\label{sec:basics-q-nonab}

We start with a brief recap of the $q$-nonabelianisation map which is defined in  \cite{Neitzke:2020jik} as an algebra homomorphism between the skein algebras 
\begin{equation}
    \psi^q_\mathcal W : \SkAlg(C, GL(2, \C)) \to \SkAlg(\Sigma, GL(1, \C)), \label{eq:q-nonab-def}
\end{equation}
that reduces to the ordinary nonabelianisation map~$\psi_\mathcal{W}$, defined in~\cref{eq:nonab-def}, when $q = 1$.  

The $G-$skein algebra\footnote{Here, we consider a simplification of the $G$-skein algebra called the HOMFLYPT skein algebra as in~\cite{Neitzke:2020jik}. In general, the definition of a $G$-skein algebra is more involved: the skeins may further be labelled by $G-$representations, and the interaction among skeins labelled by different $G-$representations is captured by the braided tensor category $\mathrm{Rep}_q(G)$ \cite[\S2]{Gunningham:2024hpp}. Unless explicitly stated, all the skeins (links and open paths) we consider, either on the base or the cover, are labelled by the defining representation.} 
\begin{equation}
    \SkAlg(C, G) := \Z[q^{\pm 1}] \langle \pi_1^\textrm{fr}(C \times I )\rangle/ \{ G-\text{skein relations}\}
\end{equation}
is defined as the free $\Z[q^{\pm 1}]-$module generated by isotopy classes of framed, oriented links in the three-dimensional space $C \times I$, modulo skein relations determined by the group $G$. The equivalence classes inside $\SkAlg(C, G)$ are known as \emph{skeins}. For $N\ge 2$, the $GL(N, \C)-$skein relations are given by the $U(N)$ HOMFLYPT relations
\begin{equation}
    \begin{tikzpicture}
        \draw[thick, -latex] (-0.5,-0.5)--(0.5,0.5);
        \draw[thick] (0.5,-0.5)--(0.1,-0.1);
        \draw[thick, -latex] (-0.1,0.1)--(-0.5,0.5);
        \draw[xshift=1cm] (-0.1,0)--(0.1,0);
        \draw[xshift=2cm, thick] (-0.5,-0.5)--(-0.1, -0.1);
        \draw[xshift=2cm, thick, -latex] (0.1, 0.1)--(0.5,0.5);
        \draw[xshift=2cm, thick, -latex] (0.5,-0.5)--(-0.5,0.5);
        \draw[xshift=3cm] (-0.1,0.05)--(0.1,0.05);
        \draw[xshift=3cm] (-0.1,-0.05)--(0.1,-0.05);
        \node[xshift=4cm] at (0.5,0) {$(q-q^{-1})$};
        \draw[xshift=6cm, -latex, thick] (-0.5,-0.5)--(-0.5,0.5);
        \draw[xshift=6.5cm, -latex, thick] (-0.5,-0.5)--(-0.5,0.5);
        \node[xshift=7cm] at (0,0) {and};
        \draw[xshift=9cm, thick] (0,0) arc(0:360:0.5cm) node [currarrow, pos = 0.9, sloped, xscale = 1] {};
        \draw[xshift=9.5cm] (-0.1,0.05)--(0.1,0.05);
        \draw[xshift=9.5cm] (-0.1,-0.05)--(0.1,-0.05);
        \node[xshift=10.5cm] at (0,0) {$\frac {q^N - q^{-N}}{q - q^{-1}}$};
        \node[xshift=11.3cm] at (0,0) {,};
    \end{tikzpicture} \label{fig:GLN-skeins}
\end{equation}
depicted here with blackboard framing, together with the "twist" relation
\begin{equation}
    \begin{tikzpicture}
        \draw[thick] (-0.5,-0.5)--(-0.5,0);
        \draw[thick] (-0.5,0) arc(180:-120:0.15);
        \draw[thick, -latex] (-0.49,0.17)--(-0.49,0.6);
        \draw[xshift=0.75cm] (-0.1,0.05)--(0.1,0.05);
        \draw[xshift=0.75cm] (-0.1,-0.05)--(0.1,-0.05);
        \node[xshift=1.5cm] at (0,0) {$q^N$};
        \draw[xshift=2.5cm, thick, -latex] (-0.5,-0.5)--(-0.5,0.6);
        \node[xshift=2.2cm] at (0,0) {.};
    \end{tikzpicture}
    \label{fig:GLN-framing}
\end{equation}

The abelian skein algebra $\SkAlg(\Sigma, GL(1, \C))$ is a $q$-deformation of the homology of~$\Sigma$ modified in the presence of branch-points. The $GL(1, \C)-$skein relations are thus simply given by
\begin{equation}
    \begin{tikzpicture}
        \draw[thick, -latex] (-0.5,-0.5)--(0.5,0.5);
        \draw[thick] (0.5,-0.5)--(0.1,-0.1);
        \draw[thick, -latex] (-0.1,0.1)--(-0.5,0.5);
        \draw[xshift=1cm] (-0.1,0.05)--(0.1,0.05);
        \draw[xshift=1cm] (-0.1,-0.05)--(0.1,-0.05);
        \node[xshift=2cm] at (-0.5,0) {$q$};
        \draw[xshift=2.25cm, -latex, thick] (-0.5,-0.5)--(-0.5,0.5);
        \draw[xshift=2.75cm, -latex, thick] (-0.5,-0.5)--(-0.5,0.5);
        \draw[xshift=2.75cm] (-0.1,0.05)--(0.1,0.05);
        \draw[xshift=2.75cm] (-0.1,-0.05)--(0.1,-0.05);
        \node[xshift=3.75cm] at (-0.5,0) {$q^2$};
        \draw[xshift=4cm, thick] (-0.5,-0.5)--(-0.1, -0.1);
        \draw[xshift=4cm, thick, -latex] (0.1, 0.1)--(0.5,0.5);
        \draw[xshift=4cm, thick, -latex] (0.5,-0.5)--(-0.5,0.5);
        \node[xshift=4.6cm] at (0,0) {,};

        \draw[xshift=6cm, thick] (0,0) arc(0:360:0.5cm) node[currarrow, pos = 0.9, sloped, xscale = 1] {};
        \draw[xshift=6.5cm] (-0.1,0.05)--(0.1,0.05);
        \draw[xshift=6.5cm] (-0.1,-0.05)--(0.1,-0.05);
        \node[xshift=7.1cm] at (0,0) {$1$};
        \node[xshift=7.7cm] at (0,0) {, and};

        \draw[xshift=8.5cm, -latex, thick] (0,-0.5)--(0,0.5);
        \draw[xshift=8.75cm, thick, orange] (0.05,-0.05)--(-0.05,0.05);
        \draw[xshift=8.75cm, thick, orange] (-0.05,-0.05)--(0.05,0.05);
        \draw[xshift=9.25cm] (-0.1,0.05)--(0.1,0.05);
        \draw[xshift=9.25cm] (-0.1,-0.05)--(0.1,-0.05);
        \draw[xshift=9.5cm, thick] (0,0)--(0.25,0);
        \draw[xshift=10cm, thick, orange] (0.05,-0.05)--(-0.05,0.05);
        \draw[xshift=10cm, thick, orange] (-0.05,-0.05)--(0.05,0.05);
        \draw[xshift=10.25cm, -latex, thick] (0,-0.5)--(0,0.5);
        \node[xshift=10.5cm] at (0,0) {.};
    \end{tikzpicture}
    \label{fig:GL1-Skein}
\end{equation}

There is a canonical isomorphism between the abelian skein algebra $\SkAlg(\Sigma, GL(1, \C))$ and the quantum torus $\mathbf T_\Gamma$, with $\Gamma = H_1(\Sigma; \mathbb{Z})$. The latter is the $\mathbb{Z}[q,q^{-1}]$-algebra with the basis $\{ X_\gamma \}_{\gamma \in \Gamma}$ and the product rule
\begin{align}
    X_\gamma X_{\gamma'} = (-q)^{-\langle \gamma, \gamma'\rangle} X_{\gamma + \gamma'}. \label{eq:q-commute}
\end{align} 
(The minus sign on the RHS is necessary to encode the branch-point relation in the abelian skein algebra.) The isomorphism is specified as follows: for a link $\tilde L \in \Sigma \times I$, let $\gamma(\tilde L) \in H_1(\Sigma; \Z)$ denote its projection to homology. Then, we define the quantum torus generator corresponding to $\tilde L$ to be \cite[\S3.5]{Neitzke:2020jik} 
\begin{equation}
    X_{\tilde L} := (-1)^{n'(\tilde L)}q^{n(\tilde L)}X_{\gamma(\tilde L)}. \label{eq:skein-torus-iso}
\end{equation}
Here, $n(\tilde L)$ is the number of signed self-intersections (also known as the writhe of $\tilde L$) in the projection of $\tilde L$ to $\Sigma$, whereas $n'(\tilde L)$ is the number of points in the projection of $\tilde L$ to $\Sigma$ that are not self-intersections in $\Sigma$ but become self-intersections when further projected down to $C$. Simply put, the isomorphism is given by taking the homology class of $\tilde L$ while keeping track of the $GL(1, \C)-$skein relations.

Through this isomorphism, the $q$-nonabelianisation map $\psi^q_\mathcal{W}$ defines a homomorphism from the nonabelian skein algebra into the quantum torus $\mathbf{T}_\Gamma^{\mathcal W}$ with generators $X_\gamma^\mathcal{W}$. The generators $X^\mathcal{W}_\gamma$ 
should be interpreted as a \emph{quantisation} of the spectral coordinates (\cref{eq:holonomy-function})
\begin{equation}
    \mathcal X^\mathcal{W}_\gamma : \nabla^\textrm{ab} \mapsto \C.
\end{equation}
Indeed, note that the product~\eqref{eq:q-commute} reduces to the semi-classical product~\eqref{eq:Xalgebra} when setting $q=1$, while changing the sign $\tau(\hat{\gamma}) \mapsto -\tau(\hat{\gamma})$. In particular, this implies that $q$-nonabelianisation with respect to an Fenchel-Nielsen type network $\mathcal W^\text{FN}$ gives a new definition of quantum FN coordinates $X^\text{FN}_\gamma$.\footnote{At the Fenchel-Nielsen phase, the skein algebra is mapped into a localised quantum torus algebra in order to handle the infinite sums that may appear \cite{Detcherry_2025}. We will not be too pedantic about this distinction in this paper.}  To compare this definition with the discussion of these coordinates in \S\ref{sec:quantumFN}, we need to be able to represent the formal variables $X_\gamma^\mathcal{W}$ as operators acting on conformal blocks. We will explain how this works in \S\ref{sec:CStheory}.

Although $q$-nonabelianisation has so far been formulated for the $GL(2, \C)$-skein algebra, it restricts to the $SL(2, \C)$-skein algebra upon projecting the image to the quantum subtorus generated by $\Gamma^\textrm{odd} = H_1^\textrm{odd}(\Sigma; \Z)$. Generally, this amounts to the condition that
\begin{equation}
    X_{\, \sigma_*  \gamma} = X_{-\gamma} \quad \text{ for all }~\gamma \in H_1(\Sigma; \Z), \label{eq:odd-torus-projection}
\end{equation}
where $\sigma$ is a covering involution on $\Sigma$. That $q$-nonabelianisation still leads to an algebra homomorphism from $\SkAlg(C, SL(2, \C))$ into $\mathbf T_{\Gamma^\textrm{odd}}$ was conjectured in \cite[\S9]{Neitzke:2020jik}, and proven in \cite{panitch20243dquantumtracemap}. Since our interest lies in the quantisation of the moduli space of flat $SL(2, \C)$-connections, henceforth, we shall implicitly assume the above restriction of our $q$-nonabelianisation map. Note that this restriction reduces to the equivariance condition on $GL(1, \mathbb{C})$-connections in ordinary $SL(2,\mathbb{C})$-nonabelianisation.

\subsubsection{Rules} \label{subsubsec:q-nonab-rules}
\begin{figure}[h]
    \centering
    \includegraphics[width=0.35\linewidth]{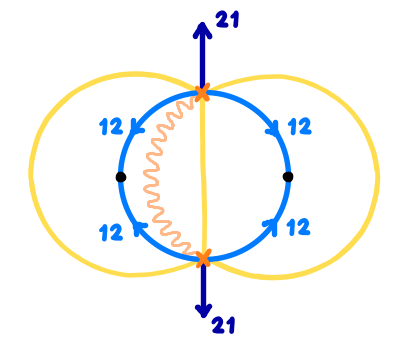}
    \caption{Leaf space (in yellow) for a simple FG spectral network (in blue) on the three-punctured sphere.}
    \label{fig:leaf-space}
\end{figure}

Whereas ordinary nonabelianisation only depends on the isotopy class of the spectral network $\mathcal W$, $q$-nonabelianisation makes use of the full WKB foliation -  introduced in~\S\ref{sec:spectral_network} as the set of \emph{all} $ij-$trajectories (and not just the critical ones). The WKB foliation can be conveniently captured in terms of a trivalent graph $\mathcal W^L$ on $C$, known as the \emph{leaf space}, which intersects every $ij-$trajectory exactly once. As an example, \cref{fig:leaf-space} illustrates the leaf space of the FG network in \cref{fig:FG-example}, which is isotopic to the FN network in \cref{fig:FN-example}. Taking the Cartesian product of the spectral network $\mathcal W$ (and its leaf space $\mathcal W^L$) with the interval $I$, we get a 3d spectral network $\mathcal W_{3d} \subset C \times I$ (and its leaf space $\mathcal W^L_{3d}$). We refer the reader to \cite[\S5]{Neitzke:2020jik} for nice illustrations of 3d leaf spaces.

By making small perturbations if necessary, we assume that the projection of $L$ onto the 3d leaf space~$\mathcal W^L_{3d}$ only has finitely many points of self-intersection and that $L$ is generically transverse (i.e.~not parallel) to the fixed height slices $C \times \{t\}$. (This is all to avoid infinite contributions from exchanges.) The $q$-nonabelianisation of $L$ is then defined as
\begin{align}
    \psi^q_\mathcal W(L) = \sum_{\tilde L} \,\alpha(\tilde L) \, X_{\tilde L}^\mathcal{W} \quad \text{with} \quad \alpha(\tilde L) \in \mathbb{Z}[q^{\pm 1}],
\end{align}
where $\tilde L$ are \emph{closed} lifts of $L$ to the cover $\Sigma \times I$, and the $X^\mathcal W_{\tilde L}$ (defined in \cref{eq:skein-torus-iso}) are formal variables valued in the (odd) quantum torus $\mathbf T_\Gamma^\mathcal W$.

\begin{figure}[h]
    \centering
    \includegraphics[width=0.75\linewidth]{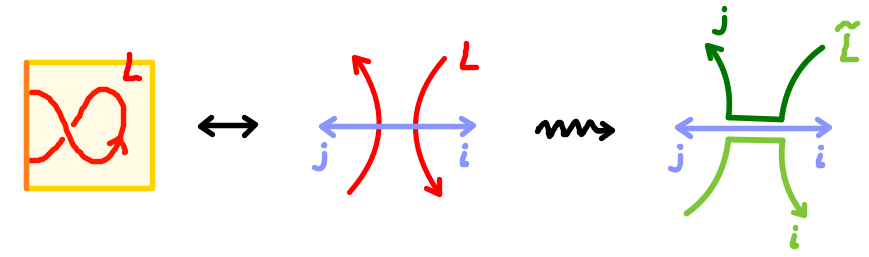}
    \caption{On the left: an exchange occurs when a link $L$ (in red) self-intersects upon projection to the 3d leaf space $\mathcal W^L_{3d}$ (in yellow). In the middle: this is equivalent to two different segments of $L$ crossing the same $ij$-trajectory (in light blue). On the right: in such case, depending on the orientations, the lifts $\tilde{L}$ (in light and dark green) of the two segments may exchange their strands.}
    \label{fig:exchanges-eg}
\end{figure}

The lifts $\tilde L$ are composed of three local pieces: direct lifts, detours and exchanges. We are already familiar with direct lifts and detours. Exchanges are a novel feature of $q$-nonabelianisation: when two segments of $L$ intersect a single $ij-$trajectory at a fixed height, i.e.~when the projection of $L$ onto $\mathcal W^L_{3d}$ self-intersects, the lifts of the two segments may exchange their strands along the $ij-$trajectory. This is illustrated in \cref{fig:exchanges-eg}. As we shall see in the next subsection, exchanges will not play a role in this paper.

\begin{figure}[h!]
     \centering
     \includegraphics[width=0.65\linewidth]{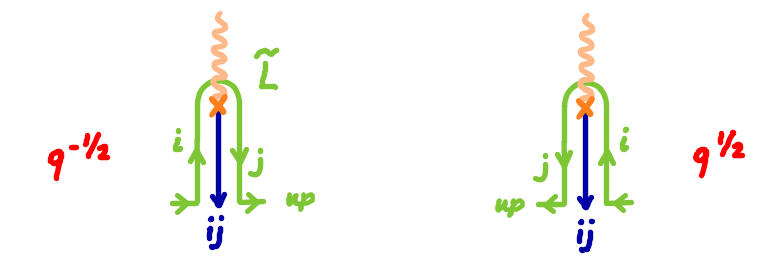}
     \caption{If an upward-travelling detour $\tilde L$ (in light green) undergoes a clockwise (on the left), or anti-clockwise (on the right), detour along a critical $ij-$trajectory (in dark blue), it is accompanied by a factor of $q^{-1/2}$ or $q^{1/2}$, respectively. For a downward-travelling detour, the $q$-factors are reversed.}
     \label{fig:framing-detour}
 \end{figure}

Both detour paths and exchanges are accompanied by an additional $q$-factor; direct lifts carry no additional $q$-factors. All these $q$-factors contribute to the coefficients $\alpha(\tilde L)$. Depending on the orientation of a detour path around the corresponding branch-point, i.e.~clockwise or anti-clockwise with respect to the orientation of $\tilde{L}$ in the height-direction, a detour contributes a \emph{detour factor} $q^{\pm 1/2}$. This is illustrated in \cref{fig:framing-detour}. Exchanges are instead accompanied by a factor proportional to $(q-q^{-1})$.

\begin{figure}[h]
    \centering
    \includegraphics[width=0.7\linewidth]{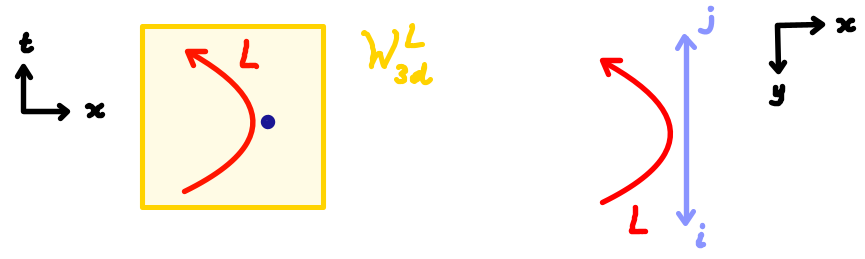}
    \caption{Locally, in a neighbourhood of a point of tangency, parametrise the base $C$ by (real) coordinates $x,y$, in such a way that the WKB foliation is parallel to $\partial_y$, and parametrise the height-direction by $t$. Then the projection of the link $L$ to $C$, at this point of tangency, must be tangential to an $ij$-trajectory.}
    \label{fig:tangency}
\end{figure}

There is one other source of $q$-factors, the \emph{winding factors}. To explain these, we consider the restriction of the canonical vector field $\partial_t$ on $C \times I$, for $t \in I$,  to the leaf space $\mathcal W_{3d}^L$. By making small perturbations if necessary, a link $L \subset C \times I$ can be positioned such that its projection to $\mathcal W^L_{3d}$ is tangential to the canonical vector field only at finitely many points. We refer to these points as \emph{points of tangency}. By making more small perturbations if necessary, we can ensure that only the projection of $L$ to $\mathcal W_{3d}^L$, and \emph{not} $L$ itself, is tangential to the vector field $\partial_t$ at all points of tangency. This implies that the projection of $L$ to $C$ is tangential to the WKB foliation at any point of tangency. This is illustrated in \cref{fig:tangency}. 

\begin{figure}[h]
    \centering
    \includegraphics[width=0.45\linewidth]{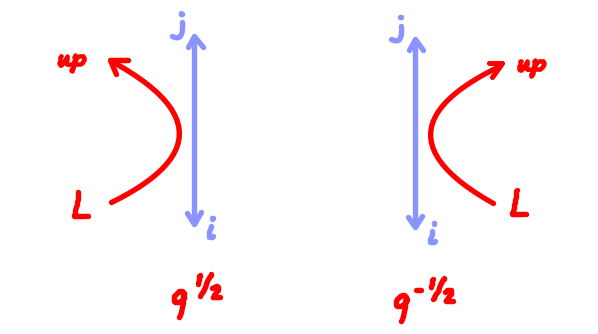}
    \caption{Any point of tangency of a link $L$ (in red) contributes an overall winding factor, which depends both on the orientation of the link with respect to the $ij-$trajectory (in light blue) and on the tangency (up or down) of $L$. The illustrated $q$-factors are assigned to upward traveling links, they are reversed for downward traveling links.}
    \label{fig:overall-q-factor}
\end{figure}

\begin{figure}[h]
    \centering
    \includegraphics[width=0.45\linewidth]{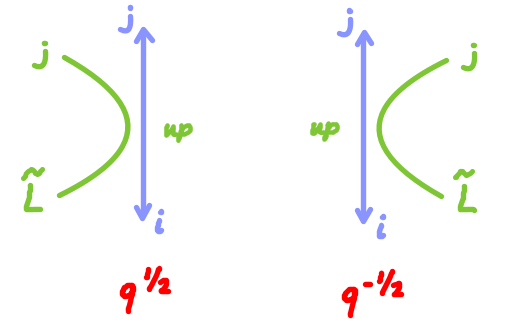}
    \caption{A point of tangency of the projection of a lift $\tilde{L}$ (in light green) contributes a sheet-dependent $q$-factor, which depends only on the orientation of the winding with respect to the $ij$-trajectory and the tangency (up or down) of $\tilde{L}$. The illustrated factors are assigned to upward travelling links, they are reversed for downward travelling links. }
    \label{fig:sheet-q-factor}
\end{figure}

The points of tangency contribute $q$-factors to $\alpha(\tilde L)$. In \cite{Neitzke:2020jik}, these contributions were separated into two local pieces: an overall winding factor and a sheet-dependent winding factor: 

\begin{itemize}
    \item Each point of tangency of the link $L$ receives an overall winding factor of $q^{\pm 1/2}$, depending on the orientation of the winding of $L$, at this point, with respect to the canonical vector field $\partial_t$. This is illustrated in \cref{fig:overall-q-factor}. (It is called an overall factor because both lifts of~$L$ are accompanied by the same factor.) 
    
    \item Each point of tangency of the projection of any lift $\tilde L$ to $C \times I$ receives a sheet-dependent winding factor. This is illustrated in \cref{fig:sheet-q-factor}. Note that the projection of~$\tilde L$ to $C \times I$ may have more points of tangency than $L$. This happens precisely when $\tilde L$ undergoes an exchange or takes a detour. 
\end{itemize}

\begin{figure}[h]
    \centering
    \includegraphics[width=0.72\linewidth]{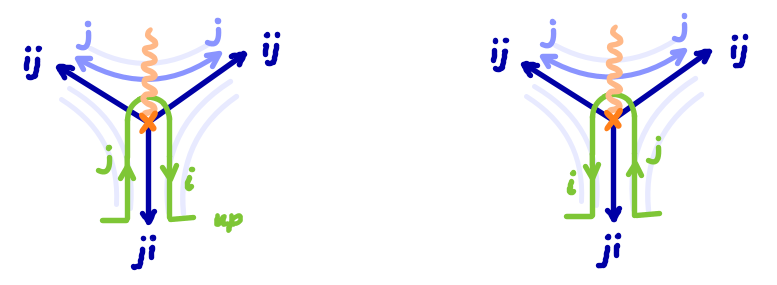}
    \caption{Locally, detours either turn clockwise or anti-clockwise in the vicinity of the branch-point. In this figure we illustrate that a detour is always followed by a point of tangency. Using the rules of $q$-nonabelianisation, we verify that the two contributions to the total framing factor precisely cancel each other.}
    \label{fig:total-detour-q-factor}
\end{figure}

Note that any detour path winds around the branch-point and that its projection to $C \times I$ necessarily has a point of tangency. This implies that there are two $q$-factors associated with any detour: the detour-factor and the sheet-dependent winding-factor. The clockwise (or anti-clockwise) detour introduces a factor of $q^{-1/2}$ (or $q^{1/2}$) into $\alpha(\tilde L)$.  Using the rules of \cref{fig:sheet-q-factor}, the point of tangency contributes $q^{1/2}$ (or $q^{-1/2}$, respectively), which therefore cancels the detour-factor exactly. This is illustrated in \cref{fig:total-detour-q-factor}. We may therefore effectively ignore the $q$-factors associated to any detour path in our computations. (This argument only depends on the local foliation structure around the branch-points, and the conclusion is thus valid for any WKB foliation at a generic phase.)

\begin{figure}[h!]
    \centering
     \includegraphics[width=0.55\linewidth]{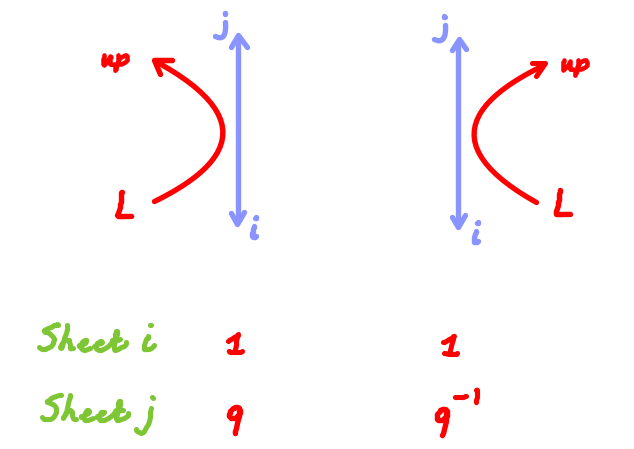}
    \caption{The overall and the sheet-dependent winding factor can be combined into a single sheet-dependent $q$-factor. The illustrated factors are associated with upward travelling links; they are reversed for a downward travelling links.}
    \label{fig:new-wind}
\end{figure}

This implies that, in the absence of exchanges, we may safely neglect all winding contributions from points of tangency that are not common to both the base link $L$ and its lifts $\tilde L$. This means that we can repackage the overall and the sheet-dependent winding factors into a single $q$-factor. This is illustrated in \cref{fig:new-wind}.  \emph{In conclusion, it is sufficient to only assign $q$-factors to direct lifts.} These $q$-factors only depend on the relative orientation between the $ij$-trajectory and the sheet of residence of a lift $\tilde L$.

Finally, note that the $q$-factors associated with exchanges vanish in the limit $q=1$, whereas all other $q$-factors simply become $1$. Together with the specialisation $X_L^\mathcal{W} \to \mathcal{X}_L^\mathcal{W}$ at $q=1$, this implies that $q$-nonabelianisation indeed reduces to ordinary nonabelianisation in the limit $q=1$

\subsubsection{\texorpdfstring{Open paths}{Open paths}} \label{subsubsec:open-paths}

The skein algebra $\SkAlg(C, SL(2, \C))$ is a quantisation of the algebra of functions on the moduli space $\mathcal{M}^\text{flat}(C, SL(2, \C))$ of flat $SL(2, \C)$ connections on $C$, so that
\begin{equation}
    \SkAlg(C, SL(2, \C))\big|_{q=1} \cong C^{\infty}_\textrm{reg}\big(\mathcal{M}^\text{flat}(C, SL(2, \C))\big).
\end{equation}
This implies that $q$-nonabelianisation essentially quantises trace functions on the moduli space of flat connections.

However, in this paper we are more interested in quantum parallel transport along open paths rather than closed paths. Ordinary nonabelianisation allows us to formulate the parallel transport of a flat $SL(2,\C)$-connection along open paths $\mathfrak{p} \subset C$ in terms of matrices valued in abelian parallel transports on the cover $\Sigma$. Analogously, we will define $q$-parallel transport in terms of matrices valued in abelian $q$-parallel transports on $\Sigma$. In a nutshell, $q$-parallel transport is defined by applying the rules of $q$-nonabelianisation to open paths inside $C \times I$. In the remainder of this section, we make precise what we mean by this. 

We formally define an \emph{open path} inside $C \times I$ as a link $\frakp \subset C \times I$, with non-empty boundary $\partial \frakp \ne \emptyset$, that is monotonously increasing (or decreasing) in the height-direction. In the skein literature, any link $\mathfrak{p}$ with non-empty boundary is called a tangle. We emphasize that open paths are, in addition, monotonically increasing (or decreasing) in the height-direction.

We denote the incoming and outgoing boundary components of an open path~$\frakp$ by 
\begin{align}
\partial \frakp^\textrm{in} = \partial\frakp\big|_{C \times \{0\}} \quad \text{and} \quad  \partial \frakp^\textrm{out} = \partial \frakp \big|_{C \times \{1\}},
\end{align}
respectively, and allow the boundary points to be part of a fixed finite set of points $\mathbb X \subset C$. For our purpose, it is convenient to take $\mathbb X$ to be the set of base-points of the path groupoid used in the nonabelianisation procedure. Note that since the links are framed, each boundary point is also equipped with a framing vector. 

We then consider the $\Z[q^{\pm 1}]-$module spanned by all upwards travelling open paths (i.e.~monotonically increasing along the height-direcion) stretching between $C \times \{0\}$ and $C \times \{1\}$, up to isotopy and $SL(2,\C)-$skein relations. We call this the space of \emph{open $SL(2,\C)-$skeins}
\begin{equation}
    \OpSk\big(C\times I, \mathbb X, SL(2, \C)\big) = \mathrm{Span}_{\Z[q^{\pm 1}]}\{\text{open paths } \frakp \subset C \times I : \partial \frakp \subset \mathbb X \}/\sim. \label{eq:def-open-skein-module}
\end{equation}
We may define a similar space of open skeins for downwards travelling open paths (i.e.~monotonically decreasing along the height-direction).

\begin{remark} 
Formally, underlying this definition is a category called the skein category. See \cite[\S2]{Gunningham:2024hpp} for an instructive primer on skein theory. Below we collect a few facts that are relevant for the construction in this section.

Let $\mathbb X^\textrm{in}$ and $\mathbb X^\textrm{out}$ be two finite sets of (framed) points inside $C$. Then, the space spanned by \emph{all} tangles $\frakp$ (not necessarily monotonous in the height direction) with boundaries $\partial\frakp^\textrm{in} = \mathbb X^\textrm{in}$ and $\partial \frakp^\textrm{out} = \mathbb X^\textrm{out}$, modulo isotopy and $SL(2, \C)-$skein relations, is called a relative skein module
\[\SkMod\big(C \times I, \mathbb X^\textrm{in}, \mathbb X^\textrm{out}, SL(2, \C)\big). \]
Unlike the previously defined space of open skeins, the relative skein module additionally contains tangles whose connected components may be closed links. In particular, if $\mathbb X^\textrm{in} = \mathbb X^\textrm{out} = \emptyset$, the relative skein module recovers the usual $SL(2,\C)-$skein algebra.

From these definitions, one can build a linear category whose simple objects are subsets of $\mathbb X$, and whose Hom spaces, between any two objects $\mathbb X^\textrm{in}$ and $\mathbb X^\textrm{out}$, are the relative skein modules $\SkMod(C \times I, \mathbb X^\textrm{in}, \mathbb X^\textrm{out}, SL(2, \C))$. This is a subcategory of the skein category $\SkCat(C, SL(2, \C))$ defined in \cite{Gunningham:2024hpp}. 

In terms of this formulation, the space of open skeins $\OpSk(C\times I, \mathbb X, SL(2,\C))$ is the subspace of the Hom space
\begin{align}
\SkMod\big(C \times I,\bigoplus_{\mathbb X^\textrm{in}}\mathbb X^\textrm{in}, \bigoplus_{\mathbb X^\textrm{out}}\mathbb X^\textrm{out}, SL(2, \C)\big) = \bigoplus_{\mathbb X^\textrm{in}, \mathbb X^\textrm{out}}\SkMod\big(C \times I,\mathbb X^\textrm{in}, \mathbb X^\textrm{out}, SL(2, \C)\big), 
\end{align}
for all subsets $\mathbb X^\textrm{in},\, \mathbb X^\textrm{out} \subset \mathbb X$, that only consists of upwards travelling open paths, and no additional closed links.
\end{remark}

One of the key features of open paths is that they are composable. Suppose we have two open paths $\mathfrak{p}_1$ and $\mathfrak{p}_2$ such that
\begin{equation}
    \partial \mathfrak{p}_1\big|_{C \times \{1\}} = \partial \mathfrak{p}_2 \big|_{C \times \{ 0\}},
\end{equation}
then we can compose them to get a new open path $\mathfrak{p}' = \mathfrak{p}_1 \circ \mathfrak{p}_2$ with boundary $\partial \frakp_1^\textrm{in} \cup \,\partial\frakp_2^\textrm{out}$. This induces a natural composition structure on open skeins as
\begin{equation}
    [\,\mathfrak p_1\,] \circ [\,\mathfrak p_2\,] = [\,\mathfrak p'\,] \in \OpSk\big(C \times I, \mathbb X, SL(2, \C)\big). \label{eq:composition-skmod}
\end{equation}
By this, we have traded the stacking product in the skein algebra for the composition structure of open skeins.\footnote{We may equip the space of open skeins with the product structure induced by composition: the product between non-composable skeins is defined to be zero. This turns the space of open skeins into an open skein algebra.} Note that this is natural from the perspective of a quantum parallel transport.

Likewise, we may associate to the cover $\Sigma$ the space of open $GL(1,\C)-$skeins
\begin{equation}
    \mathbf A:= \OpSk\big( \Sigma\times I, \pi^{-1}(\mathbb X), GL(1, \C)\big)
\end{equation}
where the boundaries of the open paths are restricted to lie inside $\pi^{-1}(\mathbb X)$, the lift of $\mathbb X$ to the cover. Whereas the abelian skein algebra is a $q$-deformation of the homology $H_1^\textrm{odd}(\Sigma; \Z)$, the space of abelian open skeins is essentially a $q$-deformation of the relative homology $H_1^\textrm{odd}(\Sigma, \pi^{-1}(\mathbb X); \Z)$. More precisely, the abelian skein relations ensure that for any open path $\widetilde \frakp \in \mathbf A$, we can define a formal variable $Y_{\widetilde p}$ as in \cref{eq:skein-torus-iso} -- by taking the relative homology class of $\widetilde p $ inside $\big(\Sigma \times I, \mathbb X \big)$ while keeping track of the abelian skein relations -- that satisfies
\begin{equation}
     X_{\eta} \, Y_{\widetilde \frakp} = (-q)^{-\langle \eta, \,\widetilde\frakp\rangle} Y_{\eta + \widetilde\frakp} \label{eq:open-paths-commutator}
\end{equation}
for all homology cycles $\eta \in H^\textrm{odd}_1(\Sigma; \Z)$. This shows that $\mathbf A$ is a two-sided module for the $GL(1,\C)$-skein algebra.

We shall refer to the $Y_{\widetilde \frakp}$ as an \emph{open path generator}. Paralleling the discussion below \cref{eq:skein-torus-iso}, the semi-classical limit of the open path generator $Y_{\widetilde{ \frakp }}$ is the abelian parallel transport along the projection $\widetilde \frakp_\Sigma$ of $\widetilde \frakp$ to $\Sigma$. By the monotonicity constraint, for every $\widetilde \frakp_\Sigma \subset \Sigma$, we can find a unique open path lift $\widetilde \frakp \subset \Sigma \times I$. Therefore, we denote the semi-classical limit of a generator $Y_{\widetilde \frakp}$ also by $\mathcal X_{\widetilde \frakp}$ when unambiguous. Recall though that the abelian parallel transport along an open path between arbitrary base-points is only defined up to abelian gauge transformations at its end-points; only the abelian parallel transport between marked points (where the vector bundle has been trivialised) is unambiguously defined.\footnote{Note that the product of the semi-classical limits of $Y_{\widetilde \frakp}$ and $Y_{\sigma_* \widetilde \frakp}$ is still fixed to be $1$ by $SL(2, \C)$-equivariance for all $\widetilde \frakp \in \mathbf A$, so that the semi-classical limit of quantum generators for open paths in the even relative homology is  trivial, as it should be.} This implies that the open path generators corresponding to open paths between marked points (which form a subset of the set of all base-points) play a special role. We will return to this in \S\ref{subsec:quantum-FN-coordinates-2}.

In this set-up, we define \emph{$q$-parallel transport} as the linear map
\begin{equation}
    P^q_\mathcal W : \OpSk\big(C \times I , \mathbb X, SL(2, \C)\big) \to \OpSk\big(\Sigma \times I, \pi^{-1}(\mathbb X), GL(1, \C)\big), \label{eq:q-parallel-def}
\end{equation} 
obtained by restricting $q$-nonabelianisation to open paths inside $C \times I$. To compute $P^q_\mathcal W[\mathfrak p]$ on any open $SL(2,\C)$-skein $[\mathfrak p]$, we apply the rules in \S\ref{subsubsec:q-nonab-rules} to a representative $\frakp \in [\mathfrak p]$ that is strictly monotonically increasing in the height direction. In this process, we consider all possible lifts of the open path $\frakp$ whose incoming and outgoing boundaries, when projected down to $C \times I$, coincide with the incoming and outgoing boundaries of $\frakp$, respectively. Each lift in the image of $P^q_\mathcal W[\mathfrak p]$ is called an \emph{abelian $q$-parallel transport}. Note that unlike the usual $q$-nonabelianisation map $\psi^q_\mathcal W$ which only selects the lifts that are closed on the cover, the $q$-parallel transport map $P^q_\mathcal W$ retains all possible lifts of the open path $\frakp$.

The same arguments which show that $q$-nonabelianisation is a skein algebra homomorphism also show that $q$-parallel transport intertwines with the composition structure \cite[\S8]{Neitzke:2020jik},
\begin{equation}
    P^q_{\frakp_1} \, P^q_{\frakp_2} = P^q_{\frakp'}.
\end{equation}
This especially means that $q$-parallel transports along long, complicated paths can be composed from $q$-parallel transports along smaller, simpler paths. In that sense, $q$-parallel transport is a \emph{local} operation.

\paragraph{Simplifications} 

For the purposes of this paper, it is sufficient to limit ourselves to $q$-parallel transports for connected open paths. We therefore restrict ourselves to open paths $\frakp \subset C \times I$ with a single incoming boundary component $z^\textrm{in}$ and a single outgoing boundary component $z^\textrm{out}$. This implies that all resulting abelian $q$-parallel transports $\widetilde \frakp^{ij} \subset \Sigma \times I$ stretch from $z^{\textrm{in}, (i)}$ on sheet $i$ to $z^{\textrm{out}, (j)}$ on sheet $j$, for some $i, j = 1, 2$. We can thus formulate the $q$-parallel transport map in terms of a matrix $P^q \in \mathrm{Mat}_{2 \times 2} (\mathbf A)$ whose $ij$-th entry contains all abelian $q$-parallel transports that start on sheet $i$ and terminate on sheet $j$ (similar to ordinary nonabelianisation \cref{eq:non-ab-par-transport}). Clearly, composition of $q$-parallel transports is compatible with the multiplication structure in $\mathrm{Mat}_{2 \times 2} (\mathbf A)$.

Furthermore, we slightly declutter our notation for lifts of connected open paths. When unambiguous, we denote the open path generator corresponding to a lift $\widetilde \frakp^{ij}$ by
\begin{equation}
    Y_{\frakp}^{ij} :=  Y_{\widetilde \frakp^{ij}}
\end{equation}
to signify that the generator corresponds to a lift of $\frakp$ going between sheet $i$ and sheet $j$. We emphasise that, unless explicitly stated, $Y^{ij}_{\frakp_1} Y^{jk}_{\frakp_2}$ means the composition
\begin{equation}
    Y^{ij}_{\frakp_1} Y^{jk}_{\frakp_2} := Y^{ij}_{\frakp_1} \circ Y^{jk}_{\frakp_2} = Y^{ik}_{\frakp'},
\end{equation}
and not the generator corresponding to the disconnected open path $\frakp_1 + \frakp_2$.

\medskip

Since $q$-parallel transport is defined by applying the rules of $q$-nonabelianisation to open paths that are monotonically increasing in the height-direction, there cannot be any self-intersections in the projection of the open paths to the leaf space. Consequently, $q$-parallel transport receives no contributions from exchanges. This simplification will enable us to define $q$-parallel transports with respect to FN-type spectral networks. Moreover, $q$-parallel transport and ordinary nonabelianisation now share the same basic ingredients, \textit{vis. a vis.} direct lifts and detours. In particular, the semi-classical limit of $q$-parallel transport (i.e.~setting $q=1$ and replacing all open path generators with their semi-classical limits) recovers the ordinary parallel transport (defined in \cref{eq:non-ab-par-transport}). It is in this sense that $q$-parallel transport quantises the space of $SL(2, \C)$-parallel transports on $C$.

Finally, we remark that it is important to carefully keep track of the framing in the presence of links with boundaries. For closed links, the framing factor is an invariant. By this we mean that without ``cutting'' the link there is no way to change the framing of the link. However, this is no longer true for open paths. We can change the framing of an open path by changing the boundary framing. Thus, it becomes essential to fix the boundary framing of all open paths  and ensure compatibility of the framing while gluing. This is implicitly assumed in the rest of this section.

\subsubsection{\texorpdfstring{3d Chern-Simons theory}{3d Chern-Simons theory}}\label{sec:CStheory}

In this paragraph, we briefly remind the reader about the physical relevance of $q$-nonabelianisation and $q$-parallel transport. Recall that the AGT correspondence states that the supersymmetric partition function of a 4d $\mathcal N=2$ class S $SU(2)$ theory, constructed by compactifying the 6d $(2,0)$-theory of type $A_1$ on a Riemann surface $C$, in the squashed 4-sphere background can be identified with the Liouville correlation function on $C$. We obtain a slightly different perspective by considering the squashed 4-sphere as a non-trivial fibration of the squashed 3-sphere over an interval $I_R$,
$$S^3_b \hookrightarrow S^4_b \to [-R, R]=: I_R,$$
that degenerates at the boundary of the interval. Then, similar to the AGT correspondence (i.e.~the 4d-2d correspondence), there exists a 3d-3d correspondence \cite{Dimofte:2011ju, Cordova:2013cea} between a 3d $\mathcal N = 2$ $SU(2)$ gauge theory on the squashed 3-sphere $S^3_b$ and the $SL(2, \C)$-Chern-Simons theory on $C \times I_R$.\footnote{Recall that complex Chern-Simons theory depends on two parameters: the quantised level $k \in \Z$ and a continuous parameter $s \in \R$ or $\mathrm i\R$. The above duality holds for $k=1$ and $s = \frac {1-b^2}{1 + b^2}$.} The interval $I_R$ (after rescaling) defines the height direction in $q$-nonabelianisation. 

The degenerating fibres at the boundary points define non-trivial boundary conditions for the Chern-Simons theory at the ends of the interval. These boundary conditions only depend on the Riemann surface $C$. Suppose we denote the incoming and outgoing boundaries by $C^\textrm{in} = C \times \{ -R\}$ and  $C^\textrm{out} = C \times \{ R \}$, respectively. Then, the space of boundary conditions on $C^\textrm{in}$ is the space of chiral Liouville blocks on $C$, and the space of boundary conditions on $C^\textrm{out}$ is the space of anti-chiral Liouville blocks on $C$. The Chern-Simons theory partition function defines a pairing between the two spaces:
$$ \mathcal Z^{CS} : \mathrm{CB}(C^\textrm{in}) \times \mathrm{CB}(C^\textrm{out}) \to \C.$$
Leaving aside potential complications associated with the dualisability of these spaces, we can write this as an endomorphism on the space of chiral Liouville blocks:
$$\mathcal Z^{CS} : \mathrm{CB}(C^\textrm{in}) \to \mathrm{CB}(C^\textrm{in}). $$

Since the 4d $\mathcal N=2$ theory on the squashed 4-sphere is trivial away from the poles, it follows that the Chern-Simons theory is trivial along the interval $I_R$ \cite{Hama:2012bg}. We thus conclude that the pairing defined by the Chern-Simons partition function simply reproduces the Liouville correlation function. This is the standard AGT correspondence, and can be directly verified in the 4d theory through supersymmetric localisation \cite{Hama:2012bg}.

Alternatively, we could engineer a scenario where the 3d theory is not quite trivial along the interval direction. For instance, suppose we introduce a compact Wilson loop $L_\gamma$ supported on a loop $\gamma$ inside $C \times I_R$. If the loop $\gamma$ is isotopic to a loop inside a fixed height slice $C \times \{t\}$\footnote{Line defects $L_\gamma$ that are not isotopic to loops in a fixed height slice are called \emph{fat line defects} in \cite{Neitzke:2020jik}.}, the Chern-Simons pairing
$$\mathcal Z^{CS}[L_\gamma] : \mathrm{CB}(C^\textrm{in}) \to \mathrm{CB}(C^\textrm{in})$$
represents the action of a Verlinde loop operator, supported on $\gamma$, on the space of Liouville blocks on $C$.

Dually, in the 3d supersymmetric gauge theory, the Wilson loop $L_\gamma$ describes a non-abelian line operator $L_\gamma^{\mathcal N=2}$ in the UV. The easiest way to study this line operator is to pass to the Coulomb branch, wherein the non-abelian line operator decomposes into a sum of abelian line operators $\widetilde L^{\mathcal N =2}$. That is,
\begin{equation}
    L_\gamma^{\mathcal N =2} = \sum_{\widetilde L} \alpha(\widetilde L) \, \widetilde L^{\mathcal N=2}.
\end{equation}
Dualising again, the above UV-IR map describes a sum of compact Wilson loop $\widetilde L$ in a $GL(1, \C)-$Chern-Simons theory on $\Sigma \times I_R$. From a 2d perspective, the abelian Wilson loop is an abelian Verlinde operator. Basically, this is the underlying physical interpretation behind the $q$-nonabelianisation map. A more detailed explanation of the 3d UV-IR map, and various other related facts, can be found in \cite[Section 1]{Neitzke:2020jik}.

There is yet another way to define a non-trivial endomorphism on the space of conformal blocks. In the 3d theory, the boundary CFT operators are lifted to non-compact Wilson lines $L^\text{open}$ starting on $C^\textrm{in}$ and terminating on $C^\textrm{out}$. For a trivial bulk theory, the non-compact Wilson lines are stationary along the $C$ direction, so that they do not interact with each other in an interesting way. However, we can modify the bulk theory by allowing for the braiding and fusion of the non-compact Wilson lines. This is simply described by the Moore-Seiberg transformations on the space of conformal blocks. In other words, the Chern-Simons pairing
$$\mathcal Z^{CS}[L^\text{open}] : \mathrm{CB}(C^\textrm{in}) \to \mathrm{CB}(C^\textrm{in})$$
describes the action of the mapping class group on the space of Liouville blocks on~$C$.

We know that these transformations are incredibly complicated for generic vertex operators in the CFT. However, we also know that the situation is much more manageable for degenerate operators. Let us therefore fix all the non-compact Wilson lines for generic representations to be spectators (these correspond to the insertions at the punctures), and only allow the braiding and fusion of non-compact Wilson lines for degenerate operators. Identifying the latter with the open skeins on $C$, the UV-IR map applied to such a configuration of non-compact Wilson lines inside the Chern-Simons theory gives precisely the $q$-parallel transport map.\footnote{More precisely, we identify a non-compact Wilson line corresponding to a $(1,2)$ degenerate operator with an open path in $C\times I$ labelled by the defining representation of $SL(2, \C)$ (also see \cref{eq:degenerate-sl2-weights}).} Note that the marked points in $\mathbb X$ are all the possible (inequivalent) insertion points of degenerate operators in $C$ in the presence of the other generic operators.

For example, the Moore-Seiberg fusion move from s-channel to t-channel (c.f.~\cref{fig:F-move})
$$
\begin{tikzpicture}
        \draw[thick] (0,0)--(2,0);
        \node[yshift = -0.2cm] at (0,0) {$\alpha_4$};
        \node[yshift = -0.2cm] at (2,0) {$\alpha_1$};
        \node[yshift = 0.2cm] at (0.5,1) {$\alpha_3$};
        \node[yshift = 0.2cm, red] at (1.5,1) {$\alpha_2$};
        \node[yshift = -0.2cm] at (1,0) {$\alpha$};
        \draw[thick] (0.5,0)-- (0.5,1);
        \draw[red, thick] (1.5,0)--(1.5,1);
        \draw [-latex] (2.1,0.5)--(2.9,0.5);
        
        \draw[xshift = -1cm, thick] (4,0)--(6,0);
        \draw[xshift = -1cm, thick] (5,0.5)-- (4.5,1);
        \draw[xshift = -1cm, red, thick] (5,0.5)--(5.5,1);
        \draw[xshift = -1cm, thick] (5,0)--(5,0.5);
        \node[yshift = -0.2cm, xshift = 3.1cm] at (0,0) {$\alpha_4$};
        \node[yshift = -0.2cm, xshift = 3cm] at (2,0) {$\alpha_1$};
        \node[yshift = 0.2cm, xshift = 3cm] at (0.5,1) {$\alpha_3$};
        \node[yshift = 0.2cm, xshift = 3cm, red] at (1.5,1) {$\alpha_2$};
        \node[yshift = 0.25cm, xshift = 3.25cm] at (1,0) {$\alpha'$};
\end{tikzpicture}
$$
with $\alpha_2$ degenerate can be depicted in three dimensions as the following braid diagram of non-compact Wilson lines (suppressing internal momenta labels for brevity):
\begin{equation*}
    \begin{tikzpicture}
        [braid/.cd, number of strands = 5, strand 1/.style = {MidnightBlue, very thick}, strand 2/.style = {MidnightBlue, very thick}, strand 3/.style = {red, very thick}, strand 4/.style = {transparent}, strand 5/.style = {MidnightBlue, very thick}]
        \pic at (0,0) {braid = 1 s_3 1};
        \draw[-latex, very thick, MidnightBlue] (0, -2.5);
        \draw[-latex, very thick, MidnightBlue] (1, -2.5);
        \draw[-latex, very thick, red] (3, -2.5);
        \draw[-latex, very thick, MidnightBlue] (4, -2.5);

        \node[yshift = -0.3cm, MidnightBlue] at (0, -3.5) {$\alpha_4$};
        \node[yshift = -0.3cm, MidnightBlue] at (1, -3.5) {$\alpha_3$};
        \node[yshift = -0.3cm, red] at (3, -3.5) {$\alpha_2$};
        \node[yshift = -0.3cm, MidnightBlue] at (4, -3.5) {$\alpha_1$};
        
        \draw[thick] (-1,0)--(5,0);
        \draw[thick] (-1,-3.5)--(5,-3.5);
        \node[xshift=0.5cm, yshift = 0.1cm] at (5, -3.5) {$C^\textrm{in}$ .};
        \node[xshift=0.5cm, yshift = 0.1cm] at (5, 0) {$C^\textrm{out}$};
    \end{tikzpicture} \label{eq:3d-S-wall}
\end{equation*}
The UV-IR map applied to this configuration describes a configuration of non-compact Wilson lines in the abelian Chern-Simons theory on $\Sigma \times I_R$. Equivalently, this is the $q$-parallel transport along a path going from the puncture labelled by $\alpha_1$ to the puncture labelled by $\alpha_3$  on the 3-punctured sphere. 

Abstractly, $q$-parallel transport is a formal sum of open path generators $Y_{\widetilde{\frakp}}$. However, the above picture allows us to \emph{evaluate} the $q$-parallel transport map on the Hilbert space of the abelian Chern-Simons theory -- the space of abelian (Heisenberg) conformal blocks on~$\Sigma$. That is, we can realise the open path generators $Y_{\widetilde \frakp}$ as operators on abelian conformal blocks that satisfy the relation \cref{eq:open-paths-commutator} with the abelian Verlinde operators.\footnote{At this point, the notion of an abelian conformal block is still rather abstract. We will attempt to make this concrete in \cite{HollandsMurugesanII}.} 

Physically, an open path generator $Y_{\widetilde \frakp}$, such that $\partial \widetilde\frakp = \{z^\textrm{in}\} \cup \{z^\textrm{out}\}$, acts on an abelian conformal block $\mathcal F^\textrm{ab}_{\vec \alpha}$ on $\Sigma$ with a degenerate insertion at $z^\textrm{in}$ as
\begin{equation}
    Y_{\widetilde \frakp} \, \mathcal F^\textrm{ab}_{\vec \alpha} \big( \widetilde V^\textrm{ab}_\textrm{deg}(z^\textrm{in}) \ldots \big) =  C(\ldots) \, \mathcal F^\textrm{ab}_{\vec \alpha'} \big( \widetilde V^\textrm{ab}_\textrm{deg}(z^\textrm{out}) \ldots \big) \label{eq:open-path-generators-action}
\end{equation}
to produce a new abelian conformal block $\mathcal F^\textrm{ab}_{\vec \alpha'}$ with the degenerate insertion now moved to $z^\textrm{out}$ while possibly modifying the labels $\vec\alpha$. The coefficients $C(\ldots)$ are given by functions of invariants of the remaining vertex operator insertions and the other parameters of the theory. Otherwise stated, the operator $Y_{\widetilde \frakp}$ computes the analytic continuation of $\mathcal F^\textrm{ab}_{\vec \alpha}(z^\textrm{in})$ along the path $\widetilde \frakp_\Sigma$. In certain cases, the UV-IR map ($q$-parallel transport map) then induces an action of the open path generators on Liouville conformal blocks. We will go through some examples in \S\ref{subsec:quantum-FN-coordinates-2}.

In principle, the $q$-nonabelianisation -- UV-IR map, allows us to consider more complicated configurations of compact and non-compact Wilson lines. Some examples include (compact) Wilson loops in the presence of (braiding and fusing) non-compact degenerate Wilson lines, and non-compact degenerate Wilson lines that start and end on the same boundary. These correspond to a more generic class of open Verlinde line operators in Liouville field theory \cite{Gaiotto:2014lma}. We will revisit abelian open Verlinde operators in the context of quantum FN coordinates in \S\ref{subsec:quantum-FN-coordinates-2}.

Whilst the rest of this paper is written in the language of 2d conformal field theories, it helps to retain the 3d picture in mind while going through the remainder of this section.

\subsection{Quantum S-matrix} \label{sec:q-FN-nonab}
Our goal in this chapter is to formulate $q$-parallel transports for FN networks. As a first step, we construct the quantum FN S-matrix.
Arbitrary $q$-parallel transports can then be obtained by composing direct lifts with the quantum S-matrices.

We define the quantum S-matrix as the $q$-parallel transport matrix along an infinitesimal path $\varepsilon \subset C \times I$ that cuts across a critical $ij$-trajectory of a spectral network. For instance, for a single $12-$wall $w$, this takes the form
\begin{equation}
    S_w^q = \begin{pmatrix}
         Y^{11}_\varepsilon &  Y^{12}_\varepsilon \\ 0 &  Y^{22}_\varepsilon
    \end{pmatrix} \in \mathrm{Mat}_{2 \times 2}(\mathbf A) \label{eq:FG-q-S-matrix}
\end{equation}
where the generators $Y^{ij}_\varepsilon$ correspond to the relative homology classes of open paths in $\Sigma \times I$ stretching from sheet $i$ to sheet $j$.

Similarly, we may define a quantum FN S-matrix. However, just like in ordinary nonabelianisation, $q$-nonabelianisation is not defined at the FN-phase. Instead, we have to take an average of the two resolutions at $\theta_\FN \pm \eta$. This was described in \S\ref{sec:nonab-pants} in the semi-classical context. Below, we define a similar average for the quantum FN S-matrix.

\begin{figure}[h]
   \centering
  \includegraphics[width=0.55\linewidth]{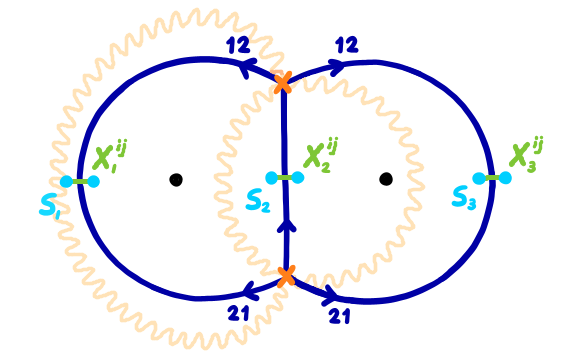}
    \caption{Fully symmetric FN network (in dark blue) on the gauged 3-sphere, together with the base-points (in light blue) placed infinitesimally away on either side of the double walls.}
    \label{fig:molecule1-nonab-2}
\end{figure}

Consider the fully-symmetric FN molecule from~\cref{fig:molecule1-nonab} on the gauged 3-sphere, which we have redrawn for convenience in \cref{fig:molecule1-nonab-2}. The quantum FN S-matrix~$S_2^{q\FN}$ is, by definition, the $q$-parallel transport matrix across an infinitesimal path crossing the double trajectory labelled by $S_2$. Inside the gauged 3-punctured sphere, this evaluates to
\begin{align}
    S^{q\FN}_2 = \begin{pmatrix}  \gamma  Y^{11}_{2, L} & \alpha  Y^{12}_{2, L} \\ - \beta'  Y^{'21}_{2, L} & \gamma  Y^{22}_{2, L} \end{pmatrix} = \begin{pmatrix}  \gamma  Y^{11}_{2, L} & \alpha  Y^{12}_{2, L} \\ - q^{-1}\beta  Y^{21}_{2, L} & \gamma  Y^{22}_{2, L} \end{pmatrix} = \begin{pmatrix}  \gamma  Y^{11}_{2, L} & q\alpha'  Y^{'12}_{2, L} \\ - \beta'  Y^{'21}_{2, L} & \gamma  Y^{22}_{2, L} \end{pmatrix}. \label{eq:quantum-FN}
\end{align}
with
\begin{align}
    \gamma &= \sqrt{\frac {(M_1 M_2 - M_3)(M_1 M_2 - M_3^{-1})}{(1- M_1^2)(1 - M_2^2)}}, \notag \\
    \alpha &= \sqrt{\frac{(1- M_1^{-1}M_2 M_3)(1-M_1M_2^{-1}M_3)}{(1-M_1^2)(1-M_2^2)}},\\
    \beta' &= -\sqrt{\frac{(1 - M_1 M_2^{-1} M_3)(1 - M_1^{-1} M_2 M_3)}{(1-M_1^2)(1-M_2^2)}}, \notag\label{eq:quantum-FN-coefficients}
\end{align}
and
\begin{equation}
    \beta = \frac{M_1 M_2 } {M_3}\, \beta' \quad \text{ and } \quad \alpha' = \frac{M_1 M_2 } {M_3} \, \alpha.
\end{equation}
Note that \cref{eq:quantum-FN} is
in agreement with the cyclically permuted ($M_1 \to M_2 \to M_3 \to M_1$) equations~\cref{eq:gammaA} and \cref{eq:alphabetaA} (computing $P^\text{FN}_b$).

\begin{figure}[h]
    \centering
    \includegraphics[width=0.7\linewidth]{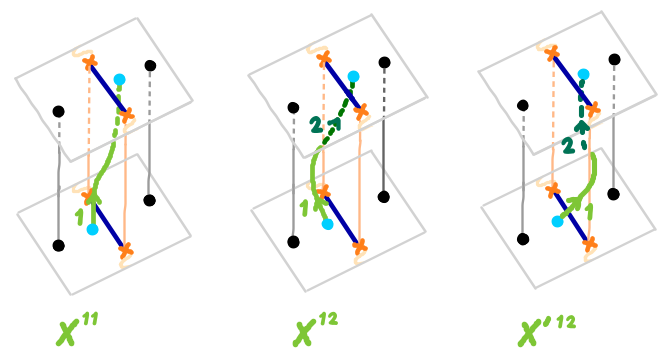}
    \caption{We depict open paths corresponding to the generators $Y^{11}_{L}$, $Y^{12}_L$ and $Y'^{12}_{L}$ (drawn as projections to $C\times I$). The paths start on the left base-point at $t=0$ and run to the right base-point at $t=1$.}
    \label{fig:open-paths}
\end{figure}

The unprimed off-diagonal open path generators $Y^{ij}_{2, L}$ (with $i \ne j$) correspond to a clockwise detour around the top branch-point, while the primed off-diagonal open path generators $Y'^{ij}_{2, L}$ correspond to an anticlockwise detour around the bottom branch-point. This is illustrated in \cref{fig:open-paths}. We explain how to go between the primed and unprimed off-diagonal generators in \S\ref{subsubsec:derive-q-FN-mat}. Following $SL(2, \C)$-equivariance, the semi-classical limits of the open path generators satisfy 
\begin{equation}
    \mathcal X^{11}_{2, L} = \frac 1 {\mathcal X^{22}_{2, L}} \quad \text{ and } \quad \mathcal X^{12}_{2, L} = - \frac 1 {\mathcal X^{21}_{2, L}},
\end{equation}
and likewise for the primed off-diagonal generators. Thus, we see that \cref{eq:quantum-FN} reproduces the correct semi-classical invariants in \cref{eq:invariants-s-matrix}. 

The inverse S-matrix is
\begin{align}
\begin{aligned}
     \left(S^{q\FN}_2\right) ^{-1} &= \begin{pmatrix}  \gamma  Y^{11}_{2, R} & -\alpha  Y^{12}_{2, R} \\  \beta'  Y^{'21}_{2, R} & \gamma  Y^{22}_{2, R} \end{pmatrix}  = \begin{pmatrix}  \gamma  Y^{11}_{2, R} & -\alpha  Y^{12}_{2, R} \\ q \beta  Y^{21}_{2, R} & \gamma  Y^{22}_{2, R} \end{pmatrix} = \begin{pmatrix}  \gamma  Y^{11}_{2, R} & -q^{-1}\alpha'  Y^{'12}_{2, R} \\  \beta'  Y^{'21}_{2, R} & \gamma  Y^{22}_{2, R} \end{pmatrix}, \label{eq:quantum-FN-2}
     \end{aligned}
\end{align}
where the subscript $L/R$ denotes that the path originates on the left/right side of the double wall (as illustrated in \cref{fig:molecule1-nonab-2}). In \cref{eq:quantum-FN} and \cref{eq:quantum-FN-2}, we have assumed that all paths are travelling upwards, in particular, when they undergo detours. The average S-matrices for downward travelling paths is obtained by replacing all $q \mapsto q^{-1}$. Indeed, note that the inverse S-matrix for a downward travelling path agrees with the na\"ive inverse of $S^{q\FN}$. The other quantum FN S-matrices are similarly obtained by a cyclic permutation of the exponentiated masses and the corresponding open path generators. 

In principle, knowledge of the quantum S-matrix is sufficient to compute $q$-parallel transports along any path inside the 3-punctured sphere. Suppose we would like to compute the $q$-parallel transport along an open path $\frakp \subset C \times I$ which crosses a double wall exactly once. Then, we may decompose $\frakp = \frakp_\iota \circ \frakp_f$ where $\frakp_\iota$ is the segment that lies before the crossing and $\frakp_f$ is the segment after the crossing. The $q$-parallel transport along the segments $\frakp_\iota$ and $\frakp_f$ only consists of direct lifts. Then, by the locality of $q$-parallel transport, we have 
\begin{equation}
    P^q_{\frakp} = P^q_{\frakp_\iota}\, S^{q\FN}\, P^q_{\frakp_f}
\end{equation}
where we compose the individual $q$-parallel transports along $\frakp_\iota$ and $\frakp_f$ with the appropriate quantum FN S-matrix. For an open path $\frakp$ that crosses a double trajectory more than once, we apply the same prescription at every crossing. We discuss the extension of this construction to arbitrary surfaces in \S\ref{subsec:gluing-q-nonab}.

Before we derive the expression \cref{eq:quantum-FN}, it is instructive to verify its consistency. One of the most important tests of consistency of $q$-nonabelianisation is invariance under two kinds of isotopies: isotopy across a critical trajectory and isotopy across a branch-point. For a single critical $ij-$trajectory, this was already established in \cite{Neitzke:2020jik}. We shall now demonstrate the same for the double wall. 

\begin{remark}
    We believe that eq.~\eqref{eq:quantum-FN} is the ``simplest'' quantisation of the semiclassical S-matrix that is compatible with $q$-nonabelianisation. Indeed, observe that the coefficients $\alpha, \alpha', \beta, \beta'$ and $ \gamma$ that appear in the S-matrices are only constituted from the exponentiated masses (i.e.~no $q$s), and thus, can be computed directly in the semiclassical limit. Moreover, if we were to start with an ansatz where each coefficient in the quantum S-matrix is identical to its semi-classical counterpart up to an overall factor $q^{a_{ij}}$, then only imposing isotopy invariance across double walls and branch-points is sufficient to fix the $q$-factors as in \cref{eq:quantum-FN}.
\end{remark}

\begin{remark}
Whereas mixing the primed and unprimed open path generators leads to relations involving the masses, such as\footnote{The composition of a detour around different branch-points leads to a closed cycle around the two branch-points. Further evaluating this closed cycle (as explained in \S\ref{subsubsec:derive-q-FN-mat}), we get the relation \cref{eq:mixed-open-relations}.}
\begin{align}
    Y_{2, L}^{'21} Y_{2, R}^{12} = - q^{-1}\frac{M_1 M_2} {M_3} Y_{2, L}^{22} Y_{2, R}^{22}, \label{eq:mixed-open-relations}
\end{align}
the identical open path generators satisfy a particularly nice set of relations (see \cref{fig:FN-isotopy-S-wall-2})
\begin{align}
    Y_{2, L}^{12} Y_{2, R}^{21} = Y_{2, L}^{'12} Y_{2, R}^{'21} = - Y_{2, L}^{11} Y_{2, R}^{11}.
\end{align}
This especially makes it attractive to work solely with either the primed or unprimed off-diagonal open path generators. In this paper, we have chosen to work with the unprimed generators throughout. 

We emphasise however that the most natural presentation of the quantum S-matrix is such that the $21-$detours follow the $21-$critical trajectory emanating from the bottom branch-point (corresponding to the primed generator), and the $12-$detours follow the $12-$critical trajectory emanating from the top branch-point (corresponding to the unprimed generator). In this form, we observe that the quantum S-matrix is devoid of any additional $q$-factors. This agrees with the form of \cref{eq:FG-q-S-matrix}.
\end{remark}

\subsubsection{Isotopy invariance across double wall} \label{subsubsec:double-wall-isotopy}

\begin{figure}[h]
    \centering
    \includegraphics[width=0.6\linewidth]{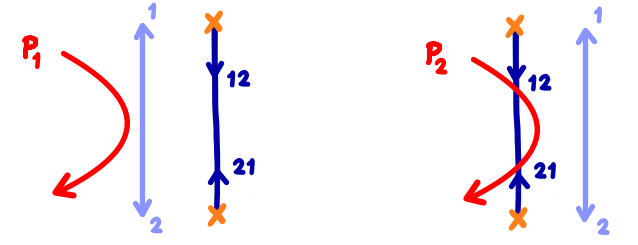}
    \caption{Isotopy move across a double wall. }
    \label{fig:FN-isotopy-S-wall-1}
\end{figure}

Consider the two paths $\mathfrak{p}_1$ and $\mathfrak{p}_2$ in $C \times I$ illustrated (in blue) in \cref{fig:FN-isotopy-S-wall-1}. We assume both are monotonically increasing in the height-direction. For the first path, only the direct lifts contribute to its $q$-nonabelianisation. Taking into account the winding factor (see \cref{fig:new-wind}), we obtain
\begin{equation}
    \begin{pmatrix}
        Y^{11}_{\mathfrak p_1} & 0 \\ 0 & q^{-1} Y^{22}_{\mathfrak p_1}
    \end{pmatrix} \label{eq:FN-isotopy-1-no-detour}
\end{equation}
for the $q$-parallel transport along this path.

The second path crosses the double wall twice. We may decompose this path into three parts labelled by $\iota, m$ and $f$, where $\iota$ labels the first part of the path that hasn't crossed the wall yet, $m$ labels the middle part in between the two crossings, and $f$ labels the final part after the crossings. Note that the middle piece $m$ comes with a winding factor. The $q$-parallel transport for the second path is thus given by
\begin{equation}\label{eq:double-wall-isotopy-w2}
    \begin{pmatrix} Y^{11}_\iota & 0 \\ 0 & Y^{22}_\iota \end{pmatrix}
    \begin{pmatrix}  \gamma  Y^{11}_{2, L} & \alpha  Y^{12}_{2, L} \\ - q^{-1}\beta  Y^{21}_{2, L} & \gamma  Y^{22}_{2, L} \end{pmatrix}
    \begin{pmatrix} Y^{11}_m & 0 \\ 0 & q^{-1} Y^{22}_m \end{pmatrix}
    \begin{pmatrix}  \gamma  Y^{11}_{2, R} & -\alpha  Y^{12}_{2, R} \\  q\beta  Y^{21}_{2, R} & \gamma  Y^{22}_{2, R} \end{pmatrix}
    \begin{pmatrix} Y^{11}_f & 0 \\ 0 & Y^{22}_f\end{pmatrix}.
\end{equation}
Ordinary nonabelianisation is insensitive to the above isotopy. This implies that the semi-classical limit of the above expression is identical to \cref{eq:FN-isotopy-1-no-detour}. Therefore, it is sufficient to only track $q$-factors on each sheet.

\begin{figure}[h]
    \centering
    \includegraphics[width=0.55\linewidth]{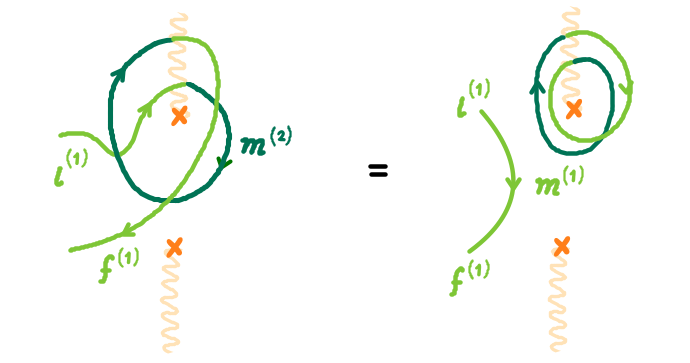}
    \caption{Equivalence between $Y^{11}_\iota Y^{12}_{2, L} Y^{22}_m Y_{2, R}^{21} Y^{11}_f$ (on the left) and $Y_\iota^{11} Y_{2,L}^{11} Y_m^{11} Y^{11}_{2,R} Y_f^{11}$ together with a closed loop around a branch-point (on the right). (We only show a single branch-cut to avoid clutter.)}
    \label{fig:FN-isotopy-S-wall-2}
\end{figure}

For instance, consider the set of lifts starting and ending on sheet $1$. There are two such lifts: the direct lift and the lift that undergoes two detours. They contribute as
\[ \gamma^2 Y^{11}_\iota  Y^{11}_{2, L}   Y^{11}_m Y_{2, R}^{11}  Y^{11}_f ~\text{ and }~ \alpha \beta Y^{11}_\iota  Y^{12}_{2, L}  Y^{22}_m Y_{2, R}^{21}  Y^{11}_f, \]
respectively, where recall that the above notation denotes composition. As illustrated in \cref{fig:FN-isotopy-S-wall-2}, we have that $Y^{12}_{2, L}  Y^{22}_m Y_{2, R}^{21} = -Y^{11}_{2, L}  Y^{11}_m Y_{2, R}^{11}$. The semi-classical result furthermore shows that $\gamma^2 - \alpha \beta  =1$. This implies that the total abelian $q$-parallel transport starting and ending on sheet $1$ is given by 
\begin{align}
Y^{11}_\iota Y^{11}_{2, L} Y^{11}_m  Y_{2, R}^{11} Y_f^{11}.
\end{align}
By composing these (non-intersecting) paths in their order of appearance, we can identify the resulting open path with the abelian $q$-parallel transport of $\mathfrak p_1$ on sheet $1$, i.e.
\begin{align}
Y^{11}_{\mathfrak p_1} = Y^{11}_\iota  Y^{11}_{2, L}  Y^{11}_m  Y_{2, R}^{11} Y_f^{11}.
\end{align}
Thus, we are able to match the abelian $q$-parallel transports of $\mathfrak p_1$ and $\mathfrak p_2$ that start and terminate on sheet $1$. Similarly,  the total abelian $q$-parallel transport starting and ending on sheet $2$ is given by 
\begin{equation}
    q^{-1} Y^{22}_{\mathfrak p_1} = q^{-1} (\gamma^2 - \alpha \beta) Y_\iota^{22} Y_{2, L}^{22} Y_m^{22} Y_{2, R}^{22} Y_f^{22}
\end{equation}
after simplifying $Y^{21}_{2, L} Y^{11}_m Y_{2, R}^{12} = -Y^{22}_{2, L} Y^{22}_m Y_{2, R}^{22}$. 

\begin{figure}[h]
    \centering
    \includegraphics[width=0.55\linewidth]{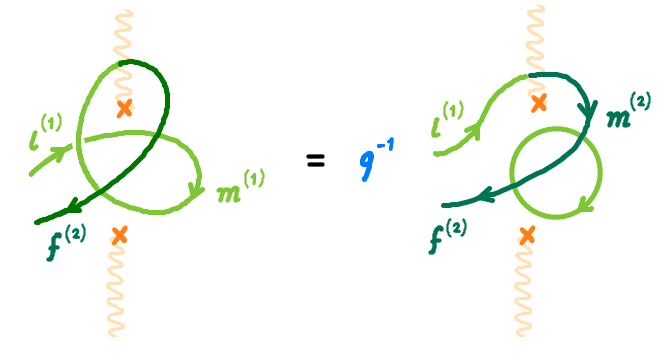}
    \caption{Equivalence between $Y_\iota^{11} Y_{2, L}^{11} Y_m^{11} Y_{2, R}^{12} Y_f^{22}$ (on the left) and $Y_\iota^{11} Y_{2, L}^{12} Y_m^{22} Y_{2, R}^{22} Y_f^{22}$ (on the right) inside $\mathbf A$. (We only show a single branch-cut to avoid clutter.) }
    \label{fig:FN-isotopy-S-wall-3}
\end{figure}

Instead, if we evaluate the off-diagonal $(12)$ component, we get
\begin{equation}
    - \alpha \gamma Y_\iota^{11} Y_{2, L}^{11} Y_m^{11} Y_{2, R}^{12} Y_f^{22} + q^{-1}\alpha \gamma Y_\iota^{11} Y_{2, L}^{12}  Y_m^{22}  Y_{2, R}^{22}  Y_f^{22} = 0, 
\end{equation}
because now $Y_{2, L}^{11} Y_m^{11} Y_{2, R}^{12} = q^{-1} Y_{2, L}^{12} Y_m^{22} Y_{2, R}^{22}$. This is illustrated in \cref{fig:FN-isotopy-S-wall-3}. The other off-diagonal component vanishes similarly. In total, we find $P_\FN^q[\mathfrak p_1] = P_\FN^q[\mathfrak p_2]$, and this establishes invariance under an isotopy across the double wall.

\subsubsection{Isotopy invariance across branch-point}
\begin{figure}[h]
    \centering
    \includegraphics[width=0.7\linewidth]{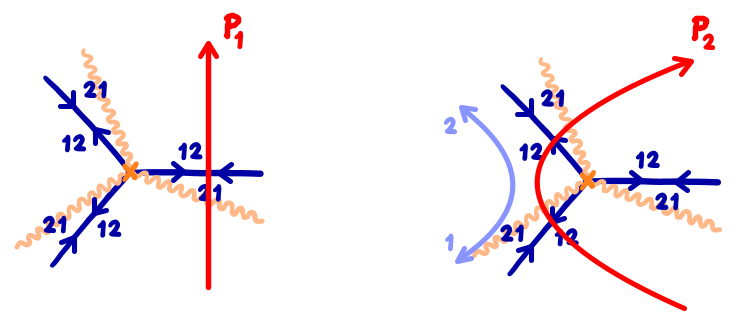}
    \caption{Isotopy move across a branch-point.}
    \label{fig:FN-isotopy-branch-point-1}
\end{figure}

\begin{figure}[h]
    \centering
    \includegraphics[width=0.8\linewidth]{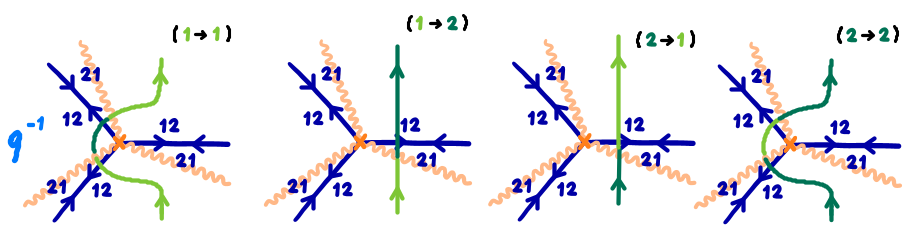}
    \caption{Projections of the lifts of the open path $\mathfrak{p}_1$ to $C$ with source and target sheets indicated. Only the lift from sheet $1$ to sheet $1$ has a non-trivial $q$ factor.}
    \label{fig:FN-isotopy-branch-point-2}
\end{figure}

We now show that $q$-nonabelianisation remains unchanged if we isotope a path across a branch-point as illustrated in \cref{fig:FN-isotopy-branch-point-1}. We will once again make use of the fact that ordinary nonabelianisation is invariant under this isotopy. So, instead of trying to explicitly evaluate the $q$-nonabelianisation of the two paths $\mathfrak{p}_1$ and $\mathfrak{p}_2$, we simply count the $q$-factors of each lift. The (projections of the) various lifts of $\mathfrak{p}_1$ are illustrated in \cref{fig:FN-isotopy-branch-point-2}. The detours of the path $\mathfrak{p}_1$ are captured by $S^{q\FN}$, and it is also easy to see that $\mathfrak{p}_1$ picks up no winding factors. So, only the lift starting and ending on sheet $1$ is accompanied by a non-trivial $q$-factor, namely, $q^{-1}$ (c.f. \cref{eq:quantum-FN}), and the rest of the lifts receive no $q$-contributions.

\begin{figure}[h]
    \centering
    \includegraphics[width=0.95\linewidth]{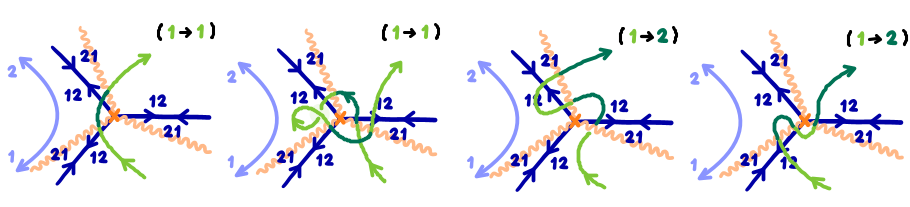}
    \caption{Projections of the lifts of the open path $\mathfrak{p}_2$ to $C$ with source and target sheets indicated.} 
    \label{fig:FN-isotopy-branch-point-3}
\end{figure}

We now compare this with the open path lifts of $\mathfrak{p}_2$.  The detours of $\mathfrak{p}_2$ are encoded in $(S^{q\FN})^{-1}$. So, in total, there are three potential sources of $q$-factors for its lifts:
\begin{enumerate}
    \item a winding factor $q^{-1}$ if the winding takes place on sheet $2$,
    \item a detour-factor $q$ for every detour from sheet $2$ to sheet $1$ (c.f. \cref{eq:quantum-FN}), and finally,
    \item a skein-factor $q^{-1}$ from resolving the self-intersection of a lift with two detours.
\end{enumerate}
The lifts of $\mathfrak{p}_2$ that start and end on sheet $1$ either undergo no detours or two detours. They are illustrated in \cref{fig:FN-isotopy-branch-point-3}. The lift that undergoes no detour is accompanied by a winding factor $q^{-1}$ that matches that of $\mathfrak{p}_1$. For the lift that undergoes two detours, both detours are from sheet $1$ to $2$, and do not contribute any $q$-factors. Furthermore, the winding factor is also $1$. But it does self-intersect, resolving which we get a factor of $q^{-1}$. 

On the other hand, the lifts of $\mathfrak{p}_2$ that start and end on sheet $2$ also undergo no detours or two detours. However, its $q$-factors are all flipped. The lift that undergoes no detours now winds on sheet $1$, and hence, carries a trivial $q$-factor. Likewise, the lift that undergoes two detours, now picks up a winding factor $q^{-1}$, a detour-factor $q^2$ and a skein factor $q^{-1}$ from resolving the self-intersection, which all effectively cancel each other.

The lifts of $\mathfrak{p}_2$ that start on sheet $1$ and end on sheet $2$ either undergo a $1\to 2$ detour at the ``first wall'' or a $2 \to 1$ detour at the ``second wall'' (see \cref{fig:FN-isotopy-branch-point-3}). First of all, there is no $q$-contribution from self-intersections in this case. Second, the former lift has no detour-factor and winding factor while the latter is accompanied by a winding factor $q^{-1}$ and a detour-factor $q$ which cancel each other. Ultimately, both lifts are accompanied by trivial $q$-factors. The same holds for lifts of $\mathfrak{p}_2$ that start on sheet $2$ and end on sheet $1$. From this, we can conclude straight away that $q$-parallel transport along $\mathfrak{p}_1$ and $\mathfrak{p}_2$ are the same. This establishes consistency of the average quantum FN S-matrices under isotopies across branch-points.

\begin{remark} \label{rem:loop-branch-point}
    The argument above also extends to a small closed loop $L$ around the simple branch-point. Since the local foliation of $C$ around such a branch-point is topologically the same for any phase, it is always possible to isotope the loop such that there are no exchanges. By a further isotopy, the loop can also be arranged so that it is travelling upwards when crossing any trajectory (see \cref{fig:closed-loop-branch-point} for an example). Let us then decompose $L = L_u \circ L_{d}$ into a component $L_u$ (shown in dark green in \cref{fig:closed-loop-branch-point}) that is an upwards-travelling open path and $L_d = L \backslash L_u$ (shown in light green). The $q$-nonabelianisation of $L$ is then given by
    \begin{equation}
        \psi^q(L) = \psi^q(L_u) \circ \psi^q(L_d) = P^q(L_u) \circ \psi^q(L_d).
    \end{equation}
    Through construction, $\psi^q(L_d)$ only consists of direct lifts, and we have already established that the $q$-parallel transport map is invariant under an isotopy across a branch-point. It then follows that $\psi^q(L)$ is also invariant under an isotopy across a simple branch-point.

    \begin{figure}[h]
        \centering
        \includegraphics[width=0.55\linewidth]{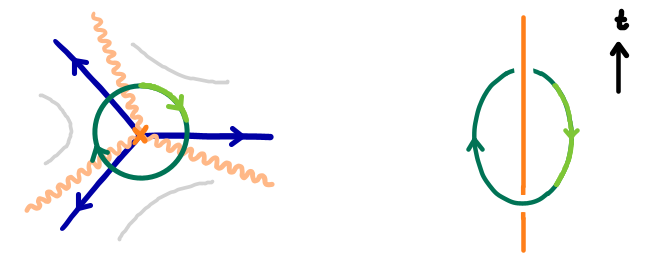}
        \caption{On the left: example of a small loop (in green) around a simple branch-point, projected onto $C$. On the right: the same configuration (without the spectral network trajectories and the branch-cuts), but now projected onto a plane in $C \times I$, perpendicular to one of the spectral network trajectories. }
        \label{fig:closed-loop-branch-point}
    \end{figure}
\end{remark}

\subsubsection{First-principle derivation}
\label{subsubsec:derive-q-FN-mat}
We would now like to derive \cref{eq:quantum-FN} starting from the rules of $q$-nonabelianisation listed in \S\ref{sec:basics-q-nonab}. Consider the infinitesimal path $\varepsilon \subset C \times I$, across the double wall marked by~$S_2$ in \cref{fig:molecule1-nonab-2}, that starts on $C \times \{0\}$ at the base-point to the left of the double wall and ends on $C \times \{1\}$ at the base-point to its right. In each resolution $\pm$, we may decompose $\varepsilon$ into three infinitesimal subpaths $\iota$, $m$ and $f$, where $\iota$ is the segment of $\varepsilon$ before it intersects the double wall, $m$ is the segment that lies in between the first and the second wall, and $f$ is the segment after the double wall. The nonabelian $q$-parallel transport along $\varepsilon$ is a sum of abelian $q$-parallel transports along all possible lifts. We can decompose it into the form
\begin{equation}
    P^{q, -}_\varepsilon = \begin{pmatrix}
        Y_{\iota}^{11, -} & 0 \\ 0 & Y_{\iota}^{22, -}
    \end{pmatrix} \begin{pmatrix}
        1 & s^{12, -} \\ 0 & 1
    \end{pmatrix} \begin{pmatrix}
        Y_{m}^{11, -} & 0 \\ 0 & Y_{m}^{22, -}
    \end{pmatrix} \begin{pmatrix}
        1 & 0 \\ t^{21, -} & 1
    \end{pmatrix} \begin{pmatrix}
         Y_{f}^{11, -} & 0 \\ 0 &  Y_{f}^{22, -}
    \end{pmatrix}
\end{equation}
in the American resolution, and
\begin{equation}
    P^{q, +}_\varepsilon = \begin{pmatrix}
         Y_{\iota}^{11, +} & 0 \\ 0 &  Y_\iota^{22, +}    \end{pmatrix} \begin{pmatrix}
        1 & 0 \\ t^{21, +} & 1
    \end{pmatrix} \begin{pmatrix}
         Y_{m}^{11, +} & 0 \\ 0 &  Y_{m}^{22, +}
    \end{pmatrix} \begin{pmatrix}
        1 & s^{12, +} \\ 0 & 1
    \end{pmatrix} \begin{pmatrix}
         Y_{f}^{11, +} & 0 \\ 0 &  Y_{f}^{22, +}
    \end{pmatrix}
\end{equation}
in the British resolution, where $Y_A^{ii, \pm}$ denotes the quantum generator corresponding to the lift of $A \in \{ \iota, m, f\}$ to the $i$th sheet in the $\pm$ resolution, and where the off-diagonal entries $s^{12, \pm}$ and $t^{21, \pm}$ encode the infinitely many detours associated with the family of $12$ and $21$-trajectories, respectively. 

\begin{figure}[h]
    \centering
    \includegraphics[width=0.8\linewidth]{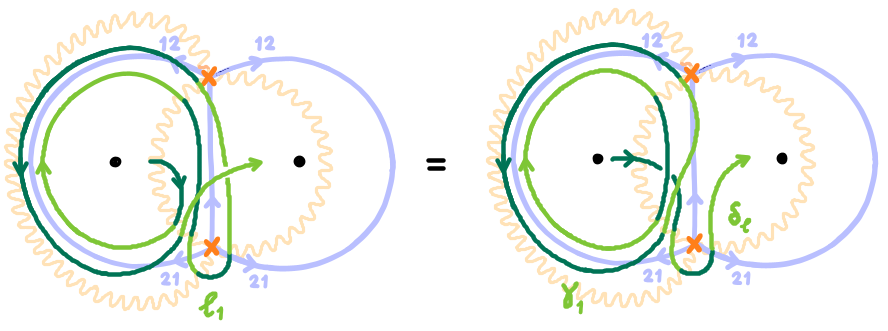}
    \caption{Decomposition of the lift $\ell_1$ (on the left) in terms of $\underline{\gamma}_1$ and $\delta_\ell$ (on the right). (The path $\ell_1$ is a detour path in the British resolution of the FN molecule drawn in the background in light blue.) }
    \label{fig:double-wall-detour-eg}
\end{figure}

Let us, for example, detail how this works for the lift $\ell_1$ shown in \cref{fig:double-wall-detour-eg}. The lift~$\ell_1$ corresponds to one of the infinitely many detours encoded in $t^+$: it is a single $(21)$-detour around the bottom branch-point in the British resolution. The decomposition of the lift $\ell_1$ in homology is simply $2 \gamma_1^\textrm{odd} + \delta_\ell$, where we recall that 
\begin{align}
\eta^\textrm{odd} = \frac 1 2 \big( \eta^{(1)} - \eta^{(2)}\big) \text{ for any } \eta \in H_1(\Sigma; \Z),
\end{align}
and $\delta_\ell$ is a detour around the bottom branch-point (see \cref{fig:double-wall-detour-eg} for notation). A similar decomposition holds inside $\mathbf A$, except that one now needs to be careful with $q$-contributions that arise from applying the $GL(1, \C)$-skein relations.

\begin{figure}[h!]
    \centering
    \includegraphics[width=0.55\linewidth]{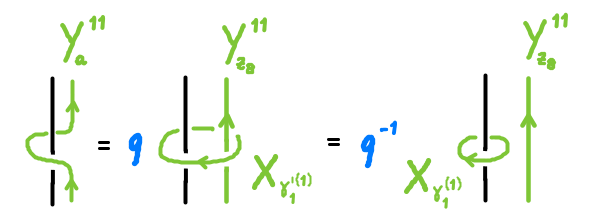}
    \caption{Two different ways to decompose the open path on the left in terms of the quantum torus generators $X_{\underline{\gamma}_1}, X_{\underline{\gamma}'_1}$ and $Y^{11}_{z_B}$. }
    \label{fig:lift-arrangement}
\end{figure}

For instance, consider the open path lift $\widetilde{\underline{\fraka}}_1^{11} \subset \Sigma \times I$ of the \emph{based} loop $\underline{\fraka}_1 \subset C$ with base-point $z_B$ as shown in \cref{fig:molecule1-nonab}. Note that the open path $\widetilde{\underline{\fraka}}_1^{11}$ is unique inside $\mathbf A$ because we have fixed a base-point for $\underline{\fraka}_1$. Then, using the abelian skein relations, we can write
\begin{equation}
    Y_{\underline{\fraka}}^{11} = q^{-1} X_{\underline{\gamma}_1^{(1)}} Y_{z_B}^{11}, \label{eq:diagonal-open-path-decomposition}
\end{equation}
where $X_{\underline{\gamma}_1^{(1)}}$ is the quantum torus generator corresponding to the closed 1-cycle $\underline{\gamma}^{(1)}_1 \subset \Sigma$ that we obtain by forgetting the base-point of $\underline{\fraka}_1$, and $Y_{z_B}^{11}$ is the generator corresponding to the open path $\{z_B^{(1)}\} \times I$. There is yet another way to write $Y_{\widetilde{\underline{\fraka}}}^{11}$ as 
\begin{equation}
    Y_{\underline{\fraka}}^{11} = q \, X_{\underline{\gamma}_1'^{(1)}} Y_{z_B}^{11}
\end{equation}
where $X_{\underline{\gamma}_1'^{(1)}}$ is the quantum torus generator corresponding to the 1-cycle $\underline{\gamma}'_1 \subset \Sigma$ illustrated in \cref{fig:lift-arrangement}. While the difference between the two closed cycles $\underline{\gamma}_1$ and $\underline{\gamma}_1'$ is apparent in this set up -- namely, $\underline{\gamma}_1$ is a closed cycle around the puncture, and $\underline{\gamma}_1'$ is a closed cycle encircling the puncture and the open path -- this distinction is subtle for a general pants cycle. We will expand on this in \S\ref{subsec:gluing-q-nonab}.

Returning to our example, we can decompose the open path $Y_{\ell_1}^+$ as
\begin{equation}
    Y_{\ell_1}^{21, +} = X_{\underline{\gamma}_1^{(1)}}^+ \, X_{-\underline{\gamma}_1^{(2)}}^+ \, Y_{\delta_\ell}^{21, +} = \big(X_{\underline{\gamma}_1^\textrm{odd}}^+\big)^2 \, Y^{21, +}_{\delta_\ell}.
\end{equation}
More generally, a generic path in $t^+$ that undergoes a single $21$-detour around the bottom branch-point while winding $k$ times on each sheet, contributes as
\begin{equation}
   Y_{\ell_k}^{21, +} =  \big(X^+_{\underline{\gamma}_1^{(1)}}\big)^{k} \, \big(X^+_{-\underline{\gamma}_1^{(2)}}\big)^{k} \, Y_{\delta_\ell}^{21, +} = \big(X^+_{\underline{\gamma}_1^\textrm{odd}}\big)^{2k} \, Y_{\delta_\ell}^{21, +} \quad \text{ for } k \ge 0. \label{eq:k-wind-decomposition}
\end{equation}

In \S\ref{sec:quantumFN}, we learned that quantum spectral coordinates $X_\gamma$ act as operators on the space of Liouville conformal blocks. In particular, the quantum generators $X_{\underline{\gamma}_k^\textrm{odd}}$, associated with odd 1-cycles $\underline{\gamma}_k^\textrm{odd}$ around the punctures $z_k$, act by multiplication by the mass eigenvalues $M_k$. We denote this by
\begin{align}\label{eq:XgammaM}
X_{\underline{\gamma}_k^\text{odd}} \rightsquigarrow M_k.
\end{align}
The same holds true in the presence of open paths (recall the 3d picture from \S\ref{sec:CStheory}). This implies that~\cref{eq:k-wind-decomposition} evaluates to
\begin{equation}
X_{\ell_k}^{21, +}    \rightsquigarrow  M_1^{2k}\, X_{\delta_\ell}^{21, +}.
\end{equation}

\begin{remark} \label{rem:mass-defn-modified}
    Since $X_{\underline{\gamma}^\textrm{odd}}$ evaluates to $M$, one might think that the $X_{\underline{\gamma}^{(i)}}$ themselves evaluate to~$M^{\pm 1}$. But this is \emph{not} true in the presence of open paths because
    \begin{align}
    \underline{\gamma}_1^{(1)} + \underline{\gamma}_1^{(2)} + \underline{\gamma}_2^{(1)} + \underline{\gamma}_2^{(2)} + \underline{\gamma}_3^{(1)} + \underline{\gamma}_3^{(2)} \ne 0 \in H_1(\Sigma - \pi^{-1}(\mathbb X); \Z).
    \end{align}
    Instead, in the example above, we find that 
    \begin{equation}
        X_{\underline{\gamma}^{(1)}} \rightsquigarrow q M, \quad  \text{ while } X_{-\underline{\gamma}^{(2)}} \rightsquigarrow q^{-1} M. \label{eq:modified-mass-def}
    \end{equation}
    More generally, the result depends on the number of open paths in the configuration.
\end{remark}

\begin{figure}[h]
    \centering
    \includegraphics[width=0.95\linewidth]{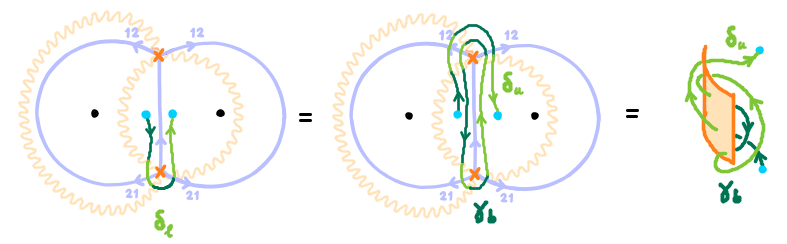}
    \caption{We can further write $\delta_\ell$ in terms of a detour $\delta_u$ around the upper branch-point. The first two figures show the 2d projections, and the last figure the 3d arrangement. We have suppressed two out of the three branch-cuts in the third figure for clarity. }
    \label{fig:double-wall-detour-eg-2}
\end{figure}

\begin{figure}[h]
    \centering
    \includegraphics[width=0.75\linewidth]{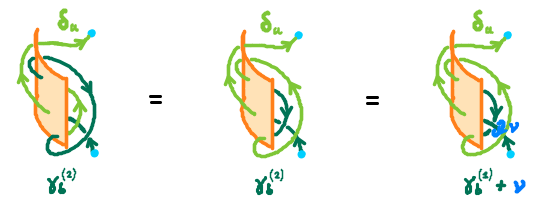}
    \caption{Two representations $\gamma_b^{(1)}$  and $\gamma_b^{(2)}$  of the 1-cycle $\gamma_b$ in the presence of the open path $\delta_u$. }
    \label{fig:gamma-ambiguity-2}
\end{figure}

It is more convenient to write the detour around the bottom branch-point $\delta_\ell$ in terms of a detour around the top branch-point $\delta_u$, as shown in \cref{fig:double-wall-detour-eg-2}. This allows us to write
\begin{equation}
    Y_{\delta_\ell}^{21, +} = X_{\gamma_\mathfrak b}^+ \, Y_{\delta_u}^{21, +} \label{eq:unprimed-generator-skein}
\end{equation}
where $\gamma_\mathfrak b$ is a closed 1-cycle that encircles both branch-points.\footnote{Note that this is also the sequence of moves that relates the primed and unprimed off-diagonal generators $Y^{ij}$ and $Y^{'ij}$ in \cref{eq:quantum-FN} and \cref{eq:quantum-FN-2}.} In order to evaluate $X_{\gamma_\mathfrak b}$, we will have to write $\gamma_\mathfrak b$ in terms of the odd mass cycles. In homology, this is simply
\[\gamma_\mathfrak b = \frac 1 2 \big(\underline{\gamma}_1^{(1)} + \underline{\gamma}_2^{(1)} - \underline{\gamma}_3^{(2)} - \underline{\gamma}_1^{(2)} - \underline{\gamma}_2^{(2)} + \underline{\gamma}_3^{(1)} \big) \in H_1^\textrm{odd}(\Sigma; \Z).\]
However, in the presence of the open path $d_u$, this relation is no longer valid inside $\mathbf A$. Using the abelian skein relations, we can equivalently write $\gamma_\mathfrak b$ as a closed cycle on sheet $1$ or sheet $2$ (also shown in \cref{fig:gamma-ambiguity-2}). Only $\gamma_\mathfrak b^{(1)}$ has a non-trivial intersection pairing with $\delta_u$. So, the above relation is modified to 
\begin{align}
    [\gamma_\mathfrak b] &= \frac 1 2 \bigg([\underline{\gamma}_1^{(1)}] + [\nu] + [\underline{\gamma}_2^{(1)}] - [\underline{\gamma}_3^{(2)}] - [\underline{\gamma}_1^{(2)}] - [\underline{\gamma}_2^{(2)}] + [\underline{\gamma}_3^{(1)}] \bigg)  \in \mathbf A \notag \\
    & = \frac 1 2[\nu] + [\underline{\gamma}_1^\textrm{odd}] + [\underline{\gamma}_2^\textrm{odd}] + [\underline{\gamma}_3^\textrm{odd}] 
\end{align}
where $\nu$ is a small clockwise loop around $d_u$. This decomposition yields
\begin{equation}
    X^+_{\gamma_\mathfrak b} \rightsquigarrow q^{-1} \frac {M_1 M_2} {M_3}, \label{eq:branch-point-closed-cycle}
\end{equation}
where we have used that $X_\nu = q^{-2}$. Recall that not all three mass eigenvalues are defined symmetrically (see \cref{eq:localmonodromy}).

This concludes the decomposition of the open path generator $Y^{21, +}_{\ell_k}$ as 
\begin{align}
Y^{21, +}_{\ell_k} \rightsquigarrow q^{-1} M_1^{2k+1} M_2 M_3^{-1} \, Y^{21, +}_{\delta_u}. 
\end{align}
Thus, the total contribution to the abelian $q$-parallel transport from $\ell_k$ for all windings $k \in \mathbb N$ is
\begin{equation}
    q^{-1} \frac {X^+_{\underline{\gamma}_1^\textrm{odd}} X^+_{\underline{\gamma}_2^\textrm{odd}} X^+_{\underline{\gamma}_3^\textrm{odd}} }{1 - X^+_{b\underline{\gamma}_1^\textrm{odd}} } \, Y_{\delta_u}^{21, +} \rightsquigarrow q^{-1} \frac {M_1 M_2 M_3^{-1}}{1 - M_1^2} \, Y_{\delta_u}^{21, +}.
\end{equation}

The rest of the exercise is similar in nature: we find a decomposition of every possible lift in our preferred basis for the space of open skeins $\mathbf A$ on $\Sigma$, and evaluate the closed generators to get masses $M_k$. This is a straightforward exercise in abelian skein theory. We find that
\begin{equation}
    P^{q, +}_\varepsilon \rightsquigarrow  \begin{pmatrix}
         Y^{11, +}_{\varepsilon} & \frac {1 - M_1^{-1} M_2 M_3}{1-M_2^2} \, Y^{12, +}_{\delta_u} \\ 
       q^{-1} \, \frac {M_1 M_2 M_3^{-1} - M_1^2 }{1 - M_1^2}  \, Y^{21, +}_{\delta_u} & \left(1 + \frac {M_1^2 - M_1 M_2 M_3 - M_1 M_2 M_3^{-1} + M_2^2}{(1 - M_1^2)(1-M_2^2)} \right)  Y^{22, +}_{\varepsilon} \end{pmatrix}.
\end{equation}
Note that the coefficients of the open path generators are identical to the semi-classical expression~\cref{eq:PTBgauged3-puncturedsphere}, up to the quantum factor $q^{-1}$ in the off-diagonal entry. Likewise, in the American resolution we find
\begin{equation}
    P^{q, -}_\varepsilon \rightsquigarrow \begin{pmatrix}
            \left(1 + \frac {M_1^2 - M_1 M_2 M_3^{-1} - M_1 M_2 M_3 + M_2^2}{(1 - M_1^2)(1-M_2^2)} \right)  Y^{11, -}_{\varepsilon}  & \frac {1 - M_1 M_2^{-1} M_3}{1-M_1^2}  \, Y^{12, -}_{\delta_u} \\ 
        q^{-1} \, \frac {M_1 M_2 M_3^{-1} - M_2^2}{1 - M_2^2} \, Y^{21, -}_{\delta_u} &  Y^{22, -}_{\varepsilon}
    \end{pmatrix}.
\end{equation}

We are now in a position to determine the average quantum FN S-matrix $S^{q\FN}$. As in the semi-classical case \ref{eq:Smatrix-average}, we define the $ij$-th entry of $S^{q\FN}$ as the entry-wise average
\begin{equation} \label{eq:quantum-FN-S-wall}
    S^{q\FN}_{ij} = \sqrt{ \left[P^{q, +}_\varepsilon \right]_{ij} \otimes \left[ P^{q,-}_\varepsilon \right]_{ij}}.
\end{equation}
This directly recovers the expressions for $\alpha, \beta$ and $\gamma$ in \cref{eq:quantum-FN}. If we furthermore define the open path generators 
\begin{align}
\begin{aligned}
& Y^{11}_{2, L} = \sqrt{Y^{11, +}_{\varepsilon} \otimes Y^{11, -}_{\varepsilon}}, \quad Y^{12}_{2, L} = \sqrt{Y_{\delta_u}^{12, +} \otimes Y_{\delta_u}^{12, -}}, \\
& Y^{21}_{2, L} = \sqrt{Y_{\delta_u}^{21, +} \otimes Y_{\delta_u}^{21, -}}, \quad Y^{22}_{2, L} = \sqrt{Y^{22, +}_{\varepsilon} \otimes Y^{22, -}_{\varepsilon}},
\end{aligned}
\end{align}
we recover the average quantum Fenchel-Nielsen S-matrix $S^{q\FN} \in \mathrm{Mat}_{2\times 2}(\mathbf A)$ defined in \cref{eq:quantum-FN}.

Instead, to write the quantum S-matrix in terms of the primed generators, we simply have to undo the relation \cref{eq:unprimed-generator-skein}, 
\begin{equation}
    Y_{\delta_u}^{21, \pm} = q \, M_1^{-1} M_2^{-1} M_3 Y_{\delta_\ell}^{21, \pm}.
\end{equation}
We now define $Y_{2, L}^{'21} = \sqrt{Y^{21, +}_{\delta_\ell}\otimes Y^{21, -}_{\delta_\ell}}$ and $Y_{2, L}^{'12} = \sqrt{Y^{12, +}_{\delta_\ell}\otimes Y^{12, -}_{\delta_\ell}}$. Then, by comparing the coefficients of the primed and unprimed generators, we immediately find
\begin{equation}
    \beta' = \frac {M_3} {M_1 M_2} \, \beta,
\end{equation}
and likewise for $\alpha'$. This concludes the derivation of the quantum FN S-matrix.

\subsection{Quantum Fenchel-Nielsen coordinates} \label{subsec:quantum-FN-coordinates-2}

The semi-classical limits of the abelian $q$-parallel transports $Y^{ij}_{L/R}$ are not gauge-invariant, and therefore do not define functions on the moduli space of flat connections and cannot be consistently quantised. In contrast, the abelian parallel transports $M_{1,2,3}$ along the lifts of the based pants cycles $\underline{\fraka}_{1, 2, 3} \subset C$, and the abelian parallel transports $T_{a,b,c}$ along the lifts of the paths $\mathfrak{p}_{a,b,c} \subset C$ (defined in \S\ref{subsubsec:gaugedsphere}) \emph{are} gauge-invariant (since we fixed a trivialisation at the marked points), and thus can be quantised. Consequently, the semi-classical limits of the generators
\begin{align}\label{eq:basisq-FN}
Y^{ii}_{\underline{\fraka}_1, \underline{\fraka}_2, \underline{\fraka}_3} \quad \text{ and } \quad  Y^{ij}_{\mathfrak{p}_a, \mathfrak{p}_b, \mathfrak{p}_c},
\end{align} 
are gauge-invariant. Moreover, since any gauge-invariant abelian parallel transport can be expressed in terms of the spectral coordinates, it follows that the open path generators~\eqref{eq:basisq-FN} form a basis for the space of all abelian $q$-parallel transports whose semi-classical limit is gauge-invariant. Thus, the (linear combination of) generators \cref{eq:basisq-FN} (corresponding to the symplectic basis of 1-cycles~\eqref{eq:ABcycles3-sphere}) have the right characteristics to be interpreted as exponentiated quantum FN length and twist spectral coordinates on the gauged 3-sphere.

\begin{figure}[h]
    \centering
    \includegraphics[width=0.5\linewidth]{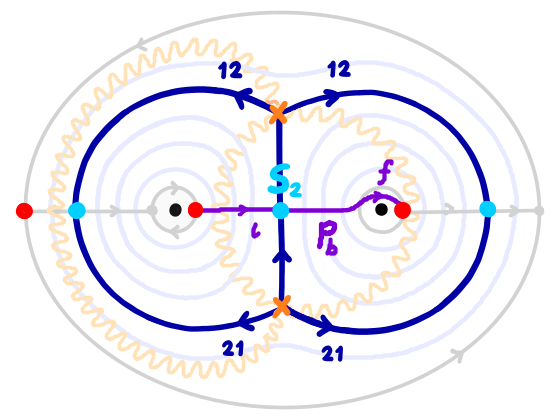}
    \caption{Since the path $\frakp_b$ extends between two marked points on $C$, the abelian parallel transports along all the lifts of $\frakp_b$ are gauge-invariant. }
    \label{fig:quantum-pb}
\end{figure}

As an example, consider the $q$-parallel transport along the (unique open path lift of the) path $\frakp_b \subset C \times I$ in \cref{eq:pathspABC}. We show that we can read off the quantum Fenchel-Nielsen twist coordinates from this $q$-parallel transport. If we decompose $\frakp_b$ into two segments $\iota$ and~$f$, where $\iota$ is the segment before crossing the double trajectory and $f$ is the segment after crossing the double trajectory (illustrated in \cref{fig:quantum-pb}), the non-abelian $q$-parallel transport along $\frakp_b$ is given by
\begin{equation} \label{eq:quantum-parallel-transport-FN}
    P^q_{b} = \begin{pmatrix}
        0 & Y^{12}_{\iota}\\ Y^{21}_{\iota} & 0
    \end{pmatrix} \begin{pmatrix}  \gamma  Y^{11}_{2, L} & \alpha  Y^{12}_{2, L} \\ - q^{-1}\beta  Y^{21}_{2, L} & \gamma  Y^{22}_{2, L} \end{pmatrix} 
    \begin{pmatrix}
        q Y^{11}_{f} & 0 \\ 0 & Y^{22}_{f}
    \end{pmatrix} = \begin{pmatrix}
        -\beta Y^{11}_{b, L} & \gamma Y^{12}_{b, L}\\
        q \gamma Y^{21}_{b, L} & \alpha Y^{22}_{b, L}
    \end{pmatrix},
\end{equation}
in terms of the open path generators $Y^{ij}_{b, L}$,
\begin{align}
\begin{aligned}
& Y^{11}_{b, L} = Y^{12}_\iota Y^{21}_{2, L} Y^{11}_f, \qquad Y^{12}_{b, L} = Y^{12}_{\iota}Y^{22}_{2,L}Y^{22}_f, \\
& Y^{21}_{b, L} = Y^{21}_\iota Y^{11}_{2,L}Y^{11}_f, \qquad Y^{22}_{b, L} = Y^{21}_\iota Y^{12}_{2,L}Y^{22}_f.
\end{aligned} \label{eq:quantum-open-coordinates}
\end{align}

Unlike the generators $Y_{2, L}^{ij}$ in the quantum S-matrix~\eqref{eq:quantum-FN}, the generators~$Y^{ij}_{b, L}$ are gauge-invariant in the semi-classical limit, and admit a decomposition in terms of the generators~\eqref{eq:basisq-FN}. We can immediately identify 
\begin{align}
Y_{b, L}^{ij}=Y_{\frakp_b}^{ij}, \quad \text{ for }~~ i \ne j,
\end{align}
while the diagonal $Y_{b, L}^{ii}$ can be expressed as the composition
\begin{equation}
    Y^{ii}_{b, L} = Y^{ij}_{-\frakp_a} \, Y^{ji}_{-\frakp_c}, \quad \text{ for }~~ j \ne i.
\end{equation}
Indeed, comparing the expression for $P^q_b$ with its semi-classical limit $P_b$ in \cref{eq:parallel-transport-FN-2}, we can simply verify that the off-diagonal $Y^{21}_{b, L}$ reduces to the twist coordinate $T_b$, whereas the diagonal $Y^{ii}_{b, L}$ reduce to a ratio of the twist coordinates $T_a$ and $T_c$.

More precisely, recall that a basis of semi-classical spectral coordinates can be given in terms of the abelian parallel transport along a basis consisting of \emph{odd} 1-cycles in the relative homology on $\Sigma$. Similarly, a basis of quantum spectral coordinates can be labelled by \emph{odd} open paths (see also the discussion around \cref{rem:mass-defn-modified}). The quantisation of the exponentiated FN twist coordinate~$T_b$ is therefore 
\begin{equation}\label{eq:oddYp}
    Y_{\frakp_b}^\textrm{odd} := Y_{\frac 1 2 \big(\widetilde\frakp^{21}_{b} - \widetilde\frakp^{12}_{b} \big)},
\end{equation}
where, in our conventions, $- \widetilde \frakp^{12} = (-\widetilde \frakp)^{21}$. We may also informally think of \cref{eq:oddYp} as the square-root of the product of the generators $Y_{b, L}^{21}$ and $Y_{b, R}^{21}$. Likewise, the quantisation of the exponentiated FN length coordinate~$M_k$ is
\begin{equation}\label{eq:average-quantum-coord}
    Y_{\underline{\fraka}_k}^\textrm{odd} := Y_{\frac 1 2 \big(\widetilde{\underline{\fraka}}_k^{11} - \widetilde{\underline{\fraka}}_k^{22} \big)}.
\end{equation}

We emphasise that this is a quantisation in terms of open skeins $Y_{\widetilde\frakp} \in \mathbf A$, and $Y_{\underline{\fraka}_k}^\textrm{odd}$ should be treated differently from the quantum torus generators (closed skeins) $X_{\underline{\gamma}_{k}^\textrm{odd}}$.\footnote{Here, we distinguish between the open path generator $Y_{\underline\fraka}$ and the quantum torus generator $X_\gamma$: the former is the quantisation of the abelian parallel transport (for a choice of base-point) around the puncture, while the latter is the quantisation of the trace of the aforementioned parallel transport. Although the two are identical in the semi-classical limit, their quantisations are elements of different quantum algebras.}

\begin{remark}
   There is a subtlety in writing down the $q$-parallel transport matrices $P^q$ in \cref{eq:quantum-parallel-transport-FN}. We have assumed that the open path approaches its end-points transversely to the WKB foliation (see \cref{fig:quantum-pb}). However, one could equally well choose a path that ends tangentially to the WKB foliation. In the latter case, the sheet-dependent winding factor is different. Consequently, special care is required when composing $P^q$-matrices to construct general $q$-parallel transport matrices.
\end{remark}

\subsubsection{Physical interpretation: action on conformal blocks} \label{subsubsec:y-coordinates-physics}

In \S\ref{sec:CStheory}, we realised the quantum torus generators $X_{\gamma}$ as abelian Verlinde operators (i.e.~Heisenberg-Verlinde operators) on the space of abelian conformal blocks on $\Sigma$. Similarly, the 3d Chern-Simons picture allows us to interpret the open path generators $Y_{\widetilde \frakp}$ as operators moving the position of an abelian degenerate operator insertion inside an abelian conformal block $\mathcal F^\textrm{ab}_{\vec \alpha}$ (c.f. \cref{eq:open-path-generators-action}),
\begin{equation}
    Y_{\widetilde \frakp} \, \mathcal F^\textrm{ab}_{\vec \alpha} \big( \widetilde V^\textrm{ab}_\textrm{deg}(z^\textrm{in}) \ldots \big) =  C(\ldots) \, \mathcal F^\textrm{ab}_{\vec \alpha'} \big( \widetilde V^\textrm{ab}_\textrm{deg}(z^\textrm{out}) \ldots \big) \: \text{ for } \: \partial \widetilde\frakp = \{z^\textrm{in}\} \cup \{z^\textrm{out}\}.
\end{equation}
In the following paragraphs, we shall detail how to compute the coefficients $C(\ldots)$ for the basis generators $Y_{\underline{\fraka}}^{ii}$ and $Y_{\frakp}^{ij}$ on the gauged 3-sphere.

This is easiest to spell out for the generators $Y_{\underline{\fraka}}^{ii}$. Recall from \cref{eq:diagonal-open-path-decomposition} that the abelian skein relations allows us to write
\begin{equation}
    Y_{\underline{\fraka}_1}^{11} = q^{-1} X_{\underline{\gamma}_1^{(1)}} Y_{z_B}^{11} \quad \text{ and } \quad  Y_{\underline{\fraka}_1}^{22} = q^{-1} \, X_{\underline{\gamma}_1^{(2)}} Y_{z_B}^{22}, \label{eq:open-a-decomposition}
\end{equation}
and likewise for the other $\underline{\fraka}_k$, where we remind the reader that $Y_{z_{B}}^{ii}$ is the identity generator associated with the open path $\{z_B^{(i)}\} \times I$. Now, we know the evaluation of the quantum torus generators (see \cref{rem:mass-defn-modified}),
\begin{equation}
    X_{\underline{\gamma}_1^{(1)}} \rightsquigarrow q M_1 \quad \text{ and } X_{\underline{\gamma}_1^{(2)}} \rightsquigarrow \frac q {M_1},
\end{equation}
in the presence of an open path.
This implies that we can write the action of the open path generator $Y_{\underline{\fraka}_1}$ as
\begin{equation}
    \begin{aligned}
        Y_{\underline{\fraka}_1}^{11} \mathcal F^\textrm{ab} \big(\widetilde V^\textrm{ab}_\textrm{deg} (z_B^{(1)}) \ldots \big) &= M_1  \, \mathcal F^\textrm{ab} \big(\widetilde V^\textrm{ab}_\textrm{deg} (z_B^{(1)}) \ldots \big), \text{ and }\\
        Y_{\underline{\fraka}_1}^{22} \mathcal F^\textrm{ab} \big(\widetilde V^\textrm{ab}_\textrm{deg} (z_B^{(2)}) \ldots \big) &= \frac 1 M_1  \, \mathcal F^\textrm{ab} \big(\widetilde V^\textrm{ab}_\textrm{deg} (z_B^{(2)}) \ldots \big),
    \end{aligned} \label{eq:evaluation-open-a}
\end{equation}
and similarly for the other $\mathfrak{a}_k$.
In this form, it is easy to see that the open path generators $Y_{\underline{\fraka}_{1,2,3}}^{ii}$ commute with the action of the quantum torus generators $X_{\underline{\gamma}_{1,2,3}}$. Further, observe that the above expressions \cref{eq:evaluation-open-a} match with the MS braiding matrices in \cref{eq:braiding-matrix-can-M}, up to an overall $q$-factor.

For the open path generators $Y_{\frakp}^{ij}$, we remind ourselves that we positioned the loops ${\underline{\gamma}_{1,2,3}^{(i)}}$ in between the punctures $z_{1,2,3}^{(i)}$ and the marked points $z_{A, B, C}^{(i)}$ (see \cref{fig:pants-cycle-ordering}). This implies that the open path generators $Y_{\frakp_{a,b,c}}^{ij}$ commute with the Heisenberg-Verlinde operators~$X_{\underline{\gamma}_{1,2,3}^{(i)}}$. Since Liouville blocks on the gauged 3-sphere are fully fixed by the masses $M_{1, 2, 3}$, and also their abelianisation as a consequence, we conclude that the open path generators $Y_{\frakp_{a,b,c}}^{ij}$ act multiplicatively on the abelian conformal block $\mathcal{F}^\text{ab}$. The multiplicative constant is the semi-classical limit of $Y_{\frakp}^{ij}$ evaluated on the corresponding BPZ connection. Thus, we can write the action of $Y^{ij}_{\frakp_{a,b,c}}$ as
\begin{equation}
    Y_{\frakp_{a, b, c}}^{ij} \mathcal F^\textrm{ab} \big(\widetilde V^\textrm{ab}_\textrm{deg} (z_{A, B, C}^{(i)}) \ldots \big) =  T^\textrm{Rie}_{a, b, c} \, \mathcal F^\textrm{ab} \big(\widetilde V^\textrm{ab}_\textrm{deg} (z_{A, B, C}^{(j)}) \ldots \big). \label{eq:evaluation-open-p}
\end{equation}
We can either read off the expressions for $T^\textrm{Rie}_{a, b, c}$ from \cref{eq:T-BPZ-1}\footnote{Recall that when expressed as in \cref{eq:MSP}, the MS parallel transports $P^\textrm{MS}_{a,b,c}$ are not $SL(2, \C)$-equivariant. So, a further basis rescaling is required at every marked point before the two quantum parallel transports $P^\textrm{MS}$ and $P^q$ can be compared (see the discussion around \cref{eq:detF}). Thus, in all our comparisons with MS parallel transports, we assume that such a basis rescaling is implemented, and the MS parallel transport has been brought in the same gauge as $\mathcal W-$nonabelianisation.} or compute them directly knowing the generating function for the subspace of opers. Note that this statement is only true up to an overall power of $q$ because it is only the evaluation of the quantum twist coordinate $Y_{\frakp_{a, b, c}}^\textrm{odd}$ that is unambiguously defined (see \cref{rem:mass-defn-modified}).

\begin{remark}
    The space of genus zero Heisenberg conformal blocks (for any number of generic vertex operator insertions with fixed momenta) is 1-dimensional. Then, any operator that commutes with the action of the Heisenberg-Verlinde operators (i.e.~doesn't modify the vertex operator insertions) must act multiplicatively. 
\end{remark}

This analysis extends even in the presence of multiple degenerate operators (inserted at any of the points in $\mathbb X$). The crucial thing to note is that all basis generators \cref{eq:basisq-FN} always commute with the action of the Heisenberg-Verlinde operators. Then, for any number of composable open path generators, not necessarily associated to connected open paths, we can immediately conclude that the action of these generators is multiplicative.

In particular, this implies that the action of the quantum FN coordinates on the gauged 3-sphere is multiplicative, and that the multiplicative constant is the corresponding abelian parallel transport evaluated on the BPZ connection. 
It is in this sense that we refer to the open path generators 
\begin{align}
Y_{\underline{\fraka}_{1, 2, 3}}^\textrm{odd}, Y_{\frakp_{a, b, c}}^\textrm{odd}
\end{align}
as (exponentiated) quantum Fenchel-Nielsen length and twist coordinates, respectively, on the gauged 3-sphere.

\subsubsection*{CFT-abelianisation} 

Below, we take a slight detour to investigate CFT-(non)abelianisation under the lens of $q$-parallel transports. We define the space of Liouville conformal blocks on the gauged 3-sphere $C$ with one marked point $z'$,
\begin{align}
    \mathrm{gCB}(C) := \bigoplus_{\alpha_1,\, \alpha_2,\, \alpha_3} \hom\bigg(\mathcal V_{1,2}(z') \otimes \mathcal V_{\alpha_1}(z_1) \otimes \mathcal V_{\alpha_2}(z_2) \otimes \mathcal V_{\alpha_3}(z_3), \, \C\bigg)^\text{Ward id},
\end{align}
as the space of gauged 3-point conformal blocks on $C$ (with unspecified generic Virasoro representations at the punctures, as in \cref{eq:gauged-conformal-block-Liouville}) and additionally one degenerate vertex operator insertion $V_{1,2}$ at the marked point. Recall that this is an infinite-dimensional vector space, but that fixing the Virasoro representations at the punctures produces a two-dimensional subspace $\mathrm{CB}(C)$. This subspace can be equivalently characterised as an eigenspace of the Liouville-Verlinde operators $\mathcal L_{\underline{\gamma}_{1, 2, 3}}$. 

Our aim in this paragraph is to find the corresponding abelianised conformal blocks. Below, through a Riemann-Hilbert type argument, we will specify their abelianisations as Heisenberg conformal blocks on the gauged 6-punctured sphere $\Sigma$ with one marked point. In the argument we assume some ingredients from CFT-abelianisation~\cite{HollandsMurugesanII} (as well as that $q$-abelianisation and CFT-abelianisation commute).

The standard choices of a basis for $\mathrm{CB}(C)$ can be characterised in terms of the position of the degenerate vertex operator. Examples are the s- and t-channel bases from \cref{fig:four-point-with-degenerate}. This position, in turn, fixes the MS parallel transport around the closest puncture to be diagonal. For instance, in the s-channel, the two-dimensional basis is spanned by the Liouville blocks
\begin{equation}
    \mathcal F_{s,\pm}^\textrm{Liou} \big(V_{1, 2} (z_B) \, V_{1}(z_1) \, V_{2}(z_2) \, V_3(z_3)\big),
\end{equation}
where $\pm$ refers to the two allowed intermediate momenta $\alpha = \alpha_1 \pm b/2$, and the MS parallel transport $P^\textrm{MS}_{\underline{\fraka}_1}$ can be written as
\begin{equation}\label{eq:Pa1Ya1}
    P^\textrm{MS}_{\underline{\fraka}_1} = q \begin{pmatrix}
        M_1 & 0 \\ 0 & \frac{1}{M_1}  \end{pmatrix} 
        = q \begin{pmatrix}
        Y_{\underline{\fraka}_1}^{11} & 0 \\ 0 & Y_{\underline{\fraka}_1}^{22} \end{pmatrix}.
\end{equation}

Meanwhile, we know from \cref{eq:evaluation-open-a} that the open path generators $Y_{\underline{\fraka}_1}^{11}$ and $Y_{\underline{\fraka}_1}^{22}$ act on abelian conformal blocks with (abelian) degenerate operator insertions at either of the lifts of $z_B$. The interplay between $q$-parallel transports and CFT-abelianisation furthermore tells us that \emph{generic} Liouville vertex operators on $C$ (at any of the three punctures) are abelianised into pairs of (conjugate) abelian vertex operators on the cover $\Sigma$ (see \cite[\S3]{HollandsMurugesanII}). Together this implies that the open path generators~$Y_{\underline{\fraka}_1}^{ii}$ must act on the Heisenberg conformal blocks
\begin{align}
    \begin{aligned}\label{eq:abelianbasisdeg}
        \mathcal F^{\textrm{ab}}_{(1)} \, (z_B) &=  \mathcal F^\textrm{ab} \bigg(\widetilde V^\textrm{ab}_\textrm{deg} (z_B^{(1)}) \, \widetilde V_1^\textrm{ab}(z_1) \,  \widetilde V_2^\textrm{ab}(z_2) \, \widetilde V_3^\textrm{ab}(z_3) \bigg), \\
        \mathcal F_{(2)}^{\textrm{ab}} \, (z_B)&= \mathcal F^\textrm{ab} \bigg(\widetilde V^\textrm{ab}_\textrm{deg} (z_B^{(2)})  \,\widetilde V_1^\textrm{ab}(z_1) \,\widetilde  V_2^\textrm{ab}(z_2) \,\widetilde V_3^\textrm{ab}(z_3)) \bigg),
    \end{aligned}
\end{align}
where $\widetilde V^\textrm{ab}(z)$ denotes the abelianisation of the Liouville vertex operator $V(z)$.

From the identity~\cref{eq:Pa1Ya1}, it then follows that CFT-abelianisation must map
\begin{align} \label{eq:abelianisation-map-deg}
\psi^\text{CFT}_{\,\text{FN}} :\quad
\begin{pmatrix}
\mathcal F_{s,+}^{\textrm{Liou}}\!\left(
V_{1,2}(z_B)\, V_1(z_1)\, V_2(z_2)\, V_3(z_3)
\right)
\\[6pt]
\mathcal F_{s,-}^{\textrm{Liou}}\!\left(
V_{1,2}(z_B)\, V_1(z_1)\, V_2(z_2)\, V_3(z_3)
\right)
\end{pmatrix}
\;\mapsto \;
\begin{pmatrix}
\mathcal F_{(1)}^{\textrm{ab}}
\\[4pt]
\mathcal F_{(2)}^{\textrm{ab}}
\end{pmatrix}.
\end{align}
In particular, this implies that we can identify the abelian blocks $\mathcal F_{(i)}^{\textrm{ab}}$ with the standard hypergeometric solutions to the BPZ equation~\eqref{eq:BPZ3pt}.\footnote{As shown in \cite{Bramley:2025zek}, these solutions can be obtained from the perspective of the exact WKB analysis as a Borel sum of the asymptotic solutions in the direction of the FN phase.}

We can similarly obtain the CFT-abelianisation of any other choice of basis for CB($C$) by displacing the position of the degenerate operator insertion to any of the other punctures.  Furthermore, more general open path generators~$Y_{\widetilde \frakp}$ (with distinct end-points) appear in the description of basis change transformations on the space of Liouville blocks as in \cref{eq:quantum-parallel-transport-FN}.

\begin{remark} \label{rem:differential-treatment-deg}
    Note that CFT-abelianisation treats generic and degenerate vertex operator insertions differently. This differential treatment can be explained through the 6d $(2, 0)$-theory in the background $\R^4 \times C$ that sources the AGT correspondence. 
    
    In the 6d theory, the punctures on $C$ are the insertion points for codimension-$2$ defects (space-filling defects inside $\R^4$) that describe the matter content of the 4d $\mathcal N =2 $ theory on $\R^4$ \cite{Gaiotto:2009we}. Under the action of the UV-IR map described in \S\ref{sec:CStheory}, the low-energy limit of the codimension-$2$ defects fixes the flavour charges of the 4d theory on the Coulomb branch. Importantly, this is a one-to-one mapping between the UV and IR limits of the 4d theories.

    The degenerate vertex operator insertions, in contrast, correspond to codimension-$4$ defects in the 6d $(2,0)$-theory that specify a canonical surface defect in the 4d $\mathcal N =2$ theory \cite{Alday:2009fs}. At low energies, this surface defect admits a discrete set of vacua (IR surface defects) which the UV-IR map selects from. That is, the UV-IR map is \emph{not} one-to-one when acting on surface defects. In particular, the two basis elements in~\cref{eq:abelianbasisdeg} can be explicitly identified with the two brane configurations engineering IR surface defects in the low-energy limit of the 4d $\mathcal N=2$ theory (see for instance \cite[Fig.~14]{Dimofte:2010tz}).
\end{remark}

\subsubsection*{For a general surface $C$} The story thus far admits a straight-forward extension to a general punctured surface $C$ (possibly with gauged punctures). Any system of FN length-twist coordinates corresponds to a symplectic basis of based loops $\widetilde{\underline{\mathfrak{a}}}_{k, \Sigma}$  and $\widetilde{\mathfrak{a}}_{p, \Sigma}$, open paths $\widetilde{\mathfrak{b}}_{k, \Sigma}$ and based closed loops~$\widetilde{\mathfrak{b}}_{p, \Sigma}$ on $\Sigma$. The corresponding open path generators 
\begin{align}
\{ Y_{\underline{\mathfrak{a}}_k}^\textrm{odd}, Y_{\mathfrak{b}_k}^\textrm{odd}, Y_{\mathfrak{a}_p}^\textrm{odd}, Y_{\mathfrak{b}_p}^\textrm{odd} \} \subset \mathbf{A}
\end{align}
define a set of quantum FN length-twist coordinates. And as before, all generators $Y^{ii}$ can be realised as operators that move the position of an abelian degenerate operator an open path.

For an arbitrary genus $g$ surface $\Sigma$, the space of conformal blocks (for fixed vertex operator insertions) depends on $g$ additional complex parameters, corresponding to the eigenvalues of the Heisenberg-Verlinde operators $X_{\gamma_p}$ for cycles $\gamma_p$ around the pants tubes. We collectively denote these parameters as $\vec \alpha$ in $\mathcal F^\textrm{ab}_{\vec \alpha}$. The generators $Y_{\underline{\fraka}_k}^{ii}$, associated with paths around the punctures $z_k$, still admit a decomposition of the form \cref{eq:open-a-decomposition}. Consequently, these generators commute with all the Heisenberg-Verlinde operators on $\Sigma$. In particular, the action of a generator $Y_{\underline\fraka_k}^{ii}$ given by
\begin{equation}
    Y_{\underline{\fraka}_k}^{ii} \mathcal F^\textrm{ab}_{\vec \alpha} \big(\widetilde V^\textrm{ab}_\textrm{deg} (z_p^{(i)}) \ldots \big) = \mathcal X_{\underline{\fraka}_k}^{ii}  \, \mathcal F^\textrm{ab}_{\vec \alpha} \big(\widetilde V^\textrm{ab}_\textrm{deg} (z_p^{(i)}) \ldots \big),
\end{equation}
where $\mathcal X_{\underline{\fraka}_k}^{ii}$ is the semi-classical limit of $Y_{\underline{\fraka}_k}^{ii} $, preserves the labels $\vec \alpha$. 

Interestingly, we find in \S\ref{subsec:gluing-q-nonab} that the basis generator $Y_{\fraka_p}$, and its dual $Y_{\mathfrak b_p}$, associated with a pants tubes (in contrast to punctures), do not commute with the Heisenberg-Verlinde operator $X_{\gamma_p}$. As a result, the basis generators $Y_{\mathfrak b_p}$ will act as operators
\begin{equation}
    Y_{\mathfrak b_p}^{ij} \mathcal F^\textrm{ab}_{\vec \alpha} \big(\widetilde V^\textrm{ab}_\textrm{deg} (\partial \widetilde {\mathfrak b}^{\textrm{in}, (i)}) \ldots \big) = \mathcal X_{\mathfrak b_p}^{ij}  \, \mathcal F^\textrm{ab}_{\vec \alpha'} \big(\widetilde V^\textrm{ab}_\textrm{deg} (\partial \widetilde {\mathfrak b}^{\textrm{out}, (j)}) \ldots \big)
\end{equation}
that shift the labels $\mathcal X_{\gamma_p} \in \vec \alpha$. We identify these operators with the quantum twist coordinates $\hat T^\textrm{BPZ}$ from \cref{eq:quantum-coordinates}.
 
Even more generally, for any system of spectral coordinates $\{\mathcal X_{\widetilde\frakp_\Sigma}^\mathcal W \}$ on an arbitrary surface~$C$ (possibly with gauged punctures), $q$-nonabelianisation produces a basis of open path generators $\{Y_{\widetilde \frakp}^\mathcal W\} \subset \mathbf A$ that quantises this system. The open paths $\widetilde \frakp \subset \Sigma \times I$ are simply the unique open path lifts of the based paths $\widetilde \frakp_\Sigma$ on $\Sigma$ that determine the basis of spectral coordinates. Note that the evaluation of $Y_{\widetilde \frakp}^\mathcal W$ for generic spectral networks (say of Fock-Goncharov type) is much more complicated though, because the space of conformal blocks does not generically admit a diagonalisation with respect to any subset of these generators (see \S\ref{sec:quantumFN}). Nonetheless, we propose a method to access this sector through CFT-abelianisation in \cite{HollandsMurugesanII}. 

\subsubsection{\texorpdfstring{Relation between $X$ and $Y$-coordinates}{Relation between X and Y-coordinates}}\label{subsubsec:xy-coordinates}
    
Although the quantum torus generators $X_\gamma$ and the open path generators $Y_{\widetilde \frakp}$ are valued in different algebras, it is possible to obtain the $X$-coordinates from the $Y$-coordinates. Semi-classically, the trace function is simply the trace of the parallel transport computed for a choice of base-point, and for an abelian connection there is no distinction between the parallel transport around a (based) loop and its trace. Likewise in the quantum theory, the open path generator $Y_{\widetilde \frakp}$ for any open path $\widetilde \frakp$, such that $\widetilde \frakp_\Sigma$ is a loop based at $z_b \in \Sigma$, can be decomposed in terms of $X_\gamma$, where $\gamma$ represents the homology class of $\widetilde \frakp_\Sigma$, and the identity open path $Y_{z_b}^{ii}$. This is precisely how we were able to read off the quantum FN coordinates from $q$-parallel transports. 
    
Alternatively, the trace can be computed as follows. Trivialize the abelian gauge bundle at two nearby points $z_b$, $z'_b \in \Sigma$ and use this trivialization to identify the fiber at $z'_b$ with the dual of the fiber at $z_b$ via the standard evaluation pairing. The semi-classical parallel transport along the closed path $p_\Sigma$ then defines an endomorphism of the fiber, and contracting its output with its input using the pairing yields the ordinary trace of the holonomy.
This is the semi-classical translation of the computation of a Heisenberg–Verlinde operator in the abelian CFT.

\begin{figure}[h]
    \centering
    \begin{subfigure}{0.4\linewidth}
        \centering
        \begin{tikzpicture}
            \draw[red, very thick] (1,0)--(1,0.75) node [currarrow, pos = 0.5, sloped, xscale = 1, red] {};
            \draw[red, very thick] (0,0)--(0,0.75) node [currarrow, pos = 0.5, sloped, xscale = -1, red] {};
            \draw[red, very thick] (0,0) arc (-180:0:0.5) node [currarrow, pos = 0.5, sloped, xscale = 1, red] {};
            \node at (1.35, 0.4) {$z_b$};
            \node at (-0.35, 0.4) {$z_b'$};
            \draw[-latex, thick] (2,0)--(2,0.5);
            \node at (2.2, 0.25) {$t$};
    \end{tikzpicture}
    \end{subfigure}%
    \begin{subfigure}{0.4\linewidth}
        \centering
        \begin{tikzpicture}
        \draw[-latex, thick] (-1,0.6)--(-1,1.1);
        \node at (-1.2, 0.85) {$t$};
        \draw[red, very thick, yshift = 0.1cm] (1,0)--(1,0.75) node [currarrow, pos = 0.5, sloped, xscale = 1, red] {};
        \draw[red, very thick, yshift = 0.1cm] (0,0)--(0,0.75) node [currarrow, pos = 0.5, sloped, xscale = -1, red] {};
        \draw[red, very thick, yshift = 0.1cm] (0,0.75) arc (180:0:0.5) node [currarrow, pos = 0.5, sloped, xscale = -1, red] {};
        \node at (1.35, 0.4) {$z_b$};
        \node at (-0.35, 0.4) {$z_b'$};,
    \end{tikzpicture}
    \end{subfigure}
    \hfill
    \caption{Geometric representation of the cup and cap maps $\iota^\text{ab}$ and $\pi^{\text{ab}}$ that describe the fusion product between an abelian degenerate vertex operator and its conjugate.}
    \label{fig:cupcap}
\end{figure}
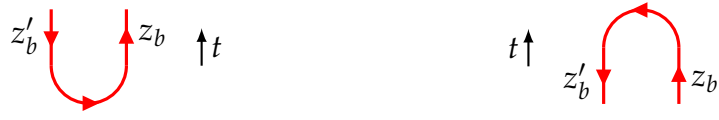

Upon quantisation, the fibre at $z_b$ is realised in terms of the identity open path $Y_{z_b}^{ii}$ and the dual fibre at~$z'_b$ in terms of the conjugate path $(Y_{z'_b}^{ii})^*$, which is simply the open path $Y_{z'_b}^{ii}$ running in the opposite direction. The pairing is then implemented by a choice of \emph{non-open} paths called the \emph{cup} and the \emph{cap}: the cup realises the identification $\iota^\text{ab}$ of the  dual fibre at $z'_b$ with the fibre at $z_b$ and the cap provides the inverse identification $\pi^\text{ab}$. See \cref{fig:cupcap} for a diagrammatic representation of the two paths. We emphasise that the above operators individually \emph{cannot} be realised as open skeins. Whereas $Y_{z_b}^{ii}$ and its conjugate are valued in different vector spaces (of upwards and downwards-travelling open paths, respectively), the cup and the cap are clearly not monotonic in the height-direction and thus not part of either spaces.

A quantum torus generator $X_\gamma$, for $[\gamma] = [\,\widetilde \frakp_\Sigma\,] \in H_1(\Sigma;\Z)$, can then be computed by first applying the cup operator $\iota^\text{ab}$, then the open path generator $Y_{\widetilde\frakp}$ (with $\partial \widetilde \frakp^\textrm{in/out} = \{z_b\}$), and finally the cap operator $\pi^\text{ab}.$ That is,
\begin{align}\label{eq:YtoXloops}
X_{\gamma} = \pi^\text{ab} \circ Y_{\widetilde \frakp} \circ \iota^\text{ab}.  
\end{align}

Compared with similar maps $\iota, \pi$ in the definition~(\ref{eq:def-verlinde-operator}) of a Liouville Verlinde operator, the cups and caps in the abelian theory are much simpler. In particular, the fusion product of an abelian degenerate operator and its conjugate is just the identity operator, unlike the fusion product~\eqref{eq:OPE12alpha} in Liouville theory. This is a simple consequence of the fact that the tensor product between two $GL(1, \C)$-representations (equivalently Heisenberg Fock spaces) $F_{\alpha_1}$ and $F_{\alpha_2}$, with weights (equivalently momenta) $\alpha_1$ and $\alpha_2$ respectively, is the representation
\begin{equation}
    \widetilde F_{\alpha_1} \otimes \widetilde F_{\alpha_2} = \widetilde F_{\alpha_1 + \alpha_2}
\end{equation}
with weight $\alpha_1 + \alpha_2$. Hence, we have that the cups and caps are defined as
\begin{equation} \label{eq:cups-caps}
    \iota^\textrm{ab} : \widetilde F_\textrm{id} \longrightarrow \widetilde F_\textrm{deg} \otimes \widetilde F_\textrm{deg}^* \quad \text{ and } \quad \pi^\textrm{ab} : \widetilde F_\textrm{deg} \otimes \widetilde F_\textrm{deg}^* \longrightarrow \widetilde F_\textrm{id}.
\end{equation}
Here, $\widetilde F_\text{deg}^*$ is the representation labelling the downwards travelling open skein.

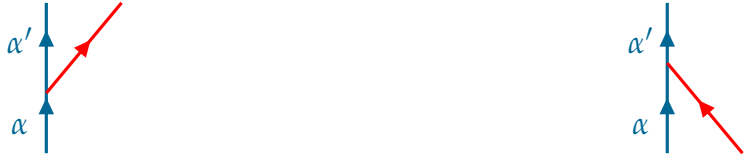
\begin{figure}[h]
    \centering
    \hfill
    \begin{subfigure}{0.4\textwidth}
        \centering
        \begin{tikzpicture}
        \draw[very thick, MidnightBlue] (0,0)--(0,2) node [currarrow, pos = 0.3, sloped, xscale = 1, MidnightBlue] {} node [currarrow, pos = 0.75, sloped, xscale = 1, MidnightBlue] {};
        \draw[very thick, red] (0,0.8)--(1,2) node [currarrow, pos = 0.5, sloped, xscale = 1, red] {};

        \node[MidnightBlue] at (-0.35,.35) {$\alpha$};
        \node[MidnightBlue] at (-0.35,1.5) {$\alpha'$};
        \end{tikzpicture}
    \end{subfigure}
    \hfill
    \begin{subfigure}{0.4\textwidth}
        \centering
        \begin{tikzpicture}
        \draw[very thick, MidnightBlue] (0,0)--(0,2) node [currarrow, pos = 0.3, sloped, xscale = 1, MidnightBlue] {} node [currarrow, pos = 0.75, sloped, xscale = 1, MidnightBlue] {};
        \draw[very thick, red] (0,1.2)--(1,0) node [currarrow, pos = 0.5, sloped, xscale = -1, red] {};

        \node[MidnightBlue] at (-0.35,.35) {$\alpha$};
        \node[MidnightBlue] at (-0.35,1.5) {$\alpha'$};
        \end{tikzpicture}
    \end{subfigure}
    \hfill
    \caption{Geometric representation of the operators $\iota^\textrm{ab}_\alpha$ and $\pi^\textrm{ab}_\alpha$ that describe the fusion product between a generic abelian vertex operator and an abelian degenerate vertex operator.}
    \label{fig:vertex-operator-fusion}
\end{figure}

In the presence of gauged punctures, in addition to the quantum torus generators $X_\gamma$ associated with closed loops $\gamma \in H_1^\textrm{odd}(\Sigma; \Z)$, we also have  quantum torus generators~$X^{ij}_{\mathfrak{b}_p}$ associated with relative homology classes $\widetilde{\mathfrak{b}}_p^{ij} \in H_1^\textrm{odd}(\Sigma, \mathbb X; \Z)$ connecting lifts of distinct marked points, which quantise the spectral coordinates~$\mathcal{X}^{ij}_{\mathfrak{b}_p}$. In order to define these \emph{open} quantum torus generators, we introduce a family of \emph{wedge operators}
\begin{equation}
    \iota^\textrm{ab}_\alpha : \widetilde F_\alpha \longrightarrow \widetilde F_\textrm{deg} \otimes \widetilde F_{\alpha'} \quad \text{ and } \quad \pi^\textrm{ab}_\alpha : \widetilde F_\textrm{deg} \otimes \widetilde F_{\alpha} \longrightarrow \widetilde F_{\alpha'}, \label{eq:wedge-op}
\end{equation}
that correspond to the fusion product of an abelian vertex operator $\widetilde V^\text{ab}_\alpha$, with generic momentum~$\alpha$, and an abelian degenerate operator $\widetilde V^\text{ab}_\text{deg}$. The corresponding skein diagrams are drawn in \cref{fig:vertex-operator-fusion}. Unlike the cup $\iota^\textrm{ab}$ and cap $\pi^\textrm{ab}$, the wedge operators \emph{can} be realised inside $\mathbf A$.

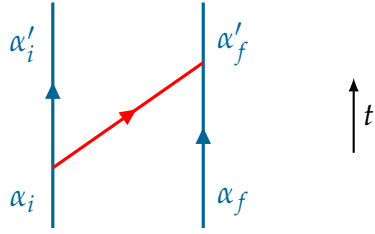
\begin{figure}[h]
    \centering
    \begin{tikzpicture}
        \draw[very thick, MidnightBlue] (0,0)--(0,3) node [currarrow, pos = 0.6, sloped, xscale = 1, MidnightBlue] {};
        \draw[very thick, MidnightBlue] (2,0)--(2,3) node [currarrow, pos = 0.4, sloped, xscale = 1, MidnightBlue] {};
        \draw[very thick, red] (0,0.8)--(2,2.2) node [currarrow, pos = 0.5, sloped, xscale = 1, red] {};

        \node[MidnightBlue] at (-0.4,.35) {$\alpha_i$};
        \node[MidnightBlue] at (-0.4,2.45) {$\alpha'_i$};
        \node[MidnightBlue] at (2.4,.35) {$\alpha_f$};
        \node[MidnightBlue] at (2.4,2.45) {$\alpha_f'$};

        \draw[-latex, thick] (4,1)--(4,2);
            \node at (4.2, 1.5) {$t$};
    \end{tikzpicture}
    \caption{The skein diagram corresponding to the quantum torus generator $X_{\widetilde{\mathfrak b}_p}^{ij}$ for an open path $\widetilde{\mathfrak b}_p^{ij}$ running between punctures $p_i$ and $p_f$.}
    \label{fig:vertex-operator-fusion-open}
\end{figure}

Suppose that $\widetilde{\mathfrak b}_p$ interpolates between marked points near punctures $p_i$ and $p_f$. The quantum torus generator $X^{ij}_{\mathfrak{b}_p}$ can then be computed by first applying the wedge operator $\iota^\text{ab}_{\alpha_i}$ at the starting point, then the open path generator $Y^{ij}_{\mathfrak{b}_p}$, and finally the wedge operator $\pi^\text{ab}_{\alpha_f}$ at the end-point. That is,
\begin{align}
 X^{ij}_{\mathfrak{b}_p} = \pi^\text{ab}_{\alpha_i} \circ Y^{ij}_{\mathfrak{b}_p} \circ \iota^\text{ab}_{\alpha_f},  \label{eq:YtoX-open}
\end{align}
as shown in \cref{fig:vertex-operator-fusion-open}. Note that since the labels $\alpha_i, \alpha_f$ of the conformal block are shifted to $\alpha_i', \alpha_f'$, this implies that the generator $X_{\widetilde{\mathfrak b}_p}$ acts as a difference operator on the abelian conformal blocks. Equivalently, the generator $X_{\widetilde{\mathfrak b}_p}$ only $q$-commutes with the Heisenberg-verlinde operators $X_{\gamma_i}$ and $X_{\gamma_f}$, i.e.
\begin{equation}
    X_{\gamma_i} \overset {\alpha_i}\rightsquigarrow M_i \quad \text{ and } \quad X_{\gamma_i} \overset {\alpha_i'} \rightsquigarrow q^{-2} M_i,
\end{equation}
modulo \cref{rem:mass-defn-modified}.

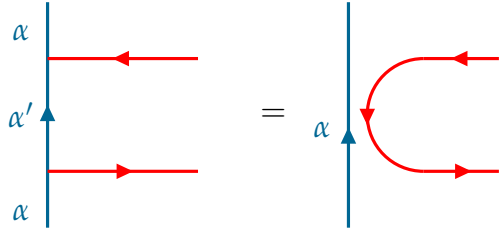
\begin{figure}[h!]
    \centering
    \begin{tikzpicture}
        \draw[very thick, MidnightBlue] (0,0)--(0,3) node [currarrow, pos = 0.5, sloped, xscale = 1, MidnightBlue] {};
        \draw[very thick, red] (0,0.75)--(2,0.75) node [currarrow, pos = 0.5, sloped, xscale = 1, red] {};
        \draw[very thick, red] (0,2.25)--(2,2.25) node [currarrow, pos = 0.5, sloped, xscale = -1, red] {};

        \node[MidnightBlue] at (-0.35, 0.2) {$\alpha$}; 
        \node[MidnightBlue] at (-0.35, 1.5) {$\alpha'$};
        \node[MidnightBlue] at (-0.35, 2.6) {$\alpha$};

        \node[ultra thick] at (3, 1.5) {$=$}; 

        \draw[very thick, MidnightBlue, xshift = 4cm] (0,0)--(0,3) node [currarrow, pos = 0.4, sloped, xscale = 1, MidnightBlue] {};
        \draw[very thick, red, xshift = 4cm] (1,0.75)--(2,0.75) node [currarrow, pos = 0.5, sloped, xscale = 1, red] {};
        \draw[very thick, red, xshift = 4cm] (1,2.25)--(2,2.25) node [currarrow, pos = 0.5, sloped, xscale = -1, red] {};
        \draw[very thick, red, xshift = 4cm] (1,2.25) arc (90:270:0.75) node [currarrow, pos = 0.5, sloped, xscale = 1, red] {};

        \node[MidnightBlue, xshift = 4cm] at (-0.35, 1.3) {$\alpha$};
    \end{tikzpicture}
    \caption{For a based loop in $\Sigma$, the operation $\pi^\text{ab}_\alpha \circ Y \circ \iota^\text{ab}_\alpha$ is the same as the operation $\pi^\text{ab} \circ Y \circ \iota^\text{ab}$. }
    \label{fig:vertex-operator-fusion-loop}
\end{figure}

Observe that if $\widetilde{\mathfrak{b}}_{p,\Sigma}$ were a based loop in $\Sigma$, then the relation in \cref{fig:vertex-operator-fusion-loop} ensures that the composition in~\cref{eq:YtoX-open} reproduces the associated Heisenberg-Verlinde operator as computed in \cref{eq:YtoXloops}, as it should.

\subsection{\texorpdfstring{On $q$-parallel transport for arbitrary Riemann surfaces}{On q-parallel transport for arbitrary Riemann surfaces}}
\label{subsec:gluing-q-nonab}
Consider an arbitrary Riemann surface $C$ equipped with a pants decomposition $\mathcal P$. Fix a path groupoid that respects $\mathcal P$ and is homotopic to the path groupoid on the (gauged) three-punctured sphere when restricted to each pair of pants. Let $\mathfrak p$ be an arbitrary open path on $C$. Since $q$-nonabelianisation is a local procedure, the $q$-parallel transport along~$\mathfrak p$ may be computed by decomposing the path $\mathfrak{p}$ as
\begin{align}
    \mathfrak{p} = \mathfrak{p}^\textrm{\maljapanese\char"4E}\mathfrak{p}^\textrm{\maljapanese\char"1A} \cdots \mathfrak{p}^\textrm{\maljapanese\char"4A}, 
\end{align}
where each $\mathfrak{p}^\textrm{\maljapanese\char"4D}$ is an (connected) open path embedded in a single pair of pants. Since the knowledge of the quantum FN S-matrix is sufficient to compute the $q$-parallel transport along any path inside the three-punctured sphere, the only remaining ingredient in the computation of the $q$-parallel transport map on an arbitrary Riemann surface is the gluing operation.

\begin{figure}[h]
    \centering
    \includegraphics[width=0.18\linewidth]{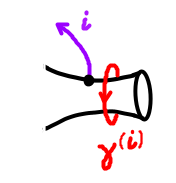}
    \caption{Ordering of the lifts of the pants-cycle $\gamma$ and an open 1-cycle around a gauged puncture on $\Sigma$.}
    \label{fig:pants-cycle-ordering}
\end{figure}

Recall the definition of a gauged puncture: we cut open a small disc around a puncture, and fix a marked point on the boundary. It is convenient to think of the thickened boundary component as an annulus, and of the marked point as lying on the inner boundary of the annulus (see \cref{fig:pants-cycle-ordering}). Lifting this configuration to the cover $\Sigma$, this allows us to define an ordering between the closed pants-cycles and the open 1-cycles with respect to the marked point. The 1-cycles are defined such that they can be collapsed onto the boundary without any obstruction from the open paths. This is illustrated in \cref{fig:pants-cycle-ordering}. 

\begin{figure}[h]
    \centering
    \includegraphics[width=0.48\linewidth]{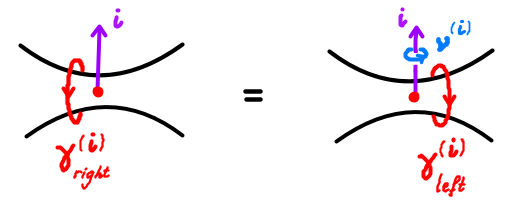}
    \caption{There is no canonical way to order the lifts of closed pants-cycles in the presence of an open path. The two possible choices differ by factor of $q^{\pm 2}$. 
    }
    \label{fig:pants-cycle-ordering-2}
\end{figure}

To glue two gauged punctures, we identify their boundary components as well as their marked points. However, this introduces the lateral ordering ambiguity illustrated in \cref{fig:pants-cycle-ordering-2}. Recall from \cref{fig:lift-arrangement} that $\gamma$ was the loop encircling the puncture while $\gamma'$ was the loop on the other side of the open path. On a tube though, there is no canonical way to distinguish between $\gamma$ and $\gamma'$. The best we can do is to fix a choice of $\gamma$ versus $\gamma'$ on each pants-tube, and stay consistent with that choice. This is the (open) skein analogue of the CFT ambiguity in the choice of internal momenta on the pants-tube with a degenerate operator, as illustrated in \cref{fig:five-point-conformal-block}.

Indeed, let us see what happens to the definition of the mass eigenvalues as we move a pants cycle across an open path. Suppose the open path originates on sheet $1$ in $\Sigma \times \{0\}$. Then the pants cycles $\gamma_\text{left}$ and $\gamma_\text{right}$ (denoted $\gamma$ and $\gamma'$ previously) on either side of the open path are related as   
\begin{equation}
    \gamma_\textrm{right}^\textrm{odd} = -\gamma_\textrm{left}^\textrm{odd} - \frac 1 2 \nu
\end{equation}
where $\nu$ is a small clockwise loop around the open path. This implies that the evaluation of the corresponding abelian Verlinde operators is related by
\begin{equation}
    M_\textrm{right} = q / M_\textrm{left}.
\end{equation}
Note that the orientation reversal of the pants cycle $\gamma_\textrm{right} \mapsto - \gamma_\textrm{left}$ across the pants tube inverts the mass on the RHS. On the other hand, if the open path originates on sheet $2$ in $\Sigma \times \{0\}$, the orientations of the closed cycles are reversed, and
\begin{equation}
    \gamma_\textrm{right}^\textrm{odd} = - \gamma_\textrm{left}^\textrm{odd} + \frac 1 2 \nu,
\end{equation}
so that
\begin{equation}
    M_\textrm{right} = q^{-1} /M_\textrm{left}.
\end{equation}

Thus, depending on whether the open path originates on sheet $1$ or sheet $2$, the mass eigenvalue is shifted by a factor of $q^{\pm 1}$. Suppose we wish to write all the expressions in terms of $M_\textrm{left}$. Then, we can collect the above action into an operator
\begin{equation} \label{eq:shift-matrix-q-nonab}
    \begin{pmatrix}
        \Delta_q^{-1} & 0 \\ 0 & \Delta_q \end{pmatrix} \bigg( \dots \bigg)^\textrm{op} = \bigg( \begin{pmatrix}
        \Delta_q & 0 \\ 0 & \Delta_q^{-1} \end{pmatrix} \dots \bigg)^\textrm{op},
\end{equation}
where
\begin{align}
\Delta_q^{\pm 1} (M) = q^{\pm 1} M
\end{align}
is a difference operator acting on the length coordinate $M$, and the superscript "$\textrm{op}$" denotes orientation reversal across the pants tube. Note that $\Delta_q$ is precisely the difference operator~$\widehat T$ that we encountered in~\cref{eq:shift-matrix-S3}. That is, $q$-parallel transports across a pants-tube is obtained by inserting the difference operator $\widehat T$.

\begin{remark}
    Suppose $\widetilde \frakp_p$ is an open path that starts and ends on a pants tube labelled by the pants cycle $\gamma_p$. In that case, we may equivalently think of the open path generator $Y_{\widetilde \frakp_p} \in \mathbf A$ as being implicitly composed with wedge operators $\iota^\textrm{ab}_{\alpha_p}, \pi^\textrm{ab}_{\alpha_p}$, and identify $Y_{\widetilde \frakp_p}$ with the open quantum torus generator $X_{\widetilde \frakp_p} \in \mathbb T$. Thus, $\Delta_q^{\pm 1}$ is also the difference operator realisation of the quantum torus generators $X_{\widetilde\frakp_p}$.
\end{remark}

\subsubsection{Example: 4-punctured sphere}

\begin{figure}
    \centering
    \includegraphics[width=0.7\linewidth]{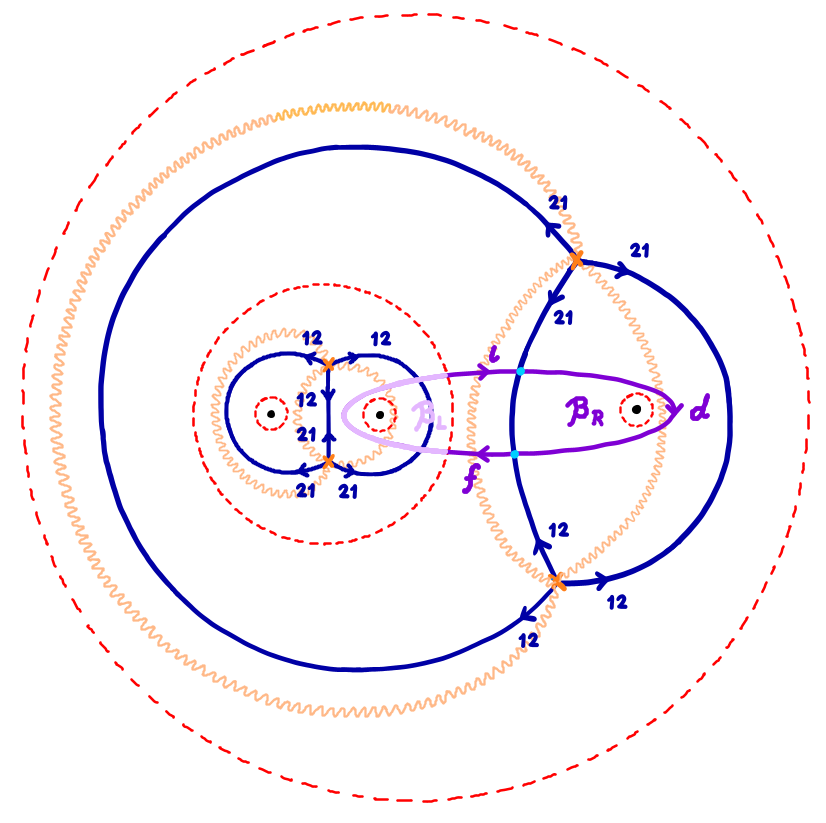}
    \caption{The path $\beta$ on the 4-punctured sphere is split into two paths $\beta_L$ and $\beta_R$, each of which is decomposed into three pieces $\iota$, $d$ and $f$. }
    \label{fig:q-PT-beta}
\end{figure}

As an example, we compare the $q$-parallel transport along the 1-cycle $\beta$ on the 4-punctured sphere, as in \cref{fig:q-PT-beta}, with the Moore-Seiberg result in~\cref{eq:qmon4ptbeta}. Let us denote the restriction of $\beta$ to the right (left) pants by $\beta_R$ ($\beta_L$). The $q$-parallel transport along both components is very similar to \cref{eq:double-wall-isotopy-w2}.\footnote{When compared to \cref{eq:double-wall-isotopy-w2}, the paths $\iota$ and $f$ cross the branch-cut here. However, this does not contribute any additional $q$-factors.} Indeed, as before, we decompose $\beta_R$ (and $\beta_L$) into three pieces $\iota, d$ and $f$ where $\iota$ labels the first part of the path that hasn't crossed the double wall yet, $d$ labels the middle part in between the walls, and $f$ labels the final part after the crossings.

\begin{figure}
    \centering
    \includegraphics[width=0.2\linewidth]{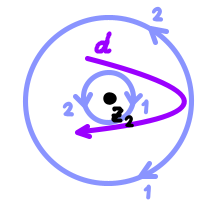}
    \caption{The winding of the diagonal path $d$ with respect to the FN foliation. }
    \label{fig:winding-diagonal}
\end{figure}

The $q$-nonabelianisation along the path $d$ is built only from the direct lifts $Y_{d^{(1)}}$ and $Y_{d^{(2)}}$. Note that we may write
\begin{equation}
    Y_{d^{(1)}} = X_{\gamma_2^{(1)}} Y_m^{11} \rightsquigarrow q M_2 Y_m^{11} \quad \text{ and } \quad Y_{d^{(2)}} = X_{\gamma_2^{(2)}} Y_m^{22} \rightsquigarrow q M_2^{-1} Y_m^{22},
\end{equation}
where $Y_m$ is the same generator that appears in \cref{eq:double-wall-isotopy-w2}. Here, we have evaluated the masses as discussed in \cref{rem:mass-defn-modified}. In addition, $Y_{d^{(1)}}$ is accompanied by a winding factor of $q$ as illustrated in \cref{fig:winding-diagonal}. Ultimately, we obtain the $q$-parallel transport along $d$ to be
\begin{equation}
    P^q_d =  q^2\begin{pmatrix}
        M_2 \, Y_m^{11} & 0 \\ 0 & q^{-1} M_2^{-1} \, Y_m^{22}
    \end{pmatrix}. \label{eq:eg-q-par-d}
\end{equation}

This form for the $q$-parallel transport matrix along $d$ is rather convenient as it allows us to reuse the conclusions from \cref{eq:double-wall-isotopy-w2}. It follows that the $q$-parallel transport along $\beta_R$ (and $\beta_L$) is of the form
\begin{equation}
    P^q_{\beta_R} = q^2\begin{pmatrix}
        \big(\ldots\big) Y^{11}_{\beta_R} & q^{-1} \big(\ldots\big) Y^{12}_{\beta_R}\\
        \big(\ldots\big) Y^{21}_{\beta_R} & q^{-1} \big(\ldots\big) Y^{22}_{\beta_R}
    \end{pmatrix}.
\end{equation}
Since both $\beta_L$ and $\beta_R$ are open paths terminating on marked points, we may evaluate the corresponding open path generators as explained in \S~\ref{subsec:quantum-FN-coordinates-2}. Upon evaluation, we can compare this with the MS parallel transport in \cref{eq:qmon4ptbeta}. Note that it is sufficient to count only the $q$-factors because both quantum parallel transports reduce to identical semi-classical parallel transports (after appropriately rescaling the MS parallel transport basis).

Although the space of open skeins is a $\Z[q^{\pm 1}]-$module, let us temporarily write the above $q$-parallel transport as
\begin{equation}
    P^q_{\beta_R} = q^{\frac 3 2}\begin{pmatrix}
        \big(\ldots\big)\, q^{\frac 1 2 } Y^{11}_{\beta_R} & \big(\ldots\big)\, q^{-\frac 1 2 } Y^{12}_{\beta_R}\\
        \big(\ldots\big) \,q^{\frac 1 2 } Y^{21}_{\beta_R} & \big(\ldots\big) \,q^{-\frac 1 2 } Y^{22}_{\beta_R}
    \end{pmatrix}. \label{eq:q-P-beta-R-split}
\end{equation}
We have already established that the MS parallel matrix inside a 3-sphere has no extra $q$-factors except for an overall factor of $q$ in the front (see \S\ref{subsec:bpz-monodromies}), i.e.
\begin{equation}
    P^\textrm{MS}_{\beta_R} = q \begin{pmatrix}
        \big(\ldots\big) & \big(\ldots\big) \\ \big(\ldots\big) & \big(\ldots\big)
    \end{pmatrix}.
\end{equation}
Comparing this expression with (the evaluation of) \cref{eq:q-P-beta-R-split}, we note that the MS parallel transport and $q$-parallel transport agree in form, up to the overall winding factor, if we redefine the open path generators by half-integral powers of $q$ as 
\begin{equation}
    \begin{aligned}
        &\widehat Y^{11}_{\beta_R} := q^{\frac 1 2 } Y^{11}_{\beta_R}, \qquad \widehat Y^{22}_{\beta_R} := q^{-\frac 1 2 } Y^{22}_{\beta_R},\\
        &\widehat Y^{12}_{\beta_R} := q^{-\frac 1 2 } Y^{12}_{\beta_R}, \qquad \widehat Y^{21}_{\beta_R} := q^{\frac 1 2 } Y^{21}_{\beta_R}.
    \end{aligned}
\end{equation}
We emphasise that this is not a universal redefinition that brings all $q$-parallel transports on any $C$ in a similar form to the MS parallel transport, but rather it is a redefinition that (at least) works in this specific example.

\begin{remark}
    It is not surprising that the winding factors of $q$-parallel transport and MS parallel transport do not coincide. Whereas $q$-nonabelianisation picks up a (sheet-dependent) winding factor from every turn with respect to the foliation, the Moore-Seiberg parallel transport only picks up a (uniform overall) winding factor around punctures. However, we do ultimately expect the two results to completely agree for closed loops (corresponding to Verlinde loop operators).

    Nonetheless, it is fairly straight-forward to translate between the two quantum parallel transports. For instance, given a (fully evaluated) $q$-parallel transport along any non-self-intersecting curve inside the 3-punctured sphere, we can recover the corresponding MS parallel transport by ``forgetting'' the additional winding factors and including an overall factor of $q^{\pm 1}$ for every turn around a puncture. In the other direction, we forget the overall $q$-factor in the MS parallel transport, and include sheet-dependent winding factors (as in \cref{fig:new-wind}) from all windings with respect to the foliation. 
\end{remark}

Yet, ignoring winding factors, we expect $q$-parallel transports and MS parallel transports to be in full agreement. Especially, the composite expressions for $P^q_\beta$ and $P^\textrm{MS}_\beta$ must agree with each other. As discussed in \S\ref{subsec:quantum-FN-coordinates-2}, we may compose the two $q$-parallel transport matrices along $\beta_R$ and $\beta_L$ by inserting the matrix~\eqref{eq:shift-matrix-q-nonab} of difference operators whenever the path crosses from one pair of pants to another. Assuming that the path starts on the left pair of pants, we find that the $q$-parallel transport along $\beta = \beta_R \circ \beta_L$ is given by
\begin{align}
\begin{aligned}
P^q_\beta &=     \begin{pmatrix}
        \Delta_q^{-1} & 0 \\ 0 & \Delta_q
    \end{pmatrix} \bigg(P^q_{\beta_R} \begin{pmatrix}
        \Delta_q^{-1} & 0 \\ 0 & \Delta_q
    \end{pmatrix} \left( P^q_{\beta_L} \right)^\textrm{op}\bigg)^\textrm{op} \\
    &= \begin{pmatrix} \Delta_q^{-1} & 0 \\ 0 & \Delta_q
    \end{pmatrix} \left(P^q_{\beta_R} \right)^\textrm{op} \begin{pmatrix}
        \Delta_q & 0 \\ 0 & \Delta_q^{-1}
    \end{pmatrix} P^q_{\beta_L},
    \end{aligned} \label{eq:gluing-q-beta}
\end{align}
where $\Delta_q^{\pm 1}(A) = q^{\pm 1} A$ is a difference operator that acts on the exponentiated internal momentum $A$, and the superscript "$\textrm{op}$" denotes the $q$-parallel transport with $A \mapsto A^{-1}$. This indeed agrees with the form
\begin{align}
P^\textrm{MS}_\beta = \widehat{\mathbf{T}}^{-1}_{\alpha} \, P^\textrm{MS}_{\beta_R, Q-\alpha} \, \widehat{\mathbf{T}}_\alpha \, P^\textrm{MS}_{\beta_L, \alpha}\, 
\end{align}
 of \cref{eq:gluing-ms-beta}. This establishes consistency between the two quantum parallel transports (up to the winding factors).

\appendix

\section*{\texorpdfstring{\Huge Appendix}{Appendix}}
\section{Nonabelianisation commutes with gluing for FN networks}\label{appendix-nonab}

We break the proof of the claim, that $\mathcal{W}$-nonabelianisation for the resolutions $\mathcal{W}^\pm_{\textrm{FN}}$ commutes with dissection the surface $C$ into pairs of pants, into two parts:

\begin{itemize}
\item In the first part of the proof, we show that the non-abelian monodromy around any pants cycle $\mathfrak{a}$ (as well as any small loop around a puncture) is simply the push-forward of the abelian parallel transport along its direct lifts. In other words, we show that all detour paths that contribute to the non-abelianisation map at finite $\eta$ ``decouple'' in the limit $\eta \downarrow 0$. (This argument requires the WKB framing from \S\ref{subsubsec:gaugedsphere}.) 

\item In the second part of the proof, we restrict ourselves to the parallel transport inside a single pair of pants. We argue that any contribution to the non-abelian parallel transport matrices $P_{a/b/c}$ from \S\ref{sec:nonab-pants} can be matched with the abelian parallel transport along a detour path present at finite $\eta$ (for small enough $\eta$), whereas any detour path that cannot be matched with a contribution to $P_{a/b/c}$  corresponds to a trajectory that traverses a pants-tube and therefore "decouples" in the limit $\eta \downarrow 0$. (Again, this argument requires the WKB framing.)
\end{itemize}

\begin{figure}[h]
   \centering
  \includegraphics[width=0.75\linewidth]{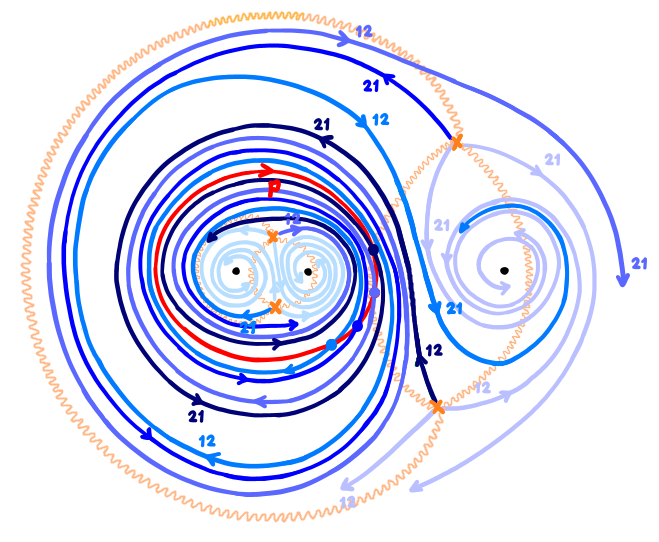}
    \caption{Cartoon of spectral network on the 4-punctured sphere for the same quadratic differential $\phi_2$ as in \cref{fig:FN-4pt-sphere-example}, but now with $\vartheta = \eta >0 $ deviating slightly from the Fenchel-Nielsen phase. Observe that four trajectories (in different shades of bright blue) cross the pants cycle $\mathfrak{a} = [\mathfrak{p}]$, which is added in red. Each of these four trajectories starts from a different branch-point, and each of them ends at a distinct puncture. The other eight trajectories of the spectral network (in light blue) remain on either side of the pants cycle.   }
    \label{fig:FN-4pt-sphere-res}
\end{figure}

First, we focus on a single pants cycle $\mathfrak{a}$ (which may as well be a small loop $\underline{\mathfrak{a}}$ around a puncture). As an illustrative example, consider the spectral network from \cref{fig:FN-4pt-sphere-example} at phase $\vartheta = \vartheta_\mathrm{FN} + \eta$, with $\eta>0$.  (The argument for $\vartheta = \vartheta_\mathrm{FN} - \eta$ is similar.) This network is illustrated in \cref{fig:FN-4pt-sphere-res}, with the pants curve~$\mathfrak{a} = [\mathfrak{p}]$ coloured red. The non-abelian parallel transport along $\mathfrak{p}$ receives contributions from direct lifts of $\mathfrak{p}$ to the cover as well as from detour paths associated with network trajectories that intersect $\mathfrak{p}$. We observe that there are just four trajectories (highlighted in bright shades of blue in \cref{fig:FN-4pt-sphere-res}) that cross the pants cycle; two in each direction. (For the loop $\underline{\mathfrak{a}}$ there would only be two intersecting trajectories.) At any finite value of~$\eta$, these trajectories wind around the pants tube (or loop around a puncture) a finite, say $N$, number of times before terminating on a puncture at the opposite side of the pants cycle. In the limit $\eta \downarrow 0$ the winding number $N$ goes to infinity.

Let us describe the relevant detour paths for the pants curve $\mathfrak{a}$ in more detail (the discussion for the loop $\underline{\mathfrak{a}}$ is similar but simpler). Fix base-points at the four intersections of the pants-tube-traversing trajectories with the path $\mathfrak{p}$. These are illustrated as blue dots in \cref{fig:FN-4pt-sphere-res}. Label the four detour paths associated with these base-points as $\delta^{L/R,u/d}$, where the superscript $L$ or $R$ refers to the left or right molecule from which the corresponding trajectory originates, whereas the superscript $u$ or~$d$ refers to whether the trajectory originates from the upper or lower branch-point in that molecule. 
Both detour paths $\delta^L$ start on sheet $1$ and end on sheet $2$ of the cover, whereas both detour paths $\delta^R$ start on sheet $2$ and end on sheet $1$. For convenience, we add this information as a subscript, so that the four relevant detour paths are named
\begin{align}
    \delta^{R,d}_{21}, \delta^{L,u}_{12}, \delta^{R,u}_{21}, \delta^{L,d}_{12},
\end{align}
in the order in which we meet them when going around the path $\mathfrak{p}$ in the clockwise direction, starting from the upper base-point $b$. 

Denote the lift to sheet $i$ of base-point $b$  by $b^{(i)}$ and the lift to sheet $i$ of path $\mathfrak{p}$  by $\mathfrak{p}^{(i)}$. Let $\mathcal{X}_A$ be the abelian parallel transport along the loop $A= \mathfrak{p}^{(2)}$ and let 
\begin{align}
a^{L/R,u/d}_{ij}
\end{align}
be the abelian parallel transport along the auxiliary path that starts at $b^{(i)}$, follows the path~$\mathfrak{p}^{(i)}$ until this hits the detour path~$\delta^{R/L,u/d}_{ij}$, then follows the detour path~$\delta^{R/L,u/d}_{ij}$, and finally returns to $b^{(j)}$ along the path $\mathfrak{p}^{(j)}$. With this notation, the nonabelian parallel transport along the path $\mathfrak{p}$, starting from the upper base-point $b$, is given by 
\begin{align}
\begin{aligned}
M_{\mathfrak{a}} = 
\begin{pmatrix}
  \mathbf{a}_{11}   & 
   \mathbf{a}_{12} \\
  \mathbf{a}_{21}  &
   \mathbf{a}_{22}
\end{pmatrix} 
\begin{pmatrix}
\mathcal{X}_A^{-1} & 0 \\ 0 & \mathcal{X}_A
\end{pmatrix}
,
 \end{aligned}
\end{align}
where
\begin{align}
\begin{aligned}
  & \mathbf{a}_{11} =
   1 + a_{12}^{L,u} a_{21}^{R,d} +  a_{12}^{L,u} a_{21}^{R,u} + a_{12}^{L,d} a_{21}^{R,u} + a_{12}^{L,u} a_{21}^{R,d} a_{12}^{L,d} a_{21}^{R,u}, \\
   &\mathbf{a}_{22} = 1+ 
   a_{21}^{R,d} a_{12}^{L,d}, \\
 &    \mathbf{a}_{12} = 
   a_{12}^{L,u} +  a_{12}^{L,d} + a_{12}^{L,u} a_{21}^{R,d} a_{12}^{L,d}, \\
&  \mathbf{a}_{21} =
  a_{21}^{R,d} + a_{21}^{R,u} + a_{21}^{R,d} a_{12}^{L,d} a_{21}^{R,u}. 
\end{aligned}
\end{align}

If we place $\mathfrak{p}$ in the middle of the pants tube, as illustrated in \cref{fig:FN-4pt-sphere-res}, all detour paths wind $N/2$ times in the anti-clockwise direction around the lift to sheet $1$ of the pants tube, and $N/2$ times in the opposite direction around the lift to sheet $2$ of the pants tube. Hence, all abelian parallel transports $a^{L/R,u/d}_{ij}$ contain a factor $\mathcal{X}_A^{N}$. 
Since we have chosen our $\mathcal{W}$-framing such that $|\mathcal{X}_A|<1$, we find that
\begin{align}
a^{L/R,u/d}_{ij} \to 0, \qquad \textrm{when}~~N \to \infty.
\end{align}

The only surviving contributions to $M_{\mathfrak{a}}$ in the Fenchel-Nielsen limit $\eta \to 0^+$ are thus given by abelian parallel transports $\mathcal{X}_A^{\pm 1}$ along the direct lifts $\mathfrak{a}^{(i)}$, i.e. 
\begin{align}
  M_\mathfrak{a} \to  \begin{pmatrix}
\mathcal{X}_A^{-1} & 0 \\ 0 & \mathcal{X}_A
\end{pmatrix}, \qquad \textrm{when}~~ \eta \to 0^+.
\end{align}
Intuitively, the pants-traversing trajectories take infinitely long to reach the puncture on the other side of the pants curve, and therefore decouple in the limit $\eta \to 0^+$. 

Even though in the above we discussed the example of the 4-punctured sphere, the local structure of the trajectories is the same for any pants cycle (as well as any small loop around a puncture) with respect to any resolution of an FN network. Therefore, we conclude that the non-abelian monodromy of any WKB-framed flat connection~$\nabla$ around any pants cycle (or any small loop around a puncture) is simply the push-forward of the abelian parallel transport along its direct lifts.

\bigskip

In the second part, we restrict to a single pair of pants and distinguish the trajectories~$w$ that originate from a branch-point \emph{inside} this pair of pants from the trajectories~$w'$ that originate \emph{outside} this pair of pants. In \cref{fig:FN-4pt-sphere-trajectories-molI.png} we have highlighted the former in bright blue for the left pair of pants on the 4-punctured sphere, as an example. Consider the nonabelian parallel transport $\widetilde{P}_{a/b/c}$ along any path $\widetilde{\mathfrak{p}}_{a,b,c}$ in the path groupoid, when restricted to this pair of pants. 

Now, any intersection of the path $\widetilde{\mathfrak{p}}_{a/b/c}$ with a trajectory~$w$ corresponds to a non-zero detour contribution to $\widetilde{P}_{a/b/c}$, while any intersection with a trajectory~$w'$ leads to a detour contribution that decouples in the limit $\eta \to 0$. The direct and detour contributions associated with the trajectories $w$ combine into a non-abelian parallel transport matrix of the form $P_{a/b/c}^{\text{FN},\pm}$ in \S\ref{sec:nonab-pants}. More precisely, any contribution to $P_{a/b/c}^{\text{FN},\pm}$ can be matched with a direct path or a detour path for a trajectory $w$, when $\eta$ is chosen small enough. 

\paragraph{Conflict of Interest}
The authors declare that there are no conflicts of interest regarding the publication of this manuscript.

\paragraph{Data Availability Statement}
Data supporting the findings of this study consist of computational derivations performed in Wolfram Mathematica. The corresponding code notebooks are available from the authors upon request.

\bibliographystyle{ieeetr}
\bibliography{ref.bib}

\end{document}